\documentclass[onecolumn,11pt,showpacs,notitlepage,nofootinbib,floatfix,superscriptaddress]{revtex4-1} 
\usepackage{graphicx}
\usepackage{amssymb}
\usepackage{amsmath}
\usepackage{amsthm}
\usepackage{amsfonts}
\usepackage{mathtools}
\usepackage{xcolor}
\usepackage{hyperref}

\newtheorem*{lem*}{Lemma}
\newtheorem*{thm*}{Theorem}
\newtheorem{thm}{Theorem}[section]
\newtheorem{lem}[thm]{Lemma}

\newcommand{\ket}[1]{|#1\rangle}
\newcommand{\bra}[1]{\langle #1|}

\newcommand{\Cres}{C_R}
\newcommand{\Ures}{U_R}

\newcommand{\eq}[1]{\hyperref[eq:#1]{Eq.~(\ref*{eq:#1})}}
\renewcommand{\sec}[1]{\hyperref[sec:#1]{Section~\ref*{sec:#1}}}
\newcommand{\secsm}[1]{\hyperref[sec:#1]{Sec.~\ref*{sec:#1}}}
\newcommand{\app}[1]{\hyperref[app:#1]{Appendix~\ref*{app:#1}}}
\newcommand{\theo}[1]{\hyperref[thm:#1]{Theorem~\ref*{thm:#1}}}
\newcommand{\algo}[1]{\hyperref[alg:#1]{Algorithm~\ref*{alg:#1}}}
\newcommand{\lemm}[1]{\hyperref[lem:#1]{Lemma~\ref*{lem:#1}}}
\newcommand{\defn}[1]{\hyperref[defn:#1]{Definition~\ref*{defn:#1}}}
\newcommand{\corr}[1]{\hyperref[cor:#1]{Corollary~\ref*{cor:#1}}}
\newcommand{\fig}[1]{\hyperref[fig:#1]{Fig.~\ref*{fig:#1}}}
\newcommand{\tab}[1]{\hyperref[tab:#1]{Table~\ref*{tab:#1}}}
\newcommand{\tabsm}[1]{\hyperref[tab:#1]{Tab.~\ref*{tab:#1}}}
\newcommand{\propos}[1]{\hyperref[prop:#1]{Proposition~\ref*{prop:#1}}}
\newcommand{\propsm}[1]{\hyperref[prop:#1]{Prop.~\ref*{prop:#1}}}
\newcommand{\rema}[1]{\hyperref[rem:#1]{Remark~\ref*{rem:#1}}}

\usepackage{soul}

\newcommand{\nocontentsline}[3]{}
\newcommand{\tocless}[2]{\bgroup\let\addcontentsline=\nocontentsline#1{#2}\egroup}

\newcommand{\face}[2]{\Delta_{#1}(#2)}
\newcommand{\facex}[2]{\Delta'_{#1}(#2)}

\newcommand{\bnd}[2]{\partial_{#1,#2}}
\newcommand{\rest}[1]{|_{#1}}

\begin{document}

\title{The cost of universality: A comparative study of the overhead\\ of state distillation and code switching with color codes}

\author{Michael E. Beverland}
\affiliation{Microsoft Quantum and Microsoft Research, Redmond, WA 98052, USA}
\author{Aleksander Kubica}
\affiliation{Perimeter Institute for Theoretical Physics, Waterloo, ON N2L 2Y5, Canada}
\affiliation{Institute for Quantum Computing, University of Waterloo, Waterloo, ON N2L 3G1, Canada}
\affiliation{AWS Center for Quantum Computing, Pasadena, CA 91125, USA}
\author{Krysta M. Svore} 
\affiliation{Microsoft Quantum and Microsoft Research, Redmond, WA 98052, USA}

\date{\today}

\begin{abstract}
Estimating and reducing the overhead of fault tolerance (FT) schemes is a crucial step toward realizing scalable quantum computers.
Of particular interest are schemes based on two-dimensional (2D) topological codes such as the surface and color codes which have high thresholds but lack a natural implementation of a non-Clifford gate.
In this work, we directly compare two leading FT implementations of the $T$ gate in 2D color codes under circuit noise across a wide range of parameters in regimes of practical interest.
We report that implementing the $T$ gate via code switching to a 3D color code does not offer substantial savings over state distillation in terms of either space or space-time overhead.
We find a circuit noise threshold of $0.07(1)\%$ for the $T$ gate via code switching, almost an order of magnitude below that achievable by state distillation in the same setting.
To arrive at these results, we provide and simulate an optimized code switching procedure, and bound the effect of various conceivable improvements.
Many intermediate results in our analysis may be of independent interest.
For example, we optimize the 2D color code for circuit noise yielding its largest threshold to date $0.37(1)\%$, and adapt and optimize the restriction decoder finding a threshold of $0.80(5)\%$ for the 3D color code with perfect measurements under $Z$ noise.
Our work provides a much-needed direct comparison of the overhead of state distillation and code switching, and sheds light on the choice of future FT schemes and hardware designs.
    \end{abstract}
\maketitle

\newpage
\tableofcontents

\clearpage
\section{Introduction}
\label{sec:background}

In this section we first provide this work's motivation and a summary of our main results in \sec{overview}, and then review some relatively standard but important background material that we will refer to throughout the paper.
In \sec{noise} we describe the noise models and simulation approaches that we use to analyze and simulate state distillation and code switching.
In \sec{color-code-basics} we provide some basic information about the color codes. 
In \sec{logicalOperations2DCC} we review how to implement logical operations using 2D color codes.
Finally in \sec{logicalOperations2DCC} we review state distillation.

\subsection{Motivation and summary of results}
\label{sec:overview}
Recent progress in demonstrating operational quantum devices~\cite{Nigg2014,Barends2014,Corcoles2013,Ofek2016,Arute2019} has brought the noisy intermediate-scale quantum era~\cite{Preskill2018}, where low-depth algorithms are run on small numbers of qubits.
However, to handle the cumulative effects of noise and faults as these quantum systems are scaled, fault-tolerant (FT) schemes~\cite{Shor1996,knill05,steane97,aharonov1997,Preskill1998,Aliferis2005,Raussendorf2007,Raussendorf2007a,Dennis01}
will be needed to reliably implement universal quantum computation.
FT schemes encode logical information into many physical qubits and implement logical operations on the encoded information, all while continually diagnosing and repairing faults.
This requires additional resources, and much of the current research in quantum error correction (QEC) is dedicated toward developing FT schemes with low overhead. 

The choice of FT scheme to realize universal quantum computing has important ramifications.
Good schemes can significantly enhance the functionality and lifetime of a given quantum computer.
Moreover, FT schemes vary in their sensitivity to the hardware architecture and design, such as qubit quality \cite{Cross2009}, connectivity, and operation speed.
The understanding and choice of FT scheme will therefore influence the system design, from hardware to software, and developing an early understanding of the trade-offs is critical in our path to a scalable quantum computer.

At the base of most FT schemes is a QEC code which (given the capabilities and limitations of a particular hardware platform) should: (i) tolerate realistic noise (ii) have an efficient classical decoding algorithm to correct faults, and (iii) admit a FT universal gate set.
In the search of good FT schemes we focus our attention on QEC codes which are known to achieve as many of these points as possible with low overhead.
Topological codes are particularly compelling as they typically exhibit high accuracy thresholds with QEC protocols involving geometrically local quantum operations and efficient decoders; see e.g. Refs.~\cite{Dennis01, Brown2014, Duclos-Cianci2010, Anwar2013, Duclos-Cianci2013, Bravyi2013a, Duivenvoorden2017, Kubica2018toom, Vasmer2020, Bravyi2014, Darmawan2018, Nickerson2017, Maskara2018, Chamberland2018, delfosse2014, kubica2019}.
Two-dimensional (2D) topological codes such as the toric code~\cite{Kitaev1997,Bravyi98} and the color code~\cite{Bombin2006} are particularly appealing for superconducting~\cite{fowler2012, Chamberland2020b, Chamberland2020} and Majorana~\cite{Karzig2017,Chao2020} hardware, where qubits are laid out on a plane and quantum operations are limited to those involving neighboring qubits.

The FT implementation of logical gates with 2D topological codes poses some challenges.
The simplest FT logical gates are applied transversally, i.e., by independently addressing individual physical qubits.
These gates are automatically FT since they do not grow the support of errors.
Unfortunately a QEC code which admits a universal set of transversal logical gates is ruled out by the Eastin-Knill theorem~\cite{Eastin2009,zeng2011,Jochym-OConnor2018}.
Furthermore, in 2D topological codes such gates can only perform Clifford operations \cite{Bravyi2013,PastawskiYoshida14,Beverland2014,Webster2020}.
There are, however, many innovative approaches to achieve universality, which typically focus on implementing non-Clifford logical gates~\cite{Bombin2007,kubica2015,Vasmer2019}, which achieve universality when combined with the Clifford gates.

The standard approach to achieve universality with 2D topological codes is known as \textit{state distillation}~\cite{Bravyi2005,knill2004a,knill2004b}.
It relies on first producing many noisy encoded copies of a \textit{$T$ state}
$\ket{\overline T} = (\ket{\overline{0}}+e^{i\pi/4}\ket{\overline{1}})/\sqrt{2}$, also known as a \textit{magic state}, and then processing them using Clifford operations to output a high fidelity encoded version of the state.
The high-fidelity $T$ state can then be used to implement the non-Clifford $T = \mathrm{diag}(1,e^{i\pi/4})$ gate.
Despite significant recent improvements, the overhead of state distillation is expected to be large in practice \cite{fowler2012,Litinski2019}.
A compelling alternative is \textit{code switching} via gauge fixing~\cite{paetznick2013,anderson2014,bombin2015,bombin2016} to a 3D topological code which has a transversal $T$ gate.
The experimental difficulty of moving to 3D architectures could potentially be justified if it significantly reduces the overhead compared to state distillation.
To compare these two approaches and find which is most practical for consideration in a hardware design, a detailed study is required.

\begin{figure}[h]
	\includegraphics[width=.95\textwidth]{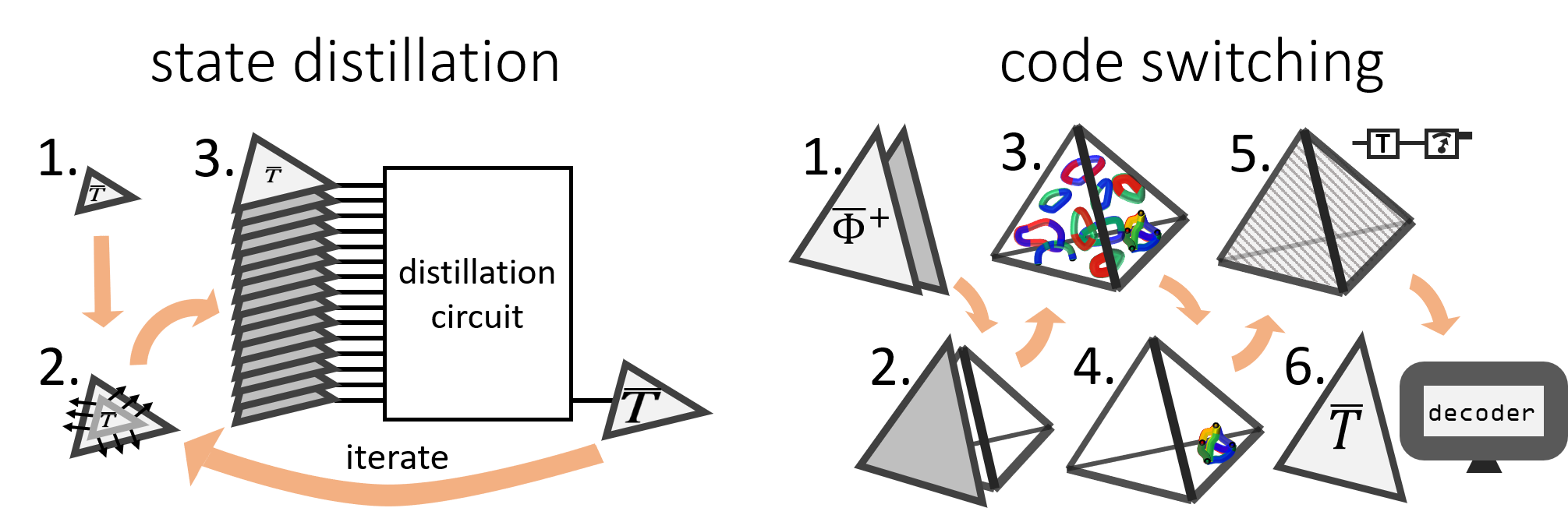}
	\caption{
	Two methods of preparing high-fidelity $T$ states encoded in the 2D color code: state distillation and code switching.
	Both approaches are implemented using quantum-local operations in 3D, i.e., noisy quantum operations are geometrically local, whereas ideal classical operations can be performed globally.
	In state distillation, many noisy encoded $T$ states are produced and fed into a Clifford distillation circuit.
	In code switching, one switches to the 3D color code, where the transversal $T$ gate is implemented.
	}
	\label{fig:DistillationAndCodeSwitchingSteps}
\end{figure}

In our work, we estimate the resources needed to prepare high-fidelity $T$ states encoded in the 2D color code, via either state distillation or code switching.
We assume that both approaches are implemented using quantum-local operations~\cite{Bombin2015b} in 3D, i.e., quantum operations are noisy and geometrically local, whereas classical operations can be performed globally and perfectly (although they must be computationally efficient).
In particular, we simulate these two approaches by implementing them with noisy circuits built from single-qubit state preparations, unitaries and measurements, and two-qubit unitaries between nearby qubits.
For state distillation, this 3D setting allows a stack of 2D color code patches, whereas for code switching it allows to implement the 3D color code; see \fig{DistillationAndCodeSwitchingSteps}.
We then seek to answer the following question: \emph{to prepare $T$ states of a given fidelity, are fewer resources required for state distillation or code switching?}

\begin{figure}[ht]
	(a)\hspace*{-5mm}\includegraphics[width=.45\textwidth]{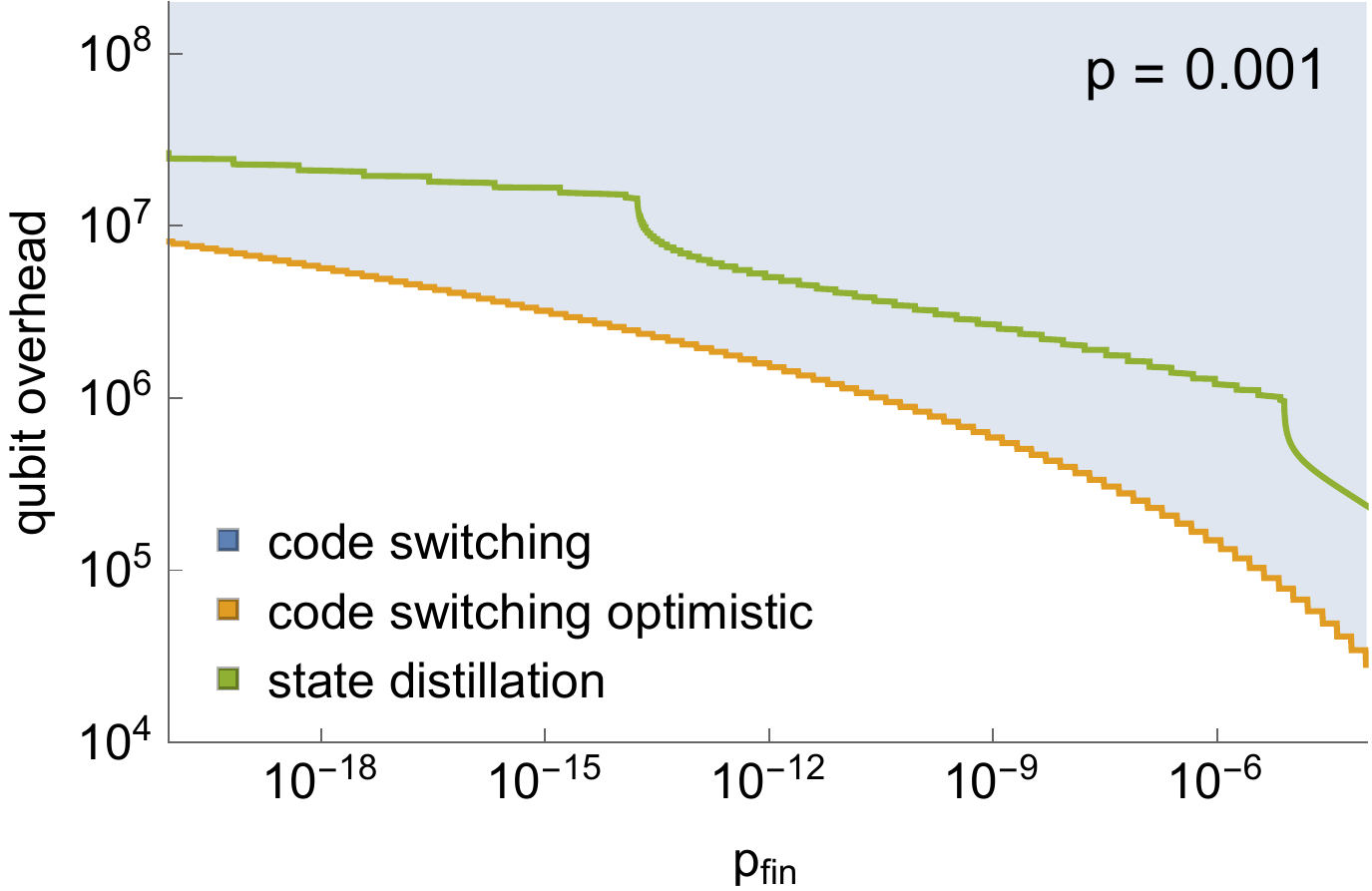}
	\quad\quad\quad
	(b)\hspace*{-5mm}\includegraphics[width=.45\textwidth]{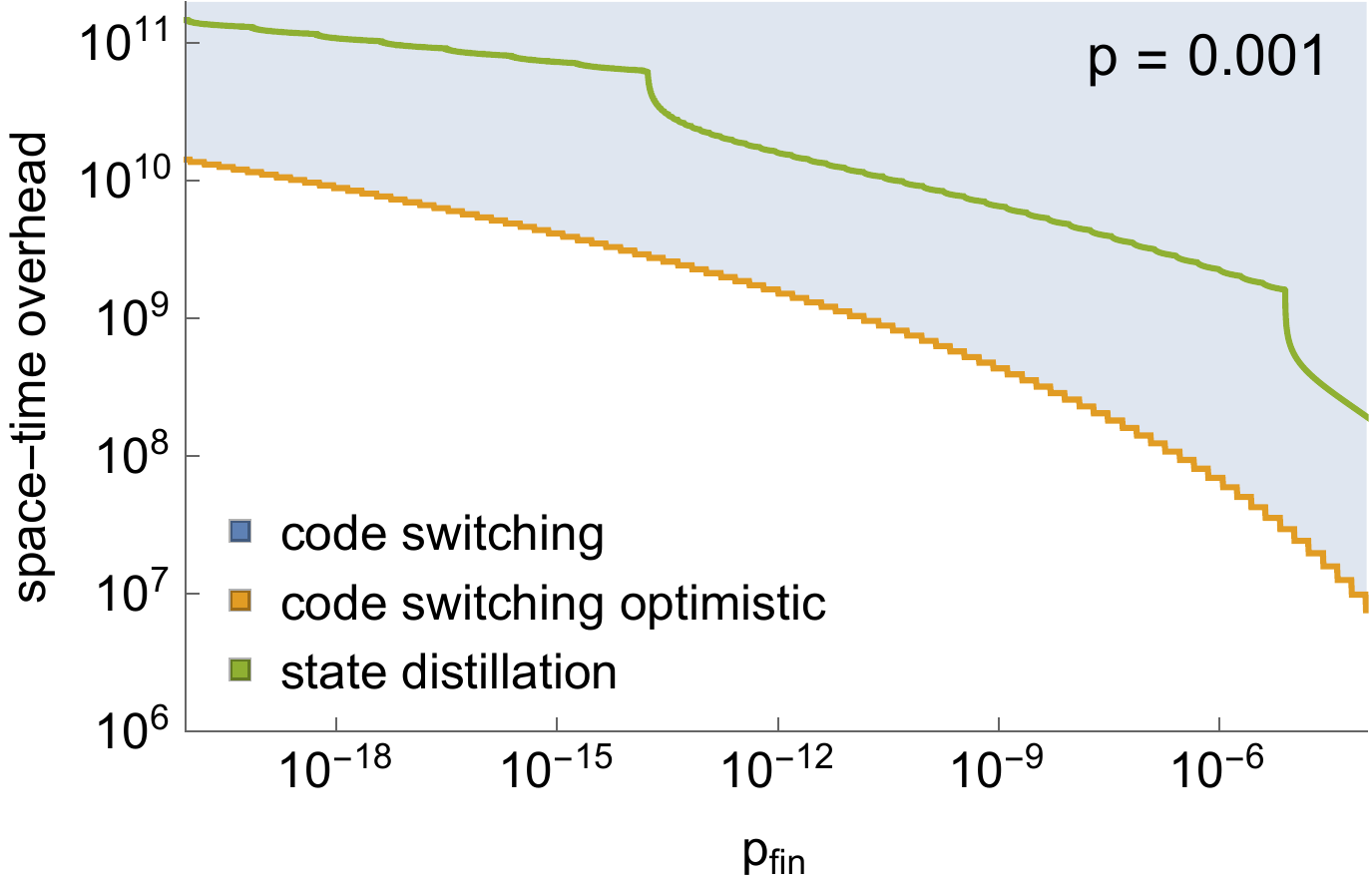}
	(c)\hspace*{-5mm}\includegraphics[width=.45\textwidth]{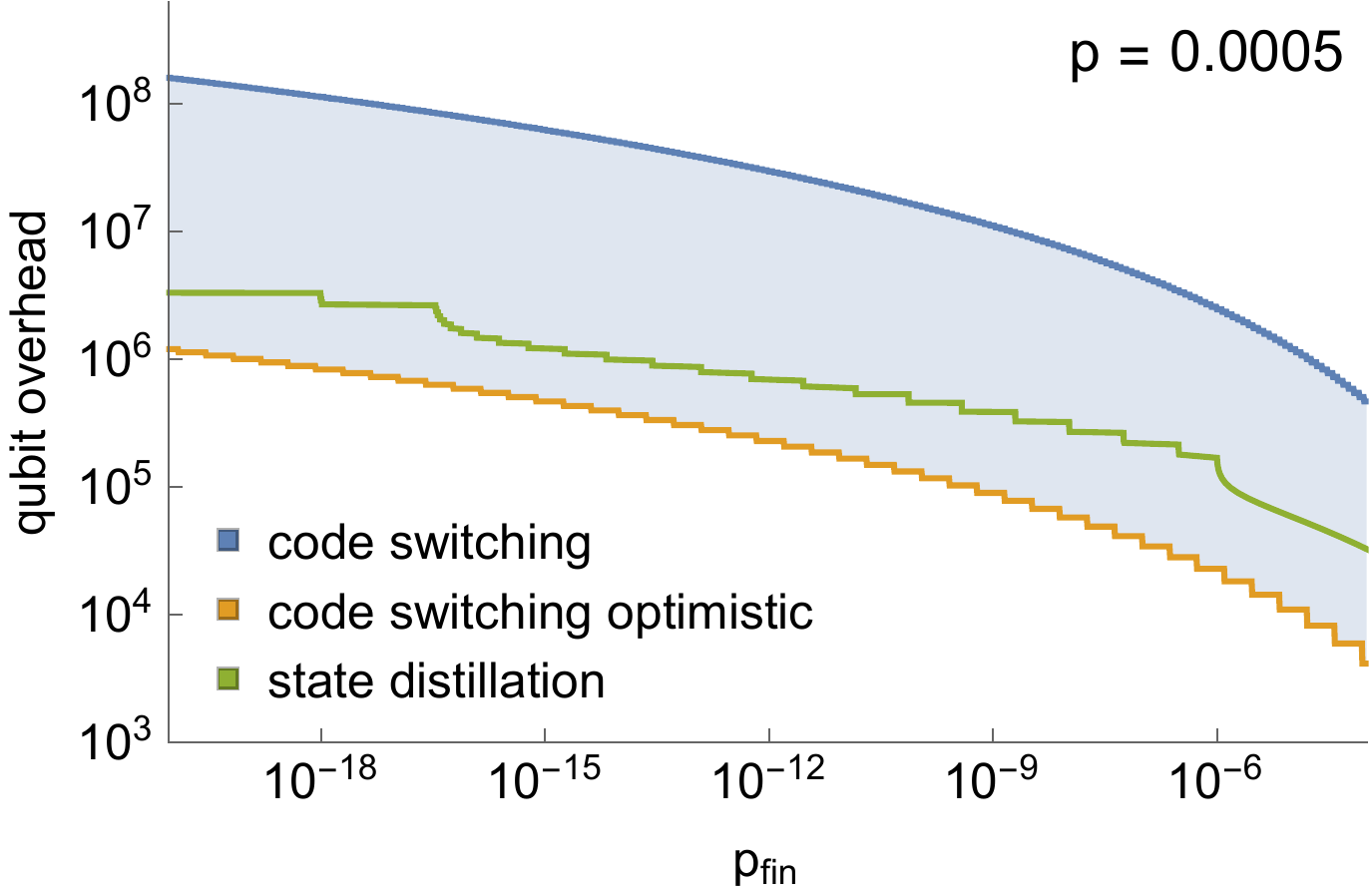}
	\quad\quad\quad
	(d)\hspace*{-5mm}\includegraphics[width=.45\textwidth]{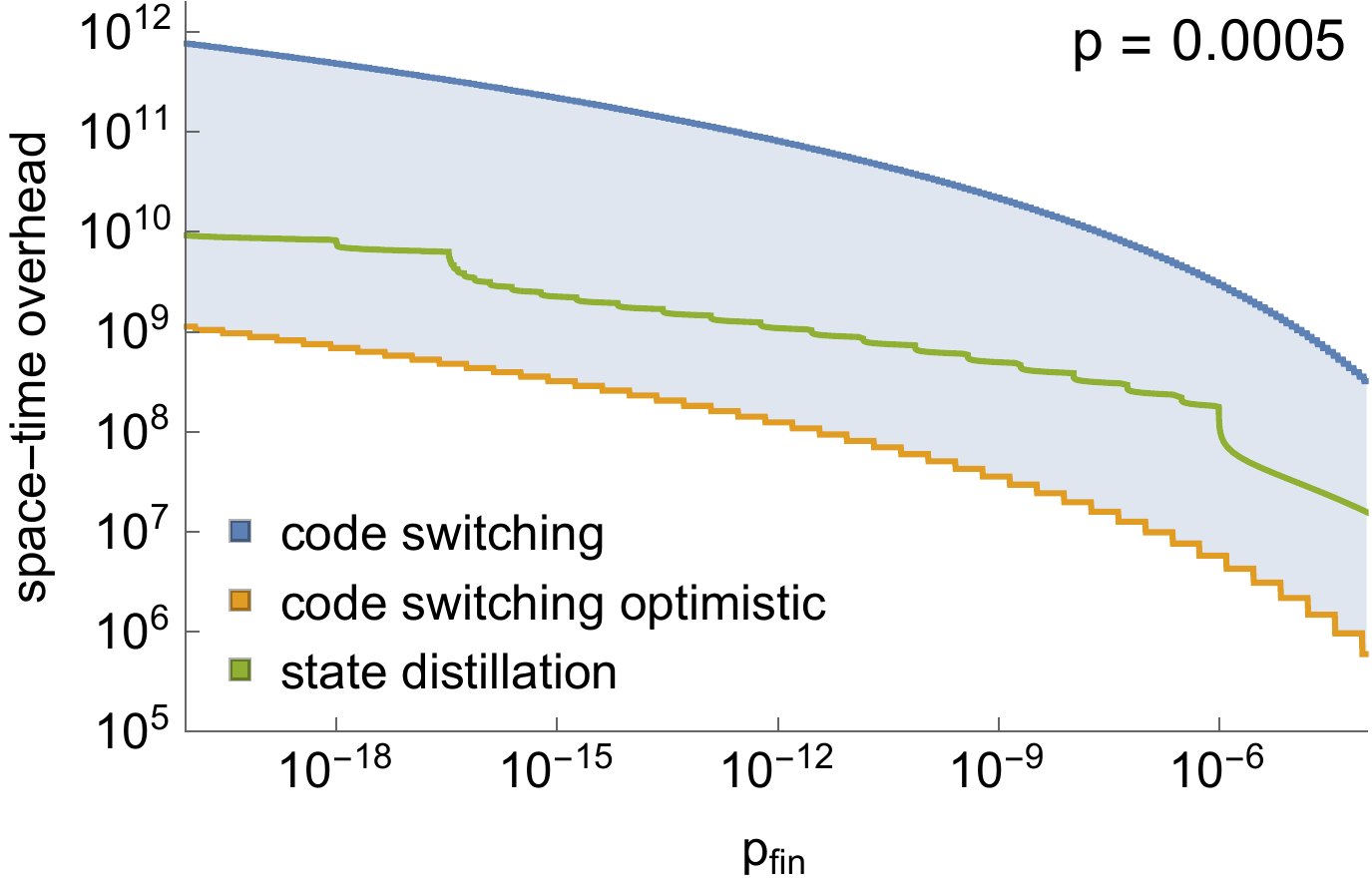}
	(e)\hspace*{-5mm}\includegraphics[width=.45\textwidth]{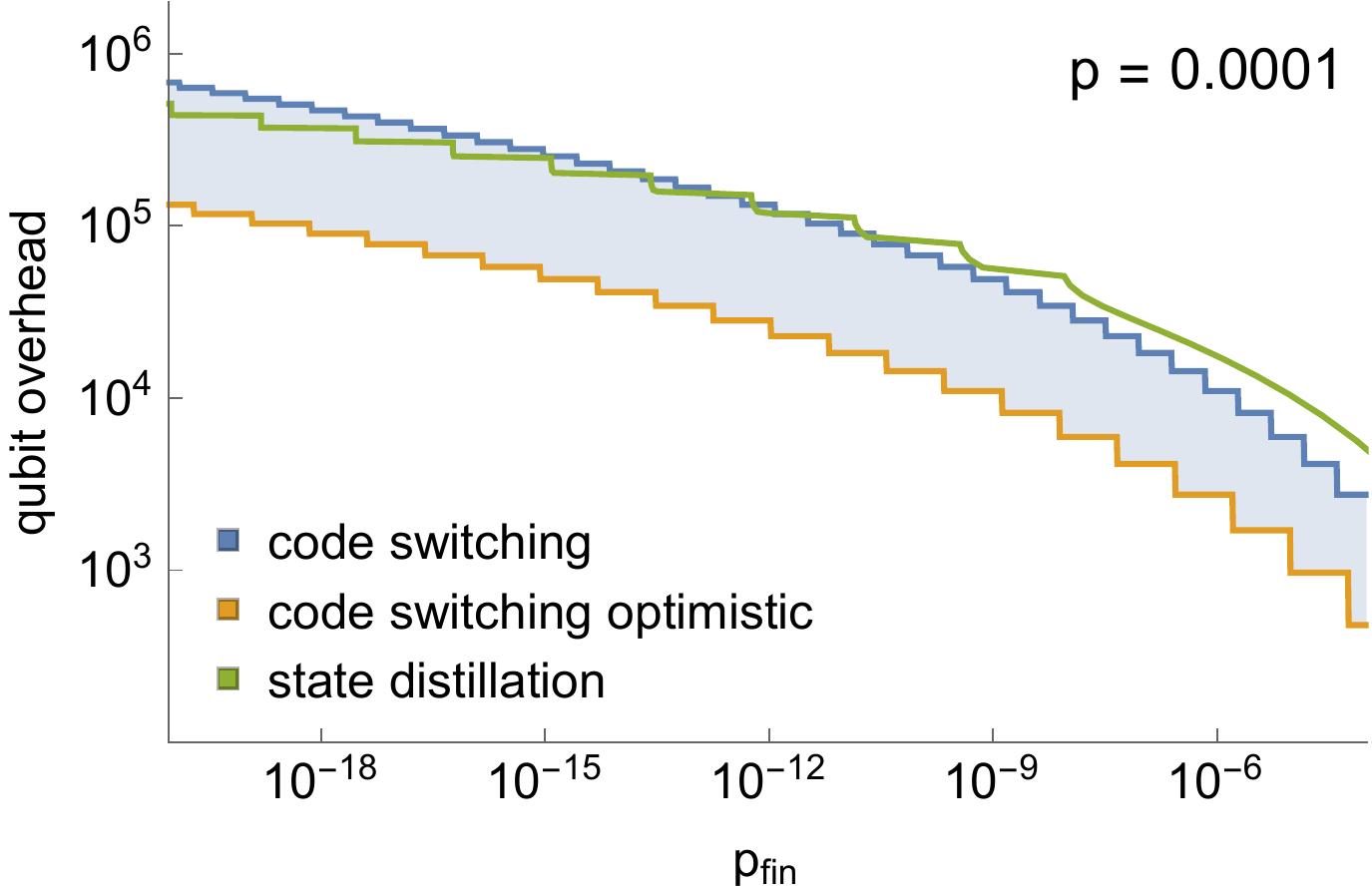}
	\quad\quad\quad
	(f)\hspace*{-5mm}\includegraphics[width=.45\textwidth]{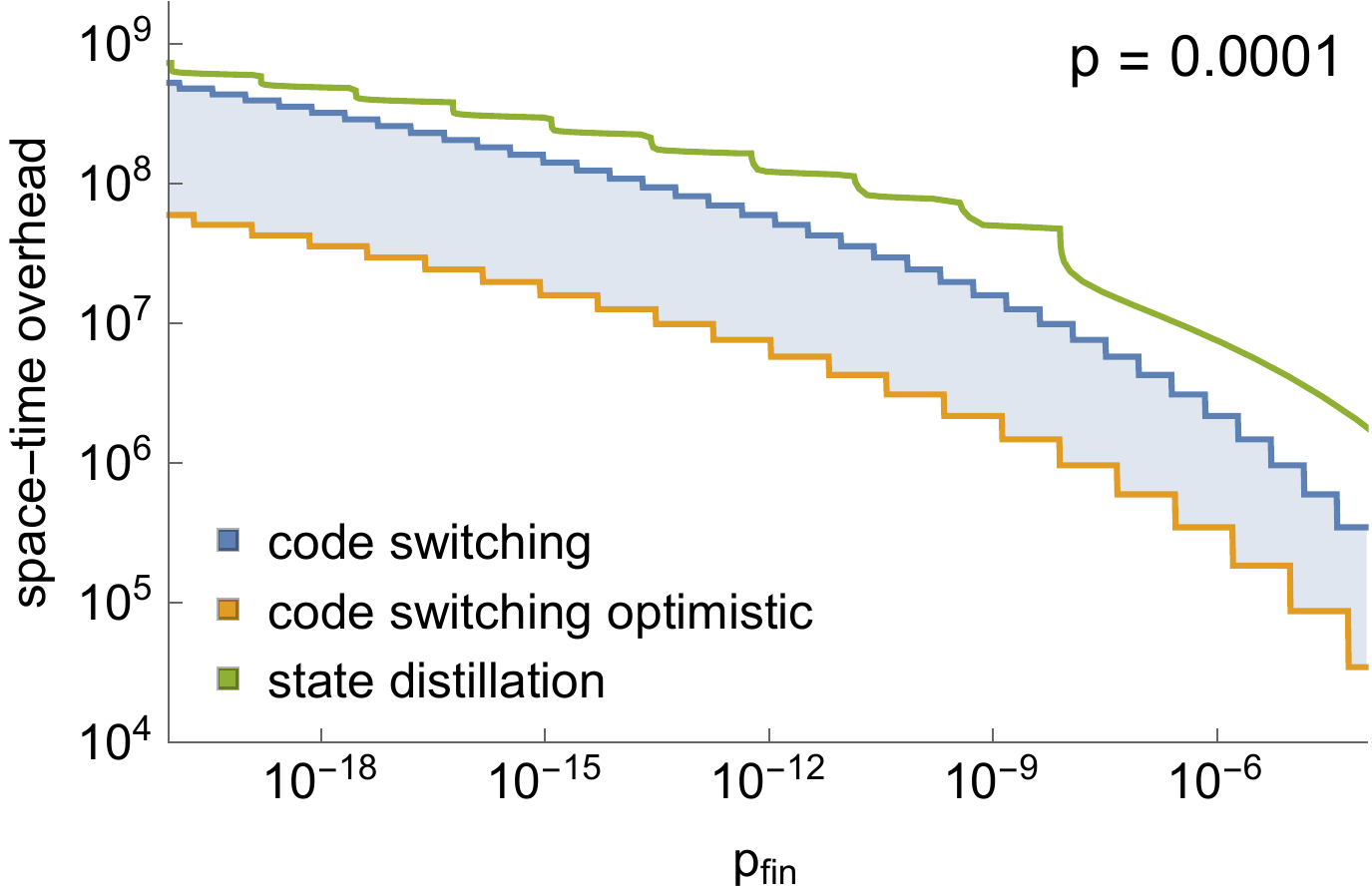}
	\caption{
	The comparison of (a) the qubit and (b) space-time overhead as a function of the infidelity $p_\text{fin}$ of the output $T$ state for state distillation and code switching.
	Possible future improvements of any steps of our code switching protocol would be included within the shaded region.
	Note there is no code switching curve for $p=0.001$ without assuming optimistic improvements to the protocol as this is higher than the observed threshold for code switching. 
	}
	\label{fig:overhead-comparison}
\end{figure}

Our main finding is that code switching does not offer substantial savings over state distillation in terms of both \textit{space overhead}, i.e., the number of physical qubits required, and \textit{space time overhead}, i.e., the space overhead multiplied by the number of physical time units required; see \fig{overhead-comparison}.
State distillation significantly outperforms code switching over most of the circuit noise error rates $10^{-4} \leq p \leq 10^{-3}$ and target $T$ state infidelities $10^{-20} \leq p_{\text{fin}} \leq 10^{-4}$, except for
the smallest values of $p$, where code switching slightly outperformed state distillation.
In our analysis we carefully optimize each step of code switching, and also investigate the effects of replacing each step by an optimal version to account for potential improvements.
On the other hand we consider only a standard state distillation scheme, and using more optimized schemes such as \cite{bravyi2012,haah2017,Haah2018a} would give further advantage to a state distillation approach.
We also find asymptotic expressions which support our finding that state distillation requires lower overhead than code switching for $p\ll 1$ and $\log p_{\text{fin}} / \log p \gg 1$.
In particular, the space and space-time overhead scale as  
$\left(\log{p_{\text{fin}}}/\log{p} \right)^{\Gamma_*}$ and $\left(\log{p_{\text{fin}}}/\log{p} \right)^{\Gamma_*+1}$ respectively, where $\Gamma_{\text{CS}}=3$ for code switching and $\Gamma_{\text{SD}} = \log_{3}15 = 2.46\ldots$ for the distillation scheme we implement.

To arrive at our main simulation results, we accomplish the intermediate goals below.

\textbf{2D color code optimization and analysis.---}In \sec{2DCCperformance}, we first adapt the projection decoder~\cite{delfosse2014} to the setting where the 2D color code has a boundary and syndrome extraction is imperfect,
as well as optimize the stabilizer extraction circuits.
We find a circuit noise threshold greater than $0.37(1)\%$, which is the highest to date for the 2D color code, narrowing the gap to that of the surface code.
We also analyze the noise equilibration process during logical operations in the 2D color code and provide an effective logical noise model.

\textbf{Noisy state distillation analysis.---}Using the effective logical noise model, we carefully analyse the overhead of state distillation in \sec{distillation}. We strengthen the bounds on failure and rejection rate by explicitly calculating the effect of faults at each location in the Clifford state distillation circuits rather than simply counting the total number of locations \cite{brooks2013, jochym2013,jones2013b,fowler2012,Litinski2019}.
We remark that we stack 2D color codes in the third dimension to implement logical operations such as the CNOT in constant time,
whereas strictly 2D approaches such as lattice surgery would require a time proportional to code distance.
The circuit-noise threshold for this state distillation scheme with the 2D color code is equal to the error correction threshold of $0.37(1)\%$. 

\textbf{Further insights into 3D color codes.---}In \sec{insights-3DCC}, we
provide a surprisingly direct way to switch between the 2D color code and the 3D subsystem color code.
Our method exploits a particular gauge-fixing of the 3D subsystem color code for which the code state admits a local tensor product structure in the bulk and can therefore be prepared in constant time. 
We also adapt the restriction decoder~\cite{kubica2019} to the setting where the 3D color code has a boundary and optimize it, which results in a threshold of 0.80(5)\%
and a better performance for small system sizes.

\textbf{End-to-end code switching simulation.---}\sec{dimjump} is the culmination of our work, where building upon results from the previous sections we provide a simplified recipe for code switching, detailing each step and specifying important optimizations.
In our simulation, we exploit the special structure of the 3D subsystem color code to develop a method of propagating noise through the $T$ gates in the system, despite the believed computational hardness of simulating general circuits with many qubits and $T$ gates.
We numerically find the failure probability of implementing the $T$ gate with code switching as a function of the code distance and the circuit noise strength, which, in turn, allows us to estimate the $T$ gate threshold to be $0.07(1)\%$.
We not only find numerical estimates of the overhead of the fully specified protocol, but also bound the minimal overhead of a code switching protocol with various conceivable improvements, such as using optimal measurement circuits, and optimal classical algorithms for decoding and gauge fixing of the 3D color code.

This work provides a much-needed comparative study of the overhead of state distillation and code switching, and enables a deeper understanding of these two approaches to FT universal quantum computation.
More generally, careful end-to-end analyses with this level of detail will become increasingly important to identify the most resource-efficient FT schemes and, in turn, to influence the evolution of quantum hardware.
Although our study focuses on color codes, we expect our main finding, i.e., that code switching does not significantly outperform state distillation, to hold for other topological codes such as the toric code as considered in Ref.~\cite{Vasmer2019}.
Furthermore, we believe that state distillation will not be outperformed by code switching exploiting either 2D subsystem codes~\cite{Bravyi2015, Jochym-OConnor2016, Jones2016} or emulation of a 3D system with a dynamical 2D system~\cite{Bombin2018a, Brown2020,Scruby2020,Iverson2020} since these schemes are even more constrained than when 3D quantum-local operations are allowed.
We remark that there are other known FT techniques for implementing a universal gate set~\cite{Hill2013,yoder2016,jochym2014,Aliferis2005,Chamberland2019,Chamberland2020c}, however they are not immediately applicable to large-scale topological codes.
Nevertheless, we are hopeful that there are still new and ingenious FT schemes to be discovered that could dramatically reduce the overhead and hardware requirements for scalable quantum computing.

\subsection{Noise and simulation}
\label{sec:noise}

The noise model we use throughout the paper is the \textit{depolarizing channel}, which on single- and two-qubit density matrices $\rho^{(1)}$ and $\rho^{(2)}$ has the action
\begin{eqnarray}
\mathcal{E}^{(1)}_p : \rho^{(1)} &\mapsto& \left(1-p\right)\rho^{(1)} + \frac{p}{3} \sum_{P\in\{X,Y,Z\}} P \rho^{(1)} P,\\
\mathcal{E}^{(2)}_p : \rho^{(2)} &\mapsto& \left(1-p\right)\rho^{(2)} + 
\frac{p}{15} \sum_{\substack{P_1, P_2\in\{I,X,Y,Z\}\\ P_1\otimes P_2 \neq I\otimes I}} (P_1 \otimes P_2) \rho^{(2)} (P_1 \otimes P_2),
\end{eqnarray}
where the parameter $p$ can be interpreted as an error probability.
The depolarizing channel leaves a single-qubit state unaffected with probability $1-p$ and applies an error $X$, $Y$, or $Z$, each with probability $\frac{p}{3}$.
Similarly, the depolarizing channel leaves a two-qubit state unaffected with probability $1-p$ and with probability $\frac{p}{15}$ applies a nontrivial Pauli error $P_1 \otimes P_2$. 

\begin{figure}[h]
	(a)\includegraphics[height=.075\textheight]{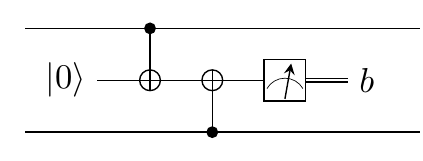}\quad\quad
	(b)\includegraphics[height=.075\textheight]{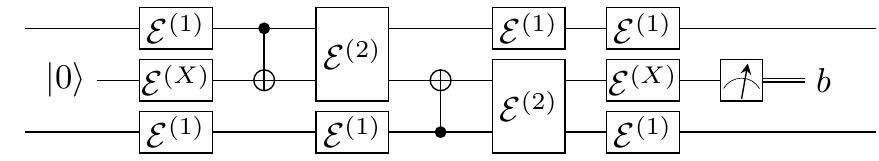}
	\qquad
	\caption{
	    (a) An example of an ideal circuit to measure weight-two operator $ZZ$.
	    We assume that every gate, as well as preparation and single-qubit measurements take one time unit.
	    (b) The noisy circuit is modeled with ideal gates followed by the depolarizing channel of strength $p$; ideal preparation can fail and produce a state orthogonal to the desired one with probability $p$, and the outcome of the ideal measurement can be flipped with probability $p$.
	    }
	\label{fig:noisy_circuit}
\end{figure}

We consider three standard scenarios,
\begin{enumerate}
    \item \textit{Depolarizing noise.---}Error correction is implemented with perfect measurements following a single time unit, during which single-qubit depolarizing noise of strength $p$ acts.
    This is often referred to in the literature as the code capacity setting.
    \item \textit{Phenomenological noise.---}Error correction is implemented with perfect measurements following each time unit, during which single-qubit depolarizing noise of strength $p$ acts.
    However the measurement outcome bits are flipped with probability $p$.
    \item \textit{Circuit noise.---}Measurements are implemented with the aid of ancilla qubits and a sequence of one- and two-qubit operations; see \fig{noisy_circuit}(a).
    One- and two-qubit unitary operations experience depolarizing noise of strength $p$. One-qubit preparations and measurements fail with probability $p$ by producing an orthogonal state or flipping the outcome; see \fig{noisy_circuit}(b).
    
    \end{enumerate}

In circuit noise, we approximate every noisy gate, i.e., Pauli $X$, $Y$ and $Z$ operators, the Hadamard gate $H$, the phase gate $T$, the controlled-not gate $\textrm{CNOT}$, and the idle gate $I$, by an ideal gate followed by the depolarizing channel on qubits acted on by the gate see \fig{noisy_circuit}.
Preparations of the state orthogonal to that intended occur with probability $p$, and measurement outcome bits are flipped with probability $p$.
We assume that all the elementary operations take the same time, which we refer to as one \textit{time unit}.

For each of the three noise models, we will use \textit{error rate} and \textit{noise strength} interchangeably to describe the single parameter $p$.

We assume a special form of noise on $T$ states, which is justified as follows. 
Consider an arbitrary single-qubit state
\begin{equation}
    \rho = \rho_{00}\ket{T}\bra{T}+\rho_{01}\ket{T}\bra{T^\perp}+\rho_{10}\ket{T^\perp}\bra{T}+\rho_{11}\ket{T^\perp}\bra{T^\perp},
\end{equation}
written in the orthonormal basis $\{ \ket{T}, \ket{T^\perp} = Z\ket{T} \}$.
Now consider a `twirling operation' consisting of randomly applying the Clifford
$XS^\dagger \propto \ket{T} \bra{T}- \ket{T^\perp} \bra{T^\perp}$, with probability $1/2$. 
This single-qubit Clifford gate can be implemented instantaneously and perfectly by a frame update.\footnote{Since we assume throughout that arbitrary one- and two-qubit physical operations are allowed, single-qubit Clifford physical operations can actually be done `offline' by tracking the basis of each physical qubit, and modifying future operations to be applied to that qubit accordingly.
They are therefore perfect and instantaneous, as they only involve classical processing.
Moreover, we will later see that logical Clifford operations can be done offline in 2D color codes such that the logical noise on encoded $T$ states can also be twirled offline.}
The state is transformed as follows
\begin{equation}
    \rho  \mapsto  \frac{1}{2} \rho + \frac{1}{2}(XS^\dagger) \rho (XS^\dagger)^\dagger = \rho_{00}\ket{T}\bra{T}+\rho_{11}\ket{T^\perp}\bra{T^\perp}.
\end{equation}
We therefore assume that the noisy $T$ state is of the form $\rho = (1-q)\ket{T}\bra{T} +q\ket{T^\perp}\bra{T^\perp}$, or equivalently that each $T$ state is afflicted by a $Z$ error with probability $q$.
Due to this simplified form of the noise, we use \textit{infidelity}, which is defined by $1-\langle T | \rho | T \rangle $, interchangeably with the noise rate and noise strength to refer to the single parameter $q$ when describing errors on $T$ states.

For the purpose of defining pseudo-thresholds later, we find it useful to define the physical error probability $p_{\text{phy}}(t)$ for a time $t$ as the probability that a single physical qubit will have a nontrivial operator applied to it over $t$ time units under this noise model.
It can be calculated as follows
\begin{eqnarray}
\label{eq:pPhysical}
p_{\text{phy}}(t) = 1-\sum_{P_1 P_2 \dots P_t =I}\text{Pr}(P_1)\text{Pr}(P_2) \dots \text{Pr}(P_t),
\end{eqnarray}
where $\text{Pr}(I)=1-p$ and $\text{Pr}(X)=\text{Pr}(Y)=\text{Pr}(Z)=p/3$.
For example $p_{\text{phy}}(1) =p$, $p_{\text{phy}}(2) =2p(1-p)+2p^2/3$, etc.
Note that $\lim_{t\rightarrow \infty} p_{\text{phy}}(t) = 3/4$.

To simulate noise, for Clifford circuits we track the net Pauli operator which has been applied to the system by noisy operations using the standard binary symplectic representation. 
When non-Clifford operations are involved, we use modified techniques which are explained throughout the text.

To estimate the statistical uncertainty of any quantity of interest $\xi$ we use the bootstrap technique, i.e., we repeat sampling from the existing data set $\mathcal I = \{I_1,\ldots, I_a\}$ to evaluate $\xi = \xi(\mathcal I)$.
In particular, for $i = 1,\ldots,b$ we
(i) randomly choose $a$ data points $I_{i(j)}$ from the data set $\mathcal I$ , where $i(j)\in\{1,\ldots,a\}$,
(ii) evaluate the quantity $\xi_i = \xi(\mathcal I_i)$ using the data set $\mathcal I_i = \{I_{i(1)},\ldots,I_{i(a)}\}$.
We remark that the same data point can be chosen multiple times in step (i).
We then estimate the quantity of interest to be
\begin{equation}
    \xi = \hat \xi \pm \sqrt{\sum_{i=1}^b \frac{(\hat \xi - \xi_i)^2}{b-1}},\quad\quad
    \hat \xi = \frac{1}{b}\sum_{i=1}^b \xi_i.
\end{equation}

Note that when reporting estimated values, we will use a digit in parenthesis to indicate the standard deviation of the preceding digit.
For example, we write $0.021(3)$ in place of $0.021 \pm 0.003$.

We will often consider the failure probability $\overline{p}_{\text{task}}$ of various tasks using a distance $d$ code and noise strength $p$.
We use the following ansatz that characterizes the generic feature that
$\overline{p}_{\text{task}}$ decreases exponentially with $d$ for any $p$ below the threshold value $p^*$, i.e.,
\begin{eqnarray}
\label{eq:ansatz_moregeneral}
\overline{p}_{\text{task}}(p,d) = \alpha(p)\beta(p)^{d},
\end{eqnarray}
where $\alpha(p)$ and $\beta(p)$ are functions of $p$ alone.
In particular, $\beta(p)$ is smaller than one for $p < p^*$.
This can be considered a generalization of the heuristic behavior of error correction failure rate $\left(p/p^{*}\right)^{(d+1)/2}$ in topological codes for error rate $p$ in the vicinity of their threshold $p^*$ \cite{fowler2012,fowler2013,landahl2011}.

\subsection{Basics of 2D and 3D color codes}
\label{sec:color-code-basics}
    
    Here we briefly review some important features of color codes, focusing on 2D and 3D.
    We also specify the lattices and some notation we use throughout the paper. 
    For a more complete review of the topics covered in this subsection, see Refs. \cite{kubica2015a,bombin2015,Kubicathesis}.

    Color codes are topological QEC codes which can be defined on any $D$-dimensional lattice composed of $D$-simplices with $(D+1)$-colorable vertices, where $D\geq 2$.
    Recall that $0,1,2,3$-simplices are vertices, edges, triangles and tetrahedra, respectively.
    Qubits are placed on $D$-simplices of the lattice; $X$- and $Z$-type gauge generators are on $(D-2-z)$- and $(D-2-x)$-simplices, and $X$- and $Z$-type stabilizer generators are on $x$- and $z$-simplices, where $x,z\geq 0$ and $x+z \leq D-2$.
    We say an operator is on a $k$-simplex when its support comprises all the $D$-simplices containing that $k$-simplex. 
\begin{figure}
\centering
	(a)\hspace*{-7mm}\includegraphics[width=.4\columnwidth]{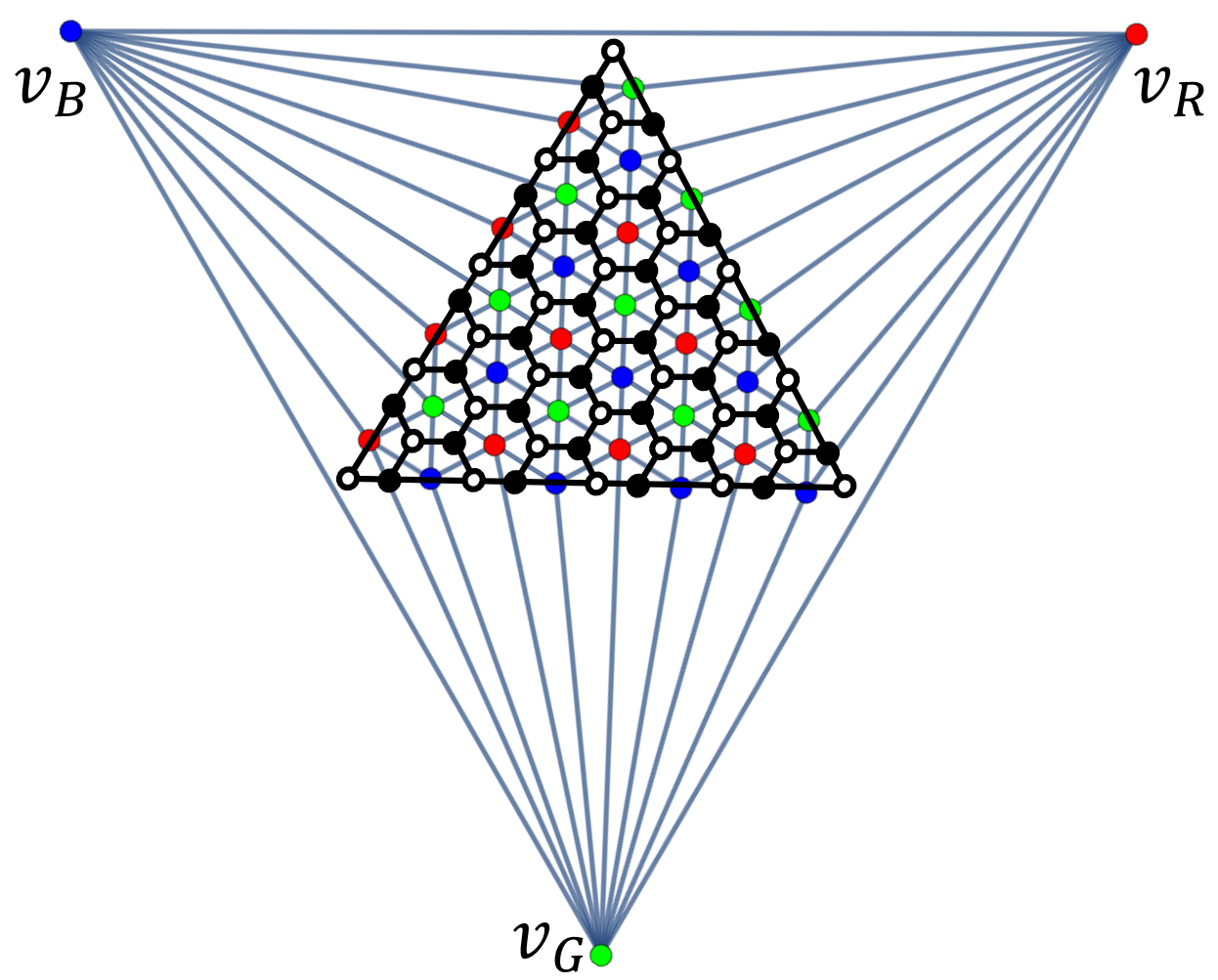}
	\quad\quad
	\includegraphics[width=.53\columnwidth]{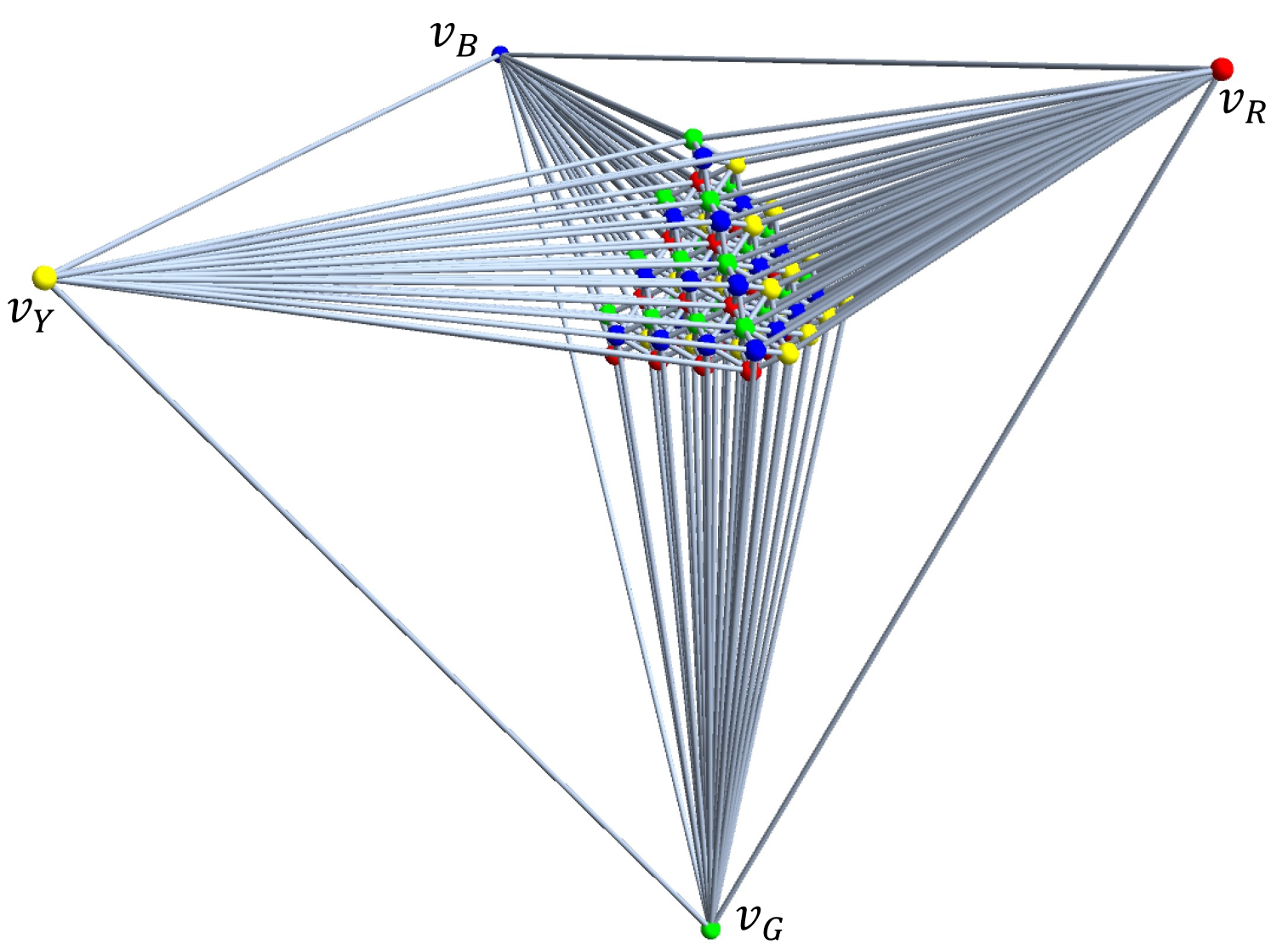}
	\hspace*{-.53\columnwidth}(b)\hspace*{.47\columnwidth}
	\caption{
	    An illustration of color code lattices $\mathcal{L}_{\text{2D}}$ and $\mathcal{L}_{\text{3D}}$ for $d=9$; see \app{lattices} for details.
	    (a) The lattice $\mathcal{L}_{\text{2D}}$ (gray edges with $R$, $G$ and $B$ vertices) and the corresponding primal lattice $\mathcal{L}^*_{\text{2D}}$ (black edges, black and white vertices).
	    Qubits are places at triangles in $\mathcal{L}_{\text{2D}}$ (vertices in $\mathcal{L}^*_{\text{2D}}$), whereas $X$- and $Z$-stabilizer generators correspond to interior vertices in $\mathcal{L}_{\text{2D}}$ (faces in $\mathcal{L}^*_{\text{2D}}$).
        (b) For the lattice $\mathcal{L}_{\text{3D}}$, qubits are identified with tetrahedra, whereas $X$- and $Z$-stabilizer generators correspond to interior vertices and interior edges, respectively.
        Note that $\mathcal{L}_{\text{2D}}$ can be obtained from $\mathcal{L}_{\text{3D}}$ by retaining only those vertices connected to the boundary vertex $v_Y$ along with the edges and faces only containing those vertices.
	   }
	\label{fig:lattices}
\end{figure}
 
    In this paper, we focus on color codes on two particular (families of) lattices: a 2D triangular lattice with a triangular boundary $\mathcal{L}_{\text{2D}}$, and a 3D bcc lattice with a tetrahedral boundary $\mathcal{L}_{\text{3D}}$; see \fig{lattices}.
    We will occasionally refer to $\mathcal{L}_{\text{2D}}$ as the \textit{triangular lattice} and $\mathcal{L}_{\text{3D}}$ as the \textit{tetrahedral lattice}.
    Most of the time we rely on context and drop the subscript, simply writing $\mathcal{L}$.
    Members of these lattice families are parameterized by 
    the code distance $d=3,5,7,\dots$ of color codes defined on them.
    We typically work with the dual lattice $\mathcal{L}$, but occasionally make use of the primal lattice $\mathcal{L}^*$.
    We remark that the graph constructed from the vertices and edges of the primal lattice $\mathcal{L}^*$ is bipartite~\cite{kubica2015a};
    see \fig{lattices}(a).
    
    Given a lattice $\mathcal{L}$, it is useful to define $\Delta_k(\mathcal{L})$ as the set of all $k$-simplices in $\mathcal{L}$. 
    We will abuse notation and write $\beta \in \face b \alpha$ (or equivalently $\beta\subseteq \alpha$) to denote that a $b$-simplex $\beta$ belongs to an $a$-simplex $\alpha$, where $0\leq b \leq a \leq D$.
    It is also useful to construct the color-restricted lattice $\mathcal{L}^\mathcal{K}$, where $\mathcal{K}$ is a set of colors, by removing from $\mathcal{L}$ all the simplices, whose vertices have colors not only in $\mathcal K$.
    For example, the restricted lattice $\mathcal{L}^{RG}$ is obtained by keeping all the vertices of color $R$ or $G$, as well as all the edges connecting them.
    We refer to the edges in $\mathcal{L}^{RG}$ as $RG$ edges; similarly for other colors.
    For any subset of edges $\gamma\subseteq\face 1 {\mathcal{L}}$ and any vertex $v\in\face 0 {\mathcal{L}}$ we define a restriction $\gamma\rest v$ as the subset of edges of $\gamma$ incident to $v$.
    We separate vertices in $\mathcal{L}$ into two types: \textit{boundary vertices} (the three and four outermost vertices in \fig{lattices}(a) and (b), respectively), and all others which we call \textit{interior vertices}.
    We call edges connecting two boundary vertices \textit{boundary edges} (there are three and six boundary edges in \fig{lattices}(a) and (b), respectively), and call all other edges \textit{interior edges}.
    More generally, we denote the sets of interior objects by $\facex  k {\mathcal{L}}$, which is the set of all $k$-simplices in $\mathcal{L}$ containing at least one interior vertex
    
    On the 2D lattice, we define the stabilizer \textit{2D color code} as follows. 
    Qubits are on triangles, and both $X$- and $Z$-stabilizer generators are on interior vertices $\facex 0 {\mathcal{L}}$.
    This code has one logical qubit and string-like logical operators.
    One can implement the full Clifford group transversally on the 2D color code. 
    
    On the 3D lattice, we can have either a stabilizer color code\footnote{There is another stabilizer code with parameters $x=1,z=0$ but we work only with the $x=0,z=1$ version here.} (with $x=0$ and $z=1$) or a subsystem color code (with $x=z=0$). In both cases, there is one logical qubit and the physical qubits are on tetrahedra. 
    For the \textit{3D subsystem color code}, $X$- and $Z$-stabilizer generators are on interior vertices $\facex 0 {\mathcal{L}}$, while $X$- and $Z$-gauge generators are on interior edges $\facex 1 {\mathcal{L}}$.
    Recall that for subsystem codes, logical Pauli operators come in two flavors: \textit{bare logical operators} which commute with the gauge generators, and \textit{dressed logical operators} which commute with the stabilizer generators.
    In the 3D subsystem color code, bare and dressed logical operators are sheet- and string-like, respectively.
    One can implement the full Clifford group transversally on the 3D subsystem color code. 
    
    For the \textit{3D stabilizer color code}, $X$- and $Z$-stabilizer generators are on interior vertices $\facex 0 {\mathcal{L}}$ and interior edges $\facex 1 {\mathcal{L}}$ respectively.
    The logical Pauli $Z$ and $X$ operators are string- and sheet-like, respectively. 
    Crucially, the $T$ gate is a transversal logical operator. 
    To implement it, we split the $n$ qubits in $\mathcal{L}$ into two groups, $(n+1)/2$ white tetrahedra and $(n-1)/2$ black tetrahedra, such that no two tetrahedra of the same color share a face.
    Applying $T$ to white qubits and $T^{-1}$ to black qubits implements the logical $\overline{T}$ gate.
    For notational convenience we write $\widetilde T = T^{\pm 1}$, determined by the color of the qubit.

    Lastly, it is useful to introduce the notions of vector spaces associated with the constituents of the lattice $\mathcal{L}$ and linear maps between them.
    We define $C_i$ to be a vector space over $\mathbb{F}_2$ with the set of $i$-simplices $\face i {\mathcal{L}}$ as its basis.
    Note that the elements of $C_i$ are binary vectors and we can identify them with the subsets of $\face i {\mathcal{L}}$.
    For any $a,b\in [D]=\{0,1,\ldots,D\}$ we define a generalized boundary operator $\bnd a b: C_a \rightarrow C_b$, which is an $\mathbb{F}_2$-linear map specified on the basis element $\alpha\in\face a {\mathcal{L}}$ as follows
    \begin{equation}
        \bnd a b \alpha =
        \begin{cases}
        \sum_{\beta \in \face b {\mathcal{L}}\mathrel{:}\beta\supseteq \alpha} \beta\quad \textrm{ if $a \leq b$,}\\
        \sum_{\beta \in \face b {\mathcal{L}}\mathrel{:}\beta\subset \alpha} \beta\quad \textrm{ if $a > b$.}
        \end{cases}
    \end{equation}
    We remark that the standard boundary operator $\partial_i: C_i \rightarrow C_{i-1}$ is a special case of the generalized boundary operator $\bnd a b$ above if we choose $a= b+1 = i$.
    These boundary maps are helpful in discussing error correction with the color code.
    In particular, since the color code is a CSS code, we can cast the decoding problem in terms of chain complexes, and treat $X$ and $Z$ errors and correction independently \cite{delfosse2014,kubica2019}.
    For the 2D color code, the boundary map $\bnd 2 0$ allows us to find for any error configuration $\epsilon\subseteq\face 2 {\mathcal{L}_{2\text{D}}}$ its point-like syndrome via $\bnd 2 0 \epsilon$, where $\epsilon$ is the support of either $X$ or $Z$ error.
    For the 3D stabilizer color code, the syndromes of $X$ and $Z$ errors correspond to loop-like and point-like objects and can be found as $\bnd 3 1 \epsilon$ and $\bnd 3 0 \epsilon$, where $\epsilon\subseteq\face 3 {\mathcal{L}_{3\text{D}}}$ denotes the support of $X$ and $Z$ error, respectively.
    In \sec{physics-gauge-flux} we discuss in detail the structure of the loop-like gauge measurement outcomes for the 3D subsystem color code.

\subsection{Fault-tolerant computation with 2D color codes}
\label{sec:logicalOperations2DCC}

\begin{figure}[ht]
\hspace*{10mm}
(a)\quad\includegraphics[height=0.13\textheight]{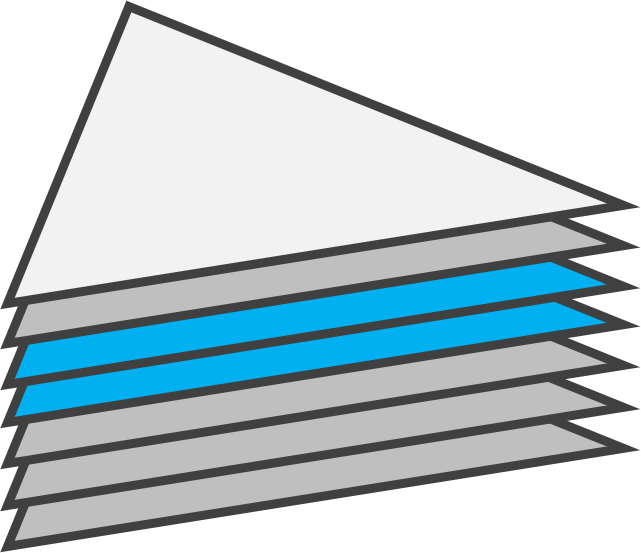}
\qquad\qquad
(b)\includegraphics[height=0.055\textheight]{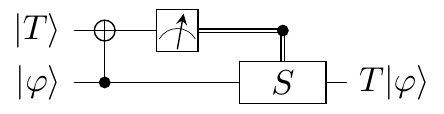}
\hspace*{10mm}
\caption{
(a) By stacking 2D color codes, one can implement transversal CNOT and SWAP operations between adjacent patches (colored in blue) with geometrically-local gates.
(b) A $T$ state can be used to implement a $T$ gate via gate teleportation using only Clifford operations.
}
\label{fig:StackAndInjectionCircuit}
\end{figure}

Here we briefly review an approach to implement quantum computation with a 3D stack of the 2D color codes
as in \fig{StackAndInjectionCircuit}(a).
Note that alternatively we could lay out the 2D color codes on a 2D plane and use lattice surgery \cite{landahl2014}, although many of the elementary logical operations, e.g. CNOT gates, would be slower than for a stack.

\textbf{Error correction and decoding.---}On each patch of the 2D color code,
QEC is continuously performed.
We use the term \textit{QEC cycle} to refer to a full cycle of stabilizer extraction circuits producing a syndrome $\sigma$, i.e., the set of measured stabilizer generators with outcome $-1$.

In a scenario in which measurements are performed perfectly, a \textit{perfect measurement decoder} is used to infer a correction $C$ for any error $E$ given the input $\sigma(E)$. The correction $C$
will return the system to the code space if applied.
The perfect measurement decoder fails if and only if $CE$ is a nontrivial logical operator.

In a scenario in which measurements are not perfect, we consider a sequence of QEC cycles $t = 1,\dots n_{\text{cyc}}$,
with error $E_t$ introduced during each cycle, and observed syndrome $\sigma_t$ for each.
For simplicity we assume that the first and last cycle have no additional error and that the syndrome is measured perfectly.
Then a \textit{faulty measurement decoder} is a decoder with takes the full history of syndromes $\sigma_1, \sigma_2, \dots \sigma_{n_{\text{cyc}}}$ and outputs a correction $C$ such that 
$C E_1 \ldots E_{n_{\text{cyc}}}$ has trivial syndrome.
The faulty measurement decoder fails if and only if $C E_1 \ldots E_{n_{\text{cyc}}}$ is a nontrivial logical operator.
We discuss decoders for the 2D color code in detail in \sec{2DCCperformance}.

\textbf{Logical operations.---}The elementary logical operations for the 2D color codes are implemented as follows.

\begin{itemize}

    \item \textit{State preparation.---}To prepare the $\ket{\overline{0}}$ state each data qubit is prepared in $\ket{0}$, then $d$ QEC cycles are performed, and the $X$-type syndrome outcomes $\sigma$ at the first cycle are inferred ($d$ cycles are needed to do so fault-tolerantly). 
    A $Z$-type fixing operator with the syndrome $\sigma$ is applied. 
    The $\ket{\overline{+}}$ state is prepared analogously. 
    
    \item 
    \textit{Idle operation.---}A single QEC cycle is performed.
    
    \item 
    \textit{Single-qubit Clifford gates.---}As the gates are transversal, these are done in software by Pauli frame updates so are instantaneous and perfect.
    
    \item 
    CNOT \textit{gate}.---This is implemented by the application of a transversal CNOT between adjacent patches in the the stack, and followed by $n_{\text{cyc}}^{\text{after}}$ QEC cycles; see \sec{2DCCOptimizedSetup} for details.
    
    \item 
    SWAP \textit{gate.---}This is implemented by swapping the data qubits on adjacent patches, and followed by a single QEC cycle.
   
    \item 
    \textit{Measurement.---}The readout of single-qubit measurements in the $\overline{X}$ and $\overline{Z}$ basis is implemented by measuring each data qubit in the $X$ and $Z$ basis. 
    The output bit string is then processed in two stages:
    (1) a perfect measurement decoder is run to correct the bit string such that it satisfies all stabilizers, and then (2) the outcome is read off from the parity of the corrected bit string restricted to the support of any representative of the logical operator.
    
\end{itemize}

The above list consists of Clifford operations, which are not by themselves universal for quantum computation. 
In addition, we consider the following non-Clifford gate.

\begin{itemize}
    \item 
    \textit{$T$ gate.} It is implemented using gate teleportation of an encoded $T$ state $\ket{\overline{T}} = (\ket{\overline{0}}+e^{i \pi/4}\ket{\overline{1}})/\sqrt{2}$ as shown in \fig{StackAndInjectionCircuit}(b).
    We will consider the production of the encoded $T$ state by both state distillation and code switching.
\end{itemize}

\subsection{State distillation}
\label{sec:basics-distillation}

Here we briefly review state distillation \cite{Bravyi2005,knill2004a}, in which Clifford operations are used with postselection to convert noisy resource states into fewer but crucially less noisy resource states.
In our analysis of the overhead of state distillation in \sec{distillation}, we consider only the standard 15-to-1 state distillation protocol.
However, given the wide range of state distillation protocols that have been proposed in recent years, we take the opportunity to attempt to consolidate the concepts behind them here.
In particular, we outline three classes of state distillation protocols, which to our knowledge include all known proposals up to small variations. 
We then go on to describe a family of protocols recently introduced by Haah and Hastings \cite{Haah2018a}, for which the 15-to-1 protocol that we consider is an special instance.
In our discussion here we will assume Clifford operations are perfect, although we will relax that assumption in \sec{distillation}. 

\begin{figure}
\hspace*{10mm}(a)\includegraphics[height=.12\textheight]{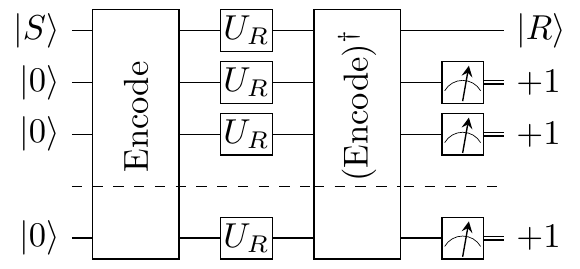}\hfill
(b)\includegraphics[height=.12\textheight]{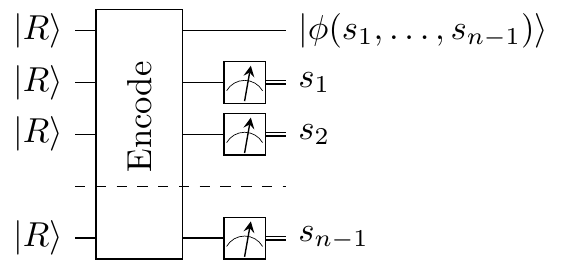}
\hspace*{10mm}
\\
\vspace*{5mm}
(c)\includegraphics[height=.14\textheight]{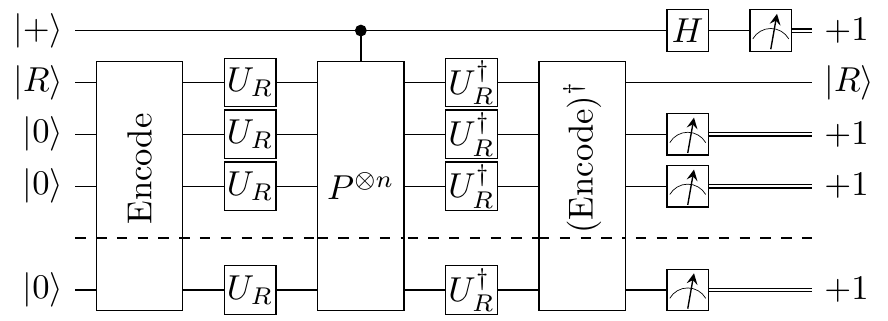}
\caption{
The three known types of state distillation schemes for a resource state $\ket{R}$.
Each wire represents a logical qubit of a QEC scheme that fault-tolerantly performs all Clifford gates.
(a) We need a code with a transversal non-Clifford gate $\Ures$, such that $\Ures\ket{S} = \ket{R}$ for some stabilizer state $\ket{S}$. 
(b) We need a code with a transversal gate $\Cres$, such that $\Cres \ket{R} = \ket{R}$. 
The measurement of the stabilizers of the code will be probabilistic even when $\ket{R}$ input is noiseless, with postselected state $\ket{\phi(+1, \dots, +1)} = \ket{R}$. (c) We need a code with a transversal gate $\Cres$, such that $\Cres \ket{R} = \ket{R}$ and $\Cres = \Ures P \Ures^\dagger$ for some Pauli operator $P$.
}
\label{fig:DistillationSchemes}
\end{figure}

Let $\Cres$ be an operator stabilizing a resource state $\ket{R}$, i.e., $\Cres \ket{R}=\ket{R}$, and $\Ures^\dag$ be a non-Clifford unitary transforming $\ket{R}$ into some stabilizer state $\ket{S}$, i.e., $\Ures^\dag\ket{R}= \ket{S}$.
We will refer to $\Cres$ and $\Ures$ as the \textit{resource stabilizer} and \textit{resource rotator}, respectively,
and throughout when we write that $\Ures$ should be applied, it can be implemented using $\ket{R}$ by gate teleportation as in \fig{StackAndInjectionCircuit}(b).
The three classes of state distillation protocols are then summarized as follows (and depicted Fig.~\ref{fig:DistillationSchemes}).
\begin{enumerate}
	\item[(a)]  \textit{Code projector then resource rotator.---}We use a code with a transversal logical gate $\overline{\Ures} = \Ures \otimes \Ures \otimes \dots \otimes \Ures$. 
	We prepare the (noisy) logical state $\ket{\overline{R}}$ by first encoding the logical stabilizer state $\ket{\overline{S}}$ and then implementing the transversal gate $\overline{\Ures}$
	(using $n$ copies of the noisy resource state $\ket{R}$).
	Unencoding $\ket{\overline{R}}$ and postselecting on the $+1$ measurement outcomes of the stabilizers of the code gives a distilled resource state with infidelity reduced from $q$ to $\mathcal{O}(q^d)$ for code distance $d$. 
	See schemes in \cite{Landahl2013,Bravyi2005,bravyi2012,Haah2018a,Haah2018b}.
	
	\item[(b)]  \textit{Resource stabilizer then code projector.---}We use a code with a transversal logical gate $\overline{\Cres} = \Cres \otimes \Cres \otimes \dots \otimes \Cres$. 
	We start with $n$ noisy resource states $\ket{R}^{\otimes n}$, which by definition satisfy $\overline{\Cres} \ket{R}^{\otimes n} = \ket{R}^{\otimes n}$, then measure the stabilizers of the code.
	If all measurement outcomes are $+1$, then the state $\ket{\overline{R}}$ has been prepared, which can then be further decoded to yield a distilled resource state $\ket{R}$ with infidelity suppressed from $q$ to $\mathcal{O}(q^d)$.
	Note that this approach seems less promising than the other two since the probability of successful postselection is less than one even for perfect resource states. See schemes in \cite{Bravyi2005,reichardt2005}.
	
	\item[(c)] \textit{Code projector then resource stabilizer.---}We use a code with a transversal logical gate $\overline{\Cres} = \Cres \otimes \Cres \otimes \dots \otimes \Cres$ and assume that $\Ures \Cres \Ures^\dagger = P$ is a Pauli operator.\footnote{
	If the unitary $\Ures$ is in the third level of the Clifford hierarchy, then $\Cres$ is a Clifford operator, but by using this approach with a code which implements a non-Clifford transversal gate $\Cres$, state distillation for higher-order schemes should be possible.}
	First, we encode a resource state $\ket{R}$ with infidelity $q_1$ in the code, giving $\ket{\overline{R}}$, and then measure the logical operator $\overline{\Cres}$ and postselect on the $+1$ outcome.
	To measure $\overline{\Cres}$ we use a measurement gadget consisting of the following three steps. 
	First, apply $\Ures \otimes \Ures \otimes \dots \otimes \Ures$ using $n$ noisy $\ket{R}$ states, each with infidelity $q_2$.
	Second, apply the Clifford gate control-$(P \otimes P \otimes \dots \otimes P)$, controlled by an ancilla state $\ket{+}$.
	Third, apply $\Ures^\dagger \otimes \Ures^\dagger \otimes \dots \otimes \Ures^\dagger$ using another $n$ noisy $\ket{R}$ states, each with infidelity $q_2$.
	After this gadget, the stabilizers are checked, and if all are satisfied, then the encoded state is decoded and kept as a distilled resource state. 
	The output has infidelity suppressed to $\mathcal{O}(q_1^2)+\mathcal{O}(q_2^d)$, but the suppression with respect to $q_1$ can be boosted by, for example, repeating the measurement gadget.
	See schemes in \cite{knill2004a,knill2004b,meier2013,jones2013,duclos2015,campbell2016,haah2017b}.
\end{enumerate}

Historically, the first state distillation protocol was proposed by Knill~\cite{knill2004a} (type c), which takes 15 input $T$ states of infidelity $q$ to produce an output of infidelity $35q^3$, with acceptance probability $1-15q$. Shortly after, Bravyi and Kitaev~\cite{Bravyi2005} proposed two schemes, the first of which is type a in our classification, but which has the same parameters as Knill's (later the two schemes were shown to be mathematically equivalent \cite{haah2018c}).
The second scheme in Ref.~\cite{Bravyi2005} is of type b and has less favorable parameters than the 15-to-1 scheme, but a higher state distillation threshold.
Another type b scheme was found by Reichardt~\cite{reichardt2005}, which could successfully distill states arbitrarily close to the stabilizer polytope.
In Ref.~\cite{bravyi2012}, type a schemes outputting multiple resource states were proposed.
Multiple outputs were also achieved for type b schemes in Refs.~\cite{jones2013,haah2017b}.
These `high rate $k/n$' schemes have promising parameters and may perform well in certain regimes.
However, they also tend to have large Clifford circuits likely inhibiting their practicality when including the effect of realistically noisy Clifford operations.
Recently, Haah and Hastings introduced yet another family of state distillation protocols of type a \cite{Haah2018a}. 
These are based on puncturing quantum Reed-Muller codes and have both interesting asymptotic properties \cite{Haah2018b} and appear to be practically favorable \cite{Haah2018a}.
We review this family here, which includes the 15-to-1 scheme that we analyze in detail in \sec{distillation}. 

\begin{figure}
\includegraphics[width=0.5\columnwidth]{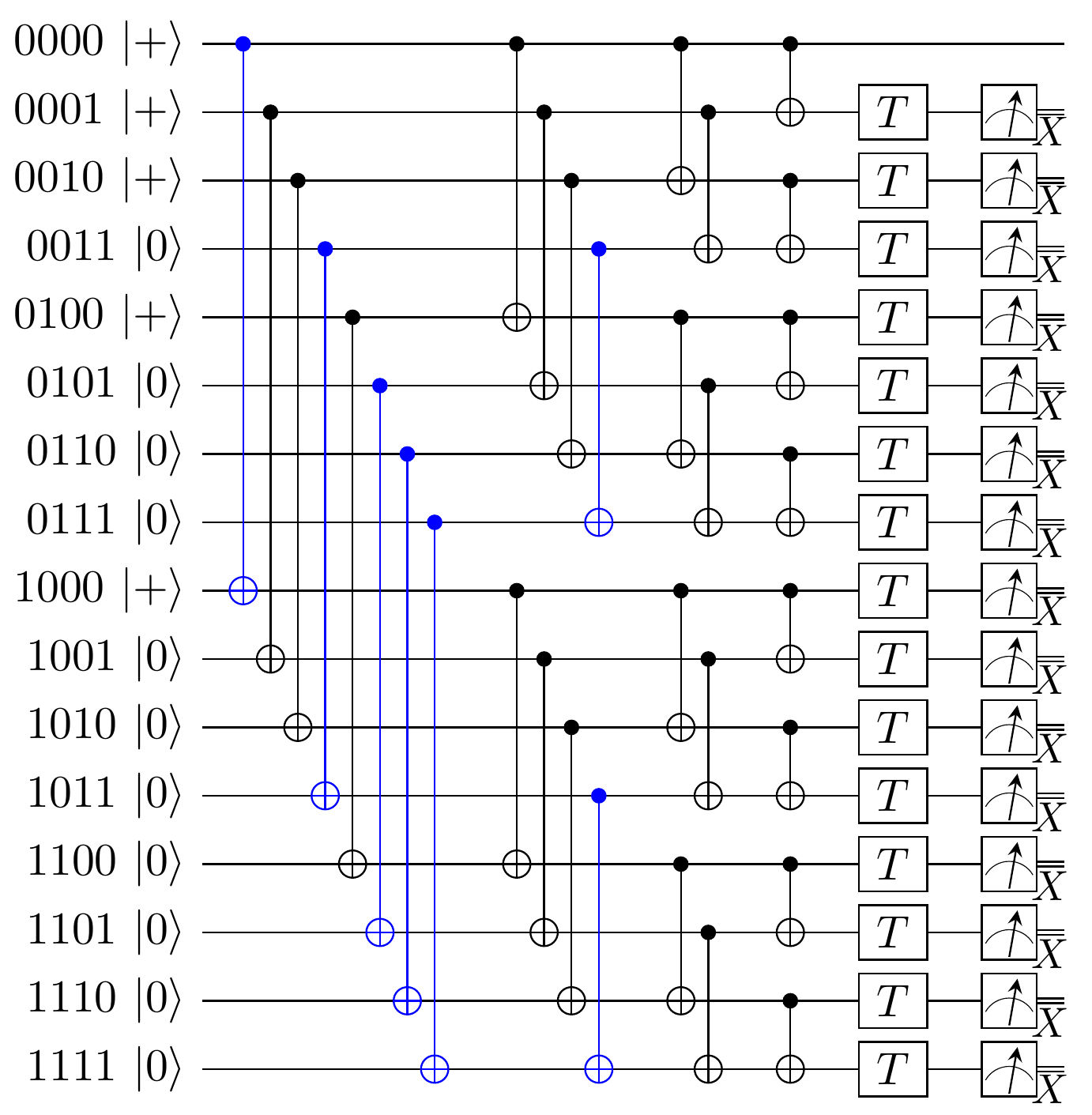}
\caption{ 
A 15-to-1 state distillation circuit as the $r=1$ instance of the Haah-Hastings construction.
Each qubit is labeled by a string of $3r+1$ bits, and if the bit string has weight at most $r$, then that qubit is prepared in the $\ket{+}$ state, otherwise in the $\ket{0}$ state. 
In the $j$th round of CNOT gates, we apply CNOTs between the pairs of qubits with bit string differing only at the $j$th position.
Note that the blue CNOTs are redundant as their control or target qubits are $\ket{0}$ or $\ket{+}$, respectively.
The resulting state is $\ket{\text{QRM}_1}\propto \ket{+}\ket{\overline{+}}+\ket{-}\ket{\overline{-}}$, where in our case $\ket{\overline{+}}$ and $\ket{\overline{-}}$ are code states of the 15-qubit Reed-Muller code.
We then apply the transversal $\overline{T}$ gate and measure every non-punctured qubit in the $X$ basis to infer the measurement outcomes $m_i$'s for $X$-stabilizers and $m_{\overline X}$ for the logical $\overline X$.
The protocol rejects if $m_i = -1$ for some $i$.
If there are no faults, then the the remaining unpunctured qubit is in the state $T^{3-2m_{\overline X}}\ket{+}$.
}
\label{fig:15to1Distillation}
\end{figure}

\textbf{Haah-Hastings state distillation protocols.---}This family of state distillation protocols \cite{Haah2018a} involves first producing what we will call a \textit{Reed-Muller state} $\ket{\text{QRM}_r}$ on $n=2^{3r+1}$ qubits, where $r =1,2,\dots$, with a Clifford circuit of depth $3r+1$ using $(3r+1)2^{3r}$ CNOTs; see \fig{15to1Distillation}. 
This state has the property that for any subset of $k$ qubits, the state can be decomposed as a set of $k$ Bell pairs between those `punctured' qubits and those which remain, namely
\begin{eqnarray}
\ket{\text{QRM}_r} = \prod_{i=1}^k \left( \frac{\ket{0}_i\ket{\overline{0}}_i+\ket{1}_i\ket{\overline{1}}_i}{\sqrt{2}}\right),
\end{eqnarray}
where the states of the $k$ punctured qubits have no bar, and where $\{ \ket{\overline{0}}_i,\ket{\overline{1}}_i \}_{i=1,...,k}$ are logical basis states for a $[[n-k,k,d_Z]]$ CSS code which is triply-even and that transversal $T$ implements a logical $\overline{T}$ on each logical qubit of the code.
Let the $Z$-distance $d_Z$ be the weight of the smallest nontrivial $Z$-type logical operator of this code.
To specify the CSS code, which we call the punctured Reed-Muller code, one can start with the $X$- and $Z$-type stabilizer generators of the $\ket{\text{QRM}_r}$ state (which, for example, could be specified by propagating the initial single-qubit stabilizers through the CNOT circuit in \fig{15to1Distillation}). 
Choosing a stabilizer generator set for $\ket{\text{QRM}_r}$ in such a way that only one $X$- and one $Z$-type generator have nontrivial support on each punctured qubit, these form logical operators and the remaining generators with no support on any punctured qubits form the stabilizer generators of the punctured Reed-Muller code. 

The next step of the protocol is to apply a (noisy) $T$ gate to each non-punctured qubit, 
\begin{eqnarray}
T^{\otimes (n-k)}\ket{\text{QRM}(r)} = \prod_{i=1}^k \left( \frac{T\ket{+}_i\otimes\ket{\overline{+}}_i+T\ket{-}_i\otimes\ket{\overline{-}}_i}{\sqrt{2}}\right).
\end{eqnarray}
To see this, note that $(I\otimes \overline{T}) \frac{\ket{0\overline{0}} + \ket{1\overline{1}}}{\sqrt{2}} = (T\otimes I) \frac{\ket{0\overline{0}} + \ket{1\overline{1}}}{\sqrt{2}} = (T\otimes I) \frac{\ket{+\overline{+}} + \ket{-\overline{-}}}{\sqrt{2}}$.
Finally, one measures the logical $\overline{X}_i$ operators of the CSS code.
If the measurement outcome of $\overline{X}_i$ is +1, then the $i$th punctured qubit is left in the state $T\ket{+}$; otherwise, the $i$th punctured qubit is in the state 
$T\ket{-} = Z T\ket{+}$, requiring a simple Pauli $Z$ fix. 
In practice this is achieved by measuring all $n-k$ non-punctured qubits in the $X$-basis, and post-processing.
Note that post-processing the measurement outcomes also tells us what the $X$-stabilizer generators for the CSS code are, which should all be $+1$ in the absence of error.
This, in turn provides, a postselection condition of the protocol.
The scheme takes $n-k$ encoded $T$ states of infidelity $q$, and outputs $k$ encoded $T$ states of infidelity $A q^{d_Z}$, where $A$ is the number of nontrivial $Z$-type logical operators of minimal weight $d_Z$.

In \fig{15to1Distillation} we show an instance of the Haah-Hastings protocol for $r=1$, which we analyze more explicitly in \sec{2DCCperformance}. 
To our knowledge, this instance was first shown in Ref.~\cite{fowler2012}, and is essentially a modification of the 15-to-1 Bravyi-Kitaev protocol, which avoids the need of the unencoding part of the circuit in \fig{DistillationSchemes}(a). 
Some larger instances of this family have very good properties for state distillation, although we do not analyze them in detail here.

\clearpage
\section{2D color code analysis}
\label{sec:2DCCperformance}

In this section, we describe an efficient and optimized implementation of the 2D color code under circuit noise.
First, in~\sec{decoder} we adapt Delfosse's color code projection decoder \cite{delfosse2014} to allow for boundaries in the lattice.
Then, we further adapt the decoder to accommodate faulty measurements in \sec{ftdecoder}.
We optimize the stabilizer extraction circuits in \sec{circuitnoise}, finding a circuit noise threshold of 0.37(1)\%.
This represents a significant improvement over the previous highest threshold value for the color code of 0.2\% in \cite{Chamberland2020},
and brings it closer to the toric code threshold of near one percent\footnote{
In Ref.~\cite{Wang2011},
a threshold above $1\%$ was reported, however follow-up work~\cite{Chamberland2020} failed to reproduce this result and reported $0.7\%$.}.
However, we stress that we are primarily interested in the finite-size rather than asymptotic performance, 
and therefore take care to accurately account for effects that are often ignored when focusing on threshold alone, such as residual error.
We believe the performance improvements we see over previous studies of the color code under circuit noise are due to our use of the hexagonal rather than the square-octagon primal lattice in the case of \cite{landahl2011,stephens2014}, and by optimizing extraction circuits and removing the restriction on the qubit connectivity in the case of \cite{Chamberland2020}.

\subsection{Projection decoder with boundaries}
\label{sec:decoder}

In this subsection, we briefly describe our adaptation of Delfosse's color code projection decoder \cite{delfosse2014} to the lattice $\mathcal{L}_{\text{2D}}$, which has boundaries.
Our adaptation is essentially the same as that presented in \cite{stephens2014}, but for the hexagonal rather than the square-octagon primal lattice.
This decoder assumes that the stabilizer measurements are perfect, and we analyse its performance under depolarizing noise.

We consider the 2D color code on the lattice $\mathcal{L} = \mathcal{L}_{\text{2D}}$ from \sec{color-code-basics}.
Since the 2D color code is a self-dual CSS code, we can decode $X$- and $Z$-errors separately and describe here the correction of $X$-errors; the correction of $Z$-errors is identical.
We denote the support of the $X$ errors by $\epsilon\subseteq\face{2}{\mathcal{L}}$, the $Z$-type syndrome by $\sigma\subseteq\facex{0}{\mathcal{L}}$ and the resulting correction by $\hat{\epsilon} \subseteq\face{2}{\mathcal{L}}$.
For each pair of colors 
$\mathcal{K}\in \{RG, RB, GB\}$ we define the set of highlighted vertices $V^\mathcal{K}$ to contain the subset $\sigma^\mathcal{K}\subseteq\sigma$ of all the vertices of color in $\mathcal{K}$ and we also include in $V^\mathcal{K}$ a boundary vertex $v_K$ for only one color $K\in\mathcal{K}$ whenever $|\sigma^\mathcal{K}|$ is odd, i.e., $V^\mathcal{K} = \sigma^\mathcal{K}$ or $V^\mathcal{K} = \sigma^\mathcal{K} + v_K$.
Note that by definition $|V^\mathcal{K}|$ is even.

\begin{figure}[ht!]
	\includegraphics[width=1.\textwidth]{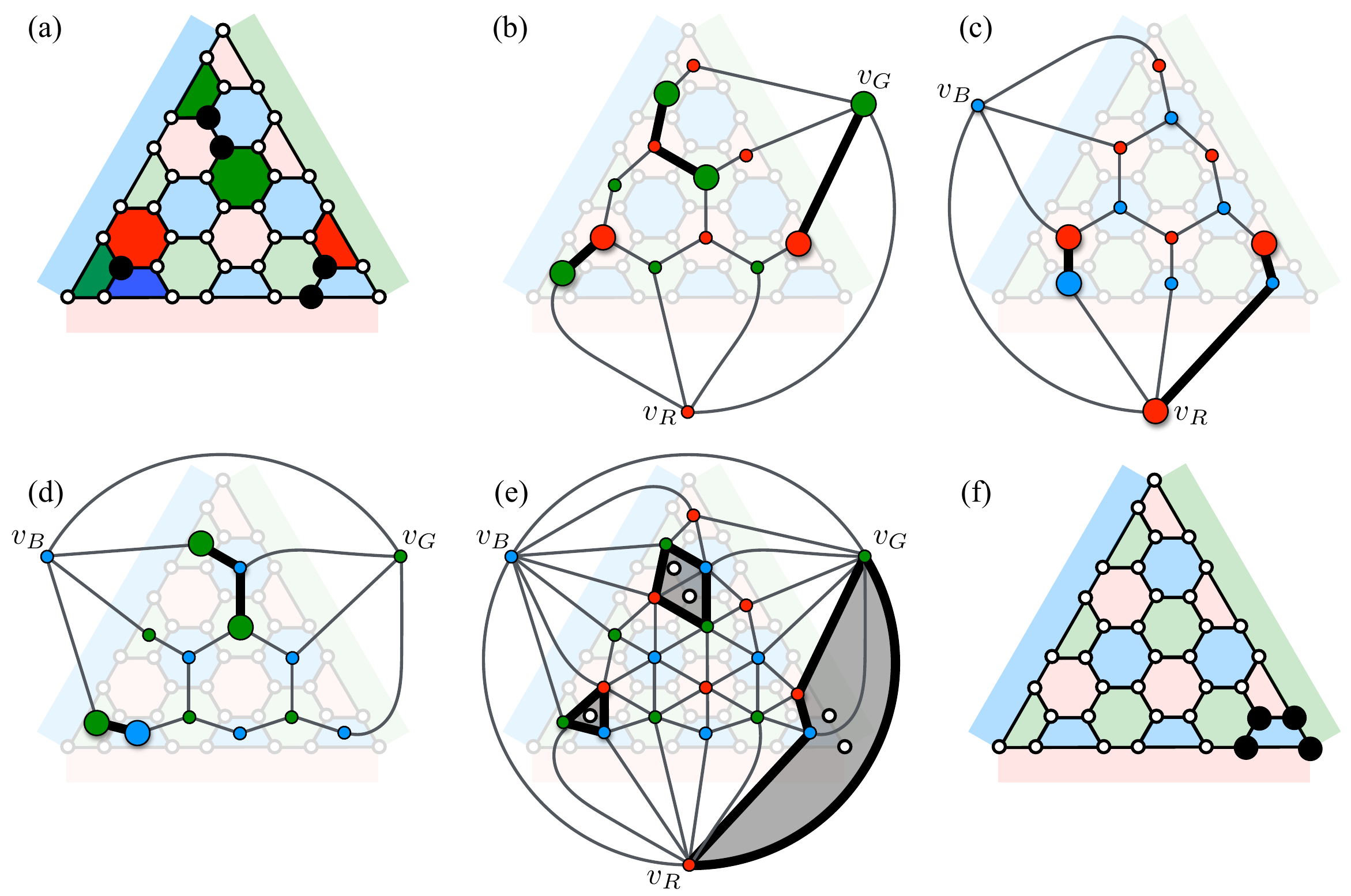}
	\caption{
		The projection decoder for the 2D color code of distance $d=7$ on the lattice $\mathcal{L}_{\text{2D}}$.
		(a) The primal lattice with errors (black dots) and the corresponding syndrome (highlighted faces).
		(b)-(d) We find pairings $E^{RG}$, $E^{RB}$ and $E^{GB}$ (thick lines) of highlighted vertices $V^{RG}$, $V^{RB}$ and $V^{GB}$ within the restricted lattices $\mathcal{L}^{RG}$, $\mathcal{L}^{RB}$ and $\mathcal{L}^{GB}$, respectively.
		Note that in (b) and (c) boundary vertices $v_G$ and $v_R$ are included as highlighted vertices.
		(e) The error estimate is the region (shaded) with boundary $E^{RG}+E^{RB}+E^{GB}$, which contains the minimal number of qubits (white dots).
		(f) The decoding succeeds since the initial error and the estimate differ by a stabilizer (black dots in the primal lattice).
	}
	\label{fig:fig_nico_illustrated}
\end{figure}

The \textit{projection decoder} (see \fig{fig_nico_illustrated}) can then be described as follows.
\begin{enumerate}
\item For each pair of colors $\mathcal{K}\in \{RG, RB, GB\}$ we use the minimum weight perfect matching (MWPM) algorithm to find a subset of edges $E^\mathcal{K}\subseteq\face{1}{\mathcal{L}^\mathcal{K}}$ which connect pairs of highlighted vertices in $V^\mathcal{K}$ within the restricted lattice $\mathcal{L}^\mathcal{K}$.
    \item The combined edge set $E=E^{RG}+E^{RB}+E^{GB}$ separates the lattice $\mathcal{L}$ into two complementary regions $\Delta \subseteq \face{2}{\mathcal{L}}$ and $\Delta^c =\face{2}{\mathcal{L}} \setminus \Delta$ and we choose the correction $\hat{\epsilon}$ to be the smaller of the regions $\Delta$ and $\Delta^c$.
\end{enumerate}

Note that step 1 can be viewed as the problem of decoding the toric code defined on the lattice $\mathcal{L}^\mathcal{K}$, and thus one could use any toric code decoder to find a pairing $E^\mathcal{K}\subseteq\face{1}{\mathcal{L}^\mathcal{K}}$.
The MWPM algorithm we chose for this step is computationally efficient.
The boundary edges are permitted in $E^\mathcal{K}$, but their edge weight is set to zero for matching.

\begin{figure}[h]
	(a)\hspace*{-5mm}\includegraphics[width=.45\textwidth]{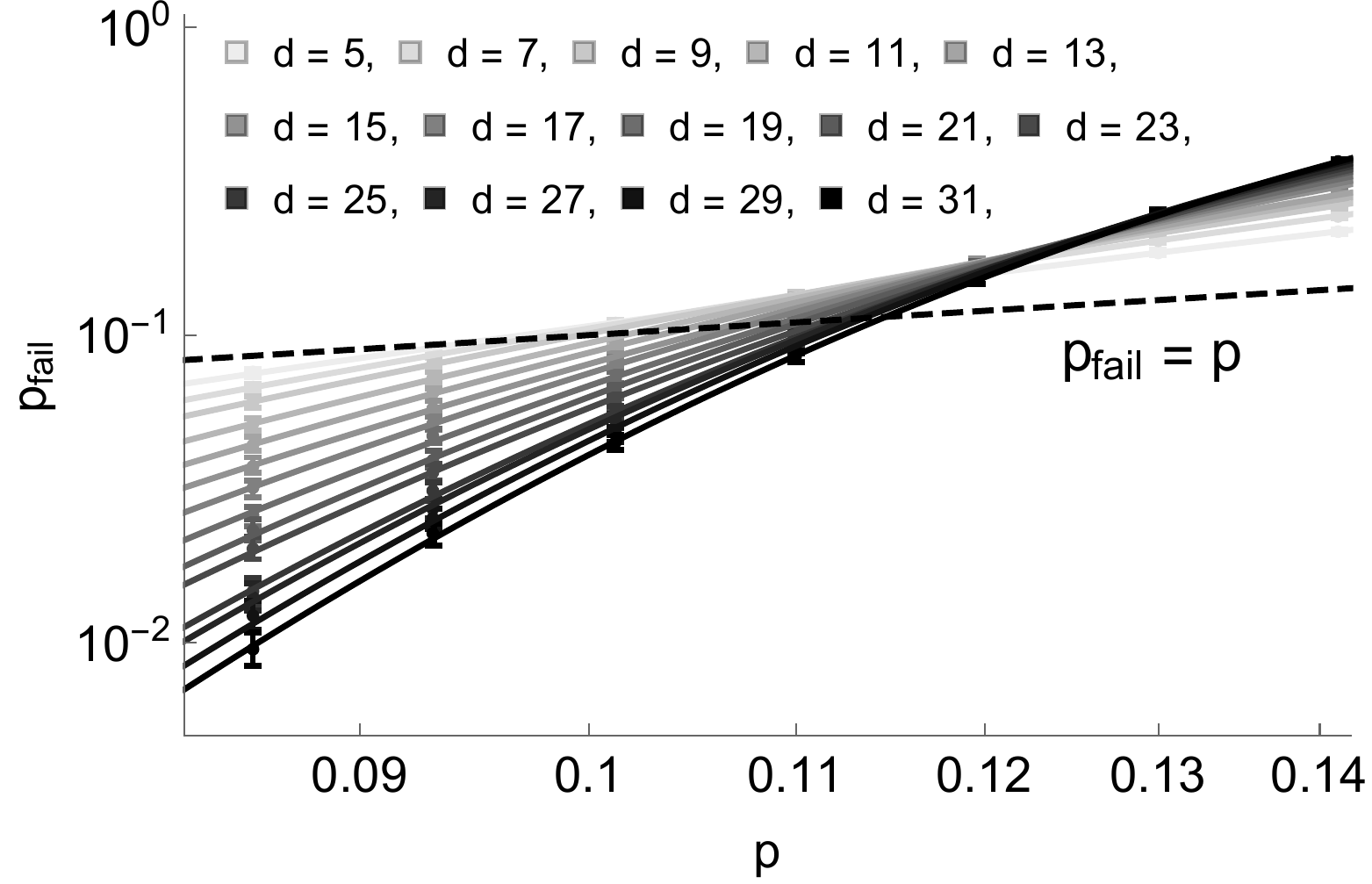}
	\quad\quad\quad
	(b)\hspace*{-5mm}\includegraphics[width=.45\textwidth]{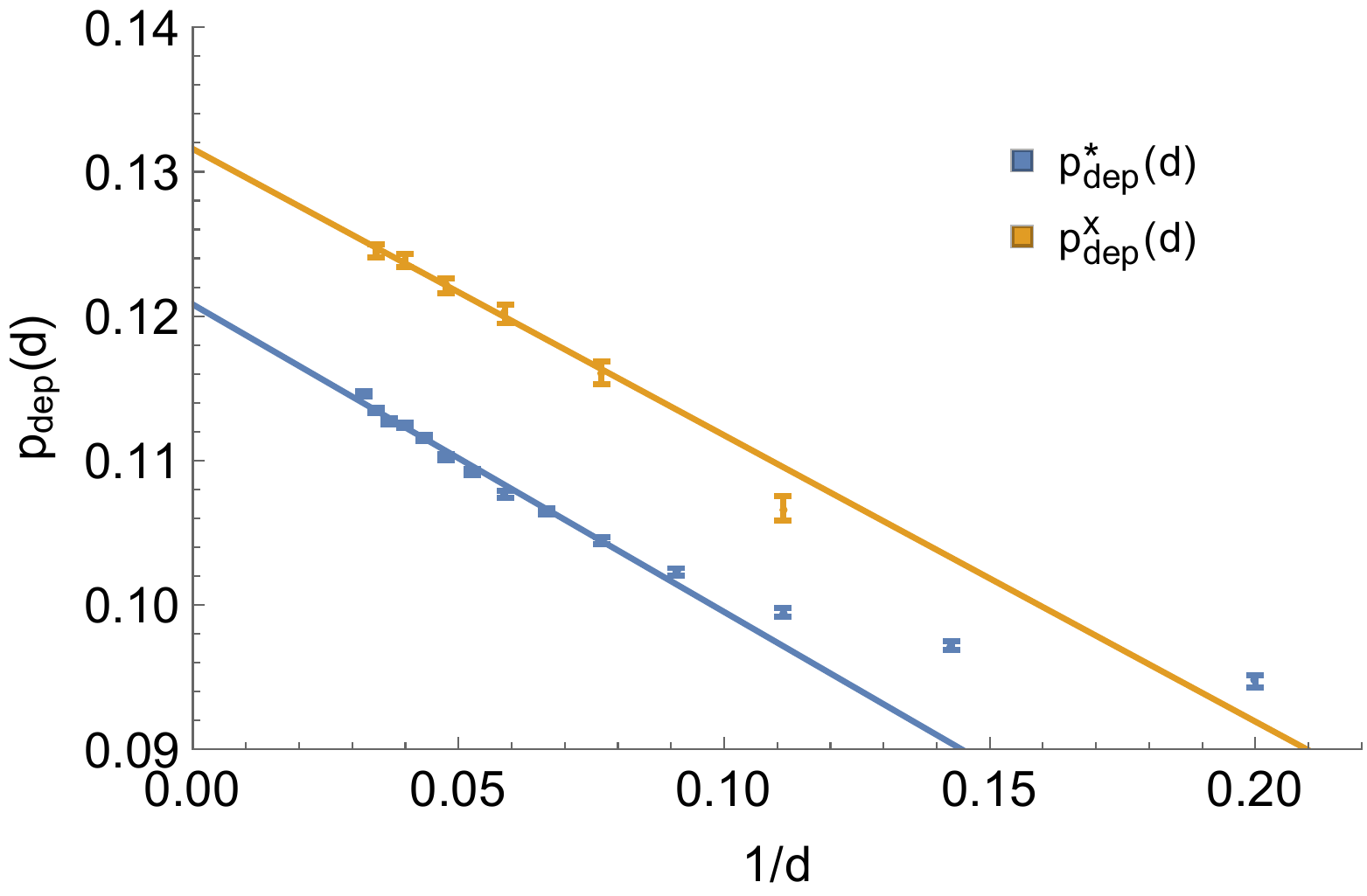}
	\caption{
	(a) Performance of the projection decoder for the 2D color code of distance $d$ under depolarizing noise and perfect measurements.
	(b) The pseudo-thresholds $p^*_{\text{dep}}(d)$ and the crossings $p^\times_{\text{dep}}(d)$ between pairs of curves corresponding to distances $d$ and $(d+1)/2$ for various $d$.
	From a linear extrapolation of the data for $d\geq 11$ we obtain intercepts of 13.16(4)\% for $p^\times_{\text{dep}}(d)$ and 12.08(4)\% for $p^*_{\text{dep}}(d)$.
	}
	\label{fig:PerfectMeasurementThreshold}
\end{figure}

In \fig{PerfectMeasurementThreshold}(a) we present the failure probability of this decoder under depolarizing noise and perfect measurements.
In \fig{PerfectMeasurementThreshold}(b) we consider two distance-dependent quantities that should both converge to the threshold as $1/d$ approaches zero. 
The first is the pseudo-threshold $p^*_{\text{dep}}(d)$, defined as a solution to $p_{\text{fail}}(p,d)=p$.
The pseudo-threshold is a good proxy to identify the regime in which the code of distance $d$ is useful. 
The second is the crossing $p^\times_{\text{dep}}(d)$ between pairs of failure curves of distances $d$ and $(d+1)/2$ for $d\equiv 1 \mod 4$.
From a linear extrapolation of the data we find intercepts $13.16(4)\%$ for $p^\times_{\text{dep}}(d)$ and $12.08(4)\%$ for $p^*_{\text{dep}}(d)$.

Note that the discrepancy in the intercept values suggests that the systems we consider are too small for a naive linear extrapolation to work reliably.
Assuming that both $p^*_{\text{dep}}(d)$ and $p^\times_{\text{dep}}(d)$ continue to change monotonically with $d$ we then take the maximum observed value of $p^\times_{\text{dep}}(d)$ at $d=29$ as a conservative estimate of the threshold under depolarizing noise, i.e., $p^*_{\text{dep}}=12.45(4)\%$ in accordance with previous threshold estimates in Refs.~\cite{Maskara2018,Chamberland2020}.

This analysis of the pseudo-thresholds and failure-curve crossings highlights two general points.
Firstly, the threshold may not be a good proxy to the finite-size performance.
For example, we see in \fig{PerfectMeasurementThreshold}(b) that our estimate of the threshold is considerably above the observed pseudo-thresholds.
Secondly, in order to find the threshold from the data for small systems one has to exert caution, as extrapolating different quantities (which nevertheless should recover the same threshold value in the limit of infinite $d$) can result in inconsistent estimates.

\subsection{Noisy-syndrome projection decoder with boundaries}
\label{sec:ftdecoder}

\begin{figure}[h]
    (a)\includegraphics[width=.33\textwidth]{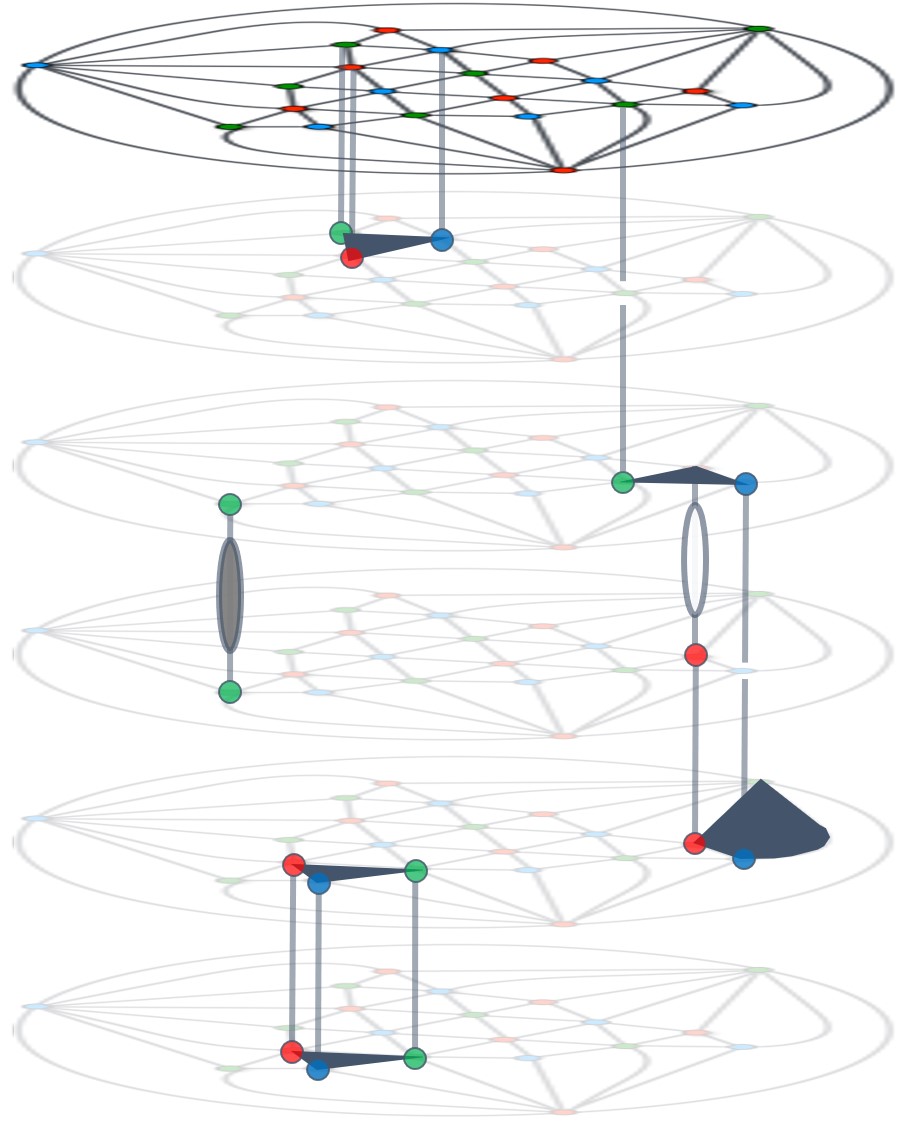}
    \quad \quad \quad \quad
	(b)\includegraphics[width=.33\textwidth]{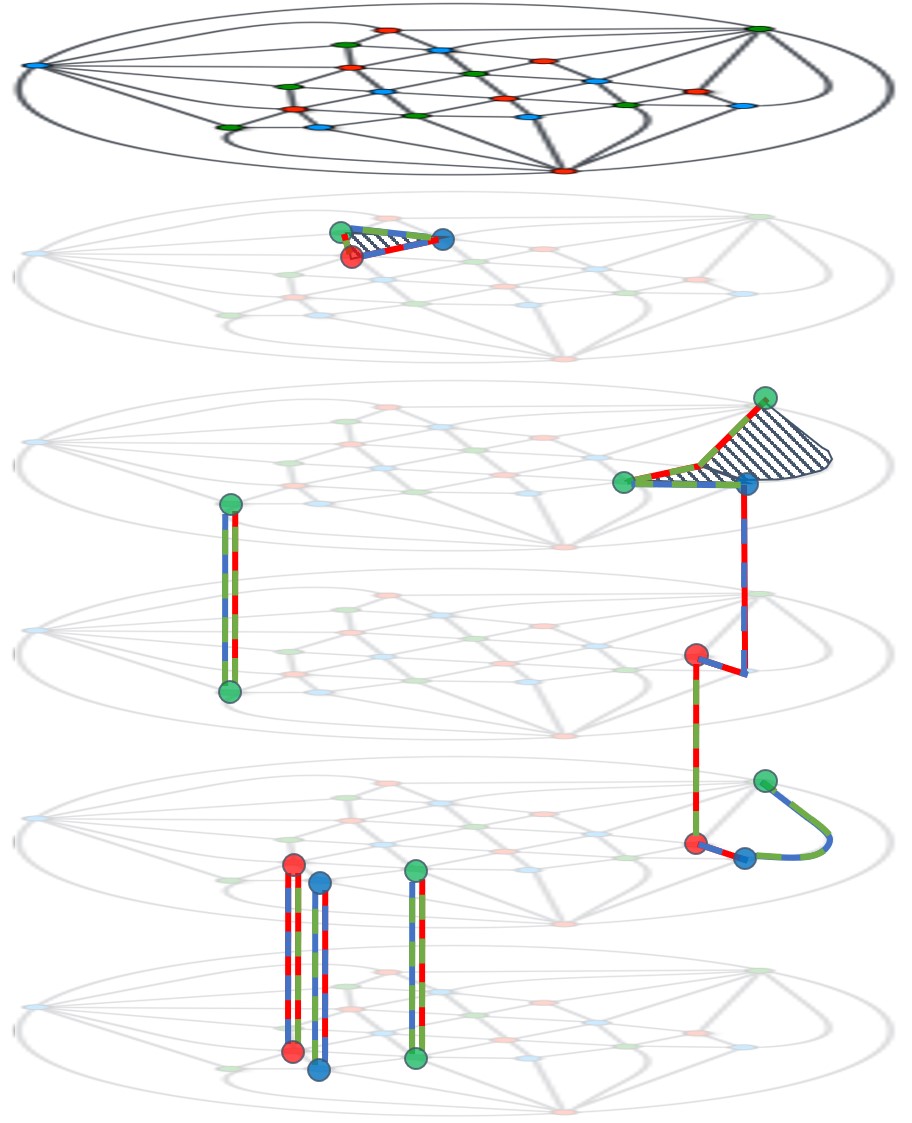}
	\caption{
	The noisy-syndrome projection decoder on the (2+1)D color code lattice $\Lambda$. 
	(a) During each QEC cycle, errors can appear on data qubits (shaded triangles), followed by stabilizer measurements which when incorrect are drawn as ellipses.
	The observed syndrome $\sigma$ (dark temporal edges and filled ellipses)
	allows us to find the set of syndrome flips $V$ (highlighted vertices).
	(b) For $\mathcal{K} \in \{RG,RB,GB\}$ we find the pairing $E^\mathcal{K}$ (colored edges) of the syndrome flips $V^\mathcal{K}$ within the restricted lattice $\Lambda$.
	Then, for every ``flattened'' connected component of $E^{RG}+E^{RB}+E^{GB}$ we find a correction $\epsilon(t)$ (hatched regions).
	}
	\label{fig:fig_2+1D}
\end{figure}

In this subsection we further generalize the projection decoder to handle phenomenological noise.
To reliably extract the syndrome and perform error correction one can repeat the stabilizer measurements multiple times.
The input of the decoder then consists of stabilizer measurement outcomes (possibly incorrect) at QEC cycles labeled by integers and can be visualized as a (2+1)-dimensional lattice $\Lambda = \mathcal{L} \times [n_\text{cyc}]$, where $[n_\text{cyc}] = \{0,1,\ldots, n_\text{cyc}\}$ and the extra dimension represents time; see \fig{fig_2+1D}.
We use a shorthand notation $\Lambda_{[t_1,t_2]}$ to denote the part of $\Lambda$ within QEC cycles $t_1$ and $t_2$.
By an QEC cycle, we simply mean a full cycle of stabilizer measurements.
Temporal edges, which vertically connect corresponding vertices in two copies of $\mathcal{L}$ at QEC cycles $t$ and $t+1$, correspond to stabilizer measurements at QEC cycle $t$.

We use the same concepts and nomenclature for the lattice $\Lambda$ as for $\mathcal{L}$ in \sec{color-code-basics}.
For instance, we say a vertex $(v,t)$ of $\Lambda$ is a boundary vertex if and only if $v$ is a boundary vertex of $\mathcal{L}$; otherwise, it is an interior vertex. 
An edge of $\Lambda$ is a boundary edge iff it connects two boundary vertices.
The sets of interior vertices and edges of $\Lambda$ are denoted $\facex{0}{\Lambda}$ and $\facex{1}{\Lambda}$, respectively.
Furthermore, we denote by $\Lambda^\mathcal{K}$ the restricted lattice of $\Lambda$ consisting of vertices of color in $\mathcal{K}$, and the edges connecting them.

The input of the \textit{noisy-syndrome projection decoder} is an observed history of the syndrome $\sigma$ consisting of the subset of temporal edges corresponding to $-1$ stabilizer measurement outcomes. 
We define the set of syndrome flips $V \subseteq \facex{0}{\Lambda}$ to be the set of all the vertices, which are incident to an odd number of edges in $\sigma$.
The decoder is then implemented using the following steps.
\begin{enumerate}
	\item For each $t \in [n_{\text{cyc}}]$ initialize $F(t)=\emptyset$. 
	\item For every $\mathcal{K} \in \{RG,RB,GB\}$ use the MWPM algorithm to find the pairing $E^{\mathcal{K}}$ of syndrome flips $V^{\mathcal{K}}$ within the restricted lattice $\Lambda^{\mathcal{K}}$.
	\item Combine the obtained pairings, $E = E^{RG} + E^{RB} + E^{GB}$, and decompose $E$ as a disjoint sum of maximal connected components, $E = \sum_i E_i$.
	\item  For every connected component $E_i$:
	\begin{enumerate}
		\item find the minimal window of QEC cycles $\tau_i = \left[t^{(1)}_i, t^{(2)}_i\right]$ enclosing $E_i$, i.e. $E_i \subset \face 1 {\Lambda_{\tau_i}}$,
		\item project $E_i$ onto $\mathcal{L}$ in order to obtain the ``flattened pairing"
		$F_i= \pi (E_i)$, where\linebreak[4] $\pi:\face{1}{\Lambda} \rightarrow \face{1}{\mathcal{L}}$ removes temporal edges and adds horizontal ones modulo two,
		\item add the edges of $F_i$ modulo two to the edge set $F\left(t^{(2)}_i\right)$ for QEC cycle $t^{(2)}_i$.
	\end{enumerate}
	\item For each QEC cycle $t$, find a correction $\epsilon(t) \subseteq\face{2}{\mathcal{L}}$ as the minimal region enclosed by $F(t)$.
\end{enumerate}
We make some additional technical remarks about the noisy-syndrome projection decoder.
In step 2, the boundary edges of the restricted lattice $\Lambda^{\mathcal{K}}$ are assigned zero weight when used for pairing.
A boundary vertex of a color in $\mathcal{K}$ (it does not matter which) is added to $V^{\mathcal{K}}$ whenever $|V^{\mathcal{K}}|$ is odd.
In step 3, we remove weight-zero edges when establishing connected components of $E$.

\begin{figure}[h]
	(a)\hspace*{-5mm}\includegraphics[width=.45\textwidth]{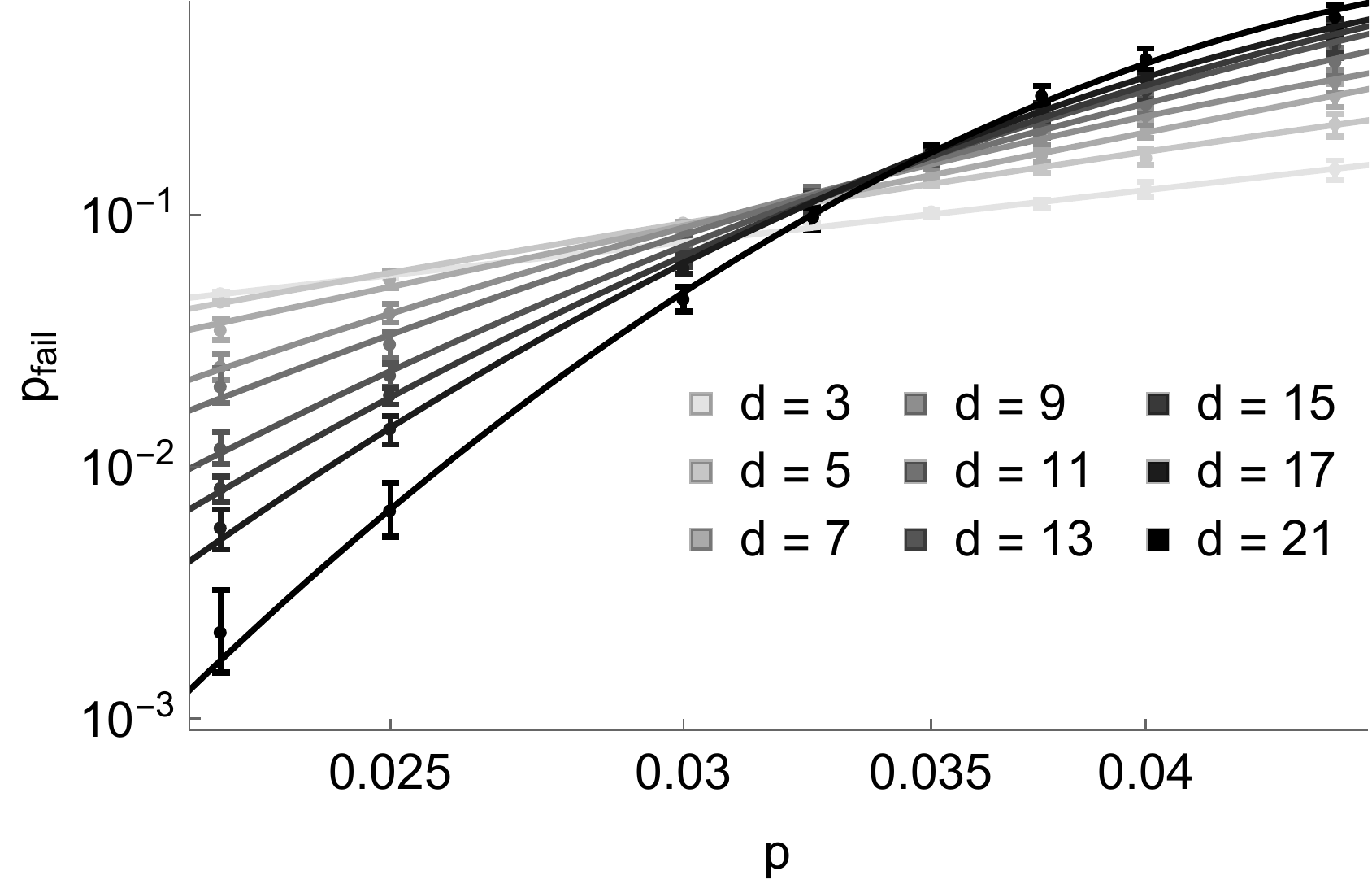}
	\quad\quad\quad
	(b)\hspace*{-5mm}\includegraphics[width=.45\textwidth]{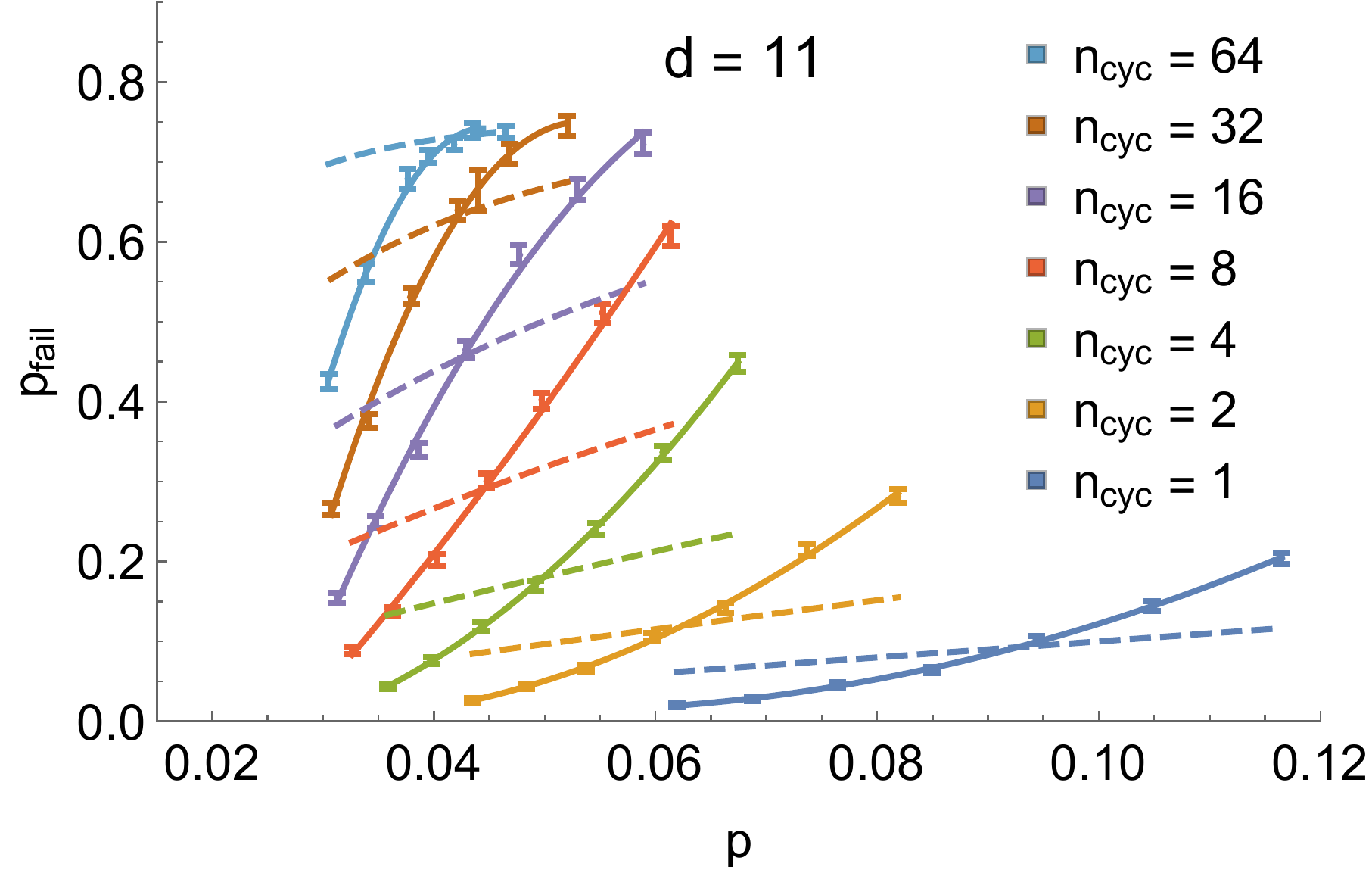}\\
	\vspace*{5mm}
	(c)\hspace*{-5mm}\includegraphics[width=.45\textwidth]{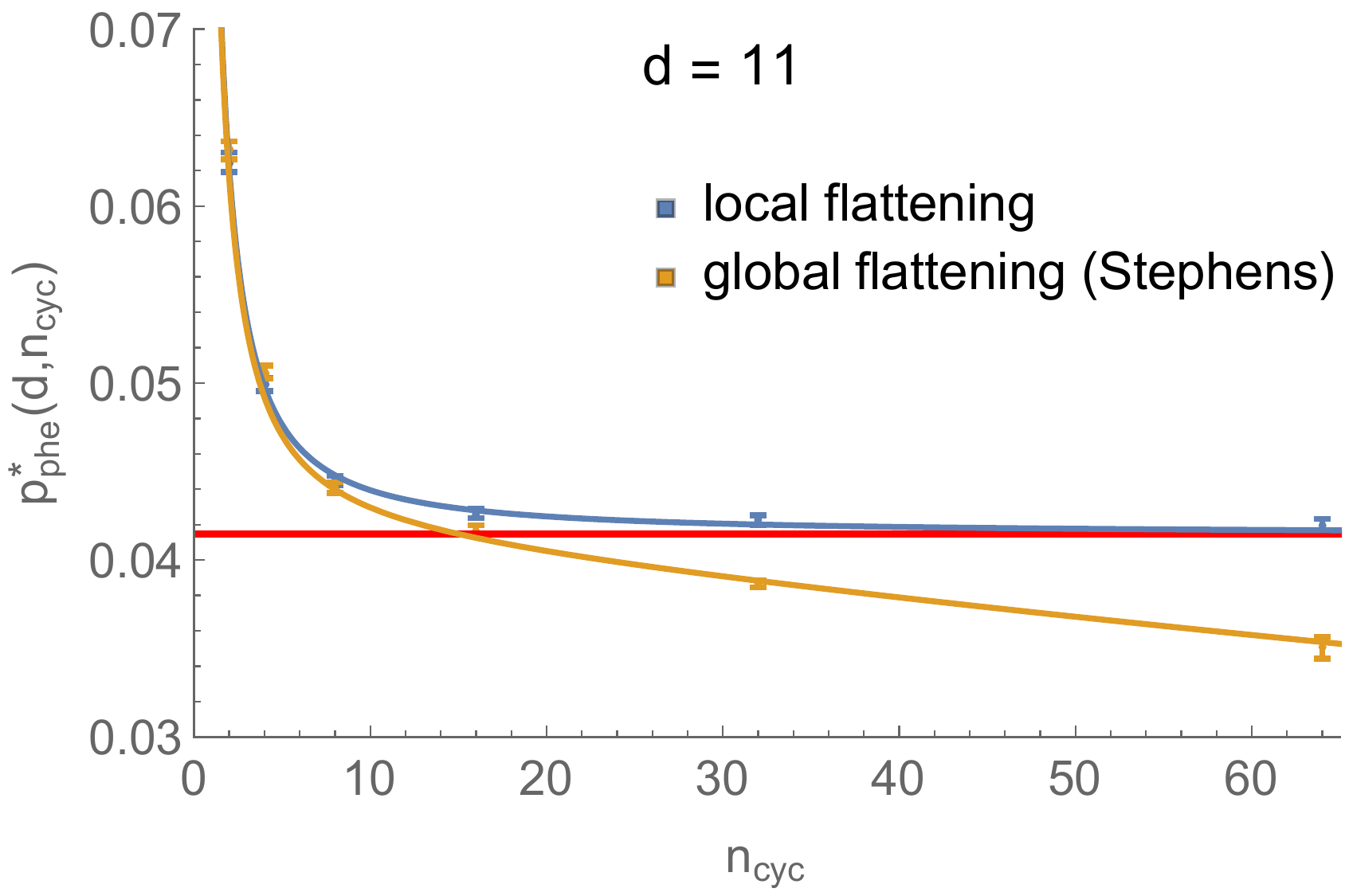}
	\quad\quad\quad
	(d)\hspace*{-5mm}\includegraphics[width=.45\textwidth]{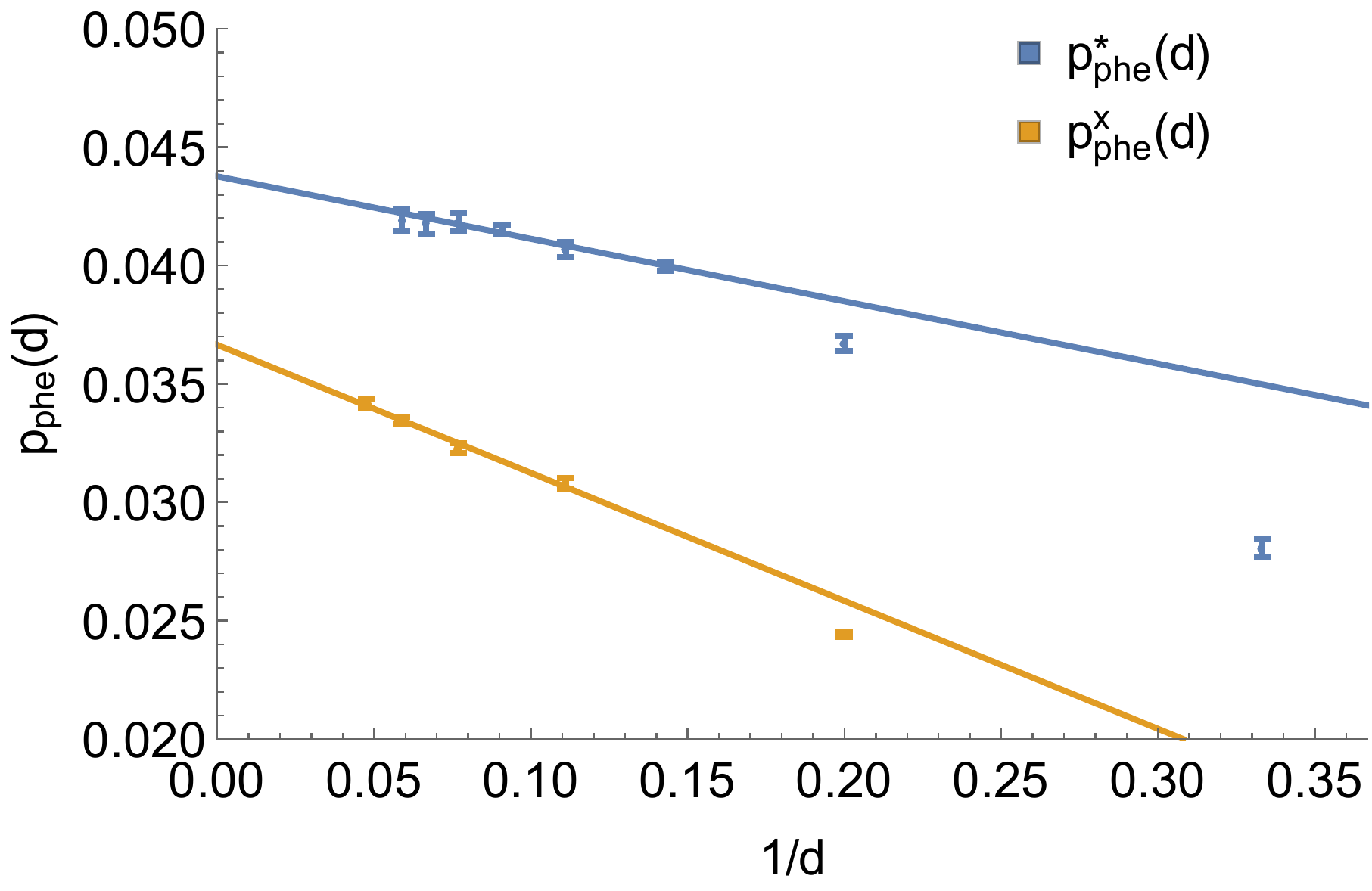}
	\caption{
	(a) The failure probability $p_\text{fail}(d,n_\text{cyc})$ of the noisy-syndrome projection decoder for $n_{\text{cyc}} = d$ time units of phenomenological noise and various distances $d$.
	(b) 
	We estimate the time-dependent pseudo-threshold $p^*_{\text{phe}}(d,n_{\text{cyc}})$ from the intersection of
	$p_\text{fail}(d,n_\text{cyc})$ with the error probability for the unencoded qubit (dashed) after $n_{\text{cyc}}$ time units for various $n_\text{cyc}$ and $d=11$.
	See \app{threshold-data} for the analysis of other distances.
	(c) We estimate the long-time pseudo-threshold $p^*_{\text{phe}}(d) = 4.15(2)\%$ for $d=11$ (red)
	by fitting $p^*_{\text{phe}}(d,n_{\text{cyc}})$ with the numerical ansatz in~\eq{time-dep-pseudothreshold}.
	Note that the Stehpens' adaptation of the decoder (yellow), which uses a global flattening, fails to stabilize.
	(d) 
	The long-time pseudo-thresholds $p^*_{\text{phe}}(d)$ (blue) and the crossings $p^\times_{\text{phe}}(d)$ (yellow) between pairs of curves corresponding to distances $d$ and $(d+1)/2$ for various $d$.
	From a linear extrapolation of the data for $d\geq 7$ we obtain intercepts of 4.38(3)\% for $p^*_{\text{phe}}(d)$ and 3.67(3)\% for $p^\times_{\text{phe}}(d)$.
	}
	\label{fig:BitflipMeasurementThresholdAndStephens}
\end{figure}

To analyze error correction thresholds in a faulty-measurement setting, it is common to study the somewhat contrived scenario of an initially perfect code state undergoing $d$ QEC cycles, followed by a single cycle of perfect measurements.
The justification for this is underpinned by the fact that in the fault-tolerant setting, the logical clock cycle (the time required to implement logical gates) requires approximately $d$ QEC cycles with lattice surgery or braiding. 
Moreover, one would expect the effects of the artificially perfect preparation and final measurement cycle to be negligible over $d$ cycles when $d$ is sufficiently large, making this scenario appropriate for estimating the threshold value (but not for estimating the actual performance of finite sizes).
In \fig{BitflipMeasurementThresholdAndStephens}(a) we find the failure probability after $d$ time units of phenomenological noise.
In \fig{BitflipMeasurementThresholdAndStephens}(d) we show crossings $p^\times_{\text{phe}}(d)$ between pairs of these failure curves corresponding to distances $d$ and $(d+1)/2$, which should converge to the threshold $p^*_{\text{phe}}$ as $1/d$ approaches to zero.

The notion of a pseudo-threshold must be revisited in the setting of faulty measurements and we cannot extract a meaningful pseudo-threshold directly from these curves as we did for the perfect measurement case in \fig{PerfectMeasurementThreshold}.
Consider the scenario in which we assume a perfect initial code state and a perfect final measurement cycle, but consider the performance over a varying number $n_{\text{cyc}}$ of noisy QEC cycles; see \fig{BitflipMeasurementThresholdAndStephens}(b).
For each $d$ and $n_{\text{cyc}}$, we define the time-dependent pseudo-threshold $p^*_{\text{phe}}(d,n_{\text{cyc}})$ as the error rate at which the encoded failure probability after $n_{\text{cyc}}$ QEC cycles matches $p_{\text{phy}}(n_{\text{cyc}})$ the unencoded failure probability for $n_{\text{cyc}}$ time units, as defined in \eq{pPhysical}.
As $n_{\text{cyc}}$ increases, $p^*_{\text{phe}}(d,n_{\text{cyc}})$ is expected to decrease due to the buildup of residual error.
However, for sufficiently large $n_\text{cyc}$ the time-dependent pseudo-threshold $p^*_{\text{phe}}(d,n_{\text{cyc}})$ should eventually stabilize to the long-time pseudo-threshold $p^*_{\text{phe}}(d)$, as can be seen in \fig{BitflipMeasurementThresholdAndStephens}(c).
The following ansatz
\begin{eqnarray}
\label{eq:time-dep-pseudothreshold}
p^*(d,n_{\text{cyc}}) = p^*(d)\left( 1 - \left[ 1 - \frac{p^*(d,1)}{p^*(d)}  \right] n_{\text{cyc}} ^{-\gamma}\right).
\end{eqnarray}
fits the data well, and allows us to extract an estimate of the long-time pseudo-threshold.
We remark that our approach of finding long-time pseudo-thresholds is similar in spirit, but not exactly the same as the one used to find the ``sustainable threshold'' \cite{Brown2015,Terhal2015}.
Also, the ansatz in \eq{time-dep-pseudothreshold} was used in \cite{kubica2018} to analyze thresholds of cellular-automata decoders for topological codes.

The long-time pseudo-thresholds $p^*_{\text{phe}}(d)$, as well as the failure curve crossings $p^\times_{\text{phe}}(d)$ should converge in the limit of infinite $d$ to the threshold under phenomenological noise.
From a linear extrapolation of the data we obtain intercepts of 4.38(3)\% for $p^*_{\text{phe}}(d)$ and 3.67(3)\% for $p^\times_{\text{phe}}(d)$; see \fig{BitflipMeasurementThresholdAndStephens}(d).
Note that $p^*_{\text{phe}}(d)$ appears to converge more quickly than $p^\times_{\text{phe}}(d)$.
Assuming that both quantities continue to monotonically change with $d$ we take the maximum observed value of $p^*_{\text{phe}}(d)$ at $d=17$ as a conservative estimate of the threshold under phenomenological noise, i.e., $p^*_{\text{phe}}=4.19(4)\%$.

We remark that the modification of the projection decoder that we have presented here to handle noisy syndrome measurements differs from that discussed by Stephens~\cite{stephens2014} in an important detail.
Namely, our ``flattening'' of pairings is local, since it occurs separately on each connected component after the matching in the (2+1)-dimensional graphs. 
In contrast, Stephens' flattening is global---all pairings are flattened together and a global correction is produced.
For any finite noise strength and for a sufficiently large number of cycles (which does not need to grow with code distance), the success probability of this global flattening will be vanishingly small.
Therefore the adaption proposed by Stephens does not have a finite error-correction threshold, and is not fault tolerant.
This did not contribute noticeably to the numerical results presented in \cite{stephens2014} since the total cycle number was fixed to $d$, for data in ranges satisfying $d < 1/p$.
However, by looking at the performance over longer times of Stephens' adaption of the projection decoder, we verify that the time-dependent pseudo-threshold for this decoder fails to stabilize for large $n_\text{cyc}$; see \fig{BitflipMeasurementThresholdAndStephens}(c).

\subsection{Optimizing stabilizer extraction and circuit noise analysis}
\label{sec:circuitnoise}

There is significant freedom in precisely which circuits are used to extract the syndrome for error correction.
We will assume that there is a separate ancilla qubit per stabilizer generator, such that there are two ancillas per face of the lattice $\mathcal{L}_{\text{2D}}$, and will not worry about the precise connectivity details, requiring only that coupled qubits are nearby.
The total number of qubits required for our implementation of the distance-$d$ 2D color code is therefore
\begin{equation}
\label{eq:Nqubits2DCC}
N_{2\text{D}}(d) = (3 d^2-1)/2.
\end{equation}
Each circuit starts by preparing an ancilla qubit in either $\ket{+}$ or $\ket{0}$ state, followed by applying CNOT gates between the ancilla qubit and all the qubits of the stabilizer generator, and finishes with measuring the ancilla qubit in the corresponding $X$- or $Z$-basis.
During each time unit new errors can appear in the system and thus it can be beneficial to parallelize as much as possible the circuits used for stabilizer measurement.
When circuits for measuring different stabilizers are interleaved, not all schedules of CNOT gates will work. 
The following conditions \cite{landahl2011} must be satisfied.
\begin{itemize}
	\item At each time unit at most one operation can be applied to any given qubit.
	\item The measurement circuit preserves the group generated by the elements of the stabilizer group and Pauli $X$ or $Z$ operators stabilizing the ancilla qubits.\footnote{We say that an ancilla prepared in $\ket{+}$ or $\ket{0}$ state is stabilized by a Pauli $X$ or $Z$ operator, respectively.}
	
\end{itemize}

\begin{figure}[h]
	(a)\includegraphics[width=.33\textwidth]{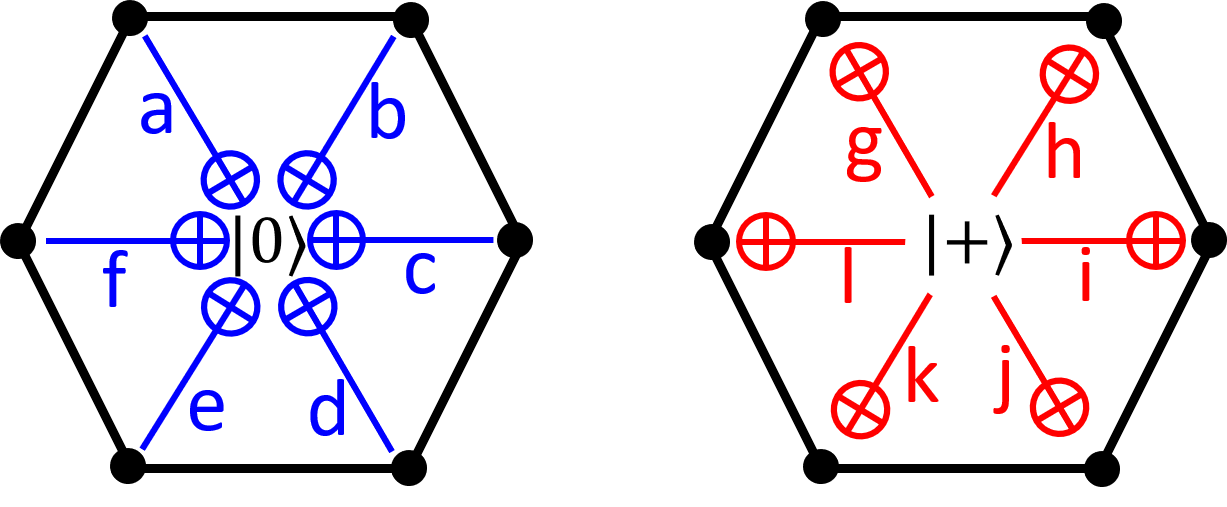}
	\quad\quad\quad
    (b)\includegraphics[width=.4\textwidth]{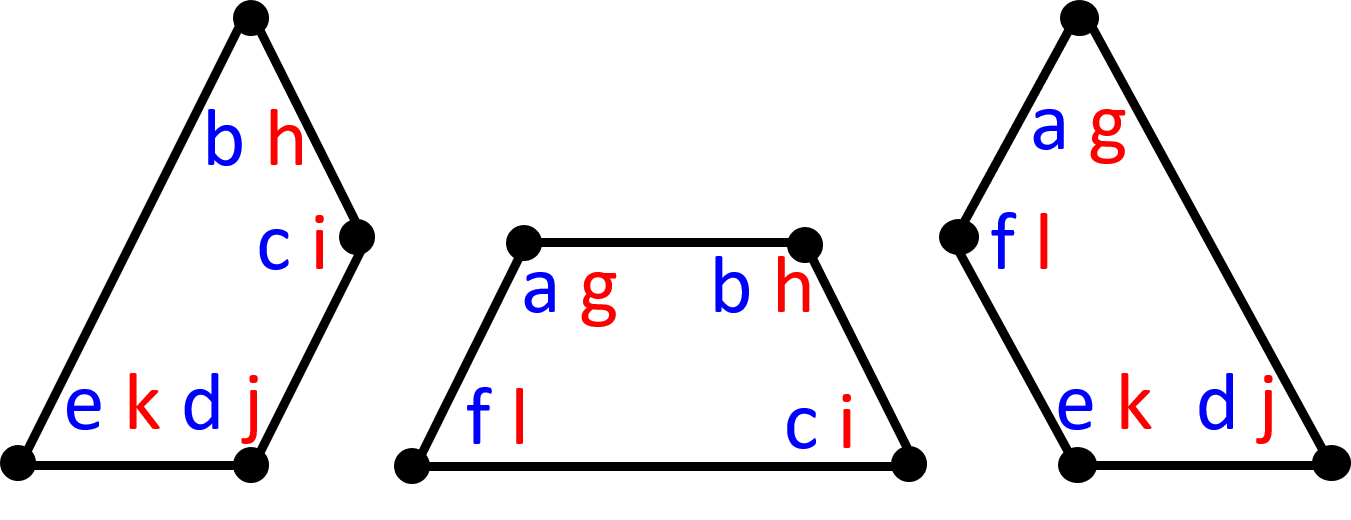}
	\caption{
	(a) The CNOT schedule $\{ a,b,c,d, e, f; g,h,i,j,k,l \}$ is a list specifying the time units when each CNOT gate is applied.
	At the end, the ancillas which are initially prepared in $\ket{0}$ and $\ket{+}$ are measured to extract the $Z$ (blue) and $X$ (red) stabilizer measurement outcomes, respectively.
	(b) The CNOT schedules for stabilizers on faces along the boundary can be viewed as restrictions of the CNOT schedule in (a).
	}
	\label{fig:color-code-stabilizer-sequence}
\end{figure}

In our optimization we assume that it suffices to specify the CNOT ordering for a single $X$ and $Z$ stabilizer generator in the bulk, as the code is translation invariant.
Moreover, the CNOT schedule for stabilizer generators along the boundary of the lattice are specified by restricting the schedule for those in the bulk; see \fig{color-code-stabilizer-sequence}.

The CNOT schedule includes twelve CNOT gates to extract both $Z$ and $X$ stabilizers.
Each CNOT gate is applied at some time unit, and thus the CNOT schedule is specified by a list of twelve non-negative integers
$\mathcal{A} = \{a,b,c,d,e,f;g,h,i,j,k,l\}$, possibly with repetitions.
We are interested in CNOT schedules which satisfy the following condition:
\begin{itemize}
    \item[1.] \textit{As short as possible.---}To ensure there is no time unit in which both ancillas in a face are idle $\mathcal{A}=\{a,b,c,d,e,f,g,h,i,j,k,l\}$ contains all numbers from $1$ to $\max\mathcal{A}$.
\end{itemize}
Note that this implies that $\max\mathcal{A}\leq 12$, and thus there are at most $12^{12}$ CNOT schedules.
However, most of them are invalid as they do not satisfy one of the following necessary conditions \cite{landahl2011}.
\begin{itemize}
    \item[2.] \textit{One operation per qubit at a time.---}The integers $a,b,c,d,e,f$ must be all distinct, as well as $g,h,i,j,k,l$, and $d,j,f,l,b,h$, and $e,k,a,g,c,i$.
    \item[3.] \textit{Correct syndrome extraction.---}To ensure the ancilla measurement after each CNOT sequence extracts the stabilizer measurement outcome, the following inequalities must hold.
    \begin{itemize}
        \item For the stabilizers in the bulk: $(a - g) (b - h) (c - i) (d - j) (e - k) (f - l) > 0$,\linebreak[4] $(e - g) (d - h) > 0$,  $(k - a) (j - b) > 0$, $(f - h) (e - i) > 0$, $(l - b) (k - c) > 0$, $(d - l) (c - g) > 0$, and $(j - f) (i - a) > 0$.
         \item For the stabilizers along the boundary: $(a - g)(b - h)(c - i)(f - l) > 0$,\linebreak[4] $(b - h)(c - i)(d - j)(e - k) > 0$ and $(a - g) (d - j) (e - k) (f - l) > 0$.
    \end{itemize}
\end{itemize}

To illustrate how we obtain the inequalities in the last condition, let us 
analyze how the Pauli $X$ operator stabilizing the $X$ syndrome extraction ancilla on a given face is spread by the CNOT schedule.
It propagates to all the data qubits on that face.
From each data qubit it may further propagate to the $Z$ syndrome extraction ancilla on the same face, and this is determined by the relative order of the CNOT gates used for the $X$ and $Z$ syndrome extraction.
We need to ensure that at the end of the CNOT schedule the Pauli $X$ operator on the $X$ syndrome extraction ancilla has not propagated to the $Z$ syndrome extraction ancilla, which is equivalent to the inequality $(a - g) (b - h) (c - i) (d - j) (e - k) (f - l) > 0$ being true.
The other inequalities are derived similarly.

We remove an ordering from the list of valid orderings if it is equivalent to another ordering in the list up to a symmetry of the lattice $\mathcal{L}_{\text{2D}}$.
No schedules with six time units satisfy all these conditions.
However, we find that there are $234$, $4854$ and $39160$ valid orderings for $7$, $8$ and $9$ time units, respectively.

\begin{figure}[h]
	(a)\hspace*{-5mm}\includegraphics[width=.45\textwidth]{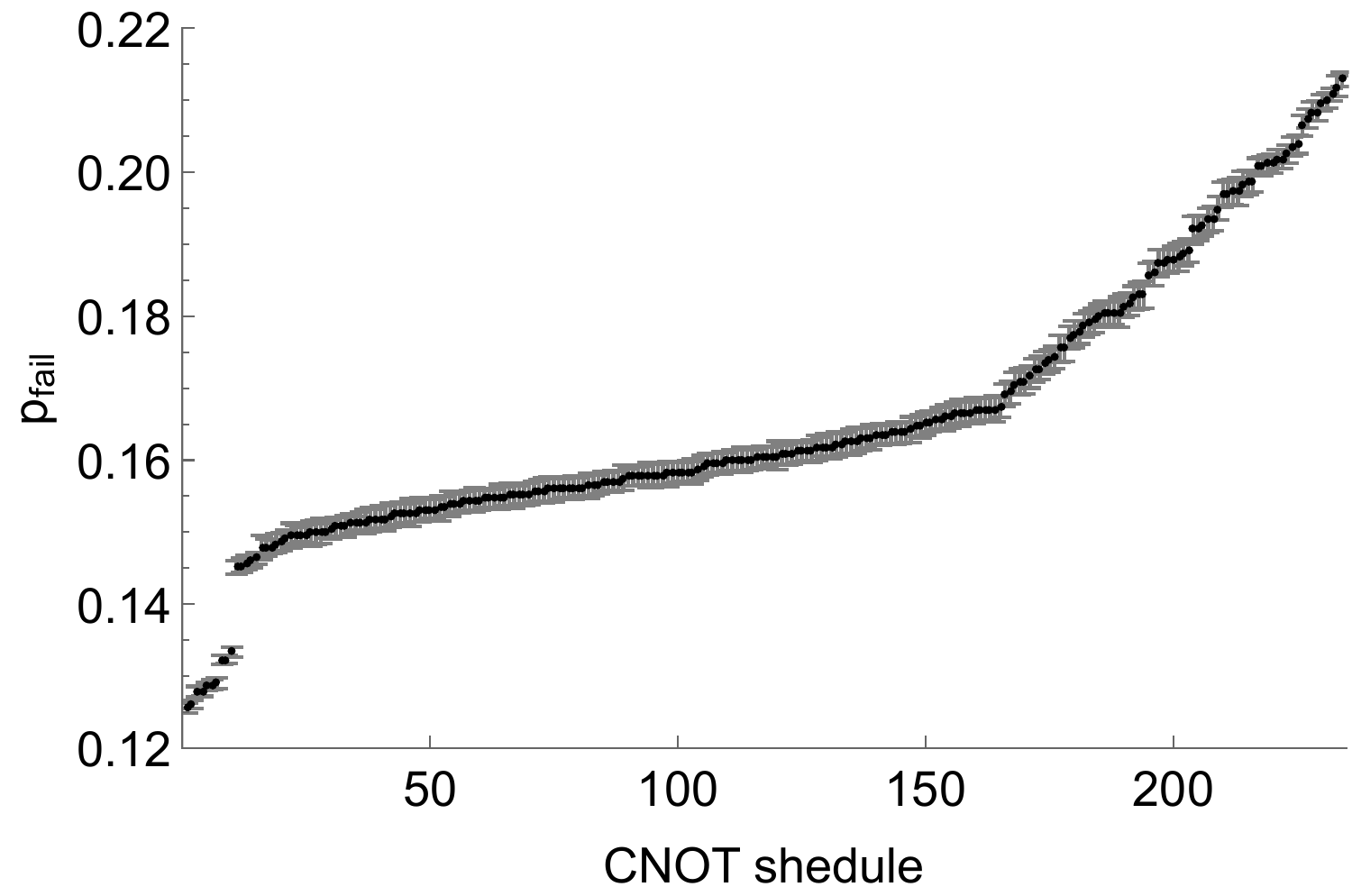}
	\quad\quad\quad
	(b)\hspace*{-5mm}\includegraphics[width=.45\textwidth]{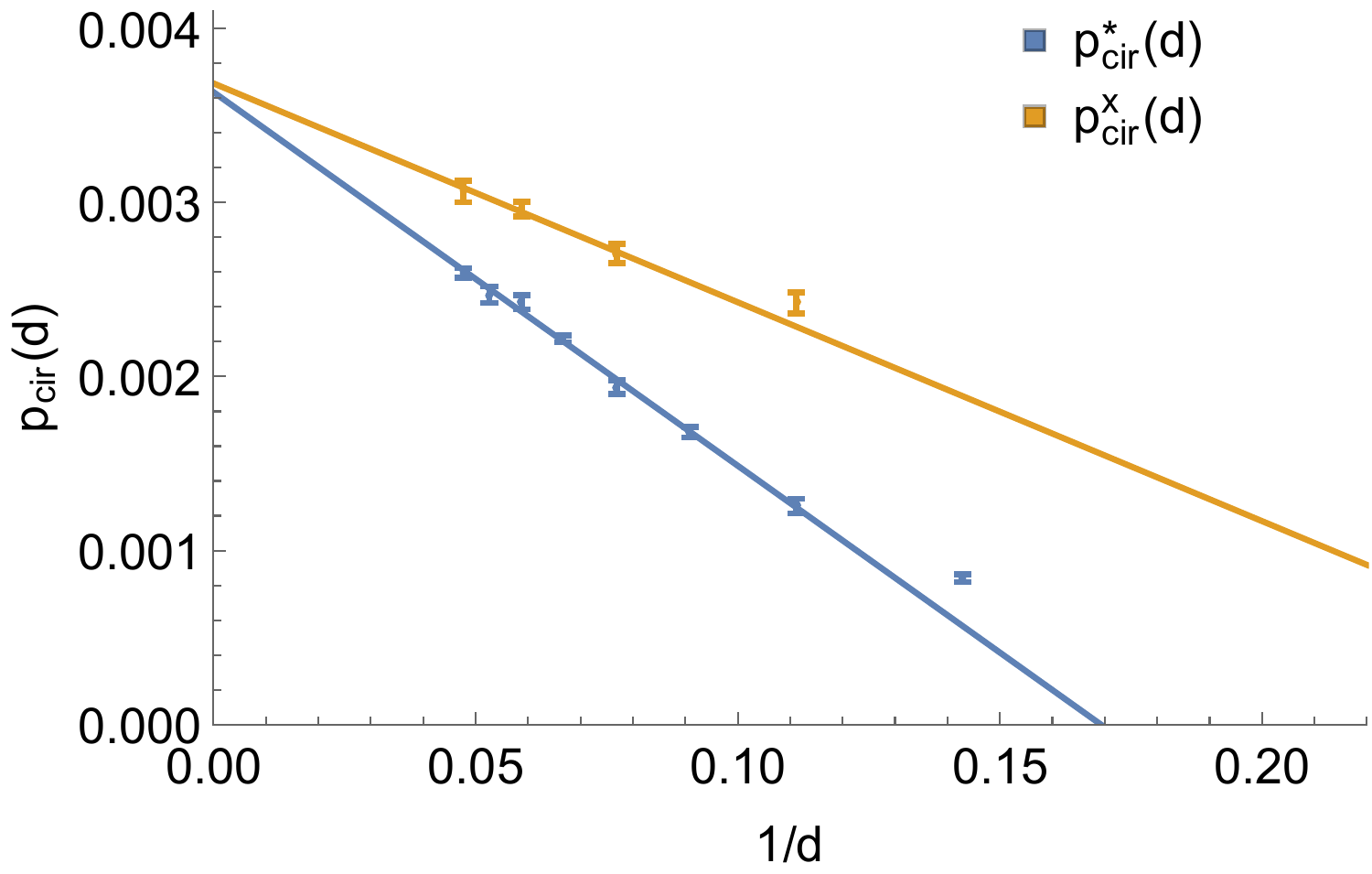}
	\caption{
	(a) The logical failure probability over $n_{\text{cyc}}=7$ cycles for distance $d=7$ and depolarizing circuit noise of strength $p=0.0035$ for each of the 234 CNOT schedules, sorted by their failure rate.
	(b) The long-time pseudo-thresholds $p^*_{\text{cir}}(d)$ and crossings $p^\times_{\text{cir}}(d)$ for the best-performing CNOT schedule 
    $\{4, 1, 2, 3, 6, 5 ; 3, 2, 5, 6, 7, 4\}$ under circuit noise; see \app{threshold-data} for further details.
	From a linear extrapolation of the data for $d\geq 11$ we obtain intercepts of 0.366(6)\% for $p^*_{\text{cir}}(d)$ and 0.37(1)\% for $p^\times_{\text{cir}}(d)$.
	}
	\label{fig:CircuitNoiseThreshold}
\end{figure}

To select a good CNOT schedule among the large number of valid ones, we focus on the $234$ shortest schedules, because fewer time units in an error correction cycle tends to translate into fewer possible faults and better performance.
We tested each using a $d=7$ code for $d$ QEC cycles with circuit noise of strength $p=0.0035$ and estimated the failure probability by sampling.
This value of $p$ was chosen because preliminary studies indicated it was close to the threshold, and distance $d=7$ was chosen as a compromise between reducing the simulation run time and limiting the impact of boundary effects.
The limited sampling resources were focused on identifying the best CNOT schedules; see \fig{CircuitNoiseThreshold}(a). 
We found that up to sampling error, 
$\{4, 1, 2, 3, 6, 5 ; 3, 2, 5, 6, 7, 4\}$ was the best-performing schedule, with a logical failure probability of $12.8(1)\%$.
For comparison, the worst-performing length-7 schedule was 
$\{4, 1, 2, 7, 6, 3 ; 1, 6, 7, 4, 5, 2\}$ and resulted in substantially worse logical failure probability of $21.1(1)\%$.
Note that the QEC cycle for the best schedule requires only 8 time units to implement by preparing and measuring the $X$ ancilla at time units 0 and 7 respectively, and preparing and measuring the $Z$ ancilla at time units 1 and 8 respectively.

Now we focus on the best-performing CNOT schedule under circuit noise and find the long-time pseudo-threshold for a range of distances in \fig{CircuitNoiseThreshold}(b) using the same approach as in \sec{ftdecoder}.
Unlike in the cases for depolarizing and phenomenological noise, the data for circuit noise appears to be in the regime where both $p^*_{\text{cir}}(d)$ and $p^\times_{\text{cir}}(d)$ can be fitted with a linear fit and their intercepts agree to within error.
We take their shared intercept as an estimate of the threshold under circuit noise $p^*_{\text{cir}}(d)=0.37(1)\%$.

To estimate the impact of this kind of circuit optimization, we compare the best and worst performing length-7 CNOT schedules and found that the long-time pseudo-threshold differs by a factor of almost two for distance $d=13$; see \app{threshold-data} for more details.

\subsection{Modelling noise in logical operations}
\label{sec:2DCCOptimizedSetup}

Now we describe an effective noise model for logical operations in the optimized 2D color code under circuit noise, which we will later use to estimate the performance of state distillation circuits.
For circuit noise strength $p$ and code distance $d$, the effective noise model is specified in terms of the overall failure probability $\overline{p}_\text{oper} (p,d)$ of each logical operation $\text{oper} = \text{prep}, \text{idle}, \text{CNOT}$, which we estimate numerically and record in \tab{logical-operations} in the form of the ansatz in \eq{ansatz_moregeneral}.

In our simulations each operation is followed by a full decoding. 
This results in the application of a logical Pauli operator, which if nontrivial is interpreted as a failure.
Our effective noise model is designed to overestimate the probability of each nontrivial logical Pauli, and assumes
each operation fails independently of others.
Specifically,
\begin{itemize}

    \item 
    for state preparation noise is modelled by preparing an orthogonal logical state to that intended with probability $\overline{p}_{\text{prep}}(p,d)$, 
    
    \item 
    for an idle operation, noise is modelled by applying $\overline{X}$ or $\overline{Z}$ each with probability $\overline{p}_{\text{idle}}(p,d)/2$ or $\overline{Y}$ with probability $\overline{p}_{\text{idle}}(p,d)/20$,
    
    \item 
    single-qubit Clifford gates are done in software by Pauli-frame updates so are noise-free,
    
    \item 
    for SWAP, noise is modelled by applying to each of the two qubits $\overline{X}$ or $\overline{Z}$ each with probability $\overline{p}_{\text{idle}}(p,d)/2$, or $\overline{Y}$ with probability $\overline{p}_{\text{idle}}(p,d)/20$,
    
    \item 
    for CNOT noise is modelled by applying $\overline{IX}$ or $\overline{ZI}$ each with probability $\overline{p}_{\text{CNOT}}(p,d)/2$, or $\overline{XI}$ or $\overline{IZ}$ each with probability $\overline{p}_{\text{CNOT}}(p,d)/4$, or other nontrivial Pauli each with probability $\overline{p}_{\text{CNOT}}(p,d)/20$,
        
    \item 
    measurements in the logical Pauli bases are assumed to be perfect.
\end{itemize}

\begin{table}
\begin{center}
 \begin{tabular}{|c || c |} 
 \hline
 logical operation &  failure probability $\overline{p}_\text{oper}(p,d)$  \\
 [0.5ex] 
 \hline\hline
   prep & 
     $\begin{array}{l}
                0.0151\cdot 0.812^{d}~~~\text{for}~ p= 10^{-3}\\
                0.0074\cdot 0.719^{d}~~~\text{for}~ p=5\cdot 10^{-4}\\
                0.0041\cdot 0.527^{d}~~~\text{for}~ p = 10^{-4}
                \end{array}$\\
    \hline 
  idle &      $\begin{array}{l}
                0.0124\cdot 0.721^{d}~~~\text{for}~ p= 10^{-3}\\
                0.0092\cdot 0.618^{d}~~~\text{for}~ p=5\cdot 10^{-4}\\
                0.0018\cdot 0.484^{d}~~~\text{for}~ p = 10^{-4}
                \end{array}$ \\
    \hline
   CNOT &  $\begin{array}{l}
                0.0972\cdot 0.894^{d}~~~\text{for}~ p= 10^{-3}\\
                0.0603\cdot 0.761^{d}~~~\text{for}~ p=5\cdot 10^{-4}\\  0.0078\cdot 0.607^{d}~~~\text{for}~ p = 10^{-4}
                \end{array}
              $ \\          \hline
   meas & \hspace*{-5.5mm} $\begin{array}{l}
   0.0011\cdot 0.690^{d}~~~\text{for}~ p= 10^{-3}
                \end{array}
              $\\ 
 \hline
\end{tabular}
 \caption{
 Failure probabilities of logical operations for the 2D color code as a function of the distance $d$.
 Since the failure probability for logical Pauli measurements is negligible, we assume measurements to be perfect in the effective noise model, and only report the measurement failure probability for $p= 10^{-3}$.
 }
 \label{tab:logical-operations}
\end{center}
\end{table}

Let us make a number of remarks about this noise model.
Firstly, the conservative estimates of different types of failure are quite loose for large distances, and could be made tighter by fine-tuning the noise model.
Secondly, one may wonder why we treat the SWAP operation as a pair of idle qubits, whereas we treat the CNOT operation separately and find it has a considerable failure probability.
The reason is that the propagation of previously existing error (which is exchanged between patches by SWAP, but added across patches by CNOT) can be very significant.

\begin{figure}[h]
	\includegraphics[width=.7\textwidth]{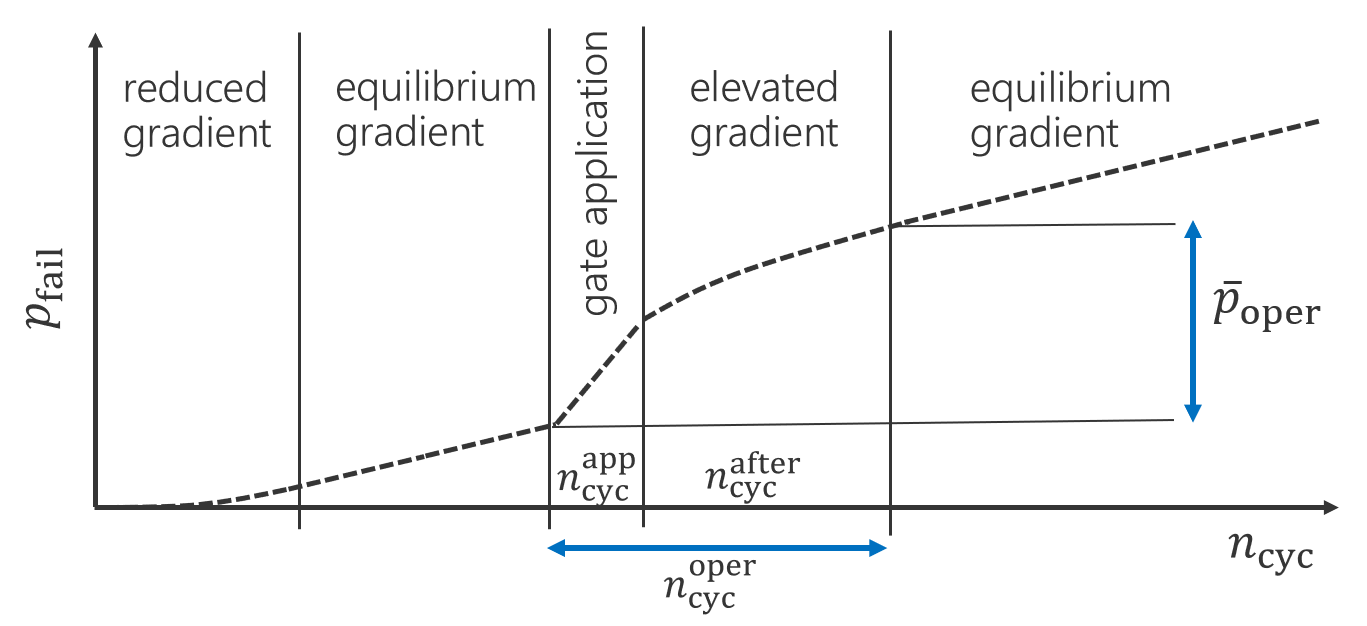}
	\caption{
	A sketch of the cumulative failure probability before, during and after a logical operation.
	Logical preparations and measurements comprise only some parts of this sketch.
    Before we apply the operation, we first make sure that the system reaches QEC equilibrium.
    The logical operation, which would take $n_{\text{cyc}}^{\text{app}}$ cycles by itself, is augmented with $n_{\text{cyc}}^{\text{after}}$ idle cycles to ensure the system has returned to QEC equilibrium afterwards.
    The logical operation failure rate  $\overline{p}_{\text{oper}}$ is the increment of $p_{\text{fail}}$ over the $n_{\text{cyc}}^{\text{oper}} = n_{\text{cyc}}^{\text{app}} + n_{\text{cyc}}^{\text{after}}$ QEC cycles.
    }
	\label{fig:QECEquilbrium}
\end{figure}

In the remainder of this section, we describe how we estimate the logical failure probabilities using distance-$d$ patches of 2D color code under circuit noise of strength $p$ followed by a perfect-measurement decoding.
To avoid the influence of boundary effects in these simulations, we ensure that the patches are close to QEC equilibrium both before and after the logical operation (or just before or after for measurement or preparation, respectively); see \fig{QECEquilbrium}.
We therefore implement $d$ QEC cycles on the patches prior to the operation and, when necessary, augment the logical operation by including additional idle QEC cycles at the end of the operation, checking that the residual noise has stabilized before the perfect measurement is implemented. 
By QEC equilibrium, we mean a regime in which the failure probability increases linearly with the number of QEC cycles.
Throughout our analysis we use the best-performing CNOT schedule from \sec{circuitnoise}, which uses one ancilla per stabilizer generator. 
This results in $(3d^2-1)/2$ qubits being needed for a distance-$d$ patch.

\begin{figure}[h]
	(a)\hspace*{-5mm}\includegraphics[width=.45\textwidth]{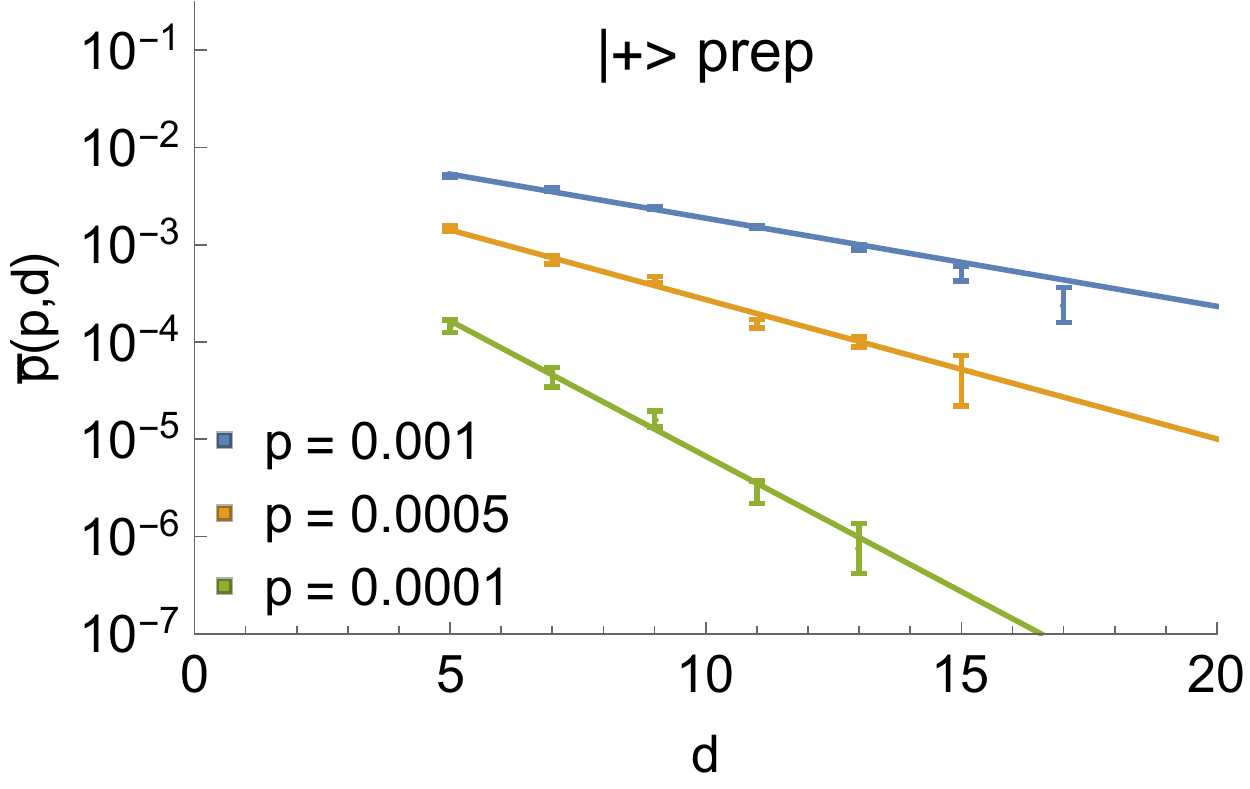}
	\quad\quad\quad
	(b)\hspace*{-5mm}\includegraphics[width=.45\textwidth]{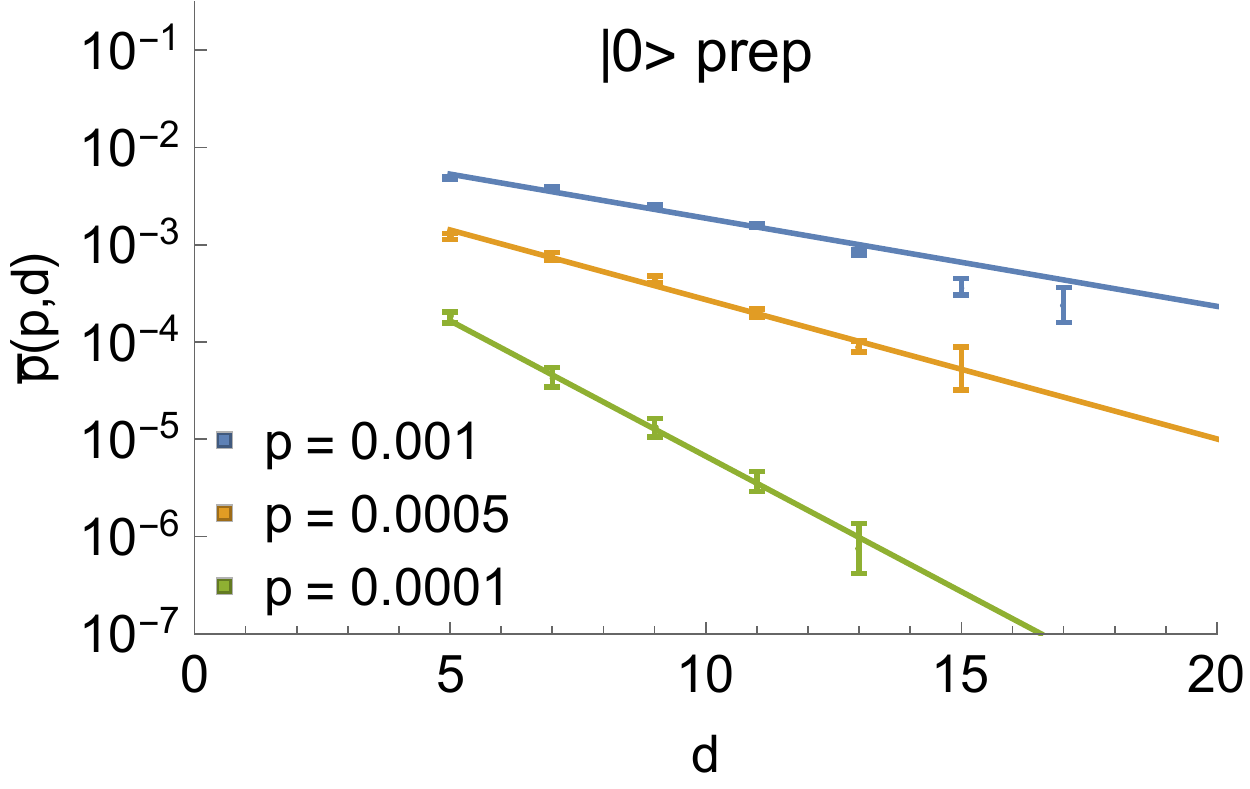}
	\caption{
    The logical failure probability for preparing the $\ket{\overline{+}}$ and $\ket{\overline{0}}$ states. 
    The ansatz in \eq{ansatz_moregeneral} is fitted, giving $\overline{p}_{\text{prep}}(p,d)$ in \tab{logical-operations}.
    Both plots can be fitted well using the same parameters, which justifies treating both $\ket{\overline{+}}$ and $\ket{\overline{0}}$ preparation identically in the noise model.
	}
	\label{fig:CircuitNoiseAnalysisPrep}
\end{figure}

\textbf{Logical state preparation.---}Recall that $\ket{\overline{0}}$ is prepared by initializing data qubits in $\ket{0}$ and then measuring the stabilizers for $d$ QEC cycles under circuit noise of strength $p$, and applying a $Z$-type operator intended to fix the $X$ stabilizers.
We simulate this followed by the perfect-measurement decoder and estimate the failure probability from the proportion of trials in which the a logical $\overline{X}$ is applied.
The analogous procedure is used to identify the failure probability for preparing $\ket{\overline{0}}$.
The data is presented in \fig{CircuitNoiseAnalysisPrep} and fitted with the ansatz in \eq{ansatz_moregeneral} using the same parameters.
This fit provides the entry for $\overline{p}_{\text{prep}}(p,d)$ in \tab{logical-operations}.

\begin{figure}[h]
	(a)\hspace*{-5mm}\includegraphics[width=.45\textwidth]{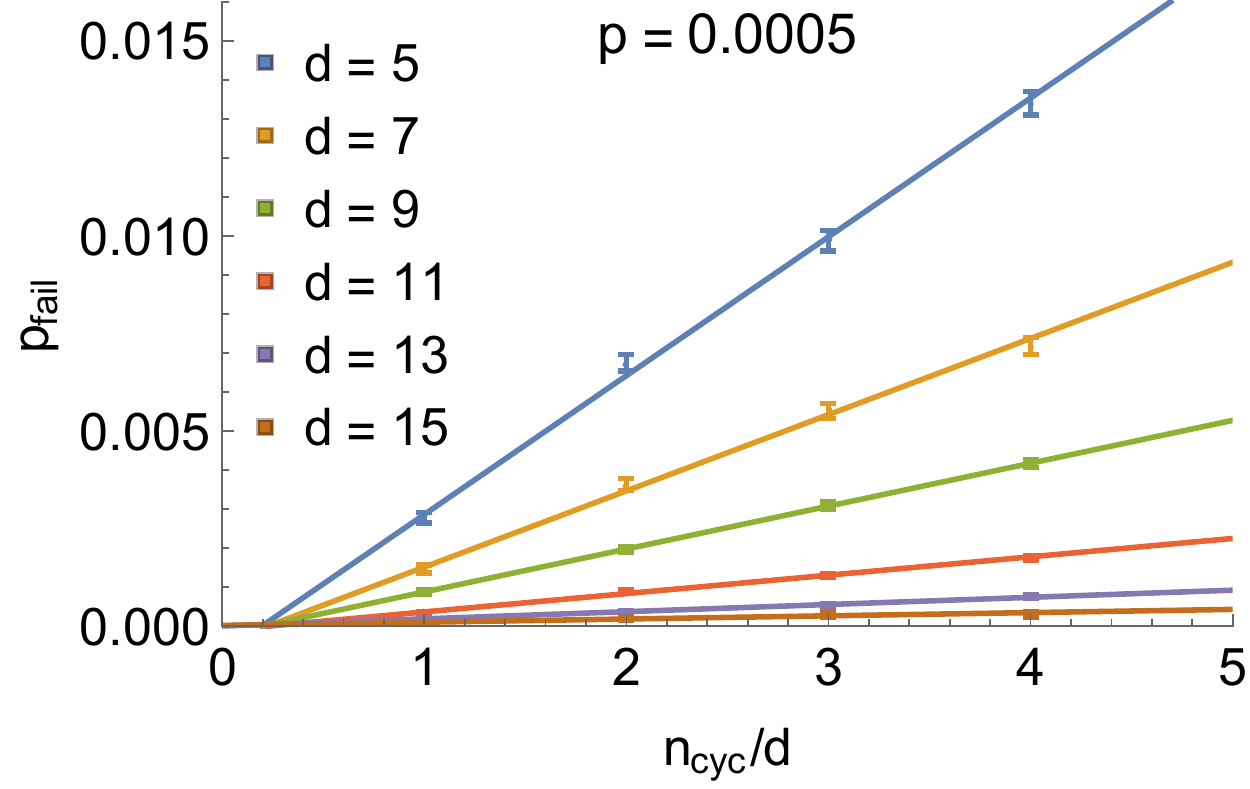}
	\quad\quad\quad
	(b)\hspace*{-5mm}\includegraphics[width=.45\textwidth]{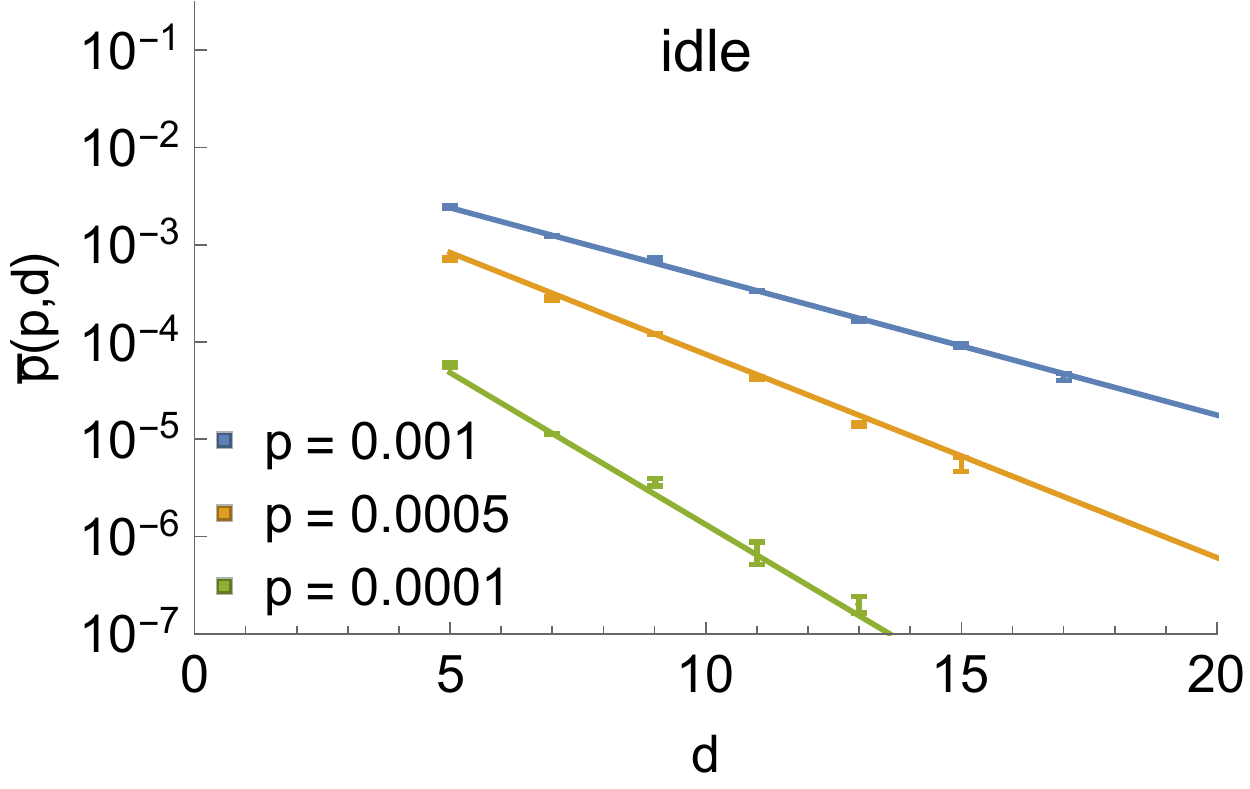}\\
	\vspace*{5mm}
	\includegraphics[width=.31\textwidth]{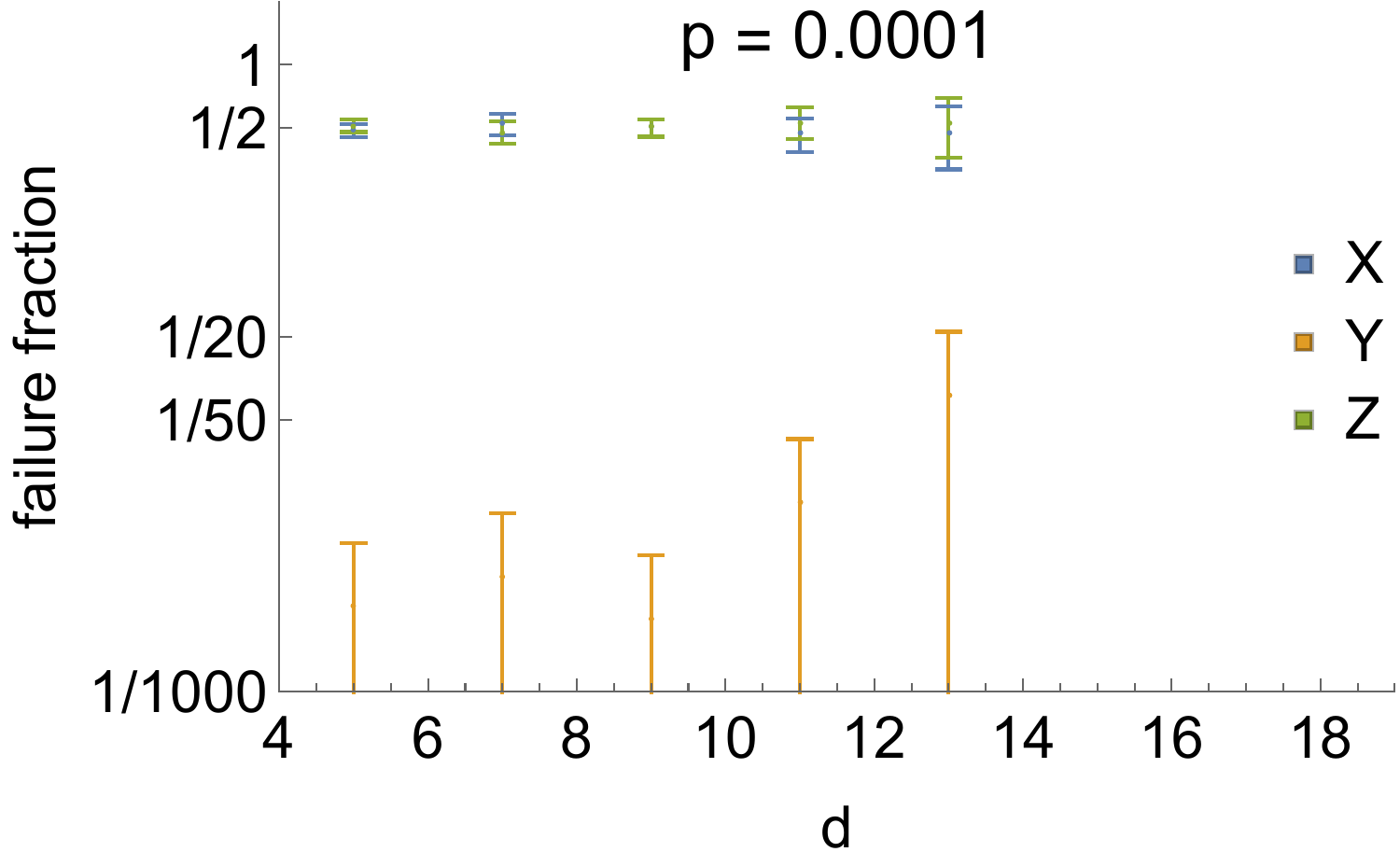}
	\quad
	\includegraphics[width=.31\textwidth]{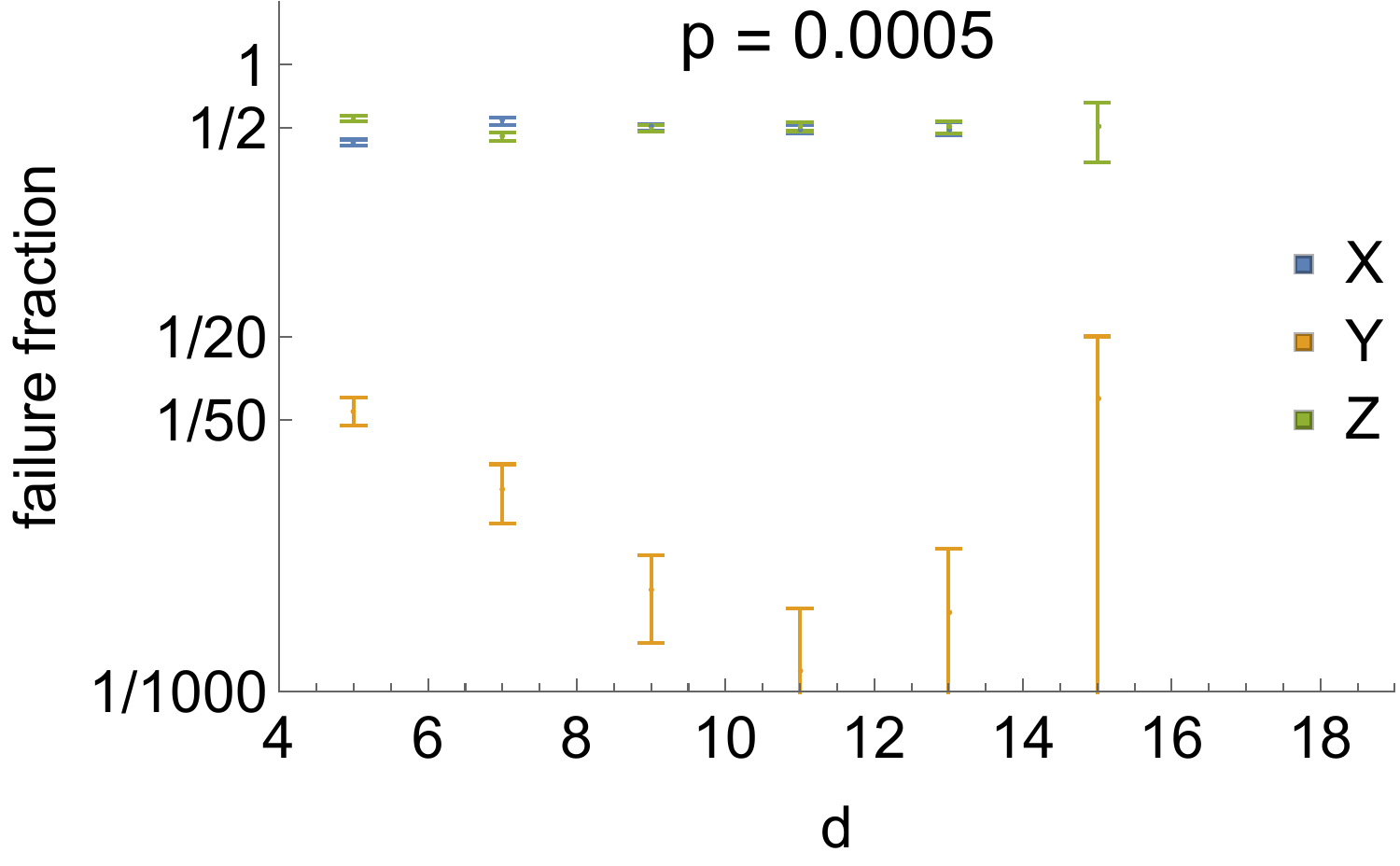}
	\quad
	\includegraphics[width=.31\textwidth]{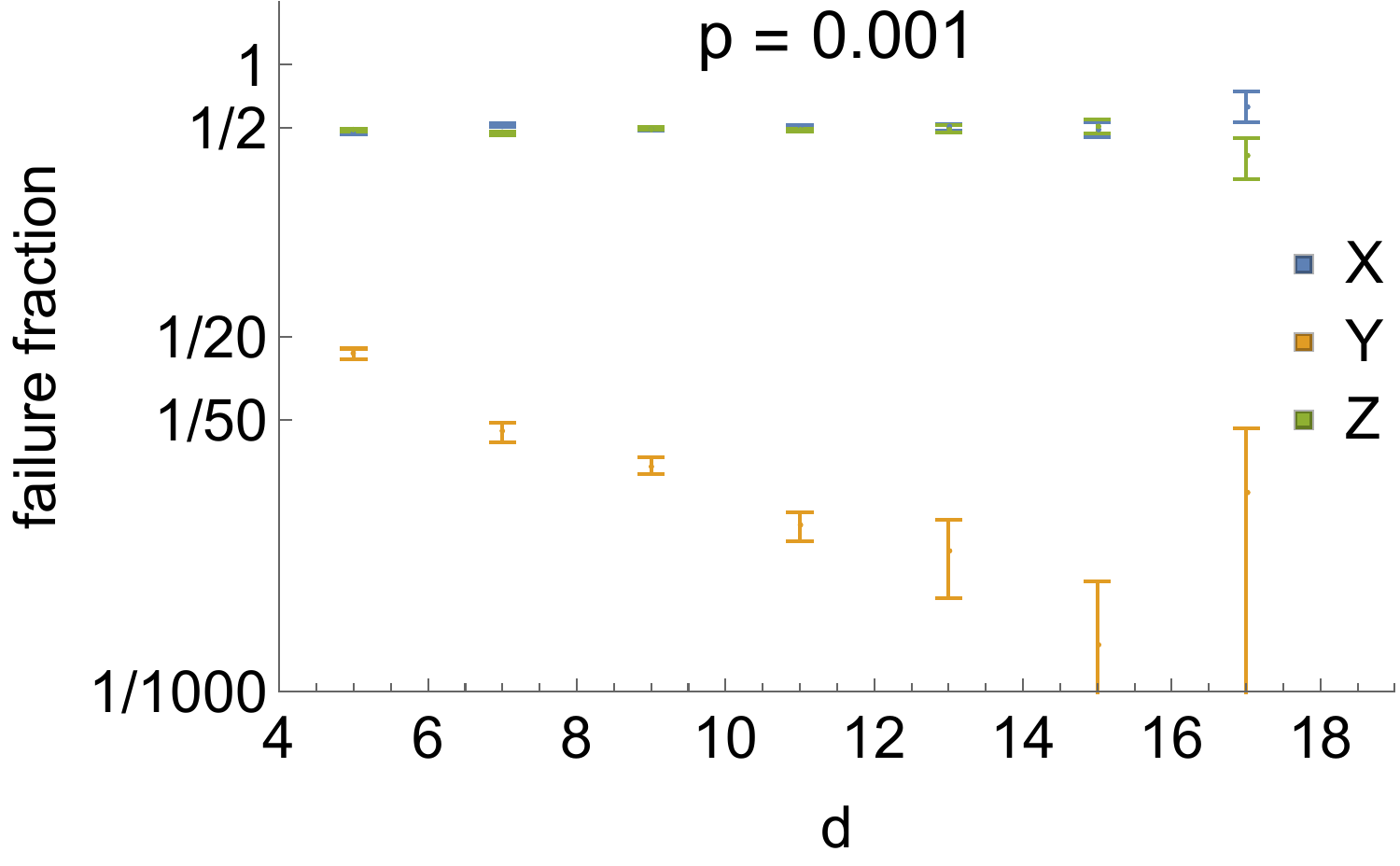}
	\caption{
	Analysis of the logical idle operation.
	(a) The failure probability using the best-performing CNOT schedule as a function of $n_{\text{cyc}}/d$ for $p=5\cdot 10^{-4}$.
	We use the ansatz in~\eq{ansatz_failure} to find the logical failure rate $\overline{p}(p,d)$.
	We observe that QEC equilibration occurs by the $d$th cycle; see \app{threshold-data} for additional data.
	Not discarding the initial equilibration period would result in an underestimate of the failure probability.
	(b) For each value of $p$, we use the ansatz in \eq{ansatz_moregeneral}
	to extract the parameters for $\overline{p}_{\text{idle}}(p,d)$ in \tab{logical-operations}.
	(bottom panels) The fraction of overall failures corresponding to different logical Pauli errors.
	The points whose data range falls below the horizontal axis correspond to no observed failures.}
	\label{fig:CircuitNoiseAnalysisIdle}
\end{figure}

\textbf{Idle logical qubit.---}We extract $\overline{p}_{\text{idle}}(p,d)$, the failure probability of an idle logical qubit for a single QEC cycle, by considering the failure probability of a single patch of distance-$d$ 2D color code over $n_{\text{cyc}}$ QEC cycles, given a perfect initial state and a perfect final QEC cycle.
After a few initial QEC cycles, we expect the residual error in the system to reach an equilibrium, and that thereafter the failure rate should increase linearly with $n_{\text{cyc}}$ in the regime of small $p_{\text{fail}}(d, n_{\text{cyc}})$.
We can use linear growth of $p_{\text{fail}}(d, n_{\text{cyc}})$ with time as a hallmark of the system being in QEC equilibrium.
To estimate $\overline{p}_{\text{idle}}(p,d)$ for fixed $p$ and $d$, we fit the failure probability $p_{\text{fail}}(d, n_{\text{cyc}})$ data, such as that presented in \fig{CircuitNoiseAnalysisIdle}(a) for $p= 10^{-3}$, with the following linear function of $n_\text{cyc}$, i.e.,
\begin{eqnarray}
\label{eq:ansatz_failure}
p_{\text{fail}}(d, n_{\text{cyc}}) &=& \overline{p}_{\text{idle}}(p,d) \cdot n_{\text{cyc}} + c(p,d),
\end{eqnarray}
where $c(p,d)$ is a constant.
We plot this and the fit in \fig{CircuitNoiseAnalysisIdle}(a) for a particular value of $p$, and use the gradient to estimate $\overline{p}_{\text{idle}}(p,d)$ for various $p$ and $d$ in \fig{CircuitNoiseAnalysisIdle}(b).
We fit \eq{ansatz_moregeneral} to the data, giving
the entry in \tab{logical-operations}.
Note that in \fig{CircuitNoiseAnalysisIdle}(a) the $y$-intercept is negative since the simulation begins with a perfect code state and so the probability of failure during the early QEC cycles is artificially reduced. 
For later logical operations we will assume that $d$ QEC cycles prior to the logical operation is sufficient to ensure the system has reached QEC equilibrium.

In our noise model, we claim that $\overline{p}_{\text{idle}}(p,d)/2$, $\overline{p}_{\text{idle}}(p,d)/2$ and $\overline{p}_{\text{idle}}(p,d)/20$ are conservative estimates of the probability of $\overline{X}$, $\overline{Z}$ and $\overline{Y}$ respectively. 
In \fig{CircuitNoiseAnalysisIdle}(c) and (d), we justify this claim by finding the fraction of failures which occur for each Pauli operator. 
Despite the symmetry of the depolarizing noise, we observe that $\overline{Y}$ failures are much less frequent than $\overline{X}$ or $\overline{Z}$ failures due to the independent detection and decoding of $X$ and $Z$ errors.

\begin{figure}[h]
	(a)\hspace*{-5mm}\includegraphics[width=.45\textwidth]{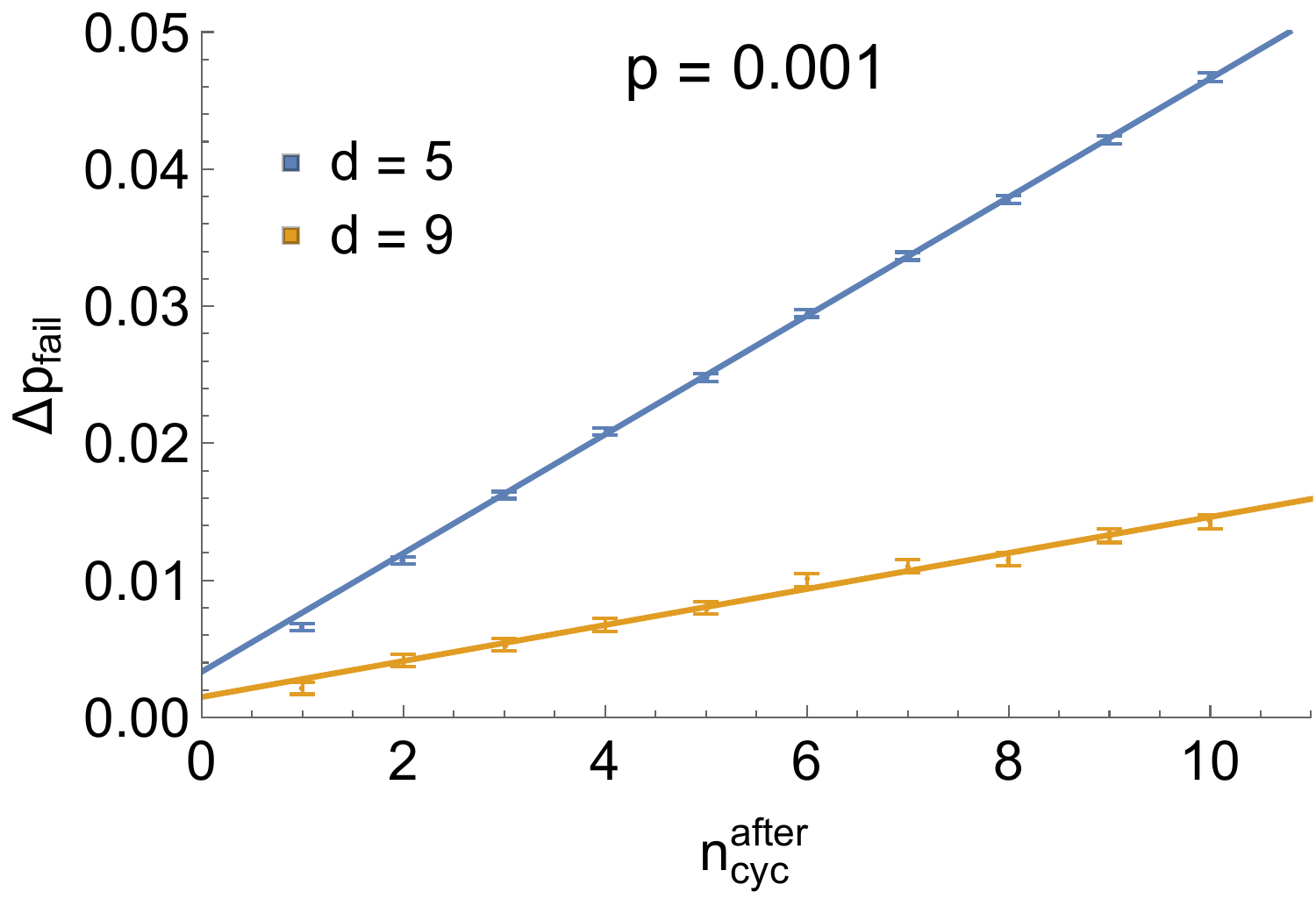}
	\quad\quad\quad
	(b)\hspace*{-5mm}\includegraphics[width=.45\textwidth]{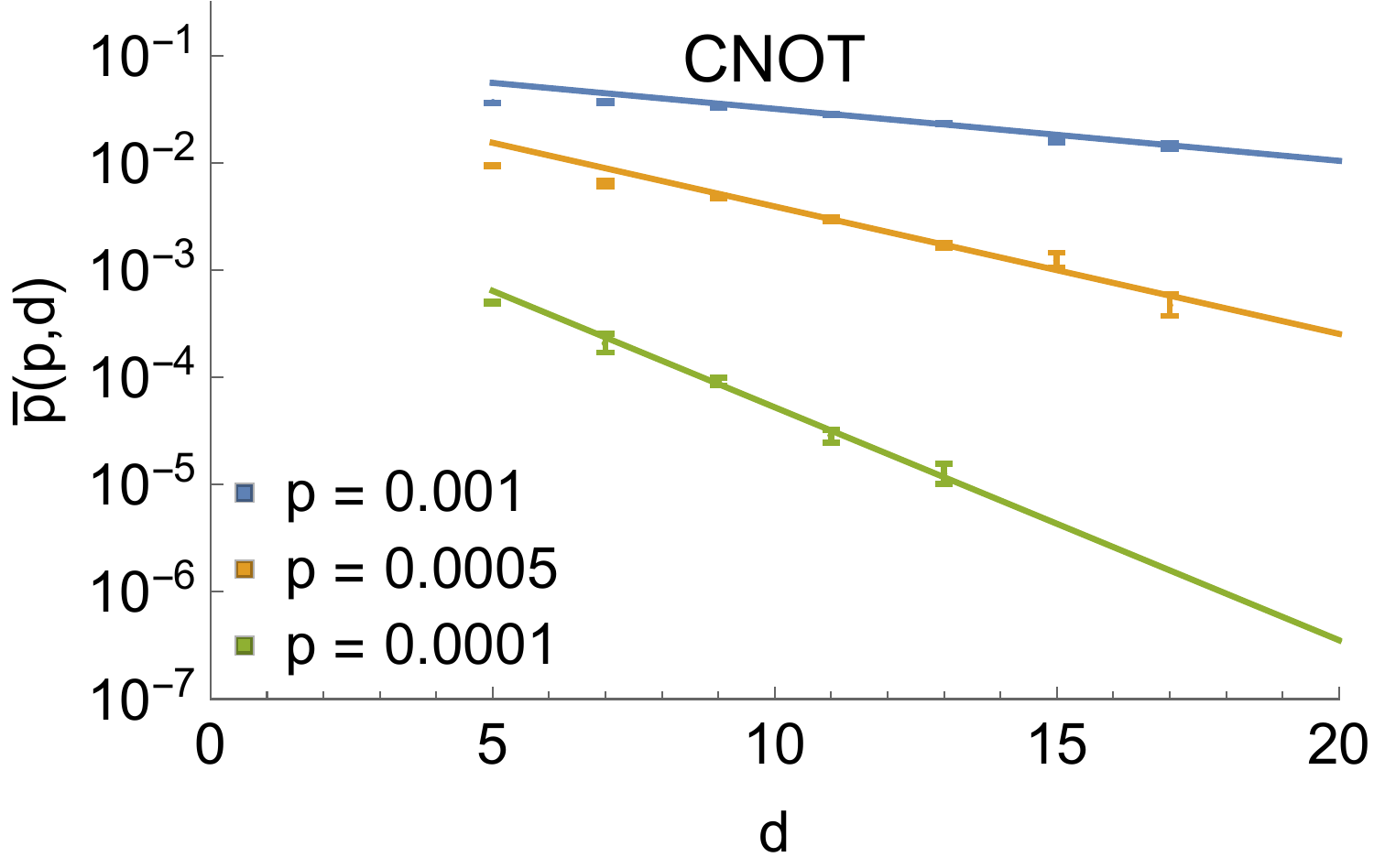}\\
	\vspace*{5mm}
	\includegraphics[width=.31\textwidth]{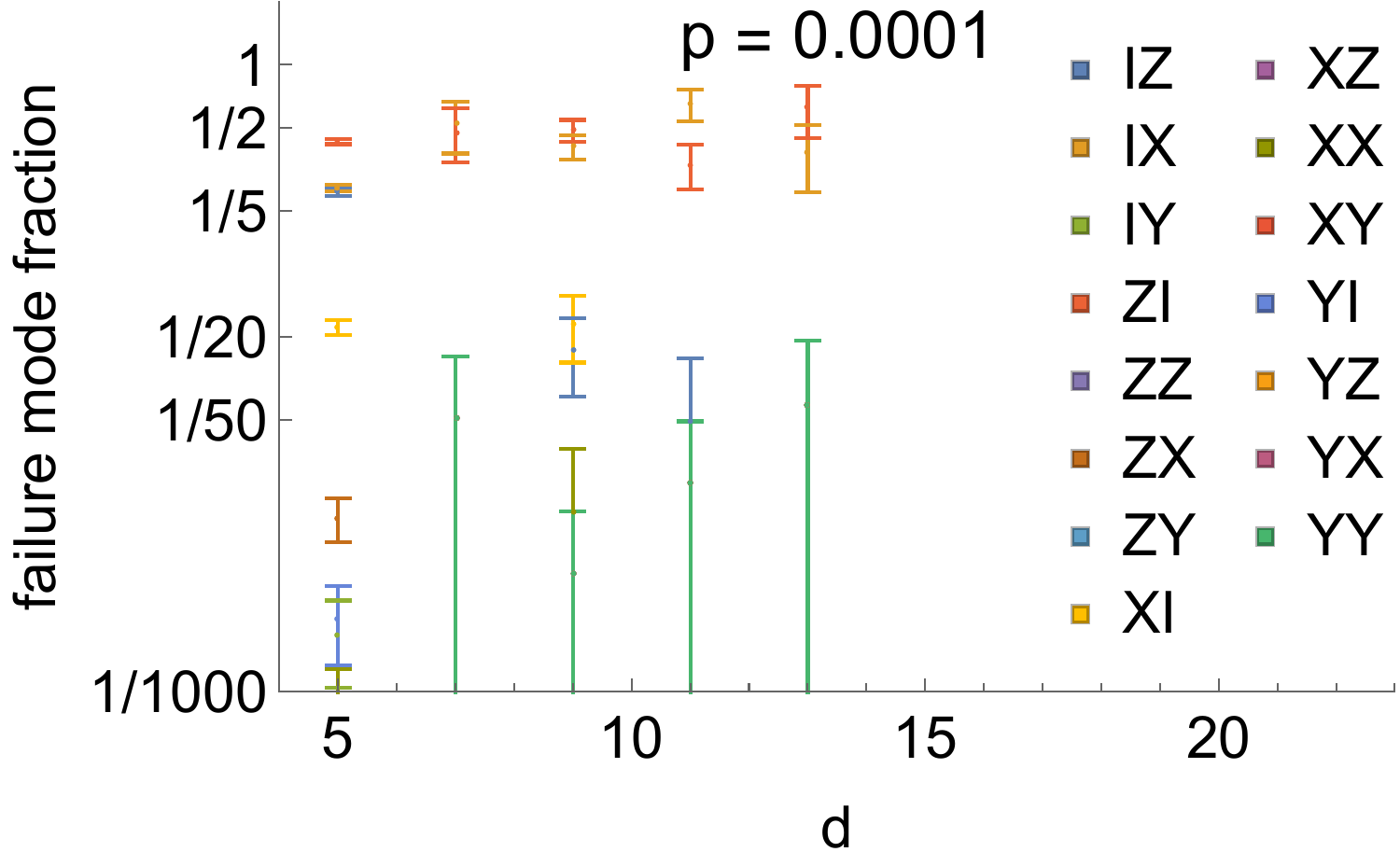}
	\quad
	\includegraphics[width=.31\textwidth]{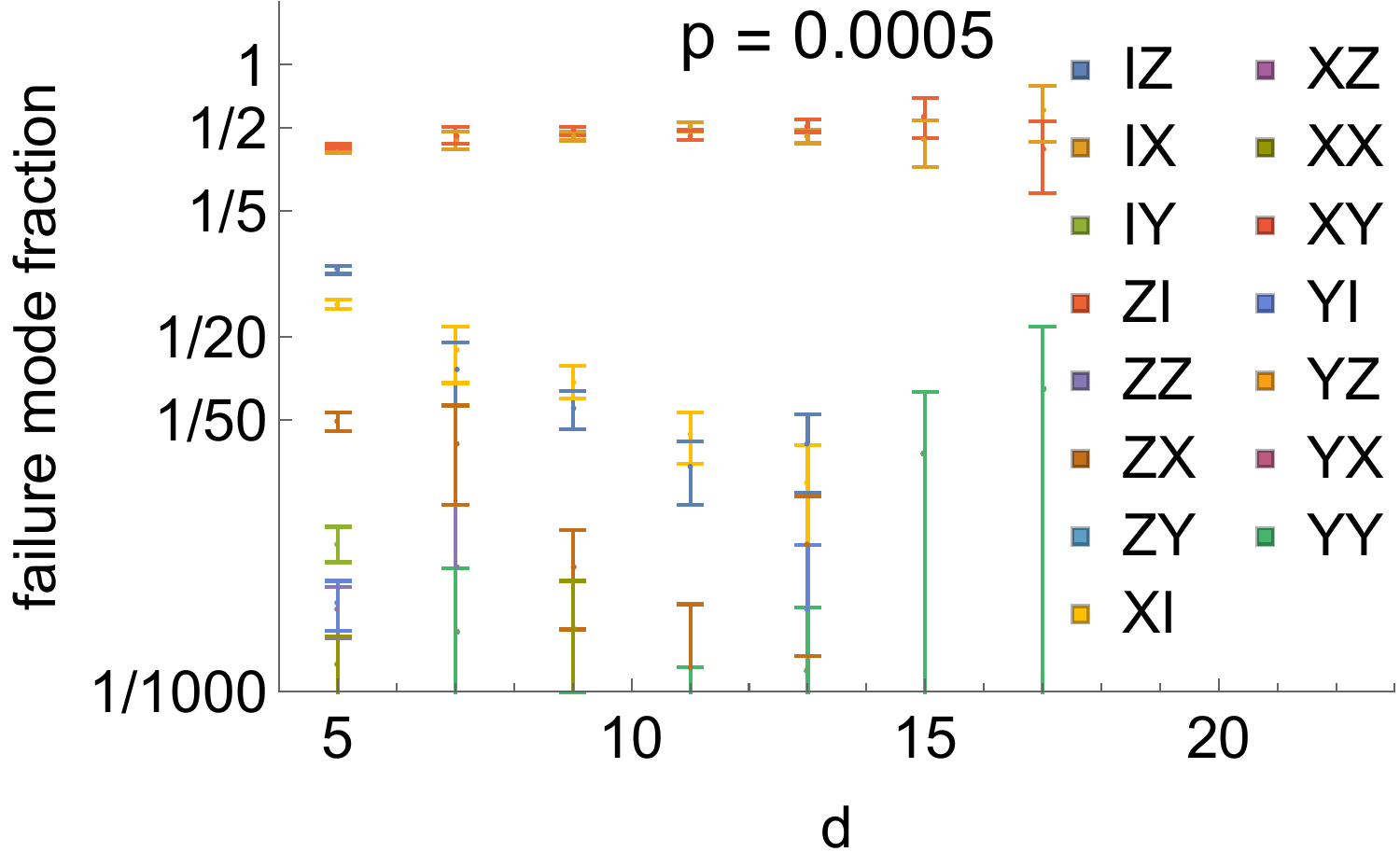}
	\quad
	\includegraphics[width=.31\textwidth]{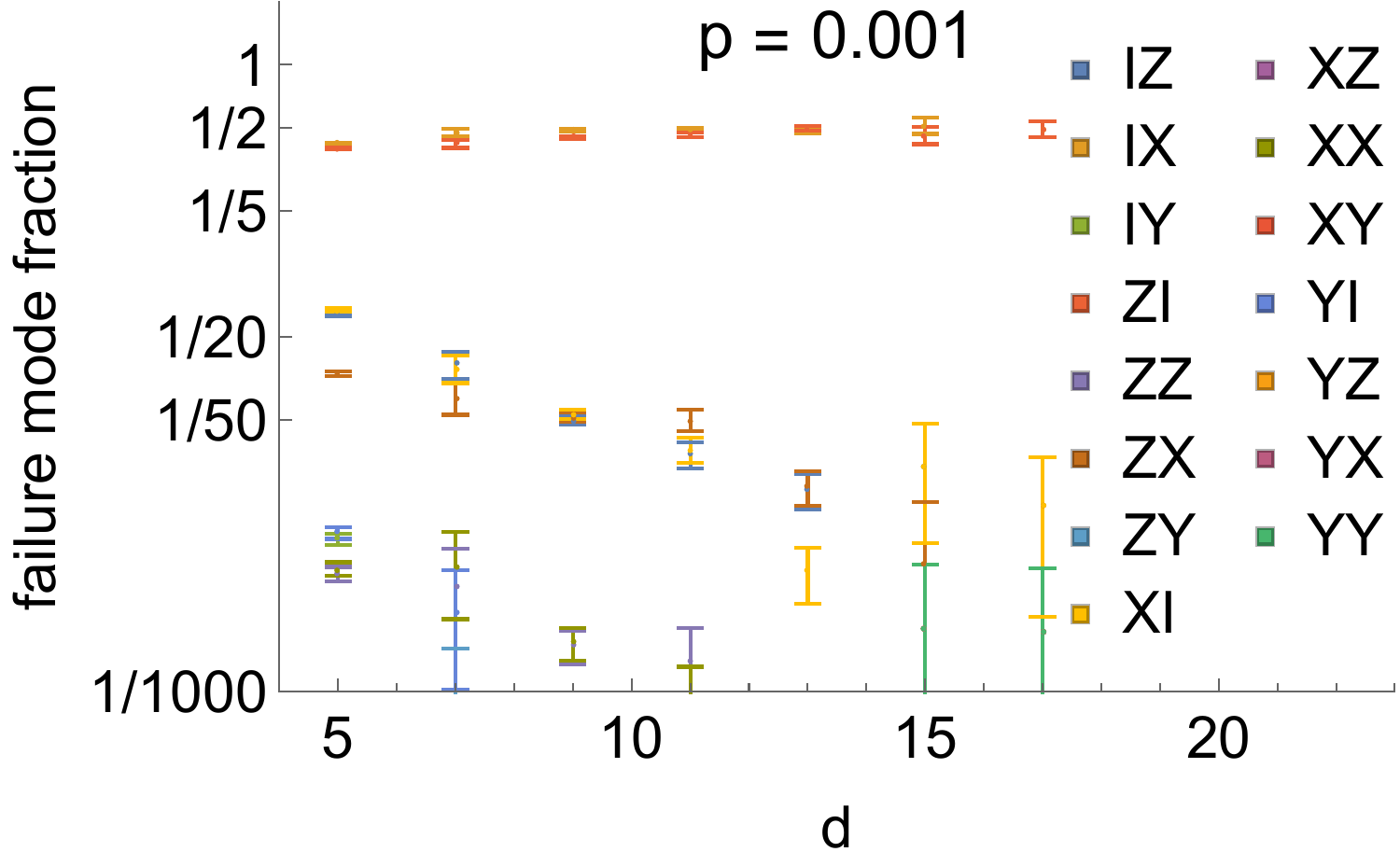}
	\caption{Analysis of the logical CNOT operation.
	(a) The change in failure probability between the time immediately after the transversal application of CNOT gates and $n_{\text{cyc}}^{\text{after}}$ QEC cycleslater.
	We  observe  that  QEC equilibration  occurs after two QEC cycles.
	(b) For each value of $p$, we use the ansatz in \eq{ansatz_moregeneral} to extract the parameters for $\overline{p}_{\text{CNOT}}(p,d)$ in \tab{logical-operations}.
    (bottom panels) The fraction of overall failures corresponding to different logical Pauli errors.
    The dominant errors are $\overline{ZI}$ and $\overline{IX}$, since the logical CNOT propagates $X$ errors onto the second patch and $Z$ errors onto the first patch.
    Other noticeable errors include $\overline{IZ}$ and $\overline{XI}$, which correspond to the failure over two QEC cycles, but become negligible for large $d$.
    The points whose data range falls below the horizontal axis correspond to no observed failures.
    }
	\label{fig:CircuitNoiseAnalysisCNOT}
\end{figure}

\textbf{Transversal logical CNOT.---}The logical CNOT is implemented transversally between a pair of 2D color code patches.
We assume the system is in QEC equilibrium before the gate, which is ensured in the simulation by running $d$ QEC cycles on an initially error-free state.
Immediately after the CNOT gates are applied, the system's residual noise is elevated since the CNOT propagates $X$ errors from the control to the target and $Z$ errors from the target to the control. 
To allow the system to return to QEC equilibrium, we include $n_{\text{cyc}}^{\text{after}}$ QEC cycles after the gate is applied in the logical CNOT operation.
We conclude from \fig{CircuitNoiseAnalysisCNOT}(a) that $n_{\text{cyc}}^{\text{after}}=2$ is sufficient, which we then incorporate into the logical CNOT operation. 
We analyze the overall failure probability of the logical CNOT in \fig{CircuitNoiseAnalysisCNOT}(b), and the fraction of failures resulting in each logical Pauli in the panels at the bottom of \fig{CircuitNoiseAnalysisCNOT}.
Note that in contrast with the phenomenon observed in \fig{CircuitNoiseAnalysisIdle}(a) in which the initial system equilibrated from a state of lower noise, here the system equilibrates from a state of higher noise. 

To decode each patch, we use the combined syndrome history---since the CNOT propagates $X$ errors from the first patch to the second one, we add the $Z$-type stabilizer history from the first $d$ QEC cycles from the first patch to that of the second before decoding $X$ errors in the second patch, and similarly for $Z$ errors in the first patch.
To isolate the contribution to the logical operator from the CNOT alone, we find and remove the contribution to the logical operator from the initial $d$ QEC cycles by applying a perfect-measurement decoding on the system immediately after the $d$ QEC cycles, and propagate that through the logical CNOT gate.\footnote{Note that it is possible to considerably improve this implementation of the CNOT gate. Namely, one first decodes the error separately for the patch which is the source of the copied error (i.e., the control patch for $X$ error and the target patch for the $Z$ error) and then applies the correction to the destination of the copied error before correcting the residual error there \cite{fowler2020}.}

\textbf{Logical readout.---}To fault-tolerantly read off $\overline{Z}$, one measures each data qubit in the $Z$ basis.
The resulting bit string can be fixed using a perfect-measurement decoder so that it satisfies all $Z$-type stabilizers, and then the outcome of the $\overline{Z}$ logical operator can be read off from any representative.
One can fault-tolerantly measure $\overline{X}$ similarly by measuring each data qubit in the $X$ basis.

\begin{figure}[h]
	(a)\hspace*{-5mm}\includegraphics[width=.45\textwidth]{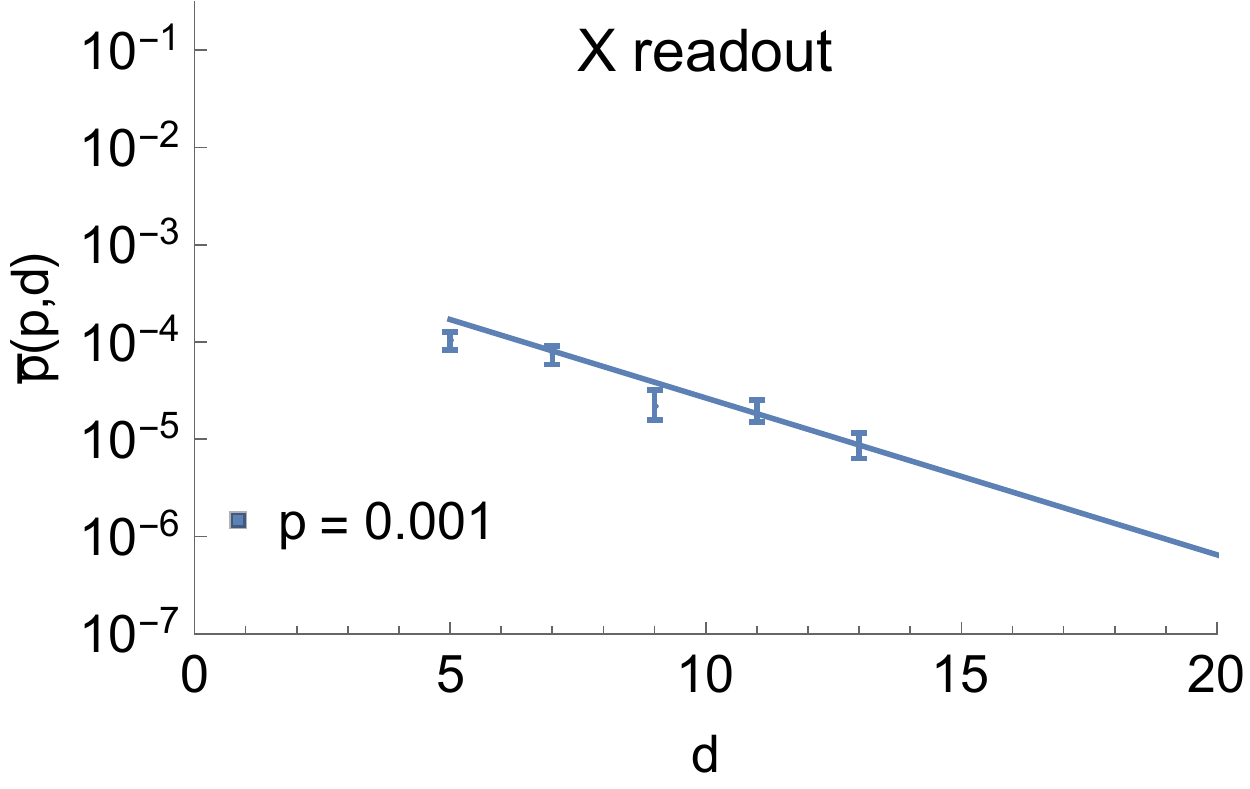}
	\quad\quad\quad
	(b)\hspace*{-5mm}\includegraphics[width=.45\textwidth]{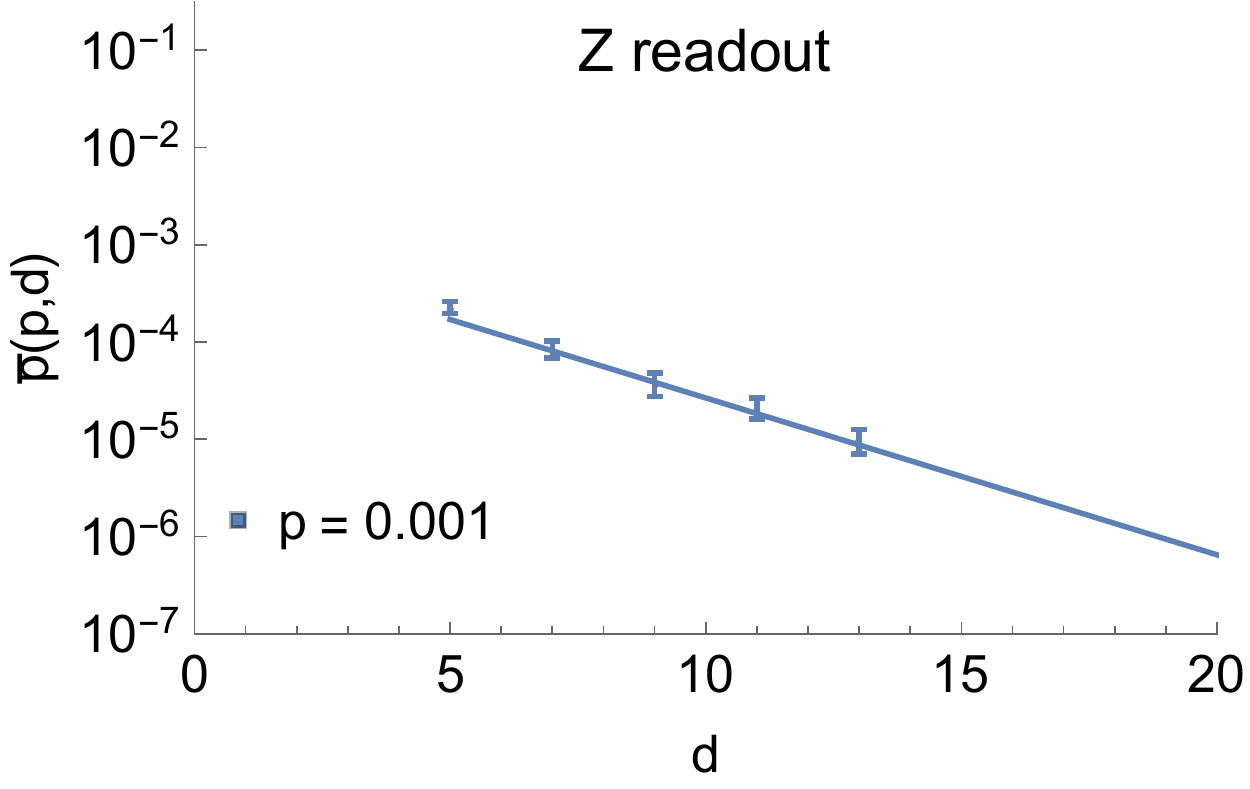}
	\caption{
	The readout failure probability for $\overline{X}$ and $\overline{Z}$.
    Note the value is orders of magnitude lower than for the other logical operations and that we only show that for $p=0.001$, since the smaller values of $p$ are so low that they are difficult to observe using Monte Carlo.
    We use the ansatz in \eq{ansatz_moregeneral} to extract the parameters for $\overline{p}_{\text{meas}}(p,d)$ in \tab{logical-operations}, although in our noise model we neglect noise in the measurement.
    }
    \label{fig:CircuitNoiseAnalysisReadout}
\end{figure}

To simulate the logical $\overline{X}$ readout we first run the system for $d$ QEC cycles to ensure equilibration has occurred, then measure each qubit in the $X$ basis under circuit noise of strength $p$, followed by perfect-measurement decoding.
To isolate the contribution to the logical operator from the logical measurement alone, we find and remove the contribution to the logical operator from the initial $d$ QEC cycles by applying a perfect-measurement decoding on the system immediately after the $d$ QEC cycles.
We estimate the failure probability and mode fraction for logical measurement of $\overline{X}$ and $\overline{Z}$ in \fig{CircuitNoiseAnalysisReadout}.
We fit \eq{ansatz_moregeneral} to the data for both $\overline{X}$ and $\overline{Z}$ measurements, which agree with one another and this fit provides the entry for $\overline{p}_{\text{meas}}(p,d)$ in \tab{logical-operations}.

Note that the noise contributed by readout is orders of magnitude below the other logical operations, and we only take data for $p = 10^{-3}$ since a prohibitively large number of samples would be required for $p = 10^{-4}$ and $p = 5\cdot 10^{-4}$. 
This justifies that we neglect contributions from readout in our effective noise model.

\clearpage
\section{State distillation analysis}
\label{sec:distillation}
In this section we carefully analyze the performance and estimate the overhead of state distillation of $T$ states using the standard 15-to-1 scheme. 


\subsection{Creating the $T$ state via state distillation in 3 steps}
\label{sec:TStateInitialization}

\begin{figure}[ht!]
	\includegraphics[width=.7\textwidth]{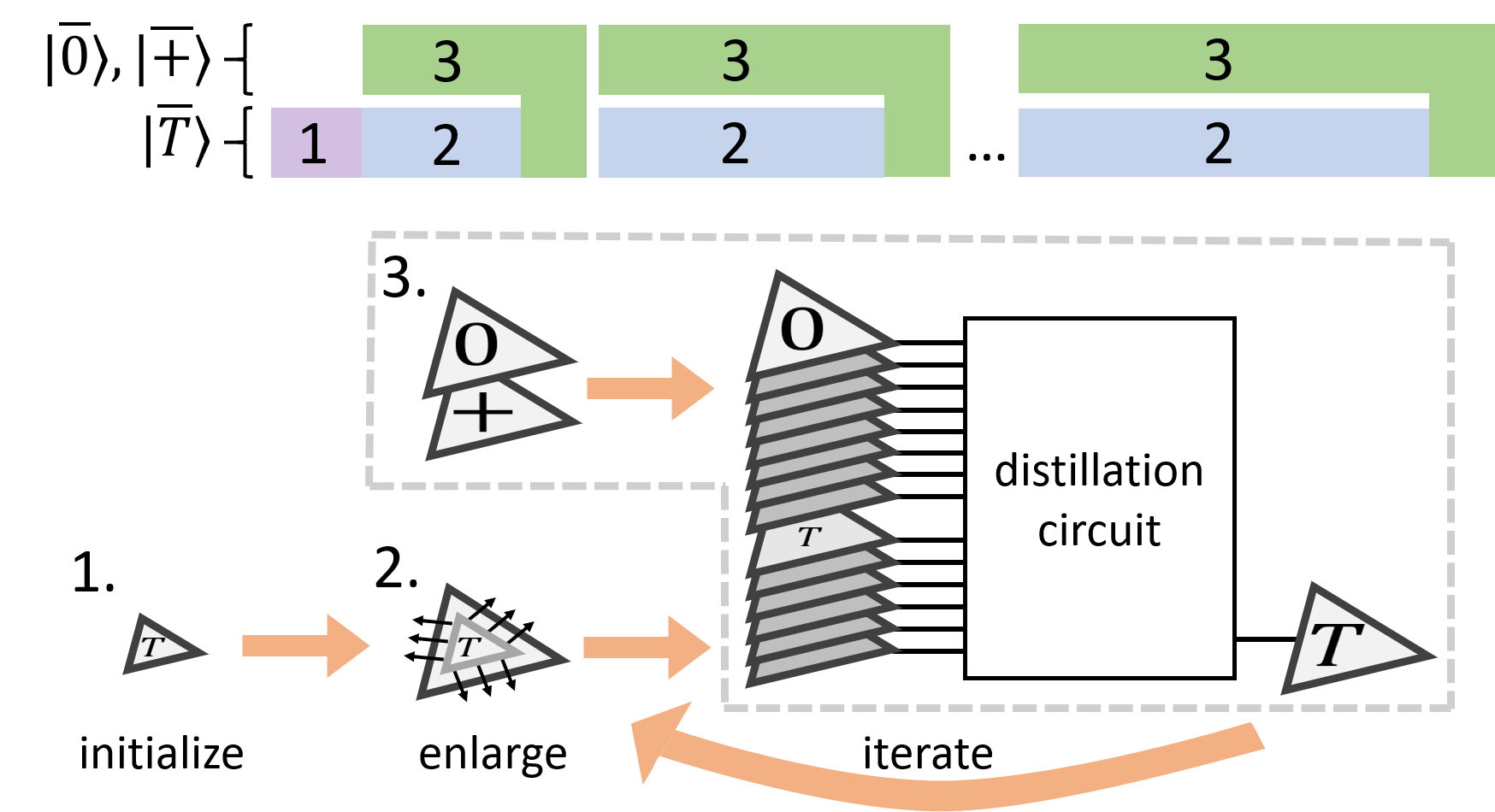}
	\caption{
	The protocol to create an encoded $T$ state via state distillation, with a qualitative timeline of each step.
	In step 1, noisy encoded $T$ states are prepared non-fault-tolerantly in a small 2D color code.
	In step 2, each code patch is expanded to increase its distance, and we assume that the logical infidelity of the encoded $T$ state does not change.
	In step 3, 16 logical states, either $\ket{\overline 0}$ or $\ket{\overline +}$, are prepared and then the state distillation circuit is run on them and 15 $\ket{\overline T}$'s.
	Given successful postselection, the output is a single encoded $T$ state of higher fidelity (indicated by a larger font size).
	Steps 2 and 3 are iterated until a state of the desired infidelity is produced. 
	}
	\label{fig:StepsDistillation}
\end{figure}

A protocol to produce an encoded $T$ state using state distillation, which we illustrate in~\fig{StepsDistillation}, consists of the following steps.

\begin{enumerate}
    \item \textit{$T$ state initialization.---}We initialize encoded $T$ states in small-distance 2D color code patches, which later serve as the input to the first round of state distillation.
    
    \item \textit{Expansion and movement of patches.---}The code distance in consecutive rounds of state distillation is required to increase to protect the produced $T$ states of improving fidelity. 
    We must therefore increase the size of a patch which is output from one round, and move it to the desired location where it becomes an input for the next round.
    
    \item \textit{State distillation circuit.---}This is a circuit with every qubit encoded in a distance-$d$ 2D color code, consisting of nearest-neighbor logical Clifford operations on patches arranged in a 3D stack.
    The input consists of 15 encoded $T$ states, and the output is one higher fidelity encoded $T$ state.
    
\end{enumerate}
The second and third steps are repeated (on multiple copies of the procedure in parallel) with increasing code distances chosen to minimize the overhead until a $T$ state of the desired quality is produced.
In the following subsections, we go through each of the three steps, elaborating on the implementation and simulation details.
In our analysis, we assume circuit noise of strength $p$, and use the effective noise model presented in \sec{2DCCOptimizedSetup} to analyze the performance of logical-level circuits implemented with the 2D color code. 

\subsubsection{T state initialization}
\label{sec:magic-state-initialization}

The first step of any state distillation protocol is to produce the initial encoded $T$ states. 
The initialization protocol to do this can be crucial since the results of state distillation depend strongly on the quality of the initial $T$ states.
For example, taking the 15-to-1 scheme with perfect Clifford operations, if the starting infidelity is decreased by a factor of two (which, as we will see, can be achieved by varying the CNOT order in the initialization protocol), the output infidelity is reduced by about one, three and eight orders of magnitude over one, two and three state distillation rounds, respectively.
Although abstract state distillation protocols have received a lot of attention, there is surprisingly little research on the initialization of $T$ states as inputs for state distillation despite the enormous potential impact.
For the surface code, a non fault-tolerant scheme with the logical error of the final encoded $T$ state comparable with that of raw state was proposed by Li~\cite{Li2015}.
More recent works~\cite{Chamberland2019,Chamberland2020a} present fault-tolerant approaches to initialize encoded $T$ states and can, in some regimes, achieve higher fidelity encoded $T$ states with low overhead, but are somewhat more challenging to implement. 

\begin{figure}[h]
	(a)\includegraphics[width=.18\textwidth]{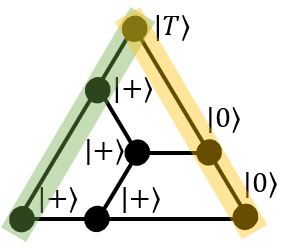}\quad\quad\quad
	(b)\includegraphics[width=.35\textwidth]{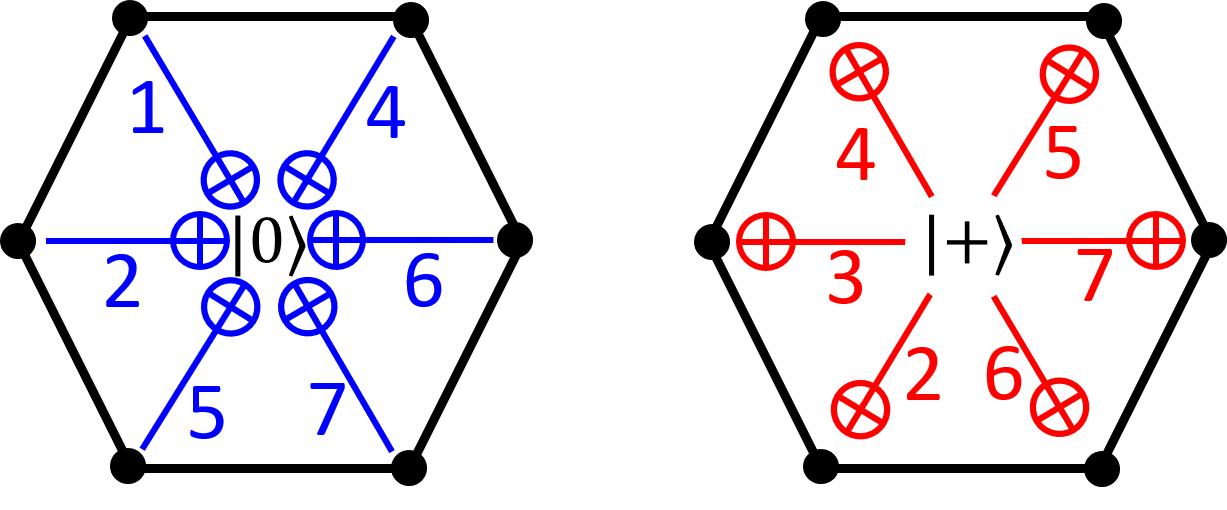}
	\caption{
	$T$ state initialization for the 2D color code with distance $d=3$.
	(a) In step 1, the top qubit is prepared in $\ket{T}$, and the remaining data qubits are prepared in either $\ket{+}$ or $\ket{0}$.
	We depict the support of logical $X$ and $Z$ representatives as shaded yellow and green strips.
	(b) In step 2, all the stabilizers are measured using the depicted CNOT order (or its restriction). 
	}.
	\label{fig:color-code-magic-initialization}
\end{figure}

Our strategy of initializing $T$ states for the 2D color code can be viewed as a generalization of the approach in Ref.~\cite{Li2015}, which consists of two main steps: (i) produce an encoded $T$ state in a distance $d_1$ code (with $d_1<d$), then (ii) enlarge the code from $d_1$ to $d$.
For judiciously chosen $d_1$, the noise added during step (ii) can be neglected because it is much less significant that the noise from step (i).
We produce an encoded $T$ state in the following steps.
\begin{enumerate}
    \item 
    Choose representatives of the logical $X$ and $Z$ operators which intersect on a single qubit, and
    prepare that qubit in $\ket{T}$.
    Prepare the remaining data qubits along the support of the logical $X$  and $Z$ in $\ket{+}$ and $\ket{0}$, respectively.
    Other data qubits are prepared in either $\ket{+}$ or $\ket{0}$; see~\fig{color-code-magic-initialization}(a).
    \item  Measure each stabilizer twice; see~\fig{color-code-magic-initialization}(b).
    If the observed syndrome is not the same or the syndrome could not have arisen without fault, the initialization procedure is restarted.
    \item Apply a Pauli operator fixing the observed syndrome.
\end{enumerate}
In our simulation, we say the procedure has succeeded in creating an encoded $T$ state if upon a single additional fault-free QEC cycle, one obtains the encoded $T$ state.
We remark that the state from step 1 is not an eigenstate of all the stabilizers measured in step 2, and thus, even in the absence of faults, we may need to apply a nontrivial Pauli operator in step 3 to ensure that all the stabilizers are satisfied.

Lastly, we optimize the initialization protocol by varying the order in which the CNOT gates are applied during the two QEC rounds in step 2.
We considered a range of system sizes, and again assume that there is a separate ancilla qubit per stabilizer generator and consider the $234$ valid schedules consisting of $7$ CNOT time units as in \sec{circuitnoise}.
We found that the $d=3$ size resulted in the best parameters, and that the worst schedule (which is not the same as that used for standard error correction) results in more than twice the lowest-order failure probability of the best schedule; see \app{haah-hastings-distillation} for more details.

The protocol takes 
\begin{equation}
\tau^{\text{init}}=19   
\end{equation}
time units:
one time unit to prepare the qubits in $\ket{0}$, $\ket{+}$ and $\ket{T}$, and two QEC cycles each lasting 9 time units.
We find that under circuit noise of strength $p$ the lowest order contributions to the output infidelity $p_{\text{fail}}^{\text{init}}$ and rejection probability $p_{\text{rej}}^{\text{init}}$ are
\begin{equation}
\label{eq:initialization}
p_{\text{fail}}^{\text{init}} = 6.07 p,\quad p_{\text{rej}}^{\text{init}} = 126 p.
\end{equation}

\subsubsection{Expansion and movement of patches}
\label{sec:movement}

We neglect any error introduced by expansion of the encoded $T$ states, i.e., while enlarging the distance of the base code from $d^{(i)}$ to $d^{(i+1)}$ between rounds $i$ and $i+1$, and while these patches are moved into the starting location for the $i+1$ round.
This is justified since errors are suppressed throughout the expansion similarly as during error correction of the distance-$d^{(i)}$ code.
We also neglect any additional time overhead introduced by the expansion and movement of patches.
This is justified since the expansion can be done within the $d^{(i)}$ QEC rounds, and the swaps needed to move the outputs each take just one QEC cycle, which we expect can be done during the $d^{(i+1)}$ QEC rounds needed to prepare the $\ket{\overline{0}}$ and  $\ket{\overline{+}}$ states at the start of the state distillation circuit; see \fig{15to1DistillationNearestNeighbor}.

\subsubsection{15-to-1 state distillation circuit}
\label{sec:noisy-distillation}

\begin{figure}
\includegraphics[width=.65\columnwidth, angle=0]{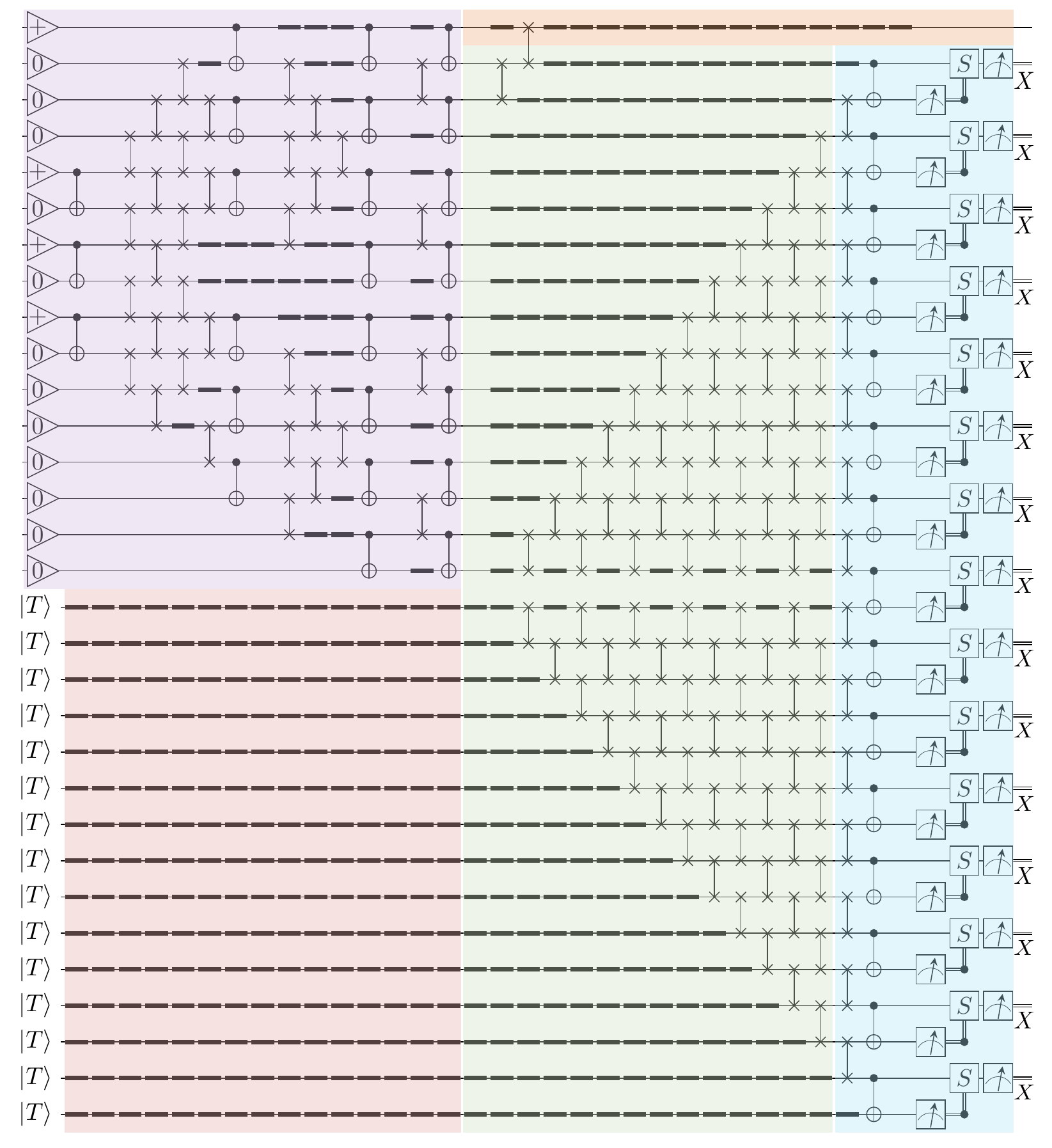}
\caption{
The 15-to-1 state distillation protocol implemented on the logical level using nearest-neighbor operations in a stack of 2D color codes.
The 31 logical qubits are encoded in distance-$d$ 2D color codes which are stacked on top of one another.
Logical idle locations (thick wire segments), SWAPs and CNOTs last 1, 1 and 2 QEC cycles respectively.
The 15 input encoded $T$ states
remain idle for $d+16$ QEC cycles (pink) while the Reed-Muller state is prepared (lilac).
We assume the logical $\ket{0}$ and $\ket{+}$ states are only prepared (indicated by a triangle) when they are needed, taking $d$ QEC cycles.
A shuffle circuit (green) which lasts 14 QEC cycles interleaves the encoded $T$ states with the qubits in the Reed-Muller state, where they undergo gate teleportation and measurement (blue) in 2 QEC cycles.
The distilled encoded $T$ state is in the top wire (orange).
}
\label{fig:15to1DistillationNearestNeighbor}
\end{figure}

Here, we analyze the 15-to-1 scheme run on logical qubits encoded in patches of 2D color code with distance $d$ arranged in a 3D stack.
The logical circuits are analysed using the effective noise model in \sec{2DCCOptimizedSetup}, which takes two parameters:
the distance $d$ of the 2D color codes used, and the strength $p$ of the underlying circuit noise.
Since we allow Clifford operations only between adjacent patches in the stack,
we have to appropriately modify the state distillation circuit; see~\fig{15to1DistillationNearestNeighbor}.
In our analysis, we only keep track of errors up to order $q^3$ and $\overline{p}$, which are of similar order of magnitude in the regime of interest, where $q$ is the infidelity of the input encoded $T$ states, and $\overline{p}$ is the noise strength in the effective noise model.
This splits the analysis into looking at error either in the encoded $T$ states alone, or in the Clifford operations alone.

{\bf Noisy $T$ states.---}First, we briefly review the effect of noise on $T$ states \cite{Bravyi2005}. 
As described in \sec{noise}, we simplify the form of noise assumed on a $T$ state by twirling, which in this case corresponds to randomly applying the Clifford $XS^\dagger \propto \ket{T} \bra{T}- \ket{T^\perp} \bra{T^\perp}$ with probability $1/2$.
Recall that single-qubit Clifford gates can be done instantaneously and perfectly by frame tracking in the 2D color code as described in \sec{logicalOperations2DCC}.
This twirling forces the noisy $T$ state to be of the form $\rho = (1-q)\ket{T}\bra{T} +q\ket{T^\perp}\bra{T^\perp}$, or equivalently that each $T$ state is afflicted by a $Z$ error with probability $q$.
The noise on the set of $15$ input noisy $T$ states is therefore represented as a $Z$-type Pauli error $E$ occurring with probability $q^{|E|}(1-q)^{15-|E|}$. 
The protocol will reject if $E$ is a detectable error for the punctured Reed-Muller code, which has distance $d=3$ for $Z$-type operators.
The protocol results in a failure if and only if $E$ is a nontrivial logical operator.
Explicit enumeration shows that there are $35$ weight-3 $Z$-type logical operators, such that the contributions $p_{\text{rej}}$ and $p_{\text{fail}}$ due to $T$ errors are
\begin{eqnarray}
    p_{\text{rej}}^{T} =  15q,\quad
    p_{\text{fail}}^{T} = 35 q^3.
    \label{eq:pRejFailT}
\end{eqnarray}

{\bf Noisy Clifford operations.---}Now we consider the effect of noise in the Clifford operations \cite{jochym2013, brooks2013} in the state distillation circuit.
First, we analyze the faults that occur during the Reed-Muller state preparation (lilac in \fig{15to1DistillationNearestNeighbor}), which takes 16 QEC cycles to complete.
We propagate each fault as a logical Pauli operator through the circuit, and assume that every other operation acts perfectly, including the $\overline{T}$ gates.
By explicitly representing the state and operations as the vector and matrices of dimension $2^{16}$ and $2^{16} \times 2^{16}$, respectively, we find that the contributions are
\begin{eqnarray}
    p_{\text{rej}}^{\text{RM}} =   
    12.3~ \overline{p}_{\text{prep}}+ 73~ \overline{p}_{\text{idle}}+ 38.2~ \overline{p}_{\text{CNOT}},\quad p_{\text{fail}}^{\text{RM}} =  
    0.875 ~\overline{p}_{\text{idle}}+ 1.93~ \overline{p}_{\text{CNOT}}.
    \label{eq:pRejFailRM}
\end{eqnarray}

Next we consider faults in the 16 idle QEC cycles of the output qubit.
Note that there is a choice of which of the 16 qubits of the Reed-Muller state to puncture in the 15-to-1 protocol.
We simulated all 16 choices, and selected the third qubit as the output since it had the lowest contribution from $\overline{p}_{\text{CNOT}}$; see \app{haah-hastings-distillation}.
Any failure in any of these locations will result in an undetected failure
\begin{eqnarray}
    p_{\text{rej}}^{\text{out}} = 0,\quad
    p_{\text{fail}}^{\text{out}} = 16~\overline{p}_{\text{idle}}.
    \label{eq:pRejFailout}
\end{eqnarray}

By explicit calculation, we find the exact contribution to $p_{\text{rej}}$ and $p_{\text{fail}}$ of every Clifford fault location in \fig{15to1DistillationNearestNeighbor} according to our effective noise model.

\begin{figure}
\includegraphics[width=.4\columnwidth, angle=0]{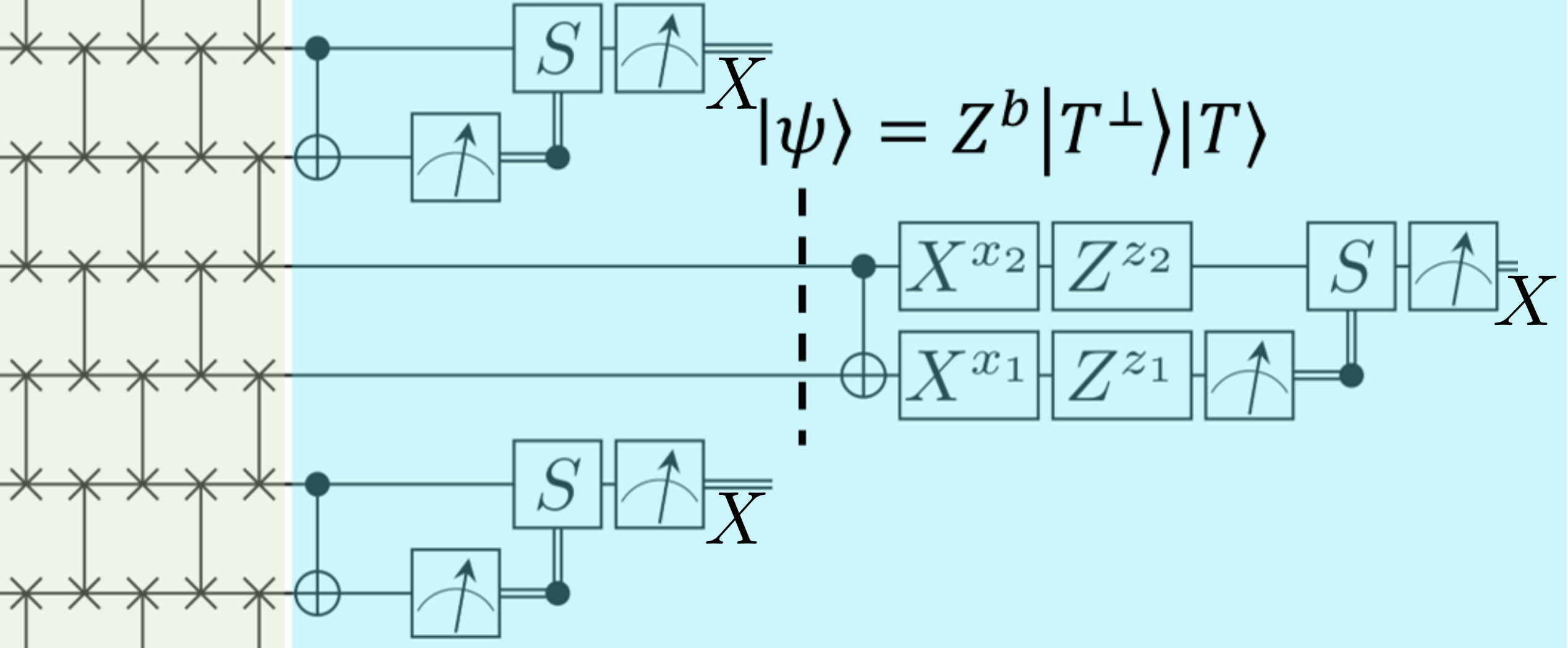}
\caption{
A part of the state distillation circuit with the inclusion of a Pauli error $X_1^{x_1}Z_1^{z_1}X_2^{x_2}Z_2^{z_2}$ on a pair of qubits.
Such an error can appear as a result of previous faults being propagated through the circuit.
Without loss of generality we assume that all the other measurements have been completed before this error, which ensures that the pair of qubits is decoupled from the rest of the system.
If the error-free outcome is $b$, then the error-free state must be $\ket{\psi} = (Z^b \ket{T^\perp}) \otimes \ket{T}$.
This simple 2-qubit circuit can then be analyzed for all 16 cases of the error $X_1^{x_1}Z_1^{z_1}X_2^{x_2}Z_2^{z_2}$.
}
\label{fig:RemainingFaults}
\end{figure}

Lastly we analyze the remaining fault locations.
These consist of the $15 (d + 16)$ idle locations involving qubits holding the $T$ states which remain idle during the production of the RM state (pink), the 420 idle and SWAP locations in the shuffle circuit (green), and the 15 CNOTs used to implement gate teleportation (blue) in \fig{15to1DistillationNearestNeighbor}.
A single fault in any of these locations will propagate to a Pauli operator $X_1^{x_1}Z_1^{z_1}X_2^{x_2}Z_2^{z_2}$ acting only on one pair of qubits as shown, where $x_1$, $z_1$, $x_2$ and $z_2$ each take values $0,1$ and the first qubit holds the $\ket{\overline{T}}$, and the second is from the Reed-Muller state.
To analyze the effect of such a Pauli operator, we imagine delaying the measurements on the affected pair of qubits until after the completion of the rest of the circuit; see \fig{RemainingFaults}.
At this point, all other measurements are completed, and if the Pauli operator were trivial, i.e., if $x_1=z_1=x_2=z_2=0$, the outcome $b\in\{0,1\}$ of the $X$ measurement would be determined by the previous outcomes since the $X$-stabilizers must be satisfied. 
Therefore, the pair of qubits must be completely unentangled with the rest of the system.
This tells us that none of these fault locations can result in a failure, but result in rejection if and only if the outcome of the $X$ measurement is modified by the Pauli operator $X_1^{x_1}Z_1^{z_1}X_2^{x_2}Z_2^{z_2}$.
We straightforwardly analyze this 2-qubit circuit with its pure initial state and for Pauli operator find the probability $p_{\text{flip}}(x_1,x_2,x_3,x_4)$ that the outcome is flipped, namely
\begin{eqnarray}
    p_{\text{flip}}(x_1,x_2,x_3,x_4)=\left\{
                 \begin{array}{ll}
                 0&~~~\text{if}~ x_1 = x_2~\text{and}~z_2=0,\\
                 1/2&~~~\text{if}~ x_1 \neq x_2~\text{and}~z_2=0,\\
                 1&~~~\text{if}~ x_1 = x_2~\text{and}~z_2=1,\\
                 1/2&~~~\text{if}~ x_1 \neq x_2~\text{and}~z_2=1.
                 \end{array} \right.
\end{eqnarray}
Then all that remains is to count the contribution to each of these Paulis from the aforementioned locations according to the effective noise model, which yields
\begin{eqnarray}
    p_{\text{rej}}^{\text{rem}} = (392 + 4.13 d)~ \overline{p}_{\text{idle}}+ 13.5~ \overline{p}_{\text{CNOT}},\quad p_{\text{fail}}^{\text{rem}} = 0.
    \label{eq:pRejFailrem}
\end{eqnarray}

The contributions from \eq{pRejFailT}, \eq{pRejFailRM}, \eq{pRejFailout} and \eq{pRejFailrem} combine to give the rejection and failure probability of the state distillation step
\begin{eqnarray}
p_{\text{rej}}^{\text{dist}} & = &  15q + 12.3~ \overline{p}_{\text{prep}}+ (466 + 4.13~ d)~ \overline{p}_{\text{idle}}+ 51.7~ \overline{p}_{\text{CNOT}},\label{eq:pRejDist}\\
p_{\text{fail}}^{\text{dist}} & = & 35 q^3 + 16.9~ \overline{p}_{\text{idle}}+ 1.93~ \overline{p}_{\text{CNOT}}. \label{eq:pFailDist}
\end{eqnarray}
The number of time units required to implement the state distillation circuit can be straightforwardly identified from \fig{15to1DistillationNearestNeighbor} and is equal to
\begin{equation}
\tau^{\text{dist}}(d) = 8 \left(d + 32 \right).    
\label{eq:distillationTime}
\end{equation}

\subsection{State distillation overhead}
\label{sec:distillation-overhead}

Here we estimate the overhead required for a $k$-round state distillation protocol using distances $\{d^{(1)},d^{(2)},\dots, d^{(k)} \}$ under circuit noise of strength $p$, as well as the infidelities output by each round $\{q^{(1)},q^{(2)},\dots, q^{(k)} \}$.
Note that $q^{(k)}$ is the infidelity of the encoded $T$ state produced by the overall protocol.

Recall that the first step of the state distillation protocol is to initialize encoded $T$ states in distance-$5$ 2D color codes; see \fig{StepsDistillation} and \sec{magic-state-initialization}.
Writing the infidelity of the state after initialization as $q^{(0)} = p_{\text{fail}}^{\text{init}}(p)$, the remaining infidelities are then calculated iteratively according to
\begin{eqnarray}
q^{(i)}&=& p_{\text{fail}}^{\text{dist}}(q^{(i-1)},p,d^{(i)}).
\end{eqnarray}

To estimate the overhead, it is useful to streamline our notation.
Recall from \eq{Nqubits2DCC}
that $N_{2\text{D}}(d) = (3 d^2-1)/2$ qubits are used to implement the distance-$d$ 2D color code.
For each state distillation round $i \in \{1,2,\dots k\}$, let $r^{(i)}=\left[1-p_{\text{rej}}^{\text{dist}}(p,q^{(i-1)},d^{(i)}) \right]$ be the acceptance probability, $R^{(i)}=15$ be the number of $\ket{\overline{T}}$s required assuming acceptance, and $\alpha^{(i)} = 31/15$ be the number of logical qubits needed per input $\ket{\overline{T}}$ in the state distillation circuit.
To account for the initialization step,
we also set $d^{(0)}=3$, $r^{(0)}=\left[1-p_{\text{rej}}^{\text{init}}(p) \right]$, $R^{(0)}=1$ and $\alpha^{(0)}=1$.

\begin{figure}[h]
	(a)\hspace*{-5mm}\includegraphics[width=.45\textwidth]{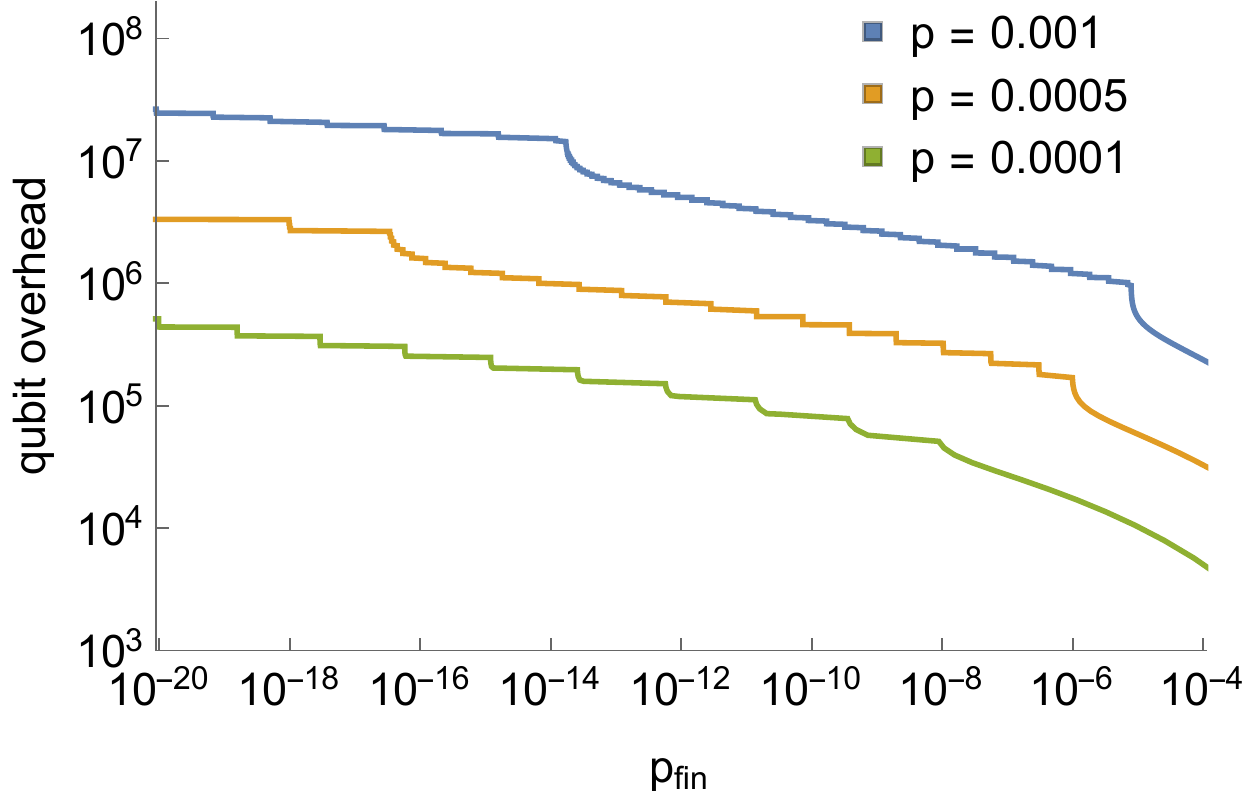}
	\quad\quad\quad
	(b)\hspace*{-5mm}\includegraphics[width=.45\textwidth]{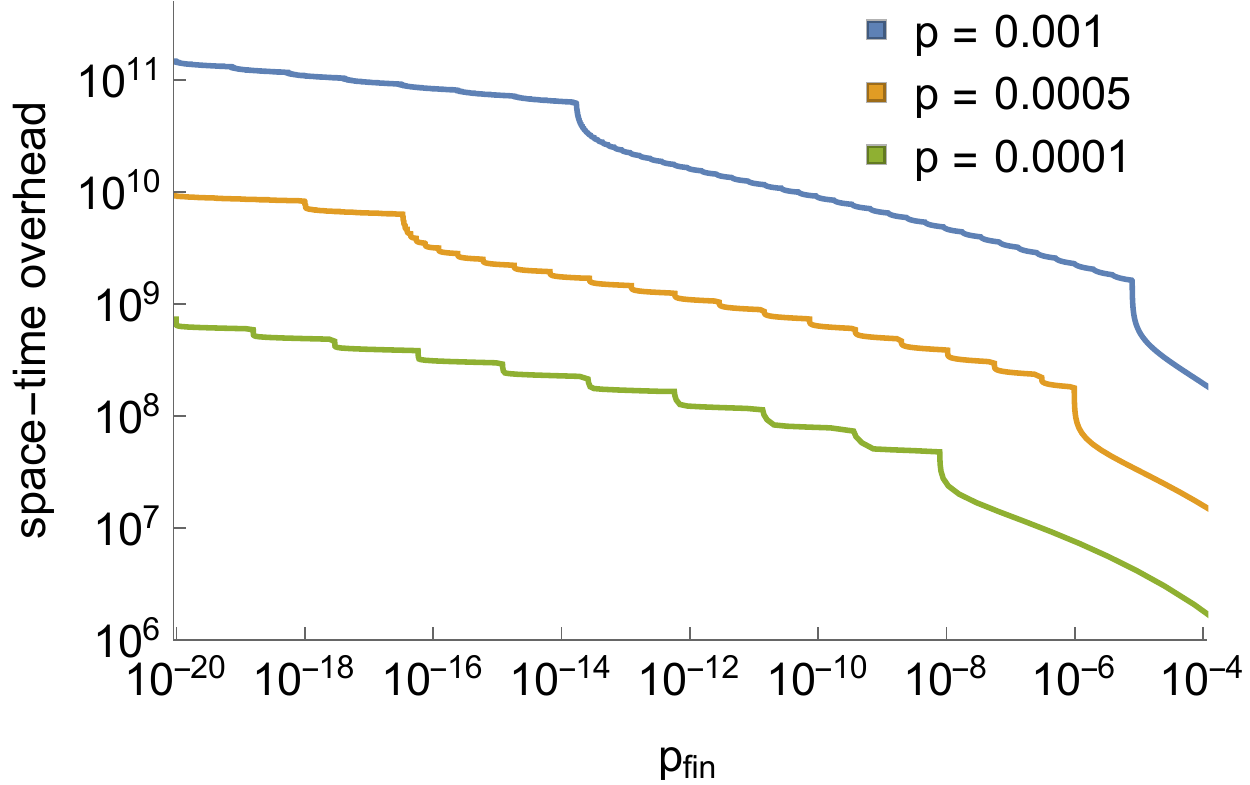}
	\caption{
	(a) The qubit and (b) space-time overhead of state distillation as a function of the infidelity $p_\text{fin}$ of the output $T$ state. 
	}
	\label{fig:overhead-distillation-15-to-1}
\end{figure}

We can calculate the expected number of physical qubits required in each round.
We imagine preparing a large number of output $T$ states, and thus we can talk about the average overhead.\footnote{
If we were to produce just one output $T$ state, then the overhead would slightly increase as we would need to guarantee that with high probability there are sufficiently many $T$ states at each level of the state distillation protocol.}
Since the state distillation protocol in the last round succeeds with probability $r^{(k)}$, on average it needs $R^{(k)}/r^{(k)}$ input $\ket{\overline{T}}$s.
We therefore require on average $\alpha^{(k)}N_{2\text{D}}(d^{(k)})R^{(k)}/r^{(k)}$ qubits for the last round.
To supply the $k$th round, the $(k-1)$th round must therefore output $R^{(k)}/r^{(k)}$ $\ket{\overline{T}}$'s on average, which requires $\alpha^{(k)}N_{2\text{D}}(d^{(k)})R^{(k-1)}R^{(k)}/(r^{(k-1)}r^{(k)})$ physical qubits, and so on.
The qubit overhead $N_{\text{SD}}$ of the state distillation protocol can then be found as the number of qubits needed in the most qubit-expensive state distillation round, namely
\begin{equation}
N_{\text{SD}} = \max_{i=1,\ldots, k} \left[\alpha^{(i)}N_{2\text{D}}\big(d^{(i)}\big)
\prod_{j=i}^k \frac{R^{(j)}}{r^{(j)}}\right].
\end{equation}
The time required for state distillation is 
\begin{equation}
\tau_{\text{SD}} = \tau^{\text{init}}+\sum_{i=1}^k \tau^{\text{dist}}(d^{(i)}).
\end{equation}
The space-time overhead is then simply $N_{\text{SD}} \tau_{\text{SD}}$.
For various values of $p$, we run a simple search over number of rounds $k \in \{1,2,3 \}$ and distances $\{d^{(1)},d^{(2)},\dots, d^{(k)} \}$ to distill a target infidelity $p_{\text{fin}}$ for a low space or space-time overhead; see \fig{overhead-distillation-15-to-1}.

Let us briefly remark on the threshold of this 15-to-1 state distillation using the 2D color code.
This threshold is the noise strength above which arbitrarily low target infidelity cannot be achieved, irrespective of the number of state distillation rounds. There are two main features that could limit the state distillation threshold.
Firstly, error correction in the 2D color code could be the limiting factor.
Namely, if $p$ is above the circuit-noise threshold of $0.37(1)\%$, one will be unable to achieve logical states with arbitrarily low infidelity.
Secondly, the first round of state distillation can be a bottleneck if the infidelity of the initial encoded $T$ state cannot be improved by it.
For simplicity, we assume that the encoded Clifford operations execute perfectly, and solve for the critical infidelity $q_c$ satisfying $35q_c^3=q_c$.
Using \eq{initialization}, we obtain a corresponding critical error rate of $1/(3.93 \cdot \sqrt{35}) \approx 4\%$. We conclude that error correction is the bottleneck, so we estimate the threshold for state distillation with this scheme to be $0.37(1)\%$.

\clearpage
\section{Further insights into 3D color codes}
\label{sec:insights-3DCC}
In this section we present some insights into 3D color codes that are useful for code switching.
In \sec{code-switching-high-level}, we describe a simple approach to switch between the 2D and 3D color codes. 
At the core of this approach is the fact that in a particular state of the gauge qubits, the 3D subsystem color code can be viewed as a collection of 2D color codes.
In \sec{physics-gauge-flux}, we describe some relevant features of the gauge operators of the 3D color code which will be relevant for gauge-fixing and also for the simulation of noise upon the application of the transversal $\overline{T}$ gate.
Our discussion closely follows material in \cite{Bombin2018}.
In \sec{3D-color-code-perfect-measurements}, we generalize the restriction decoder to correct $Z$-errors in the 3D color code with boundaries, which will later be used in our code switching protocol.

\subsection{A simple way to switch between 2D and 3D color codes}
\label{sec:code-switching-high-level}

First we recall the general procedure for \textit{gauge fixing} from a subsystem code with gauge group $\mathcal{G}$ to a stabilizer code with stabilizer group $\mathcal{S}' \subseteq \mathcal{G}$, where both codes share a set of bare logical operators.
We require that the stabilizer group $\mathcal{S}$ of the subsystem code, which is the centralizer of $\mathcal{G}$ in the Pauli group intersected with $\mathcal{G}$ modulo the phase, i.e., $\mathcal{S} = (\mathcal{Z}(\mathcal{G})\cap\mathcal{G})/\langle iI\rangle$, is contained in $\mathcal{S}'$, namely $\mathcal{S} \subseteq \mathcal{S}'$.
Consider any state $\ket{\psi}$ in the code space of the subsystem code, which by definition is a $(+1)$-eigenstate of all elements of $\mathcal{S}$. 
To switch from $\mathcal{G}$ to $\mathcal{S}'$, we first measure a generating set of $\mathcal{S}' \setminus \mathcal{S}$.
A subset of those measured generators may have $-1$ outcomes, but there must exist an element $g \in \mathcal{G}$ which anti-commutes with precisely that subset of generators.
Hence after applying $g$, all the stabilizers of $\mathcal{S}'$ will be satisfied, and since $g$ commutes with the bare logical operators, the logical state is unaffected, which completes the transfer from $\mathcal{G}$ to $\mathcal{S}'$.
This procedure is named gauge fixing since it involves measuring some (initially unsatisfied) gauge operators, and fixing them to be $+1$.

\begin{figure}
	(a)\includegraphics[height=.24\textwidth]{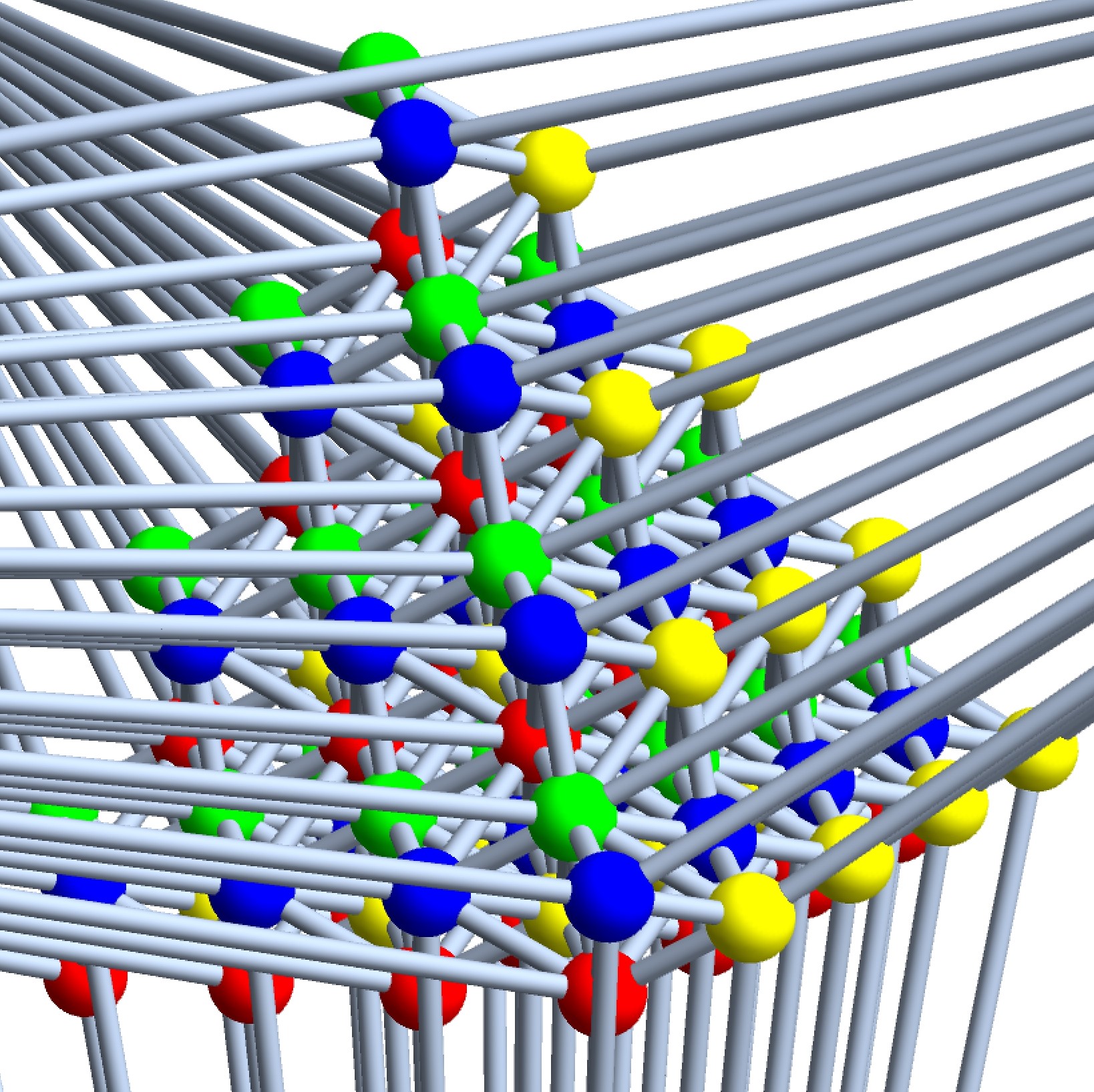}
	\quad
	(b)\includegraphics[height=.24\textwidth]{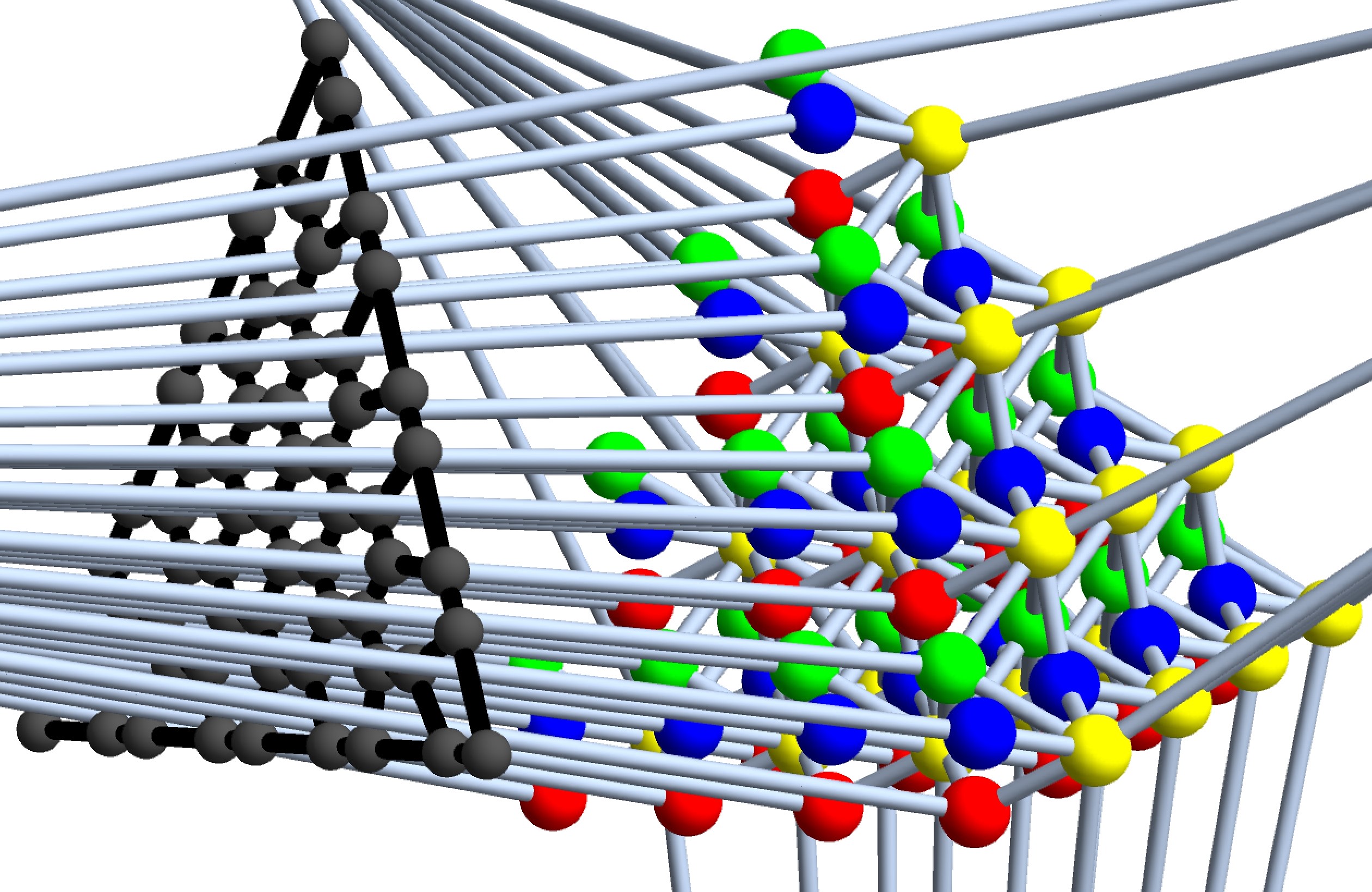}
	\quad\quad
	\includegraphics[height=.24\textwidth]{figures/2DCCSphere24Qubits}
	\hspace*{-.26\textwidth}(c)\hspace*{.24\textwidth}
	\caption{
	    (a) In one gauge fixing of the 3D subsystem color code $\mathcal{G}_\text{sub}$ on the lattice $\mathcal{L}_\text{3D}$ (shown in \fig{lattices}{b}), all $Z$-edges (depicted by light struts) are satisfied.
	    This corresponds to the 3D stabilizer color code $\mathcal{S}_\text{3D}$.
		(b) In another gauge fixing of $\mathcal{G}_\text{sub}$, all $X$- and $Z$-edges incident to $Y$ vertices in $\face 0 {\mathcal{L}_\text{3D}}$ (depicted by light struts) are satisfied.
		This corresponds to the 2D color code $\mathcal{S}_\text{2D}$ on the qubits near the boundary vertex $v_Y$
		and the 2D spherical color codes $\mathcal{S}_{\text{int}}$ on other qubits.
		We depict the primal lattice of the 2D color code on the lattice $\mathcal{L}_\text{2D}$ (shown in \fig{lattices}{a}) in black.
		(c) Around each $Y$ vertex in $\facex 0 {\mathcal{L}_\text{3D}}$, there is the 2D spherical color code, 
		whose primal lattice we depict in black.
	}
	\label{fig:3DColorCodeBoundary}
\end{figure}

Central to switching between color codes is the fact that both the 3D stabilizer color code and the 2D color code can be viewed as gauge fixings of the 3D subsystem color code; see \fig{3DColorCodeBoundary}. 
The gauge and stabilizer groups $\mathcal{G}_{\mathrm{sub}}$ and $\mathcal{S}_{\mathrm{sub}}$ for the 3D subsystem color code, and the stabilizer group $\mathcal{S}_{\text{3D}}$ for the 3D stabilizer color code are:
\begin{eqnarray}
&\mathcal{G}_{\mathrm{sub}} = \langle X(e), Z(e) \mathrel{|} \forall e \in \facex  1{\mathcal{L}_\text{3D}} \rangle,\quad \mathcal{S}_{\mathrm{sub}} = \langle X(v), Z(v) \mathrel{|} \forall v \in \facex 0 {\mathcal{L}_\text{3D}} \rangle,&\\
&\mathcal{S}_{\text{3D}} = \langle X(v), Z(e) \mathrel{|} \forall v \in \facex 0 {\mathcal{L}_\text{3D}}, e \in \facex 1 {\mathcal{L}_\text{3D}} \rangle.&
\end{eqnarray}
We can define the stabilizer group $\mathcal{S}_{\text{2D}}$ of the 2D color code within the 3D lattice $\mathcal{L}_\text{3D}$ since $\mathcal{L}_\text{2D}$ is  `contained' in $\mathcal{L}_\text{3D}$; see \fig{3DColorCodeBoundary}(b). Namely,
\begin{equation}
\mathcal{S}_\text{2D} = \langle X(e), Z(e) \mathrel{|}  \forall e \in \facex 1 {\mathcal{L}_\text{3D}}\text{: $e$ is incident to $v_Y$}\rangle,
\end{equation}
where $v_Y$ is the $Y$ boundary vertex.
Let us define the stabilizer group $\mathcal{S}_{\text{int}}$ supported on the qubits which are not near $v_Y$ as follows
\begin{equation}
\mathcal{S}_{\text{int}} = \langle X(e), Z(e) \mathrel{|}  \forall e \in \facex 1 {\mathcal{L}_\text{3D}}\text{: $e$ is incident to any interior $Y$ vertex}\rangle.
\end{equation}
We can think of $\mathcal{S}_{\text{int}}$ as the group generated by the stabilizers of the 2D spherical color codes centered around every $Y$ interior vertex of $\mathcal{L}_\text{3D}$; see \fig{3DColorCodeBoundary}(c).
Note that every 2D spherical color code encodes zero logical qubits.

It is straightforward to see that $\mathcal{G}_\text{sub}$ contains both $\mathcal{S}_\text{3D}$ and $\langle\mathcal{S}_\text{2D},\mathcal{S}_{\text{int}}\rangle$.
Moreover, $\mathcal{S}_\text{sub}\subseteq \mathcal{S}_\text{3D}$ and $\mathcal{S}_\text{sub}\subseteq \langle\mathcal{S}_\text{2D},\mathcal{S}_{\text{int}}\rangle$ since vertex operators $X(v)$ and $Z(v)$ can be formed by multiplying operators $X(e)$ and $Z(e)$ on all the edges of the same color incident to $v$.
Also note that there is a shared representation of bare logical operators for all three codes (for example $X$ and $Z$ applied to every qubit in $\mathcal{L}_\text{2D}$).
Therefore, to move from the subsystem code $\mathcal{G}_\text{sub}$ to either of the stabilizer codes, i.e., $\mathcal{S}_\text{3D}$ or $\langle\mathcal{S}_\text{2D},\mathcal{S}_{\text{int}}\rangle$, gauge switching can be used.
Switching from either of the stabilizer codes to the subsystem code requires no action,
since any state which is a $(+1)$-eigenstate of every element of $\mathcal{S}_{\text{3D}}$ or $\langle\mathcal{S}_\text{2D},\mathcal{S}_{\text{int}}\rangle$ must also be a $(+1)$-eigenstate of every element in $\mathcal{S}_\text{sub}$.

This example of gauge fixing is sometimes referred to as a \textit{dimensional jump} because the logical information is moved between codes defined on 2D and 3D lattices \cite{bombin2016}.

\subsection{Physics of the gauge flux in 3D color codes}
\label{sec:physics-gauge-flux}
Here we consider general features of the gauge operators of the 3D subsystem color code.
Our discussion closely follows material in \cite{Bombin2018}.
Suppose there is some $X$ error $\epsilon\subseteq \Delta_3(\mathcal{L})$ in the system.
We define the $Z$-type \textit{gauge flux} $\gamma\subseteq \facex 1 {\mathcal{L}}$ to be the subset of interior edges which would return $-1$ outcomes if all $Z$ edges in the system were measured perfectly.
Since each edge has one color from $\mathcal{K} = \{RG,RB,RY,GB,GY,BY\}$, we distinguish six types of the flux and write
\begin{equation}
    \gamma = \sum_{K\in\mathcal{K}} \gamma^K. 
\end{equation}

\begin{figure}
	(a)\includegraphics[width=.4\columnwidth]{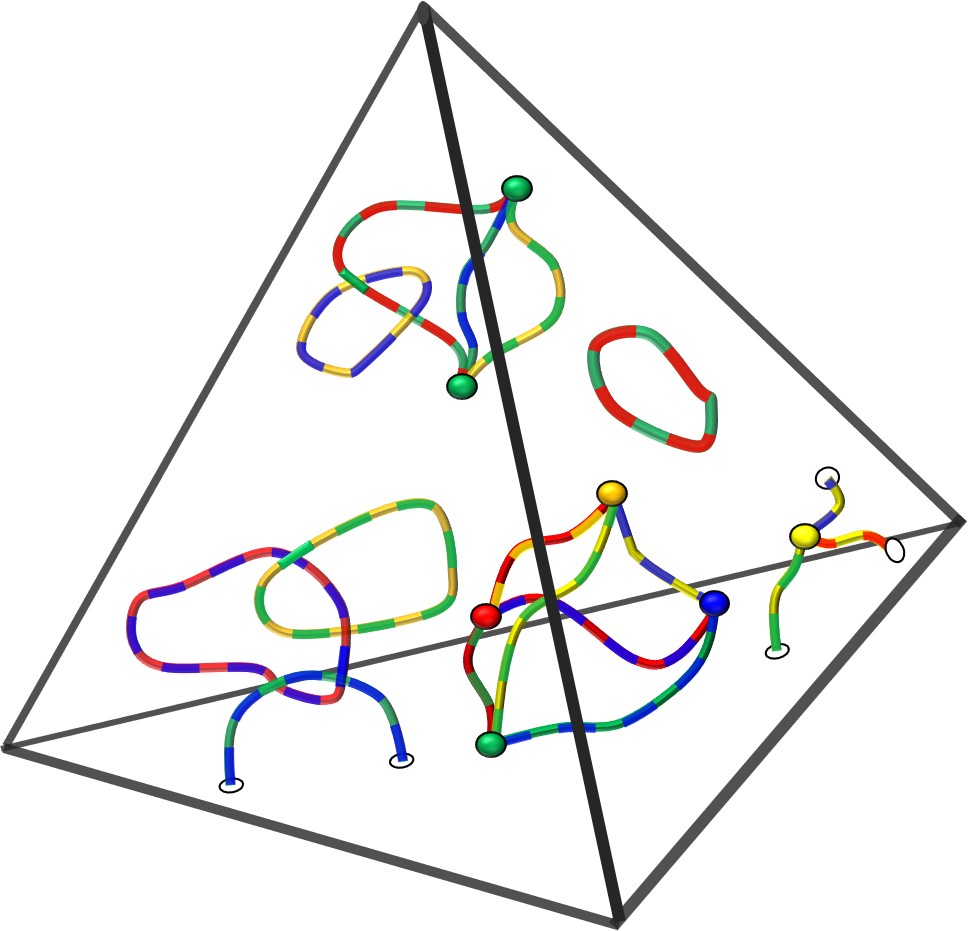}\qquad\qquad
	(b)\includegraphics[width=.4\columnwidth]{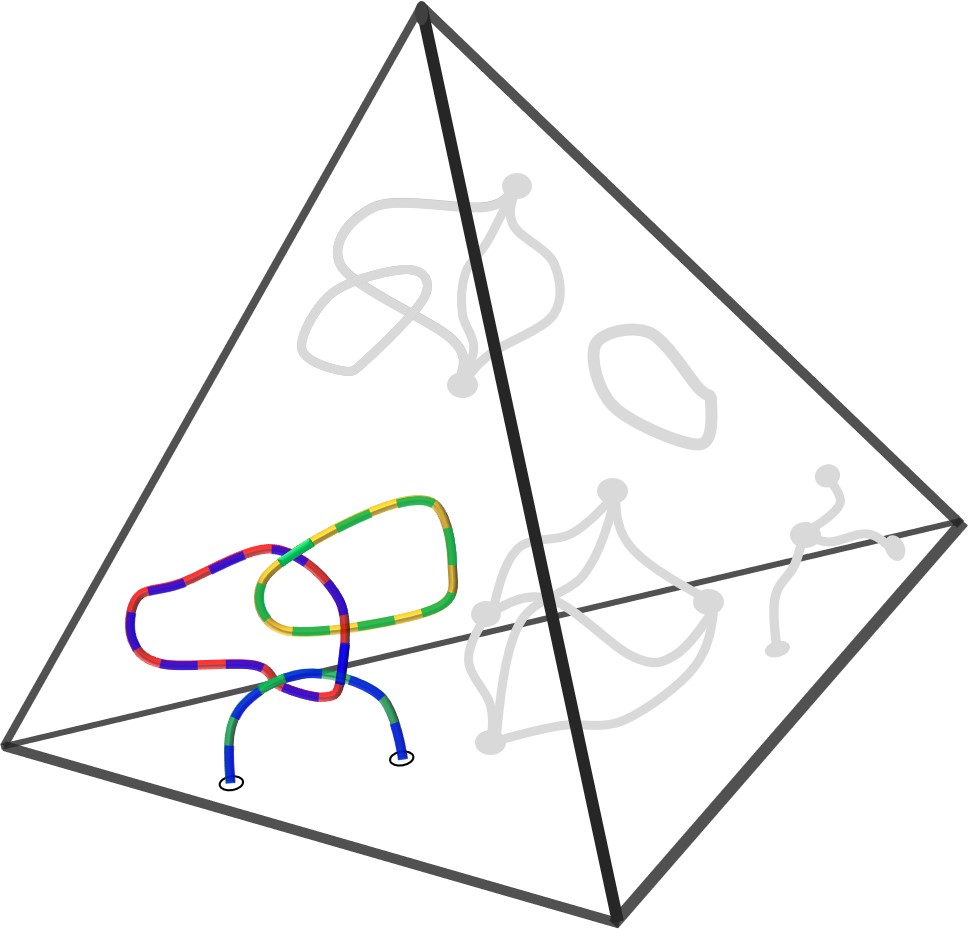}
	\qquad
	\caption{
	A schematic representation of the $Z$-gauge flux $\gamma$ in the bulk (enclosed by the black tetrahedron) of the 3D subsystem color code on the lattice $\mathcal{L}_\text{3D}$ in \fig{lattices}(b).
	(a) The flux $\gamma$ consists of strings of six different colors.
	There are seven branching points of $\gamma$ (depicted as red, green, blue and yellow vertices).
	The flux $\gamma$ has to satisfy the Gauss law in the bulk, and can terminate at any of the boundary vertices of $\mathcal{L}_\text{3D}$.
	(b) We highlight one linked component of $\gamma$, which contains three connected components of $\gamma$ that are linked.
	}
	\label{fig:gauge-flux}
\end{figure}

Although the flux $\gamma$ can be random, it has to form a collection of strings, which may branch and can terminate only at the boundary of the lattice; see \fig{gauge-flux}.
A local constraint capturing this behavior, which we call the \textit{Gauss law}, can be stated as follows.
Let $K_1,K_2,K_3,K_4\in\{R,G,B,Y\}$ be four different colors.
Then, for any vertex $v\in\facex 0 {\mathcal{L}}$ of color $K_1$, the number of edges of $\gamma$ of color $K_1 K_2$ or $K_1 K_3$ and incident to $v$ has to be even, i.e.,
\begin{equation}
\label{eq_gauss}
    \left|\gamma^{K_1K_2}\rest v + \gamma^{K_1K_3}\rest v\right| \equiv 0 \mod 2.
\end{equation}

This local constraint arises from the redundancies among gauge generators.
Namely, in order to form a stabilizer generator $Z(v)$ identified with the vertex $v$, we can multiply all the gauge generators on edges of color $K_1K_2$ incident to $v$.
Alternatively, we can obtain $Z(v)$ as the product of all gauge generators on edges of color $K_1K_3$ incident to $v$.
Since the parity of the number of $-1$ measurement outcomes among those gauge generators in both cases is the same, we recover Eq.~\eqref{eq_gauss}.

Note that when the stabilizer $Z(v)$ is violated (indicating the presence of some $X$ error), then the number of $-1$ outcomes for gauge generators on edges incident to $v$ of color $K_1K_2$ has to be odd, i.e., $\left|\gamma^{K_1K_2}\rest v \right | \equiv 1 \mod 2$.
In such a case, we call the vertex $v$ a branching point of $\gamma$, as three different flux types $\gamma^{K_1K_2}$, $\gamma^{K_1K_3}$ and $\gamma^{K_1K_4}$ meet at $v$.
We remark that to perform error correction with the 3D subsystem color code, one can use the information about the branching points of the flux $\gamma$.
If the flux $\gamma$ has no branching points at any interior vertex, then all $Z$ stabilizers are satisfied and there always exists an $X$-type gauge operator which anti-commutes with precisely those $Z$ gauge generators in $\gamma$. 

\begin{figure}[ht]
	(a)\includegraphics[width=.25\textwidth]{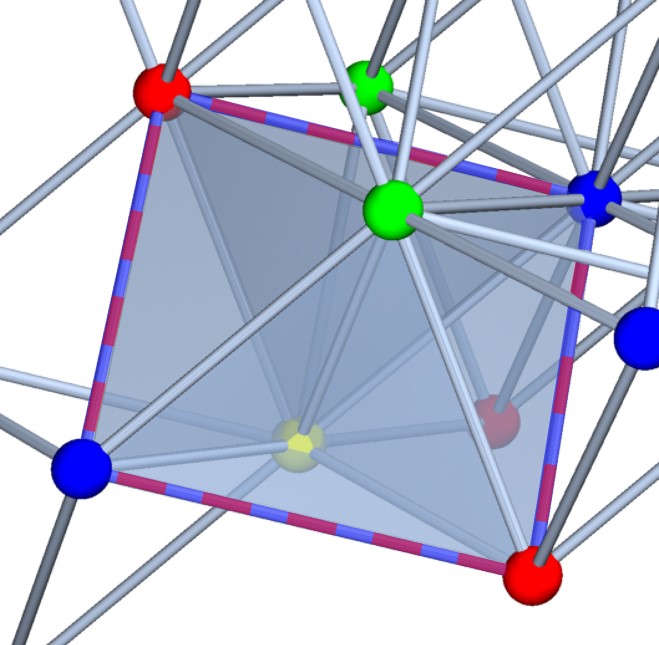}
	\quad\quad\quad
	(b)\includegraphics[width=.25\textwidth]{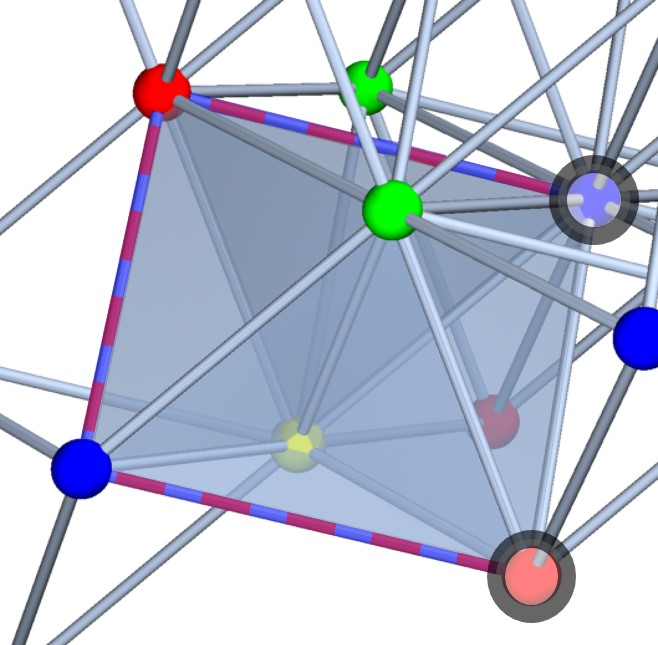}
	\caption{
	(a) The flux $\gamma$ (highlighted $RB$ edges) due to a single $GY$-edge generator (shaded four tetrahedra) contains only $RB$ edges.
	This set of edges $\gamma$ satisfies the Gauss law.
	(b) A noisy measurement of the flux may not be valid, meaning it does not satisfy the Gauss law. 
	Here $\widetilde\gamma$ has two violation points (highlighted $R$ and $B$ vertices) at which the Gauss law does not satisfied.
	}
	\label{fig:gauge-flux-validity}
\end{figure}

If an edge set satisfies the Guass law, we say that it is \textit{valid}.
An edge set which does not satisfy the Gauss law is said to be \textit{invalid}, and we refer to all vertices which violate the Gauss law as \textit{violation points}.
When the gauge flux is measured with noisy circuits, there can be errors in the reported \textit{noisy gauge flux} causing it to be invalid; see \fig{gauge-flux-validity}.

Now, we discuss the structure of the flux $\gamma$ which satisfies the Gauss law.
First, we can find a decomposition of $\gamma$ in terms of its connected components, i.e.,
\begin{equation}
    \gamma = \sum_{j=1}^a \gamma_j.
\end{equation}
By definition, different connected components are disjoint, i.e., $\gamma_j \cap \gamma_k = \emptyset$ for $j\neq k$.

Then, we define a linked component of $\gamma$ as a subset of all connected components of $\gamma$, which are linked (in the sense of knot theory); see \fig{gauge-flux}(b).
Finally, we can decompose $\gamma$ as a disjoint union of its linked components
\begin{equation}
    \gamma = \sum_{i=1}^b \lambda_i.
\end{equation}
Note that each $\lambda_i$ is the sum of some $\gamma_j$'s.

For convenience, we introduce a function $\mathrm{col}:\face 0 {\mathcal{L}} \rightarrow \mathbb{Z}^3_2$, which for each vertex $v$ returns its color, where we set $R = (1,0,0)$, $G = (0,1,0)$, $B = (0,0,1)$ and $Y = (1,1,1)$.

For any linked component $\lambda_i$ of $\gamma$ we refer to a subset $\sigma\subseteq\face 0 {\lambda_i}$ as an excitation configuration for $\lambda_i$ and call $\sum_{v\in\sigma} \mathrm{col}(v)$ the total charge of $\sigma$.
We denote \textit{the collection of excitation configurations} $\Sigma(\lambda_i)$ for $\lambda_i$ with the neutral total charge as follows
\begin{eqnarray}
\Sigma(\lambda_i) = \biggl\{ \sigma \subseteq \face 0 {\lambda_i} \mathrel{\bigg|} \sum_{v\in\sigma} \mathrm{col}(v) = (0,0,0) \biggr\}.
\end{eqnarray}
Writing the linked component $\lambda_i$ in terms of its connected components $\lambda_i = \gamma_{i_1}+\gamma_{i_2}+\dots+\gamma_{i_k}$, we also introduce \textit{the collection of excitation configurations without the linking charge}
\begin{eqnarray}
\Sigma'(\lambda_i) = \Sigma(\gamma_{i_1})\times \Sigma(\gamma_{i_2})\times \dots \times \Sigma(\gamma_{i_k}).
\end{eqnarray}
We remark that the linking charge is a charge which can be transferred between two connected components $\gamma_i$ and $\gamma_j$, which are linked; see~\cite{Bombin2018}.
Note that $\Sigma'(\lambda_i)$ is contained in $\Sigma(\lambda_i)$, i.e., $\Sigma'(\lambda_i) \subseteq \Sigma(\lambda_i)$, and they coincide whenever the linked component $\lambda_i$ consists of a single connected component.

\subsection{Restriction decoder for 3D color codes with boundaries}
\label{sec:3D-color-code-perfect-measurements}

Here we provide details on correcting $Z$ type errors in the 3D stabilizer color code with perfect measurements, which is needed for the final step of the code switching protocol.
We seek an efficient decoder for the 3D color code with good performance.
Our approach is to adapt the restriction decoder from Ref.~\cite{kubica2019}, which was originally defined for lattices with no boundary, to the tetrahedral lattice $\mathcal{L}_{\text{3D}}$.
We also apply additional minor modifications to improve the performance.
In what follows we briefly review the restriction decoder and describe our modifications.

Let $\sigma\subseteq\facex 0 {\mathcal{L}_\text{3D}}$ be the syndrome of the 3D stabilizer code on the tetrahedral lattice $\mathcal{L}_\text{3D}$.
We pick one color, say $Y$, and for each color $K\in\{R,G,B\}$ we separately analyze the restricted syndrome $\sigma^{KY} \subseteq \facex 0 {\mathcal{L}_\text{3D}^{KY}}$ within the restricted lattice $\mathcal{L}^{KY}_\text{3D}$.
If $|\sigma^{KY}| \equiv 1 \mod 2$, then we add the boundary vertex $v_Y$ to $\sigma^{KY}$.
Next, we find a subset of edges $E^{KY}\subseteq\face 1 {\mathcal{L}_\text{3D}^{KY}}$, which provides a pairing of vertices of $\sigma^{KY}$ within the restricted lattice $\mathcal{L}^{RY}_\text{3D}$.
Note that we can find a pairing of minimal weight by using the MWPM algorithm.
Also, the weight of the edge connecting the boundary vertices $v_Y$ and $v_K$ is set to zero.

After finding the pairing $E = E^{RY}+E^{GY}+E^{BY}$ we apply a local lifting procedure to every $Y$ vertex in the interior of $\mathcal{L}_\text{3D}$.
Namely, for every $Y$ vertex $v \in \facex 0 {\mathcal{L}_\text{3D}}$ we find any subset of tetrahedra $\tau(v)\subseteq \bnd 0 3 v$ in the neighborhood of $v$, whose $1$-boundary locally matches $E$, i.e., $(\partial_{3,1}\tau(v))\rest v = E|_v$.
We emphasize that all such choices of $\tau(v)$ result in operators $Z(\tau(v))$, which may differ only by a stabilizer operator.

To lift the boundary vertex $v_Y$, we need to adapt the original restriction decoder from Ref.~\cite{kubica2019}. 
Since $\mathcal{L}_\text{2D}$ is a sublattice of $\mathcal{L}_\text{3D}$ (see \fig{3DColorCodeBoundary}(b)), one can show that the problem of finding $\tau(v_Y)\subseteq \bnd 0 3 v_Y$, which satisfies $(\partial_{3,1}\tau(v_Y))\rest {v_Y} = E|_{v_Y}$, is equivalent to the problem of decoding the 2D color code defined on the facet of the tetrahedral lattice $\mathcal{L}_\text{3D}$ near $v_Y$.
We remark that different choices of $\tau(v_Y)$ may lead to operators $Z(\tau(v_Y)$ differing by a logical operator.
We use the projection decoder for the 2D color code as described in \sec{decoder}.
Finally, the correction operator is found as
$\prod_{v\in\face 0 {\mathcal{L}_\text{3D}}}Z(\tau(v))$,
where the product is over all $Y$ vertices, including the boundary vertex $v_Y$.
Note that this adaption to accommodate a lattice boundary in 3D is analogous to an adaption presented in \cite{Chamberland2020} for the 2D case.

\begin{figure}
	(a)\hspace*{-5mm}\includegraphics[width=.44\columnwidth]{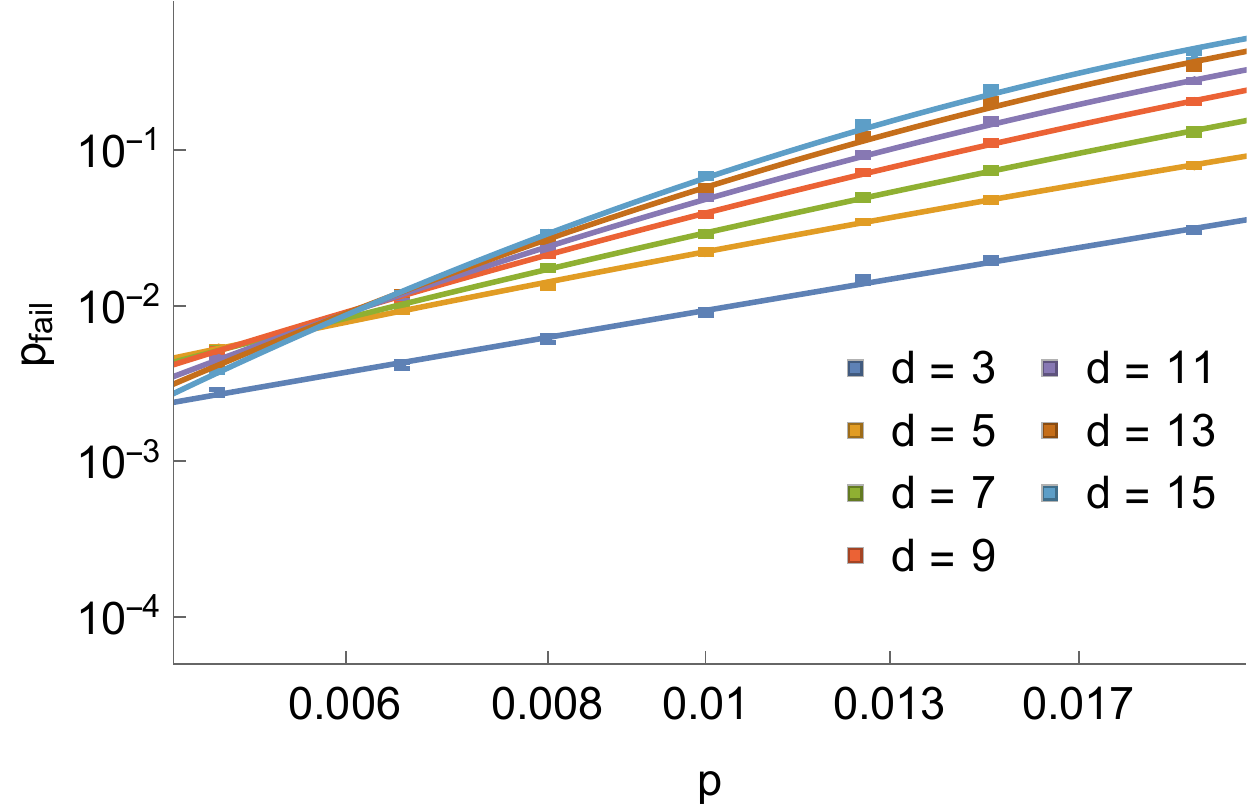}
	\quad\quad\quad
	(b)\hspace*{-5mm}\includegraphics[width=.44\columnwidth]{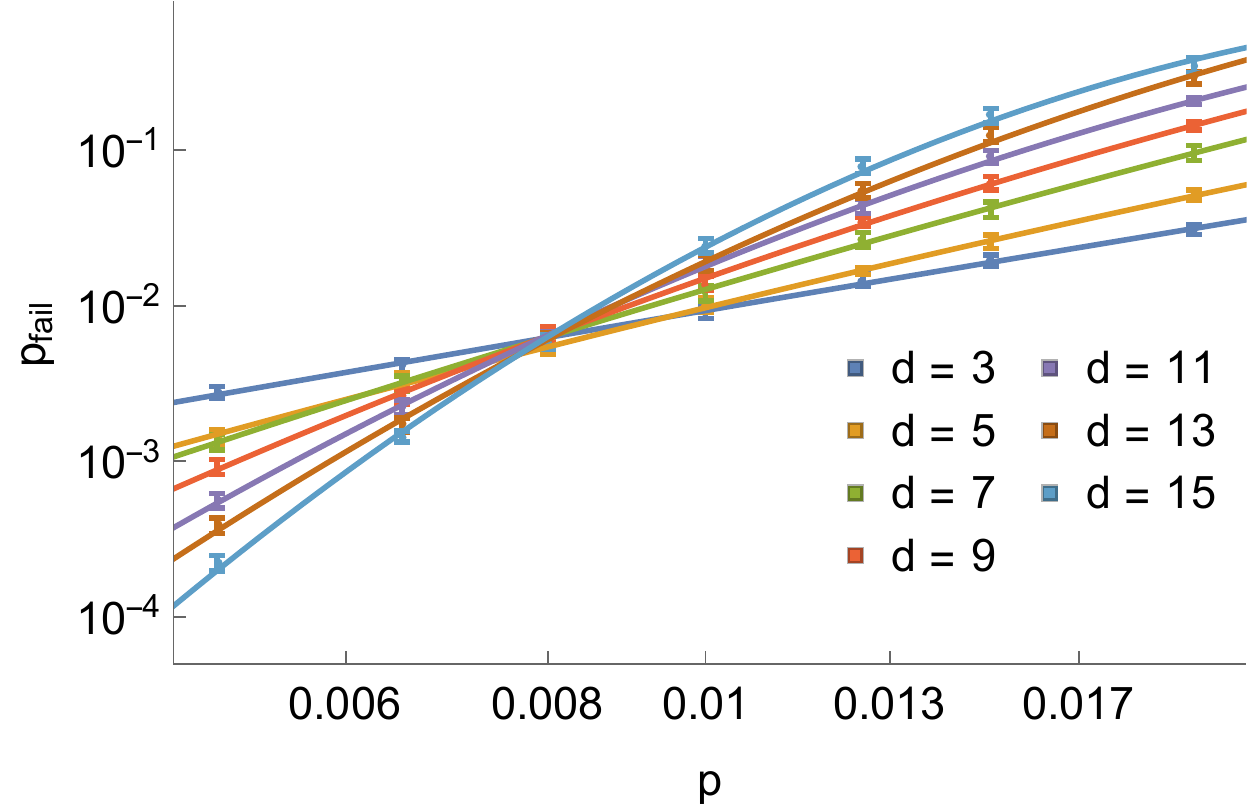}
	\qquad
	\caption{
	Performance of the restriction decoder for the 3D color code adapted to the tetrahedral lattice $\mathcal{L}_\text{3D}$ with iid $Z$ noise. Using the decoder (a) once for color $Y$, and (b) once each for $R$, $G$, $B$, and $Y$ and then selecting the lowest weight output.
	This improves the performance and raises the observed threshold from $0.55(5)\%$ to $0.80(5)\%$.
	}
	\label{fig:3DColorCodePerfectMeasurementThreshold}
\end{figure}

In \fig{3DColorCodePerfectMeasurementThreshold}(a) we show the performance of the restriction decoder adapted to the 3D stabilizer color code on the tetrahedral lattice $\mathcal{L}_\text{3D}$, finding a threshold of $0.55(5)\%$ for independent and identically distributed (iid) phase-flip $Z$ noise. This value can be contrasted with the restriction decoder threshold of $0.77\%$ for the color code on the $3$-torus reported in \cite{kubica2019}.
Also, a similar adaption of the restriction decoder to the 3D color code with a boundary was recently presented in \cite{Turner2020}, however the reported values of $0.1$--$0.2\%$ are surprisingly low.

This adapted restriction decoder on moderate system sizes is far from optimal. 
For example, some weight-2 errors can cause failure for any $d\leq 9$.
This phenomenon is not solely due to the presence of the boundaries.
Namely, we found that there are some weight-$2$ errors in the color code of distance $d=6$ on the $3$-torus, which cause the restriction decoder to introduce a logical error.

To improve the performance, we consider a very simple additional modification of the restriction decoder to improve its performance for moderate system sizes.
In addition to selecting a small weight set $\tau(v_Y)$ for the boundary vertex $v_Y$, we choose as $\tau(v)$ the set of minimal weight for every $Y$ vertex $v\in\facex 0 {\mathcal{L}_\text{3D}}$.
As mentioned above, this can only change the decoder output by a stabilizer, but the explicit representation will typically be of lower weight.
Then, we simply rerun the same decoder three times by picking other colors, i.e., $R$, $G$ and $B$ instead of $Y$, and finally select the correction which has the lowest total weight among the four for colors $Y$, $R$, $G$, and $B$. 
This simple modification yields significant improvements: the lowest distance which corrects all weight-2 errors is now $d=7$ compared to $d=11$ which was needed without the modification, and the threshold is increased from $0.55(5)\%$ to $0.80(5)\%$; 
see \fig{3DColorCodePerfectMeasurementThreshold}. 

\clearpage
\section{Code switching analysis}
\label{sec:dimjump}

In this section we describe how code switching between 2D and 3D color codes can be used to fault-tolerantly produce an encoded $T$ state in the 2D color code, and analyze the overhead of this process.

\subsection{Creating the T state via code switching in 6 steps}

\begin{figure}[h]
	\includegraphics[width=.85\columnwidth]{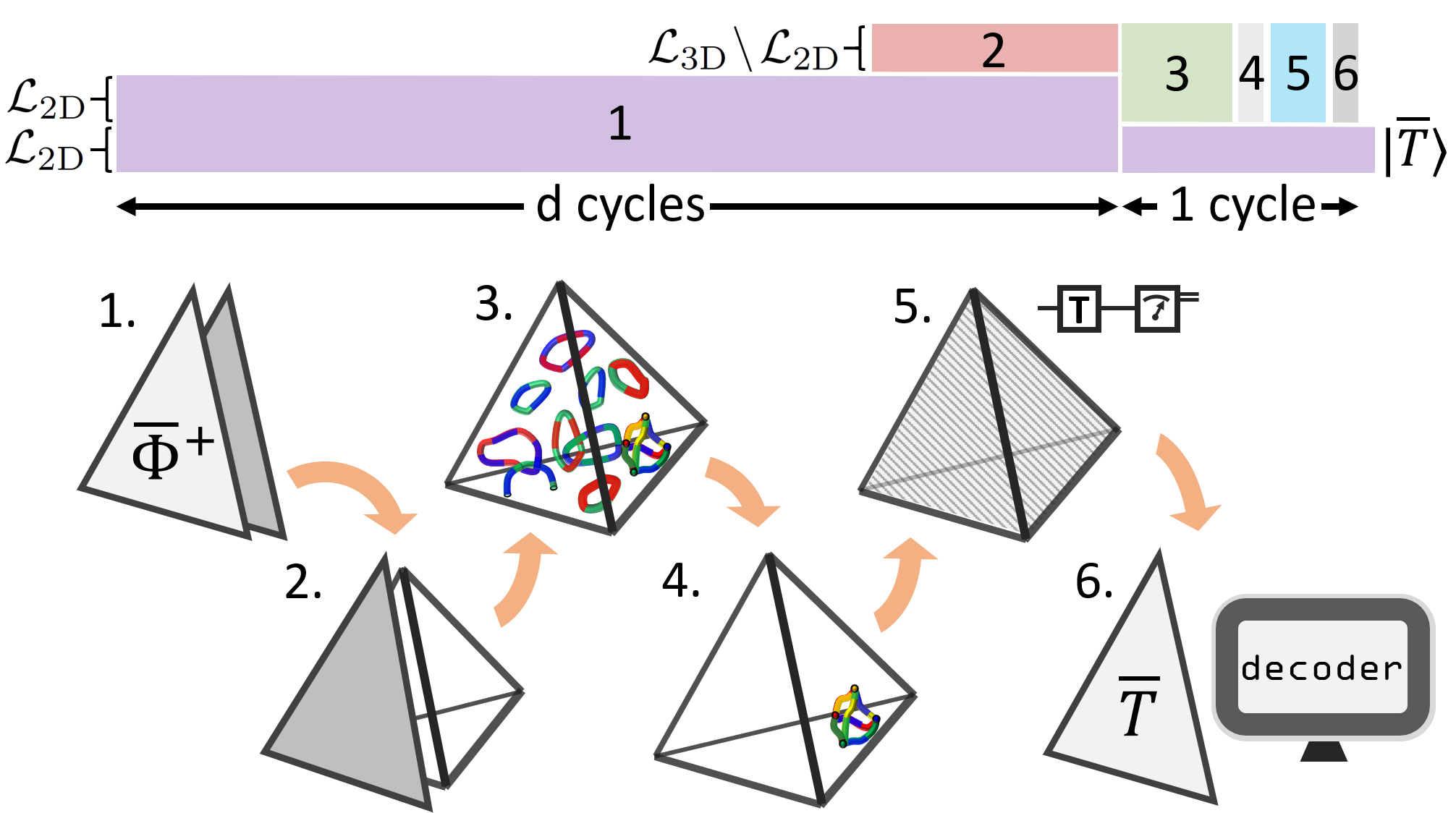}
	\qquad
	\caption{
	The protocol to create an encoded $T$ state via code switching, with the timeline of each step in units of QEC cycles for the 2D color code.
	In steps 1 and 2 we simultaneously prepare the Bell state $|\overline{\Phi^+} \rangle$ in a pair of 2D codes, and the interior of the 3D subsystem color code.
    In step 3, we measure gauge operators, before fixing them in step 4 to end up in the 3D stabilizer color code. In step 5, $\overline{T}$ is applied and all the qubits are measured.
    In step 6, a decoder is used to infer the outcome 
    of the logical $\overline X$ of the 3D stabilizer color code, and the state of the remaining 2D patch is $|\overline T\rangle$ up to a logical $\overline Z$.
    }
    \label{fig:StepsCodeswitching}
\end{figure}

Here we outline the protocol to produce an encoded $T$ state using code switching in the presence of circuit noise.
The main idea is to first produce a Bell state across a pair of 2D color codes, and switch one of the two into a 3D color code where the logical $\overline{T}$ is applied transversally.
Then, by measuring $\overline{X}$ for the 3D code, the encoded $T$ state is effectively teleported to the remaining 2D code (up to a known logical Pauli correction).
This approach avoids switching from the 3D code back to the 2D code, which we believe considerably reduces the amount of extra noise and simplifies our simulation.
More explicitly, the protocol consists of the following steps (which are illustrated in \fig{StepsCodeswitching}).

\begin{enumerate}
    \item \textit{Prepare Bell state in 2D codes.---}The encoded Bell state $(\ket{\overline{0}}_{2\text{D}}\ket{\overline{0}}_{2\text{D}}+\ket{\overline{1}}_{2\text{D}}\ket{\overline{1}}_{2\text{D}})/\sqrt{2}$ state is fault-tolerantly prepared in a pair of 2D color codes, defined on two copies of the lattice $\mathcal{L}_{\text{2D}}$.
    The second of these 2D color codes should be seen as the code defined along the 2D boundary near the yellow vertex of the 3D lattice as in \fig{3DColorCodeBoundary}(b).
    
    \item \textit{Prepare the 3D interior.---}The remaining qubits in $\mathcal{L}_{\text{3D}}$, i.e., those which are not near the boundary vertex $v_Y$, are prepared as a tensor product of unique spherical 2D color code states. 
    The logical state of the system is a Bell pair between a 2D color code and the 3D subsystem color code
    $(\ket{\overline{0}}_{2\text{D}}\ket{\overline{0}}_{\text{sub}}+\ket{\overline{1}}_{2\text{D}}\ket{\overline{1}}_{\text{sub}})/\sqrt{2}$.
    
    \item \textit{Measure gauge operators.---}All $Z$-edge operators for all edges not incident to any $Y$ vertex are measured, yielding the subset of edges $\tilde{\gamma}$ corresponding to $-1$ outcomes. 
    
    \item \textit{Gauge fix.---}We find and apply an $X$-gauge operator which seeks to fix all $Z$-edge operators to have +1 outcomes.
    The logical state is now a Bell state encoded into the 2D color code and the 3D stabilizer color code $(\ket{\overline{0}}_{2\text{D}}\ket{\overline{0}}_{3\text{D}}+\ket{\overline{1}}_{2\text{D}}\ket{\overline{1}}_{3\text{D}})/\sqrt{2}$.
    
    \item \textit{Apply $\overline{T}$, and measure.---}We apply $\widetilde{T}$ to every data qubit. 
    This, in turn, implements a logical $\overline{T}$ gate, yielding the state $(\ket{\overline{0}}_{2\text{D}}\ket{\overline{0}}_{3\text{D}}+e^{i \pi/4}\ket{\overline{1}}_{2\text{D}}\ket{\overline{1}}_{3\text{D}})/\sqrt{2}$.
    Then we measure each individual data qubit in the 3D code in the $X$ basis.
    
    \item \textit{Decode $Z$ errors in 3D code.---}We use the single-qubit $X$-basis measurements to first decode $Z$-errors, and then infer the outcome $m = \pm 1$ of the logical $\overline X$.
    Then, the state encoded in the 2D color code is $(\ket{\overline{0}}_{2\text{D}}+ m e^{i \pi/4}\ket{\overline{1}}_{2\text{D}})/\sqrt{2}$, which for $m=-1$ needs to be fixed to the encoded $T$ state by application of logical $\overline{Z}$.
\end{enumerate}

In the following subsections, we go through each of the six steps, elaborating on the implementation and simulation details.
In our analysis, we adhere to the following guiding principles.
\begin{itemize}
\item For each step, we use Monte Carlo simulations under circuit noise to estimate the performance.
We select the best error correction techniques, fault-tolerant gadgets and efficient decoding algorithms that we are aware of, and optimize measurement circuits where possible.

\item It is possible that some steps will benefit from future improvements in error correction techniques and decoders. We estimate the impact these could have on the performance of this code switching protocol by replacing those steps by a justified estimate of the best improvement one could hope for.

\item We choose the fault-tolerant error correction for the 2D color code to be the same optimized configuration we assumed for state distillation (see \sec{2DCCOptimizedSetup}) to allow for a fair comparison between code switching and state distillation. 

\item We assume a single ancilla qubit per gauge operator in the 3D color code interior. 

\end{itemize}

\begin{figure}[h]
	(a)\hspace*{-5mm}\includegraphics[width=.45\textwidth]{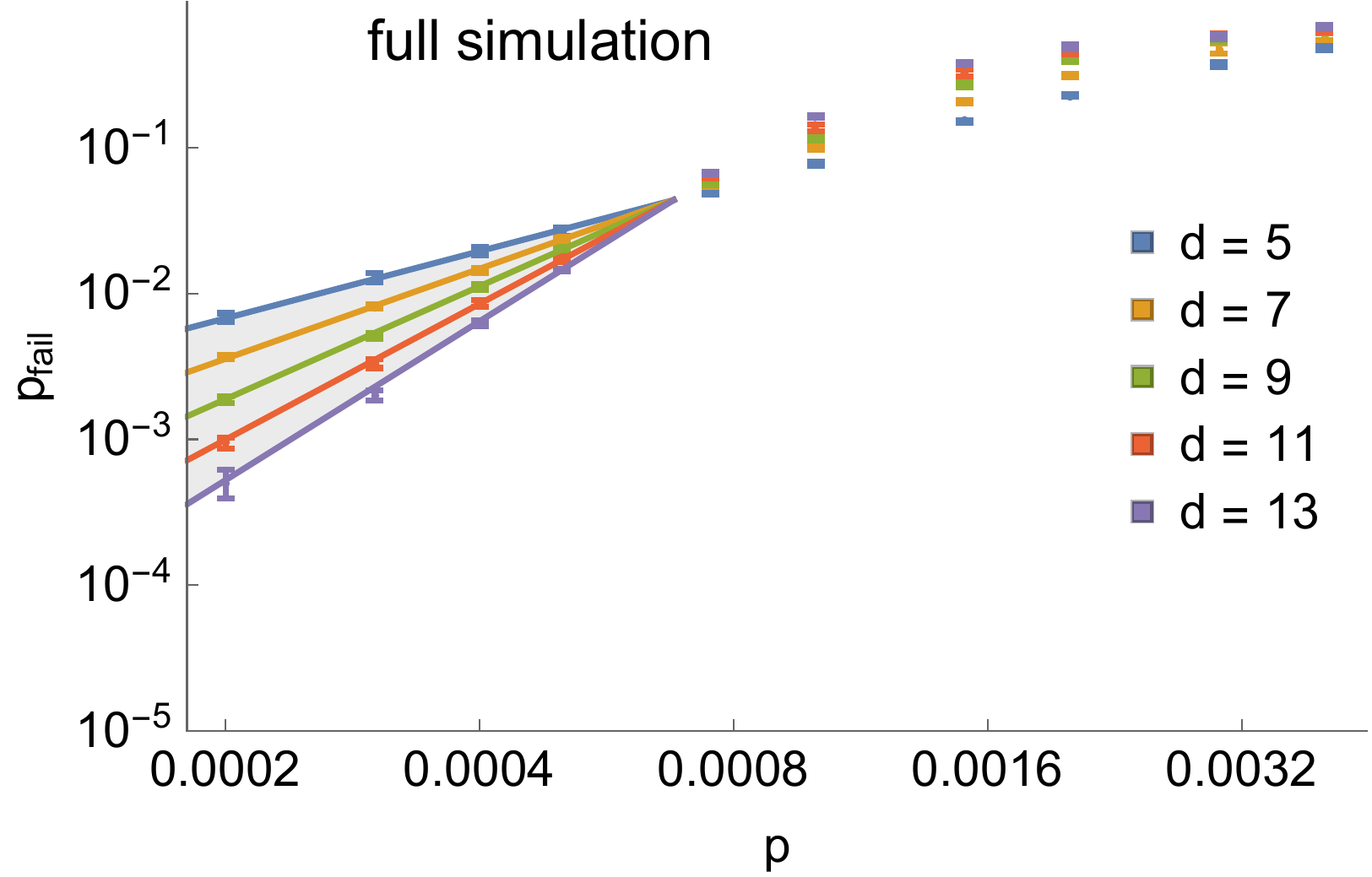}
	\quad\quad\quad
	(b)\hspace*{-5mm}\includegraphics[width=.45\textwidth]{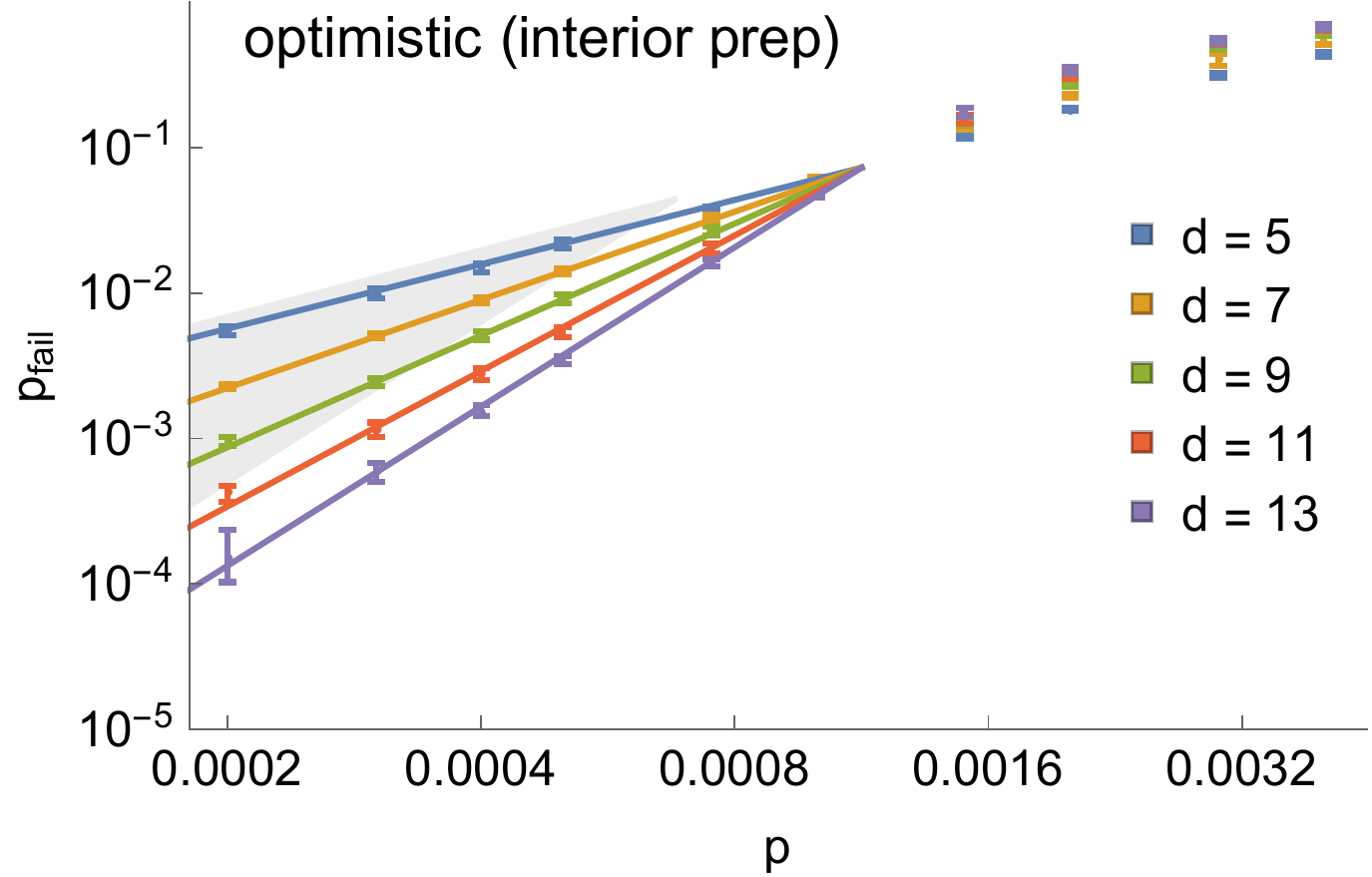}\\
	\vspace*{5mm}
	(c)\hspace*{-5mm}\includegraphics[width=.45\textwidth]{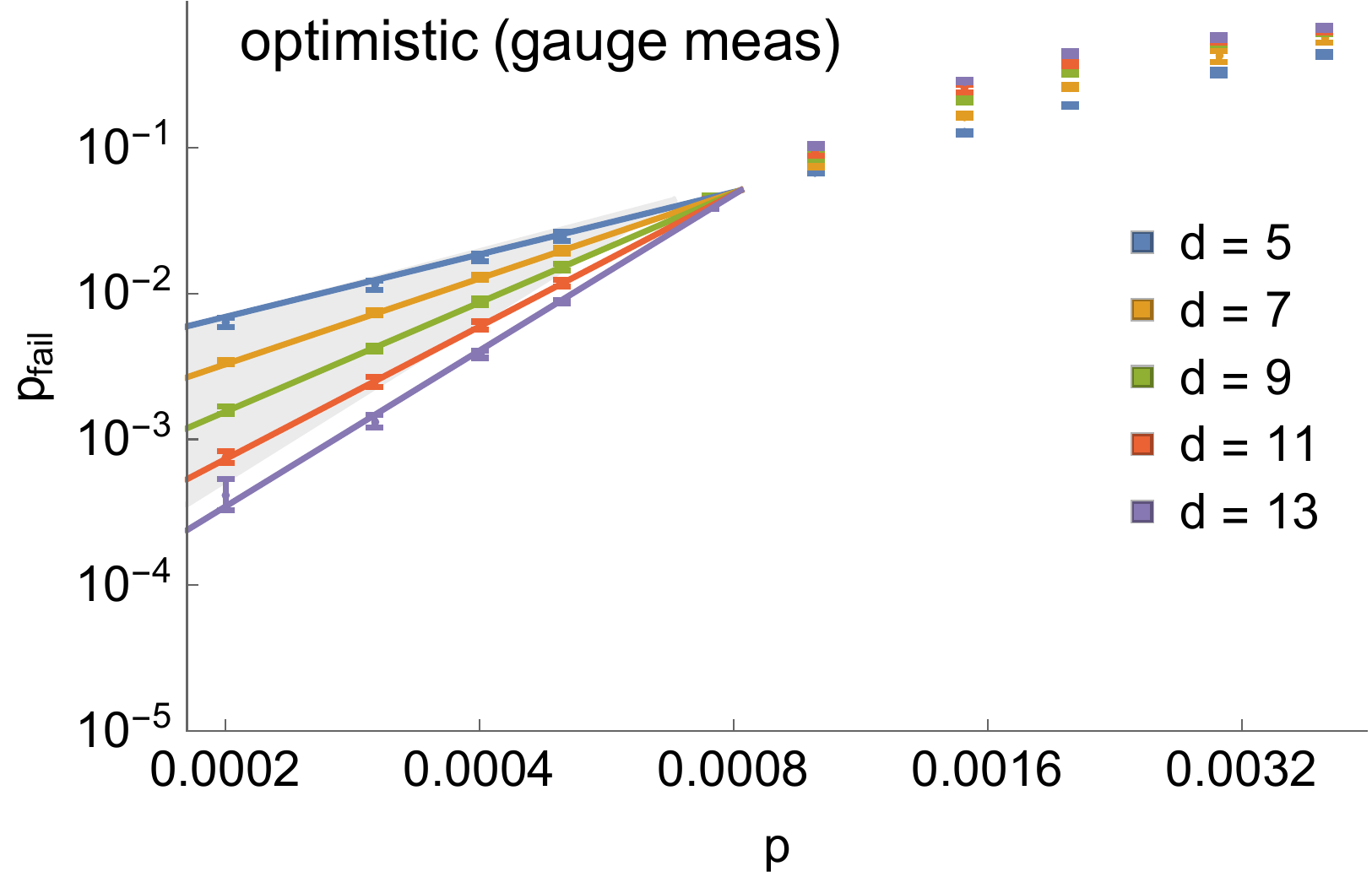}	
	\quad\quad\quad
	(d)\hspace*{-5mm}\includegraphics[width=.45\textwidth]{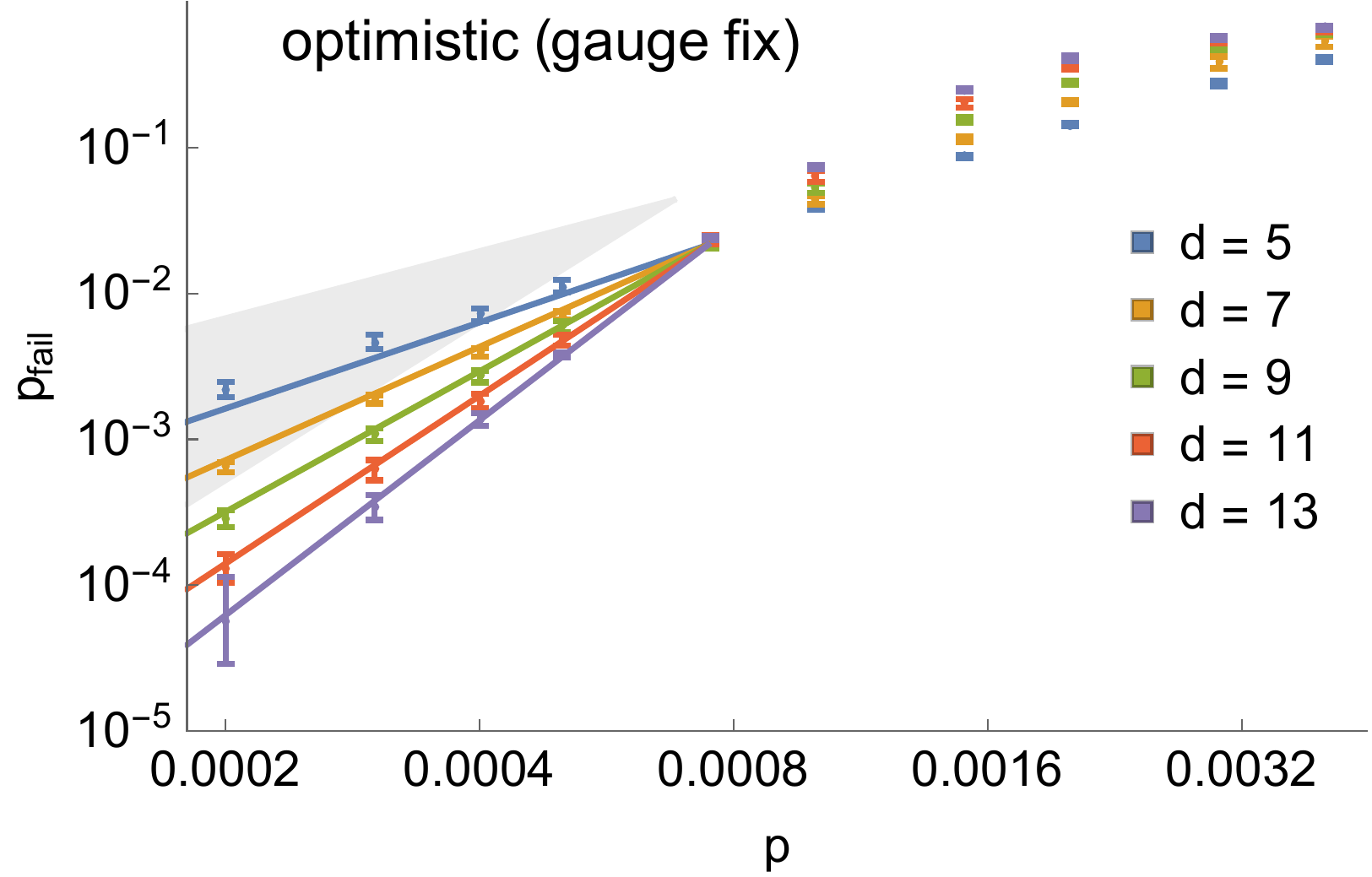}\\
	\vspace*{5mm}
	(e)\hspace*{-5mm}\includegraphics[width=.45\textwidth]{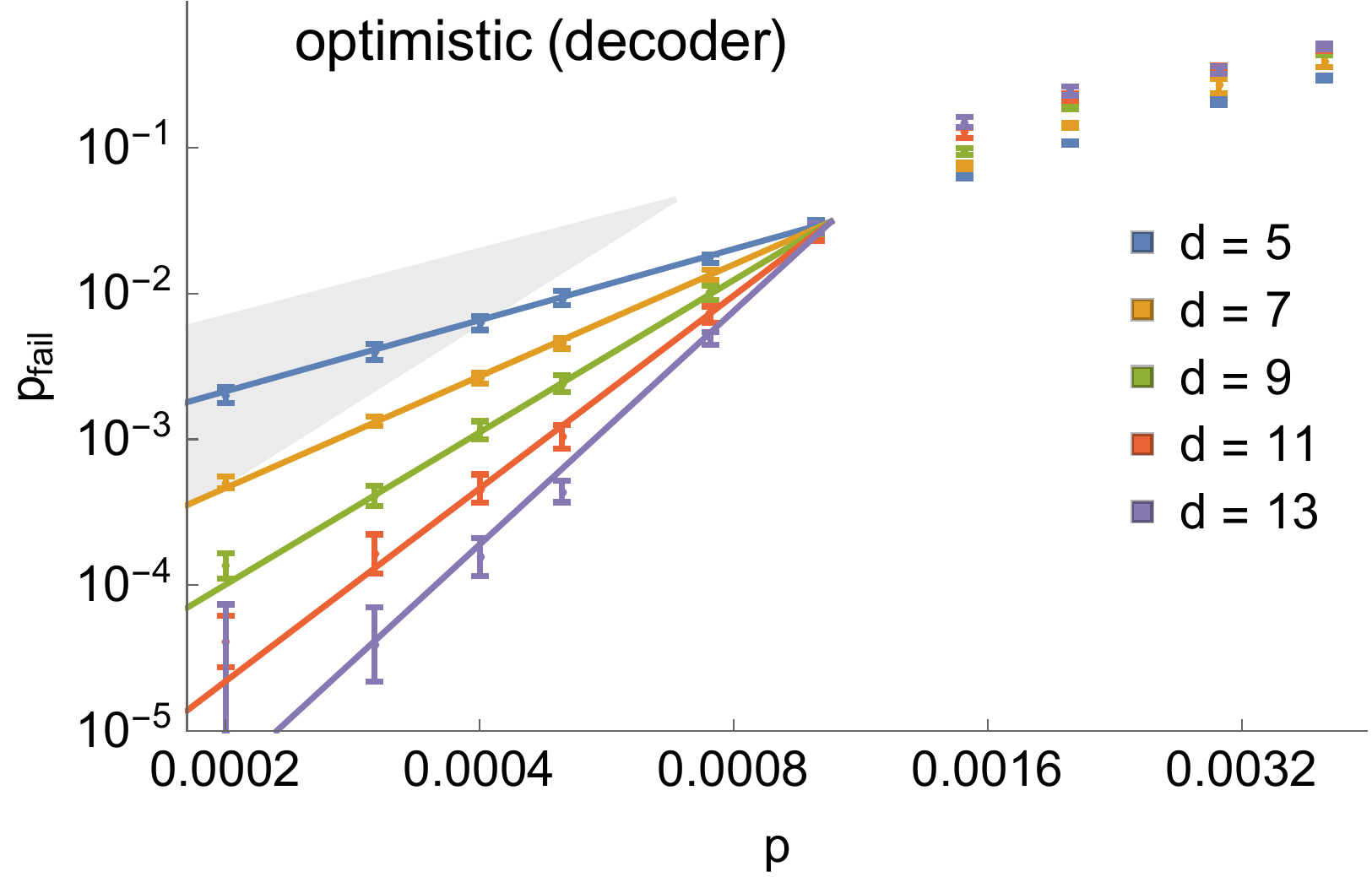}
	\quad\quad\quad
	(f)\hspace*{-5mm}\includegraphics[width=.45\textwidth]{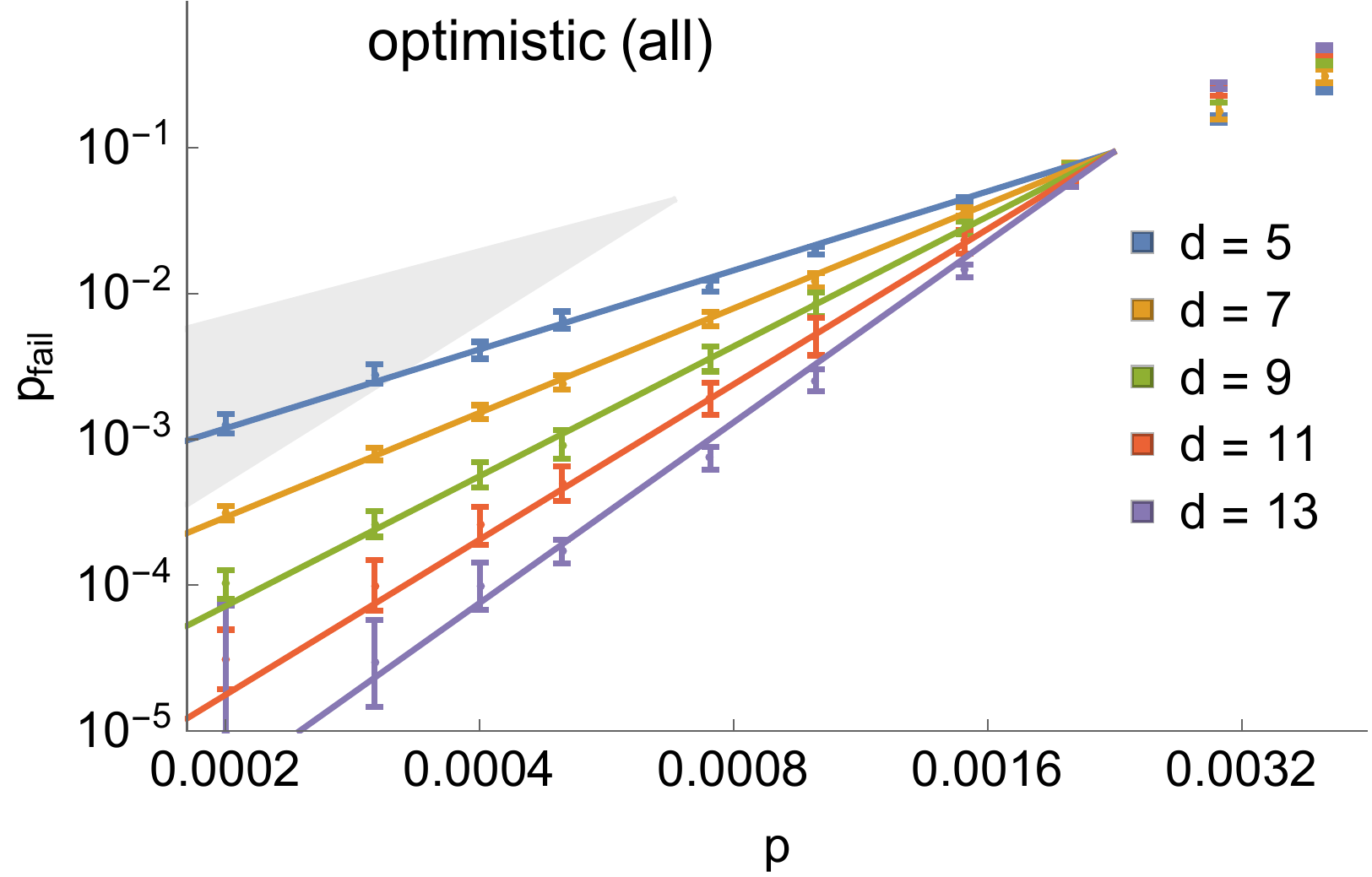}
	\caption{
	(a) The failure probability of the code switching protocol under circuit noise, for which we observe a $T$ gate threshold of $0.07(1)\%$.
	We also estimate the code switching failure probability when one step, i.e., (b) interior preparation, (c) gauge measurement, (d) gauge fixing and (e) the decoder, is replaced with an optimistic version.
	In (f), all four of these steps are replaced by their optimistic versions.
	We use a special form of the ansatz in~\eq{ansatz_moregeneral}, i.e., $p_{\text{fail}} = A \left(p/p^*_{\text{CS}} \right)^{(C d + D)}$ to fit the data up to the crossing.
	The gray region between the fit lines from (a) is superimposed onto (b)--(f) to guide the eye.
	We observe that improving the interior preparation and developing a better decoder have the most potential to improve the performance of code switching.
	If all the potential improvements are achievable, the threshold could be as large as $0.22(5)\%$.
	}
	\label{fig:EndToEndCodeSwitchingAllBounds}
\end{figure}

In \fig{EndToEndCodeSwitchingAllBounds}(a) we present our findings by showing the overall probability of failure of code switching using our simulations. 
In \fig{EndToEndCodeSwitchingAllBounds}(b)-(f) we indicate the impact of optimistic improvements of various steps in the protocol on the overall probability of failure of code switching.

\subsubsection{Preparing the Bell state in 2D codes}
\label{sec:Bell-state-prep}

To fault-tolerantly prepare the encoded Bell state $(\ket{\overline{0}}_{2\text{D}}\ket{\overline{0}}_{2\text{D}}+\ket{\overline{1}}_{2\text{D}}\ket{\overline{1}}_{2\text{D}})/\sqrt{2}$ in a pair of 2D color codes of distance $d$, we first fault-tolerantly prepare them in $\ket{\overline{+}}_{2\text{D}}$ and $\ket{\overline{0}}_{2\text{D}}$ respectively, and then transversally apply a CNOT from the first to the second.
Preparation of $\ket{\overline{+}}_{2\text{D}}$ is carried out by initializing all data qubits in the 2D color code in $\ket{+}$, and then performing $d$ QEC cycles and fixing the inferred initial syndrome of the $Z$ stabilizers.
Preparation of $\ket{\overline{0}}_{2\text{D}}$ similarly involves 
initialization of all data qubits in $\ket{0}$, followed by $d$ QEC cycles and fixing the inferred initial syndrome of the $X$ stabilizers.
Since each QEC cycle for the 2D color code requires 8 time units, the preparation of the encoded Bell state takes $8d$ time units.
After this point the second of the 2D codes undergoes a single additional QEC cycle while code switching occurs for the first 2D code patch.
Although it is arbitrary which of the two patches is used for code switching, in practice there is an asymmetry in the $X$ and $Z$ noise for the patches.
We find a marginal benefit from feeding the first patch (initialized in $\ket{+}$ prior to the CNOT) to be used for code switching.
For details, see \app{prep-basis-choice-code-switching}.

\subsubsection{Preparing the 3D interior}
\label{sec:prep-interior}

There are 2D spherical color code states to be prepared around each yellow vertex as that shown in \fig{3DColorCodeBoundary}(b). 
Implementation details and optimizations of this step are given in \app{residual-noise-sphere-color-code} and \app{prep-basis-choice-code-switching}, but we provide the key features here.

Data qubits are prepared in $\ket{+}$, then $Z$ stabilizers are measured with the shortest circuit that uses a single ancilla. 
If the syndrome extracted for a 2D color code around a yellow vertex is \textit{valid}, meaning there is an $X$ operator that will set all stabilizers to have $+1$ outcomes, then such a fix is applied, otherwise the preparation is repeated.
If the second iteration also yields an invalid syndrome, one randomly selected outcome is flipped, producing a valid syndrome which is then fixed.
The interior preparation step is started sufficiently early so that any repeated preparations have finished by the time the two 2D color code patches completes.
Note that since the 2D spherical color code encodes no logical qubits, it is not necessary to repeat measurements, in contrast to preparation of a logical state in the 2D color code patch, which requires a number of QEC cycles proportional to $d$.

\textit{Optimistic improvements.}
One may hope to improve the preparation of the 2D spherical color codes.
Although we found a schedule of shortest length to measure the stabilizers, there could be many others of the same length which could potentially lead to fewer errors.
Moreover, it may be possible to exploit the unused ancilla associated with the $RG$, $RB$ and $GB$ edges and to consider collective rather than independent preparations of the 2D color codes.
No matter how good such a preparation protocol is, it would require at least three time units as each qubit appears in three $Z$-type stabilizer generators.
Therefore, we use three idle time units with circuit noise of strength $p$ to bound potential improvements of this step in \fig{EndToEndCodeSwitchingAllBounds}(b).

\subsubsection{Measuring gauge operators}
\label{sec:measure-gauge-operators}

To change gauge to the 3D stabilizer code, the $Z$-edges of color in $\mathcal{K} = \{ RG, RB, GB \}$ are measured.
Implementation details and optimizations of this step are given in \app{gauge-measurement-circuits}.
We find a minimal length circuit to measure the gauge operators in parallel with a single ancilla qubit per edge.
Including ancilla preparation and measurement, this circuit requires 8 time units, although the preparation can be done during the last time unit of the interior preparation step. 

Since we choose to measure only $Z$-edge operators of color in $\mathcal{K}$, not all $Z$-edge operators, we only learn the restricted noisy gauge flux $\widetilde\gamma^\mathcal{K}$.
We emphasize that due to faults in the measurement process, $\widetilde\gamma^\mathcal{K}$ is likely to violate the Gauss law and differ from the restricted gauge flux $\gamma^\mathcal{K}$ in the system.

\textit{Optimistic improvements.}
It seems difficult to reduce the errors introduced by this step without increasing the space or time overhead significantly, and for example to use verified cat states to measure each edge operator.
We will neglect any space or time overhead in our estimate so that we bound the effect of potential improvements.
As each data qubit is in three measurements, a minimum of three time units would be required to implement the measurement.
Therefore, we use three idle time units with circuit noise of strength $p$ in our optimistic simulation in \fig{EndToEndCodeSwitchingAllBounds}(c).

\subsubsection{Gauge fixing}

This step corresponds to a classical algorithm which takes as its input the restricted noisy gauge flux $\tilde\gamma^\mathcal{K} \subseteq\facex 1 {\mathcal{L}_\text{3D}^\mathcal{K}}$ corresponding to some subset of edges of color in $\mathcal{K} = \{ RG, RB, GB \}$, and outputs some $X$-type gauge operator aiming to fix $\tilde\gamma^\mathcal{K}$.
The algorithm proceeds as follows.
\begin{enumerate}
    \item[(i)] Validation of the noisy flux: for any color $K\in\mathcal{K}$ we validate the noisy restricted flux $\widetilde\gamma^K$ by matching its violation points within the restricted lattice $\mathcal{L}_\text{3D}^{RG}$.
    We denote by $\lambda^K\subseteq\facex 1 {\mathcal{L}_\text{3D}^K}$ the set of edges used in the pairing of the violation points of $\widetilde\gamma^K$.
    Then, $\hat{\gamma} = \sum_{K\in\mathcal{K}}(\tilde{\gamma}^K +\lambda^K)$ is the validated flux.
    Note that $\hat{\gamma}\subseteq\facex 1 {\mathcal{L}_\text{3D}^\mathcal{K}}$.
    \item[(ii)] Gauge-fixing from the validated flux: 
    we find an operator $\prod_{e\in f(\hat\gamma)} X(e)$ consistent with $\hat{\gamma}$ and apply it.
    Such an $X$-gauge operator can be efficiently found by e.g. Gaussian elimination.
\end{enumerate}

\begin{figure}[h]
	(a)\includegraphics[width=.25\textwidth]{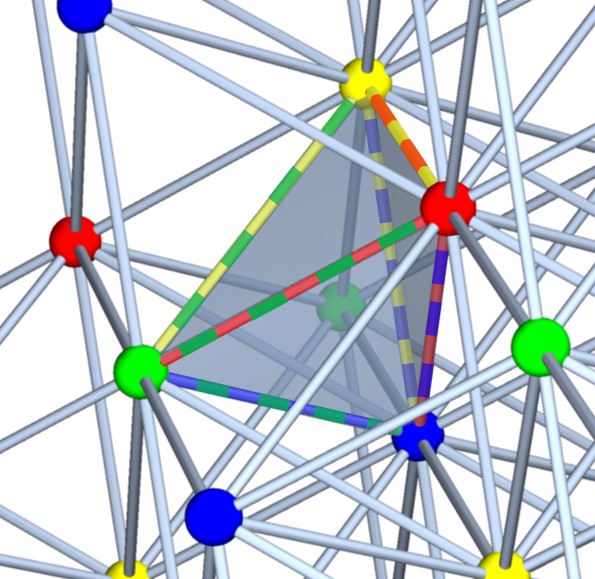}\quad\quad\quad
	(b)\includegraphics[width=.25\textwidth]{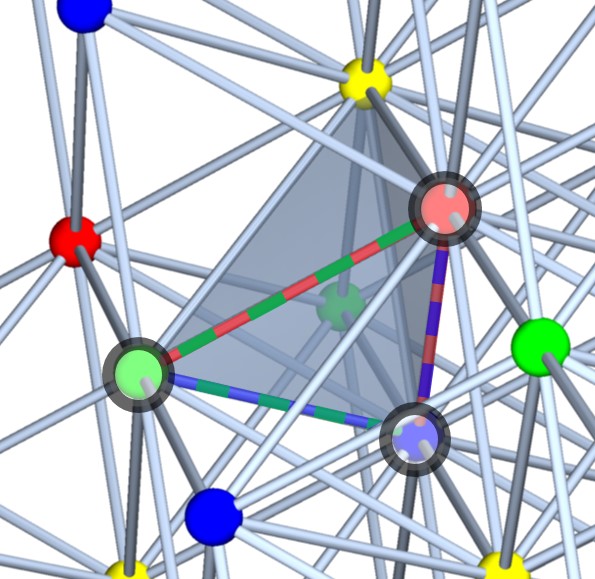}
	\caption{
	(a) A $Z$-gauge flux due to an isolated $X$ error (shaded tetrahedron) consists of six edges of different colors.
	(b) The restricted gauge flux which is measured (assuming no errors occur during the measurement) excludes edges incident to $Y$ vertices. 
	The gauge fixing algorithm in this scenario would connect the marked vertices in each sub-graph thereby removing the $RG$, $RB$ and $BG$ edges in the measured restricted gauge flux as if they were each erroneous, leaving the $X$ error present in the system following gauge fixing.
	An improved algorithm could potentially correct this error.
	}
	\label{fig:gauge-fixing}
\end{figure}

In this step, we are treating the restricted noisy gauge flux $\widetilde\gamma^{\mathcal{K}}$ as arising from a gauge flux $\hat\gamma$ with no branching points along with some measurement errors; see \fig{gauge-fixing}.
In other words, we want to find a gauge flux $\hat\gamma$ corresponding to some $X$-gauge operator just by looking at its restricted noisy gauge flux $\widetilde\gamma^{\mathcal{K}}$.
However the gauge flux can in fact have branching points due to $X$ errors in the system.
In such a case, the restricted noisy gauge flux does not even satisfy the Gauss law in the absence of measurement errors as we omit $Z$-type operators associated with $RY$, $BY$ and $GY$ edges.
For every $K\in\mathcal K$ we choose to independently pair up all the violation points of $\widetilde\gamma^K$ within the restricted lattice $\mathcal{L}^K$, which in turn allows us to find the desired $\hat\gamma$.

\textit{Optimistic improvements.}
One can ask how close to optimal the performance of the algorithm is.
For instance, it may be possible to not only estimate a gauge fix from the noisy $Z$-flux measurement, but also to correct $X$ errors in the system; see \fig{gauge-fixing}.
To bound the impact of any potential improvements of this step in our simulation, we assume that all $X$ errors in the system prior to the gauge measurements are corrected, and that the gauge itself is identified exactly; see \fig{EndToEndCodeSwitchingAllBounds}(d).
The only remaining $X$ error that is retained in this bound is therefore that which is introduced by the noisy gauge measurement circuits themselves.

\subsubsection{Applying T and measuring the 3D code's data qubits}
\label{sec:non-Clifford-noise-prop}

In this step, we first apply the $\widetilde T$ gate to each data qubit, and then measure it in the $X$ basis. 
These can be combined into a single operation requiring one time unit.
Despite being one of the simplest code switching steps to implement, it is quite challenging to simulate efficiently due to the large number of $T$ gates.
In what follows we describe how we exploit some additional structure and assumptions that render the simulation tractable.

Let the $X$-type residual error before the application of the logical $\overline{T}$ gate be supported on $\rho_X \subseteq\facex 3 {\mathcal{L}_\text{3D}}$; similarly, let $\rho_Z$ denote the $Z$-type residual error.
Note that if we could perfectly fix the gauge in the previous step and there were no additional $X$-errors, then there would be no residual $X$-error. 
Generically, there is some nontrivial residual $X$-error, which unlike the $Z$-error does not commute with the logical $\overline T$ gate.
The effect of the residual $X$-error on the measurement outcomes amounts to, roughly speaking, flipping some outcomes along the $Z$-gauge flux $\gamma = \bnd 3 1 \rho_X$ of the the residual $X$-error $\rho_X$.
This is made precise in the following lemma.
\begin{lem}
\label{lem:non-Clifford-noise}
Let $\rho_X,\rho_Z \subseteq\facex 3 {\mathcal{L_\text{3D}}}$ be residual $X$- and $Z$-errors.
Let $\gamma = \bnd 3 1 \rho_X$ be the $Z$-gauge flux and $\gamma = \sum_{j=1}^b \lambda_j$ be a decomposition of $\gamma$ into its linked components $\lambda_j$'s.
If $\widetilde{T}$ is applied to every data qubit and the $X$ stabilizers are perfectly measured, then one obtains a syndrome 
\begin{equation}
\label{eq:sampling_full}
    \sigma = \bnd 3 1 \rho_Z + \sum_{j=1}^b \sigma_j,
\end{equation}
where each excitation configuration $\sigma_j \in \Sigma(\lambda_j)$ is chosen uniformly at random from $\Sigma(\lambda_j)$.
\end{lem}
In above, we use notation introduced in \sec{physics-gauge-flux}.
To simulate this step, one can generate Pauli operators resulting in the same distribution of syndromes.
Note that there is a possibility of introducing a logical $Z$ operator at this step, which we ignore in our simulation, leading to an underestimate of the failure of code switching.
For more details and a justification of this lemma, see Ref.~\cite{Bombin2018}.

In our simulation  we make the additional simplifying assumption to sample $\sigma_j$ uniformly from within the collection of excitation configurations without the linking charge $ \Sigma'(\lambda_j)$ rather than the collection of excitation configurations $\Sigma(\lambda_j)$.
Recall that $\Sigma'(\lambda_j)\subseteq \Sigma(\lambda_j)$.
This assumption amounts to ignoring the linking charge, and we do not expect it to substantially effect the performance.
We generate the random $Z$-error $\tau_Z\subseteq\facex 3 {\mathcal{L}_\text{3D}}$ 
as
\begin{equation}
    \tau_Z = \sum_{v\in \facex 0 {\gamma}} \tau(\gamma\rest v),
\end{equation}
where $\tau(\gamma\rest v)$ is a local sampling procedure, which we now describe in detail.
Let $v\in\facex 0 \gamma$ be a vertex of color $A$, which is incident to the flux $\gamma = \bnd 3 1 \rho_X$, and $\gamma\rest v$ be the restriction of $\gamma$ to the edges incident to $v$.
Let $\mathcal{K}' = \{R,G,B,Y\}\setminus \{A\}$ denote the set of three different colors.
We find a subset $\tau(\gamma\rest v) \subseteq \bnd 0 3 v$ of tetrahedra containing $v$ as follows.
\begin{itemize}
    \item[(i)] With probability $1/2$ set $\Xi=0$; otherwise $\Xi=1$.
    \item[(ii)] For each $K\in\mathcal{K}'$, if $\Xi\prod_{K\in\mathcal{K}}\big|\gamma\rest v^{AK}\big|=0$, then choose uniformly at random a subset $E^{AK}\subseteq\gamma\rest v ^{AK}$ of even cardinality; otherwise choose uniformly at random a subset $E^{AK}\subseteq\gamma\rest v ^{AK}$ of odd cardinality.
    \item[(iii)] Find a subset of tetrahedra $\tau(\gamma\rest v)$, whose $1$-boundary locally matches $\sum_{K\in\mathcal{K}'} E^{AK}$, i.e.,
    \begin{equation}
    \label{eq:local_match}
        (\bnd 3 1 \tau(\gamma\rest v))\rest v = \sum_{K\in\mathcal{K}'} E^{AK}.
    \end{equation}
\end{itemize}

We now explain why this algorithm produces $\tau_Z\subseteq\facex 3 {\mathcal{L}_\text{3D}}$ with the correct syndrome distribution.
First, any excitation configuration $\sigma_j\in \Sigma'(\lambda_j)$ can be viewed as a sum of local excitation configurations with neutral total charge.
If the vertex $v$ is in $\facex 0 {\lambda_j}$, then step (ii) is equivalent to randomly selecting a local excitation configuration for $\lambda_j\rest v$, which is created by operators supported within the neighborhood $\bnd 0 3 v$ of $v$ and with the neutral total charge.
Since each local excitation configuration is equally likely selected, thus the resulting excitation configuration $\sigma_j$ is chosen uniformly at random from $\Sigma'(\lambda_j)$. 
Also note that the search in (iii) can be implemented by exhaustively checking which of the possible subsets of tetrahedra $\bnd 0 3 v$ containing $v$ satisfies \eq{local_match}.
This naive implementation is nevertheless efficient since $|\bnd 0 3 v|$ is bounded.
Lastly, we remark that the residual $Z$-error present in the system right before the measurement in the $X$ basis is $\rho_Z + \tau_Z$.
We use this $Z$-error in the following code switching step.

\subsubsection{Decoding Z errors in the 3D color code}
\label{sec:DecodingZ3DCC}

In the final step of the protocol, a classical decoding algorithm is run to correct $Z$ errors for the 3D stabilizer color code.
The input to the decoder is the set of $X$ measurement outcomes for each data qubit from the previous step. 
The output of the decoder will be a $Z$-type Pauli correction which, if applied, would flip all the syndromes computed given the single-qubit $X$ outcomes. 
At that point one can reliably read off the logical $\overline X$ measurement $m = \pm 1$.
Then, the state encoded in the remaining 2D color code from step 1 is $(\ket{\overline{0}}_{2\text{D}}+ m e^{i \pi/4}\ket{\overline{1}}_{2\text{D}})/\sqrt{2}$, which is the encoded $T$ state, up to the multiplication by $\overline{Z}$ for $m=-1$.

We implement a decoder based on the restriction decoder from Ref.~\cite{kubica2019}, but modified in two ways.
The first modification is to adapt the decoder to the 3D stabilizer color code on the lattice $\mathcal{L}_{\text{3D}}$ which has boundaries.
The second modification is to improve performance.
We do this by favoring low-weight corrections (which differ by a stabilizer from the output of the original decoder).
Running four independent versions of this decoder in parallel, we then simply select the lowest-weight correction from the four candidates.
We describe and analyze the performance of this modified restriction decoder in \sec{3D-color-code-perfect-measurements}.

\textit{Optimistic improvements.}
A different decoder may improve the performance of code switching.
It is difficult to give as rigorous bounds on the performance of this step as we were able to give for the previous steps. 
However, by assuming that the best decoder performs as if the phase-flip $Z$ noise was iid,
we estimate the effect of any potential performance improvements using the following steps:
\begin{enumerate}
    \item[(i)] Run the modified restriction decoder, and if it succeeds then we assume the improved version would also succeed, if it fails then continue to the next step.
    \item[(ii)] Let $w$ be the minimum of the weight of the error, and the weight of the correction produced by the modified restriction decoder.
    Let $n$ be the number of data qubits in the distance-$d$ 3D stabilizer color code.
    Choose uniformly at random a real number from $0$ to $1$, and if it is smaller than $ p^{(1)}_\text{3DCC}\left((w/n)/p^{(1)}_\text{3DCC}\right)^{\frac{d+1}{2}}$, then the correction fails; otherwise, the correction succeeds. 
\end{enumerate}

Note that $p^{(1)}_\text{3DCC} \simeq 1.9\%$ is the known threshold of the optimal decoder for the 3D stabilizer color code under iid $Z$ noise \cite{kubica2018}.
We provide more detailed justifications for this estimate in \app{3D-color-code-potential-improvements}.
The impact of potential improvements of this step on code switching is shown in \fig{EndToEndCodeSwitchingAllBounds}(e).

\subsection{Code switching overhead}
\label{sec:codeswitchingoverhead}

To calculate the space and time overhead required to produce an encoded $T$ state with failure probability $p_{\text{fin}}$, we first find the minimum distance $d(p_{\text{fin}})$ which achieves this by extrapolating the data in \fig{EndToEndCodeSwitchingAllBounds}(a).
The time to implement the code switching protocol at distance $d$ is simply $8 (d+1)$, which is the time for $(d+1)$ QEC cycles of the 2D color code to complete.
The number of qubits to implement the code switching protocol at distance $d$ is
\begin{equation}
N_{\text{CS}}(d)=(26d^3+ 51d^2+ 22d -51)/24.
\label{eq:OverheadCSdistance}
\end{equation} 
This is comprised of two 2D color code patches (each of which has $N_{2\text{D}}(d) = (3 d^2-1)/2$ qubits from \eq{Nqubits2DCC}) and the 3D interior (which has $d(d^2 + 1)/2-(1 + 3d^2)/4$ data qubits and $(d -1)(7d^2 + 10d + 15)/12 -3(d + 1)(d -1)/8$ measurement qubits).
Details of the lattices which make clear where these numbers originate from can be found in
\app{lattices}.

\begin{figure}[h]
	(a)\hspace*{-5mm}\includegraphics[width=.45\textwidth]{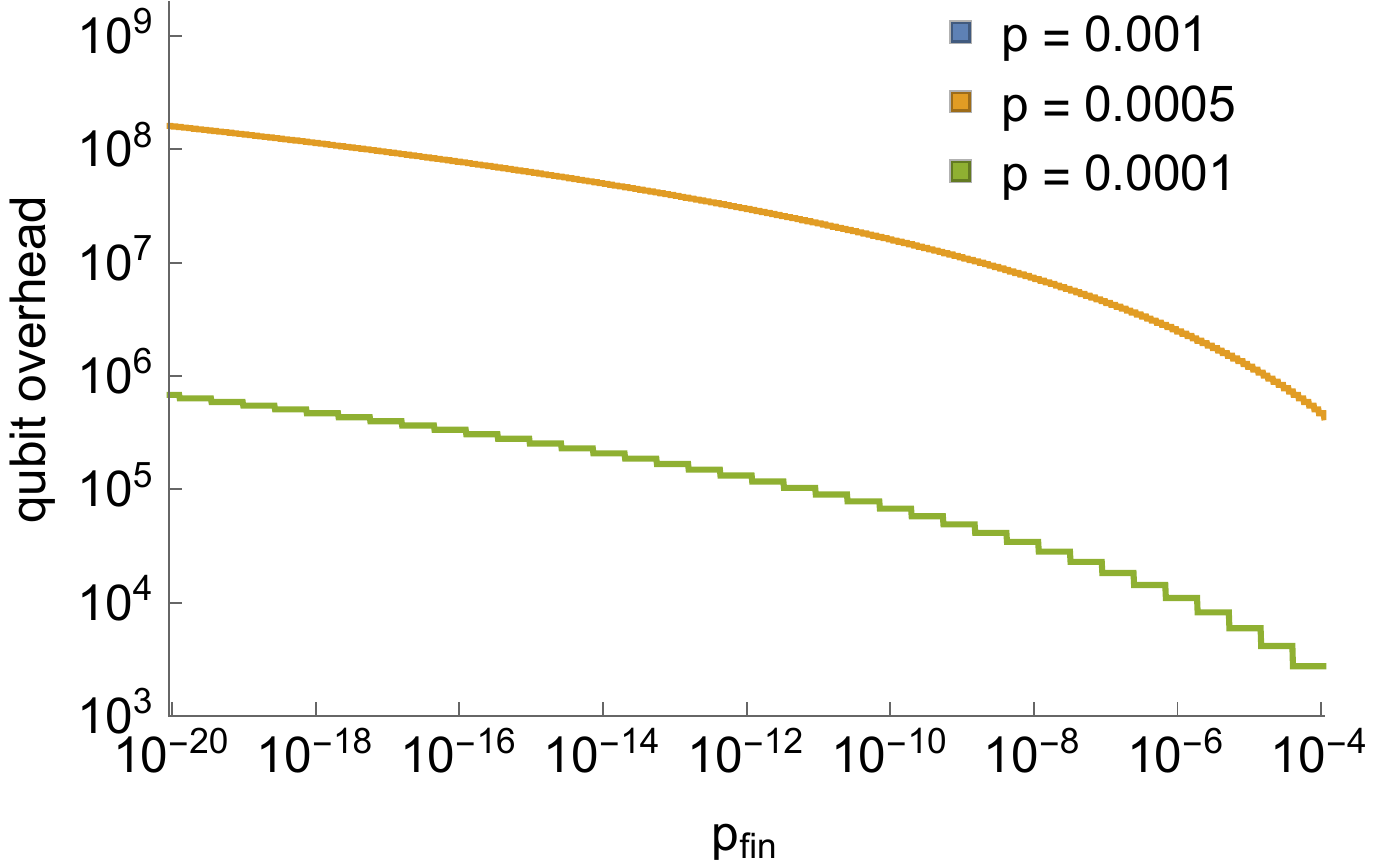}
	\quad\quad\quad
	(b)\hspace*{-5mm}\includegraphics[width=.45\textwidth]{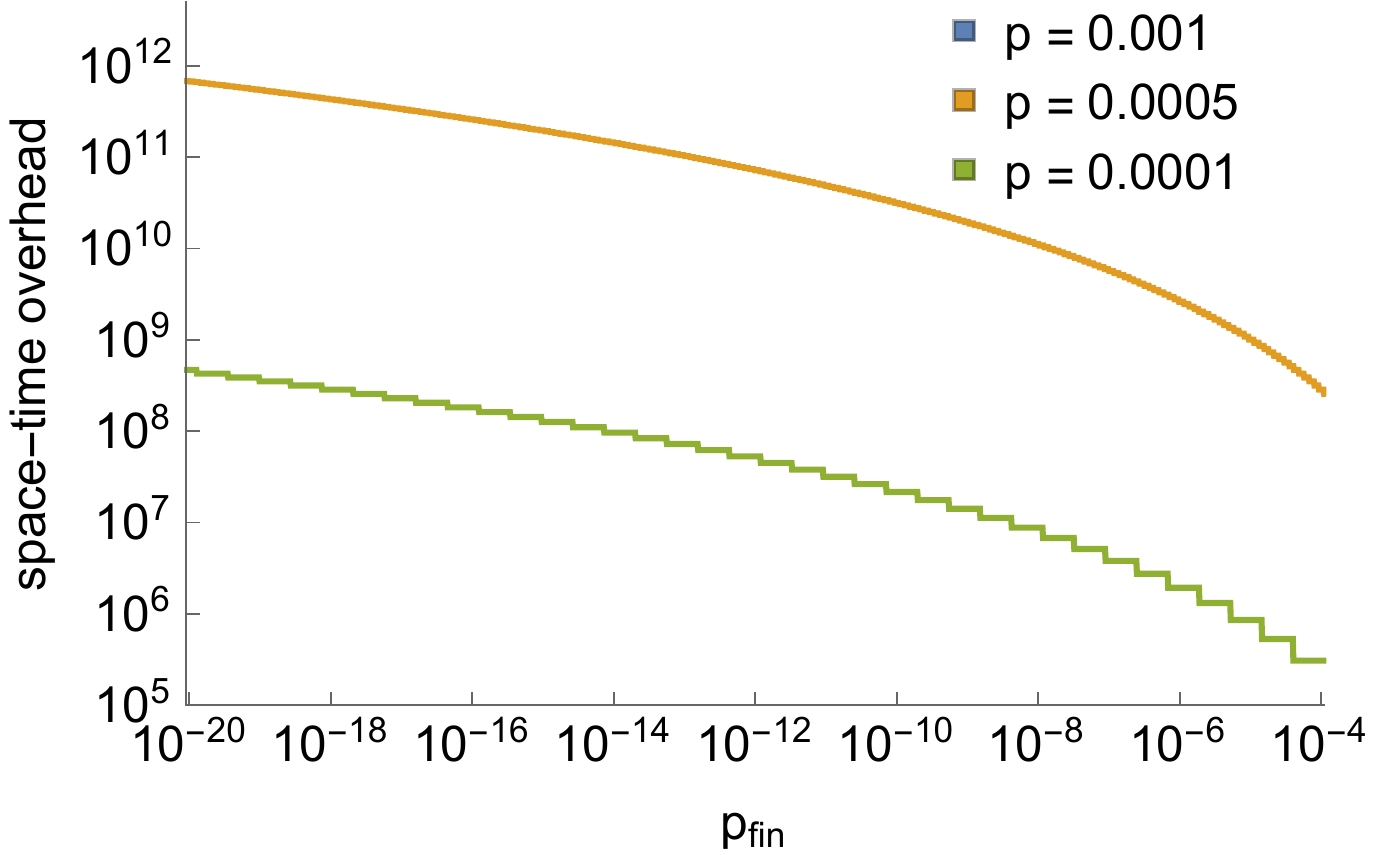}
	\caption{
	(a) The qubit and (b) space-time overhead of code switching as a function of the infidelity $p_\text{fin}$ of the output $T$ state. 
	Note there is no curve for $p=0.001$ as this is higher than the observed threshold for code switching.
	}
	\label{fig:overhead-codeswitching}
\end{figure}

In \fig{EndToEndCodeSwitchingAllBounds} we fit a special form of the ansatz in~\eq{ansatz_moregeneral}, i.e., $p_{\text{fail}} = A \left(p/p^*_{\text{CS}} \right)^{(C d + D)}$ to fit the data up to the crossing.
Setting the failure probability to be at most the target infidelity, i.e. $p_{\text{fail}}\leq p_{\text{fin}}$, and solving for $d$ we obtain 
\begin{eqnarray}
d = \frac{\log(p_{\text{fin}}/A)}{C\log\left(p/p^*_{\text{CS}} \right)} -D/C,
\end{eqnarray}
where we round the right hand side up to the closest odd integer.
Finally, in \fig{overhead-codeswitching} we substitute this into $N_{\text{CS}}(d)$ and plot the qubit overhead and the space-time overhead as a function of $p_{\text{fin}}$ for various values of $p$ for two cases: assuming no improvements, and assuming all optimistic improvements can be achieved.  

\clearpage
\section{Discussion and Acknowledgments}
\label{sec:discussion}
In this work, we have simulated concrete realizations of state distillation and code switching under circuit noise and compare the overhead that each requires to produce high-fidelity $T$ states encoded in 2D color codes.
We have focused on regimes of practical interest, with physical error rates from $10^{-4}$ to $10^{-3}$ and target logical error rates from $10^{-20}$ to $10^{-4}$; see \fig{overhead-comparison}.
For error rates of $10^{-3}$ and above, our implementation of code switching is not viable since its observed threshold was $0.07(1)\%$, considerably lower than that of $0.37(1)\%$ for state distillation.
For error rates near $5\cdot10^{-4}$, our implementation of code switching becomes viable, but requires considerably more overhead than state distillation.
However for error rates around $10^{-4}$, our code switching implementation begins to slightly outperform that of state distillation. 
Lastly, we see in \fig{overhead-comparison} that a highly-optimistic implementation of code switching (which may be very difficult if not impossible to achieve) would not provide substantial savings over our simple implementation of state distillation in the studied regimes. 

We now discuss the scaling of the space overhead $O_*^\text{S}$ and space-time overhead $O_*^\text{ST}$ of code switching and state distillation for small physical error rate, i.e., $p \ll 1$.
We further assume that $\log{p_{\text{fin}}}/\log{p}\gg 1$.
As we show in \app{scaling-calculations}, we have
\begin{eqnarray}
\label{eq:scaling-expressions}
O_*^{\text{S}} \sim c_*^{\text{S}} \left( \frac{\log{p_{\text{fin}}}}{\log{p}} \right)^{\Gamma_*},\quad\quad O_*^{\text{ST}} \sim c_*^{\text{ST}} \left( \frac{\log{p_{\text{fin}}}}{\log{p}} \right)^{\Gamma_*+1},
\end{eqnarray}
where $c_*^{\text{S}}$ and $c_*^{\text{ST}}$ are some constants, $\Gamma_{\text{CS}}=3$ for code switching and $\Gamma_{\text{SD}}=\max(2,\log_F R)$ for state distillation.
Here, $R$ is the ratio of the number of input to output magic states and $F$ is the order of error suppression in a single distillation round.
The change in the value of $\Gamma_{\text{SD}}$ can be understood as follows---if $\log_F R>2$, then the overhead is dominated by the initial distillation round; otherwise, it is dominated by the final round.
For the 15-to-1 distillation scheme we have $R=15$ and $F=3$, leading to $\log_F R \approx 2.46$ and $\Gamma_{\text{SD}} < \Gamma_{\text{CS}}$.
From \eq{scaling-expressions} we thus conclude that for $\log{p_{\text{fin}}}/\log{p}\gg 1$ state distillation will always outperform code switching.
On the other hand, if we are willing to extrapolate \eq{scaling-expressions} to more modest values of $\log{p_{\text{fin}}}/\log{p}$, the overhead of code switching is predicted to drop below that of state distillation.
The precise value of the crossover is highly sensitive to the implementation details, but this could explain why we observe code switching outperforming state distillation for low $p$ and modest $p_{\text{fin}}$ in \fig{overhead-comparison}.

Our findings point toward a number of future research directions.
First, it could be fruitful to explore the potential of code switching in the regime of very low error rates, which may be achievable in trapped ion qubits \cite{Bruzewicz2019} and topological qubits \cite{Karzig2017}.
Second, improving the 2D color code state preparation procedure, syndrome extraction, and decoding algorithms would further reduce the overhead of both code switching and state distillation.
We expect that code switching may also greatly benefit from better 3D color code decoders.
Third, our code switching protocol produces the $T$ state, but it is possible to directly apply the $T$ gate rather than injecting it which may allow one to implement code switching in constant time rather than a time which scales with the code distance $d$.
Fourth, we have seen that in contradiction with the commonly held belief that a distillation scheme overhead is dominated by that of the first round $\sim(\log 1/p_{\text{fin}})^{\log_F R}$, the last round overhead is $\sim(\log 1/p_{\text{fin}})^2$ when implemented with 2D codes, and dominates when $\log_F R<2$.
This suggests that searches and optimizations of new distillation schemes should not be restricted to those with very low $\log_F R$.


\textbf{Acknowledgments.---}
M.E.B. acknowledges relevant scientific discussions with many researchers from the QEC community, including 
Thomas Bohdanowicz,
H\'ector Bomb\'in
Ben Brown,
Earl Campbell,
Nicolas Delfosse,
Austin Fowler,
Jeongwan Haah,
Tomas Jochym O'Connor,
Cody Jones,
Andrew Landahl,
Ying Li,
Michael Newman,
John Preskill,
and
Marcus P da Silva.
He is especially grateful for the encouragement for this project from David Poulin.
M.E.B. also wishes to thank Achiamar Lee Rivera for her unwavering support throughout this research.
A.K. is deeply indebted to H\'ector Bomb\'in for many colorful discussions.
A.K. acknowledges funding provided by the Simons Foundation through the ``It from Qubit'' Collaboration.
Research at Perimeter Institute is supported in part by the Government of Canada through the Department of Innovation, Science and Economic Development Canada and by the Province of Ontario through the Ministry of Colleges and Universities.
This work was completed prior to A.K. joining AWS Center for Quantum Computing.

\clearpage
\appendix
\section*{Appendices}
\renewcommand\thesubsection{\Alph{subsection}}

\subsection{Lattice parameters and specifications}
\label{app:lattices}
Here we provide parameters of the lattices which are used throughout the paper.
\tab{2DCC-direct-lattice} and \tab{3DCC-dual-lattice} show parameters of the direct lattice and the dual lattice used to define the 2D and 3D color codes of distance $d$.
In general,
these lattices can be obtained by tessellating a $D$-sphere with $D$-simplices and then removing one vertex from that tessellation, as well as all the simplices containing that vertex \cite{kubica2015a,bombin2015}.

\begin{table}[h]
\begin{center}
 \begin{tabular}{|c|c|} 
 \hline
 \# vertices  &  \# incident faces \\
 [0.5ex] 
 \hline\hline
  $3 (d - 1)/2$ & 4 \\
  $3 (d - 3) (d - 1)/8$ & 6 \\
  3 & $d$ \\
 [1ex] 
 \hline
\end{tabular}
\caption{
The lattice $\mathcal{L}_{\text{2D}}$, which we use to define the 2D stabilizer color code of distance $d$, consists of $3(d^2+7)/8$ vertices,
$3(3d^2+5)/8$
edges and $(1+3d^2)/4$ faces.
We group vertices of $\mathcal{L}_{\text{2D}}$ into three different categories depending on the number of faces incident to them.
Note that qubits are placed on faces, whereas stabilizers are identified with vertices of $\mathcal{L}_{\text{2D}}$.
 }
 \label{tab:2DCC-direct-lattice}
\end{center}
\end{table}


\begin{table}[h]
\begin{center}
 \begin{tabular}{|c|c||c|c|} 
 \hline
 \# vertices  &  \# incident tetrahedra  &  \# edges  & \# incident tetrahedra \\
 [0.5ex] 
 \hline\hline
  4 & $(1+3d^2)/4$ & 6 & $d$ \\
  $2(d-1)$ & 8 & $(d^3 + 3 d^2 + 11 d -15)/4$ & 4  \\
  $(d-3)(d-1)/2$ & 12 & $(d-3) (d-1) (2 d+5)/6$ & 6 \\
  $(d-3)(d-1)/2$ & 18 & & \\
  $(d-5)(d-3)(d-1)/12$ & 24 && \\
  [1ex] 
 \hline
\end{tabular}
 \caption{
 The lattice $\mathcal{L}_{\text{3D}}$, which we use to define the 3D subsystem and stabilizer color codes of distance $d$, consists of $(d-1)(d+1)(d+3)/12 + 4$ vertices, $(d-1)(7d^2 +10d +15)/12 + 6$ edges, $d^3 + d +2$ faces, and $(d^3+d)/2$ tetrahedra.
 We group vertices and edges of $\mathcal{L}_{\text{3D}}$ into, respectively, five and three different categories depending on the number of tetrahedra incident to them.
Note that qubits are placed on tetrahedra, whereas stabilizer and gauge generators are identified with vertices and edges of $\mathcal{L}_{\text{3D}}$.
 }
 \label{tab:3DCC-dual-lattice}
\end{center}
\end{table}

\clearpage
\subsection{Supplementary details for 2D color code simulations}
\label{app:threshold-data}
\begin{figure}[h]
	\includegraphics[width=.24\textwidth]{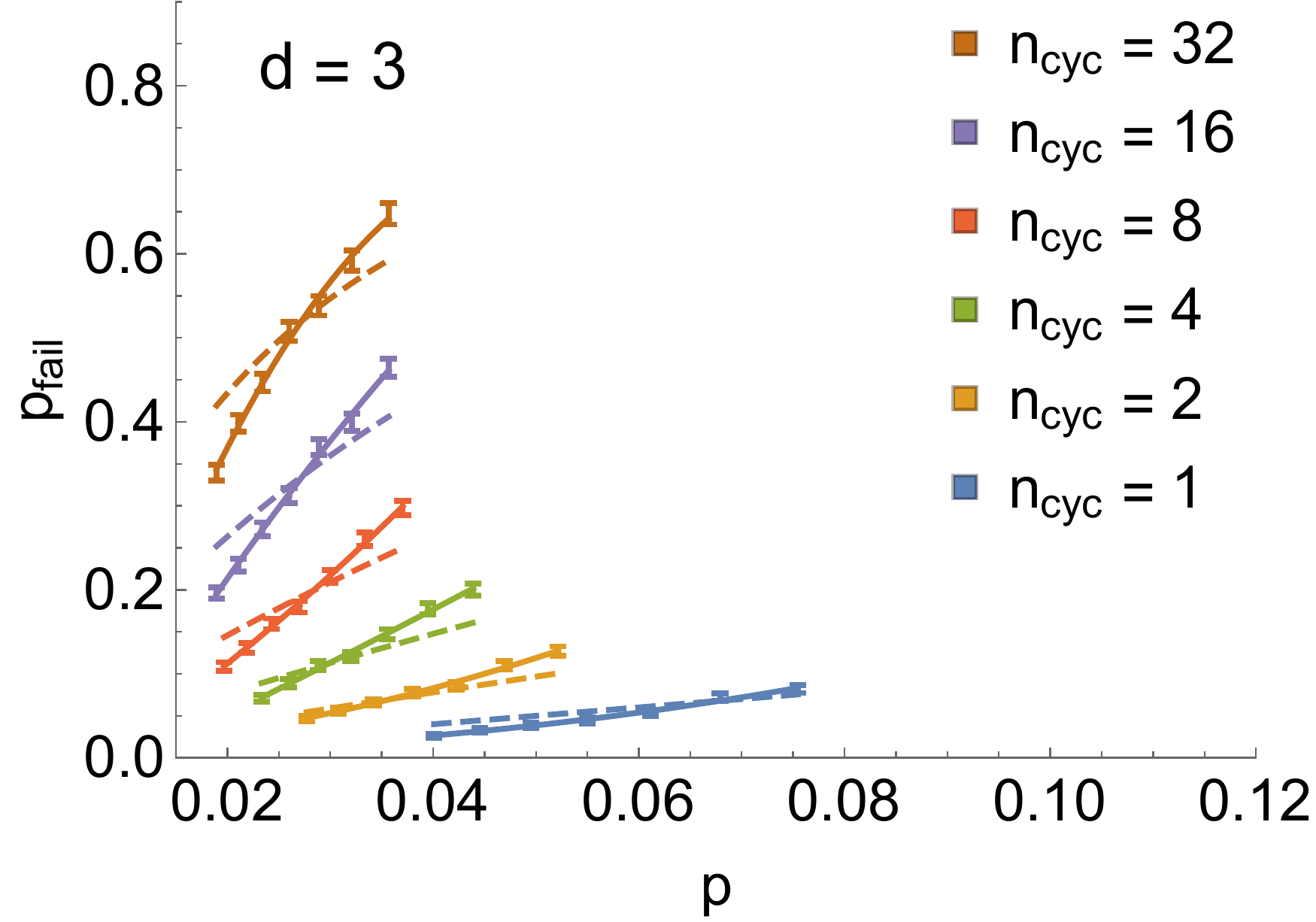}
	\includegraphics[width=.24\textwidth]{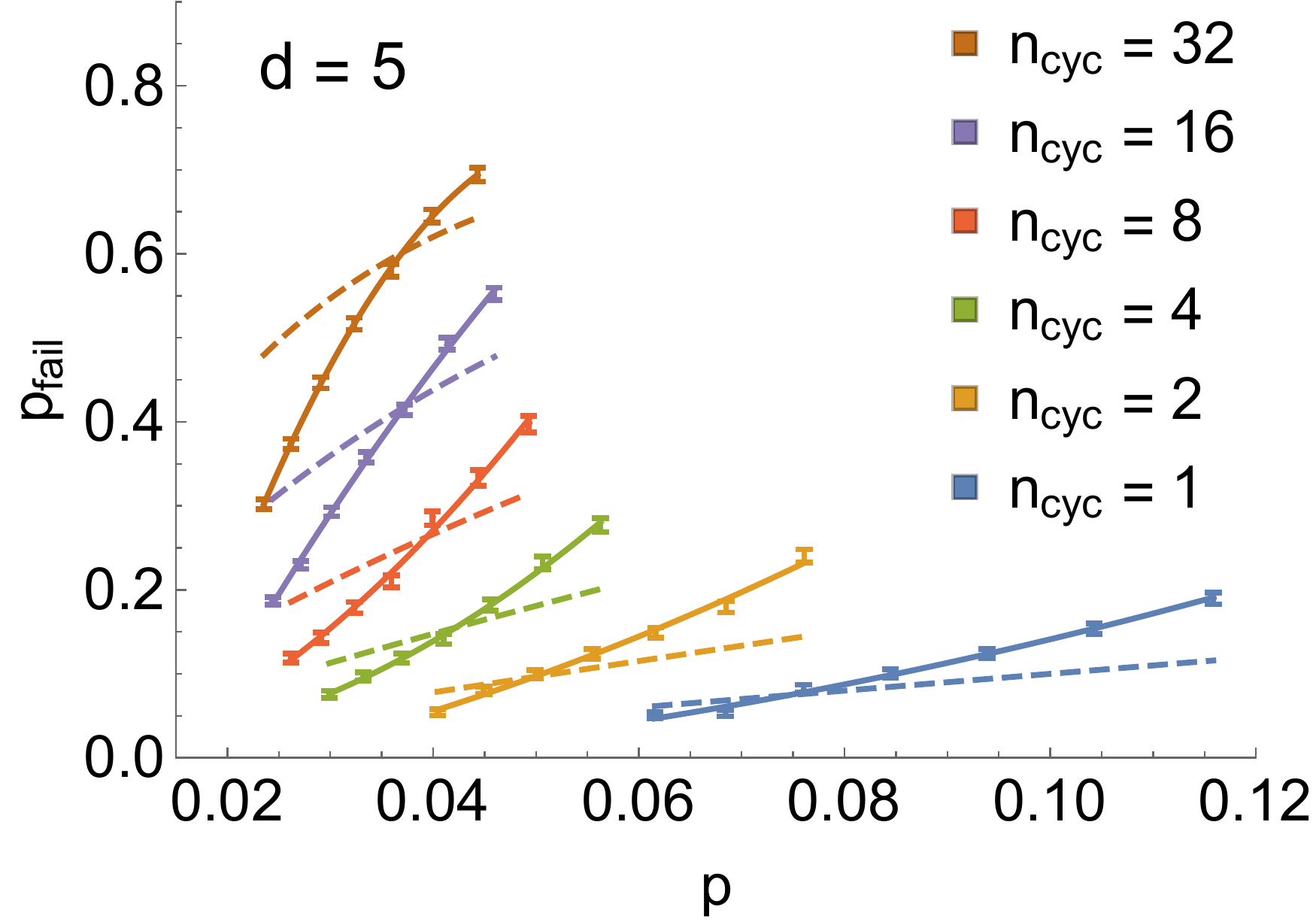}
	\includegraphics[width=.24\textwidth]{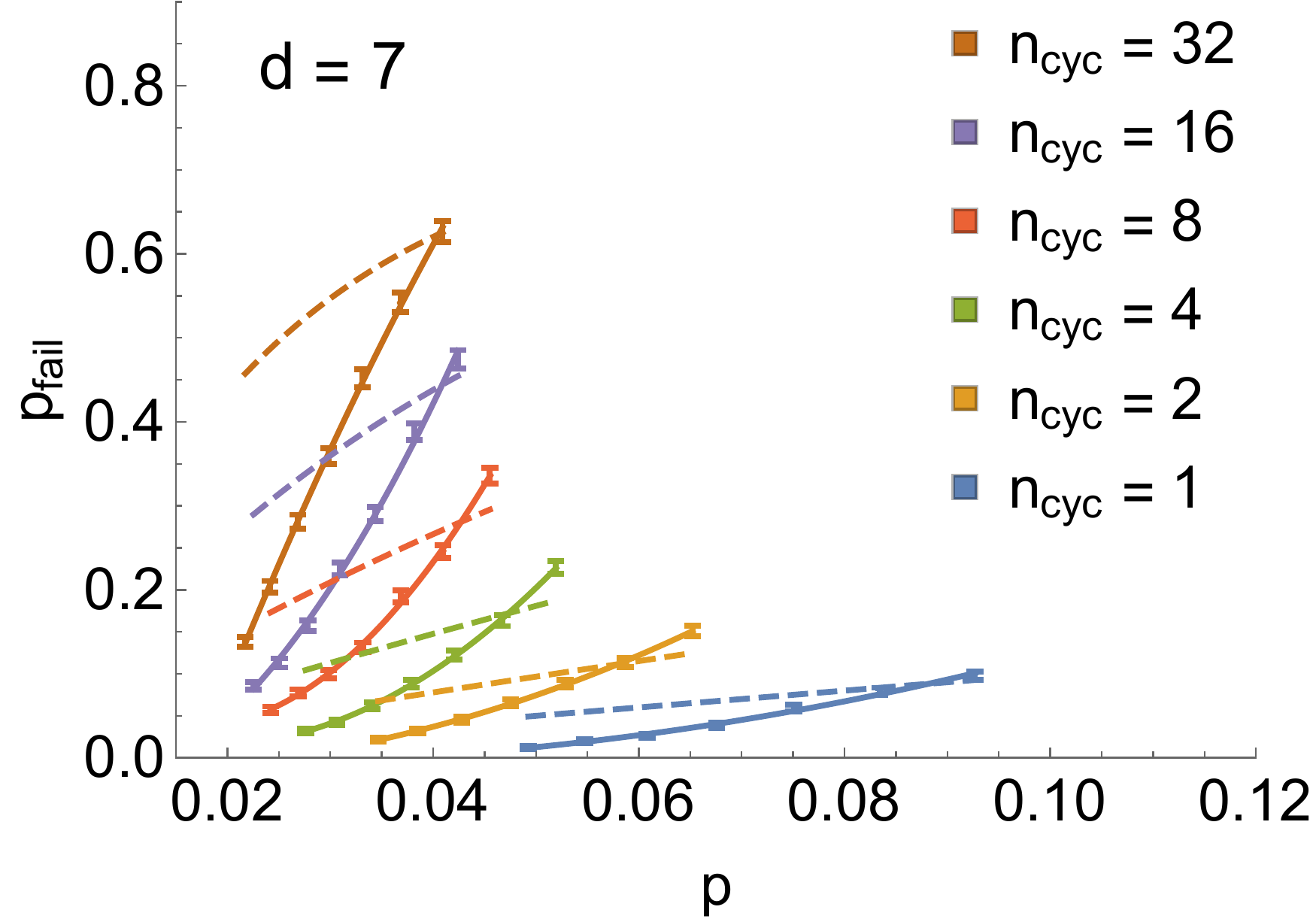}
	\includegraphics[width=.24\textwidth]{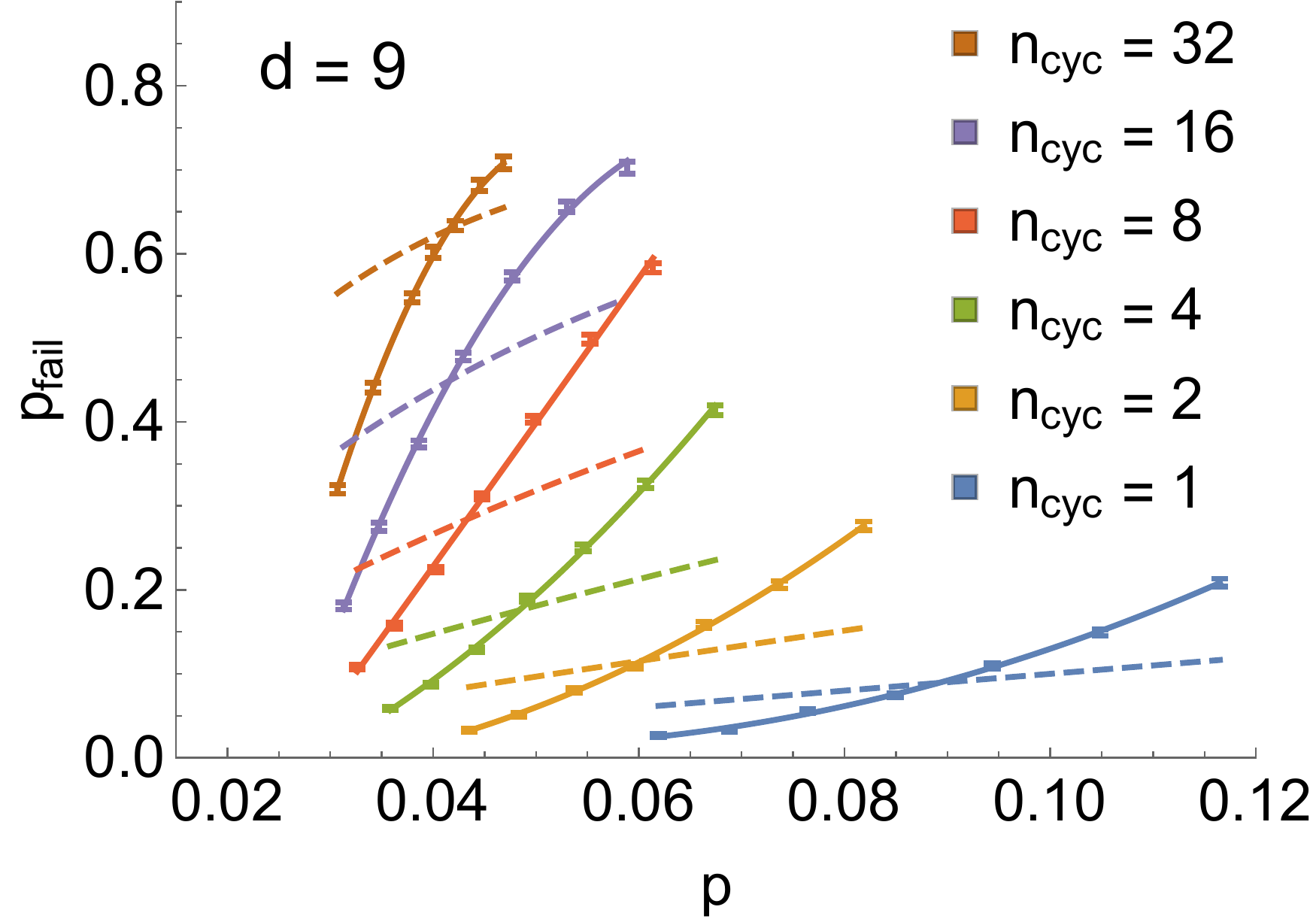}
	\includegraphics[width=.24\textwidth]{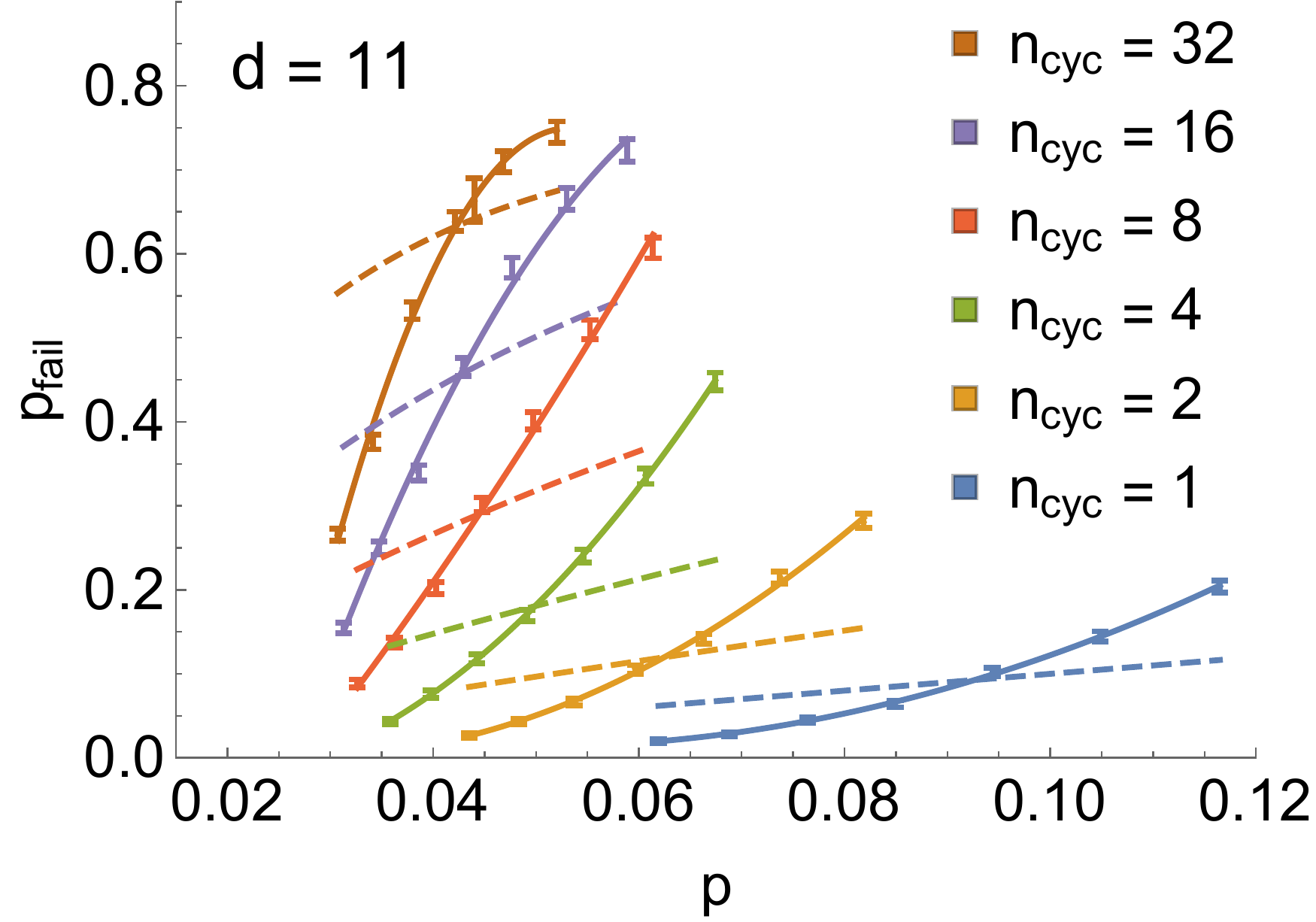}
	\includegraphics[width=.24\textwidth]{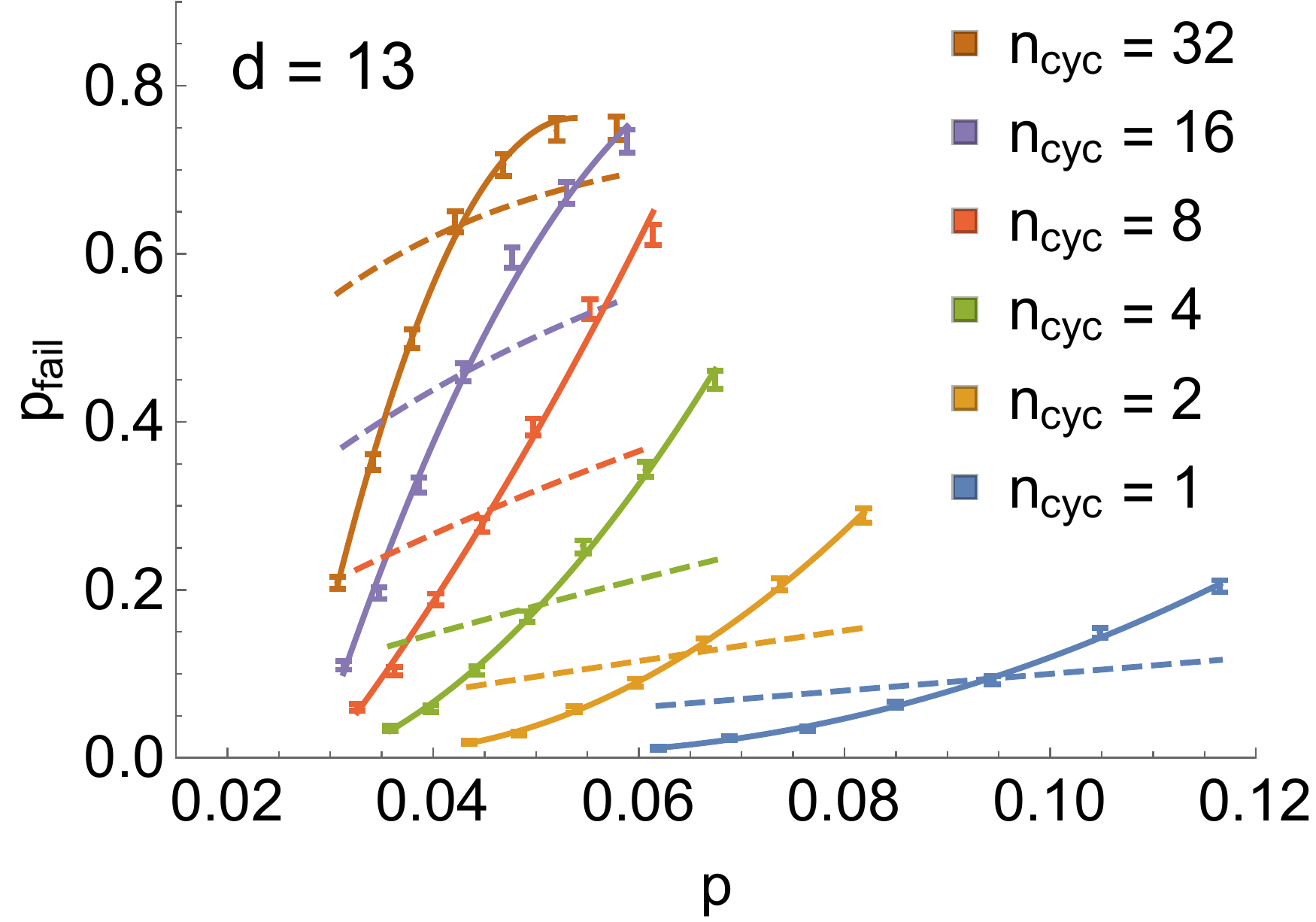}
	\includegraphics[width=.24\textwidth]{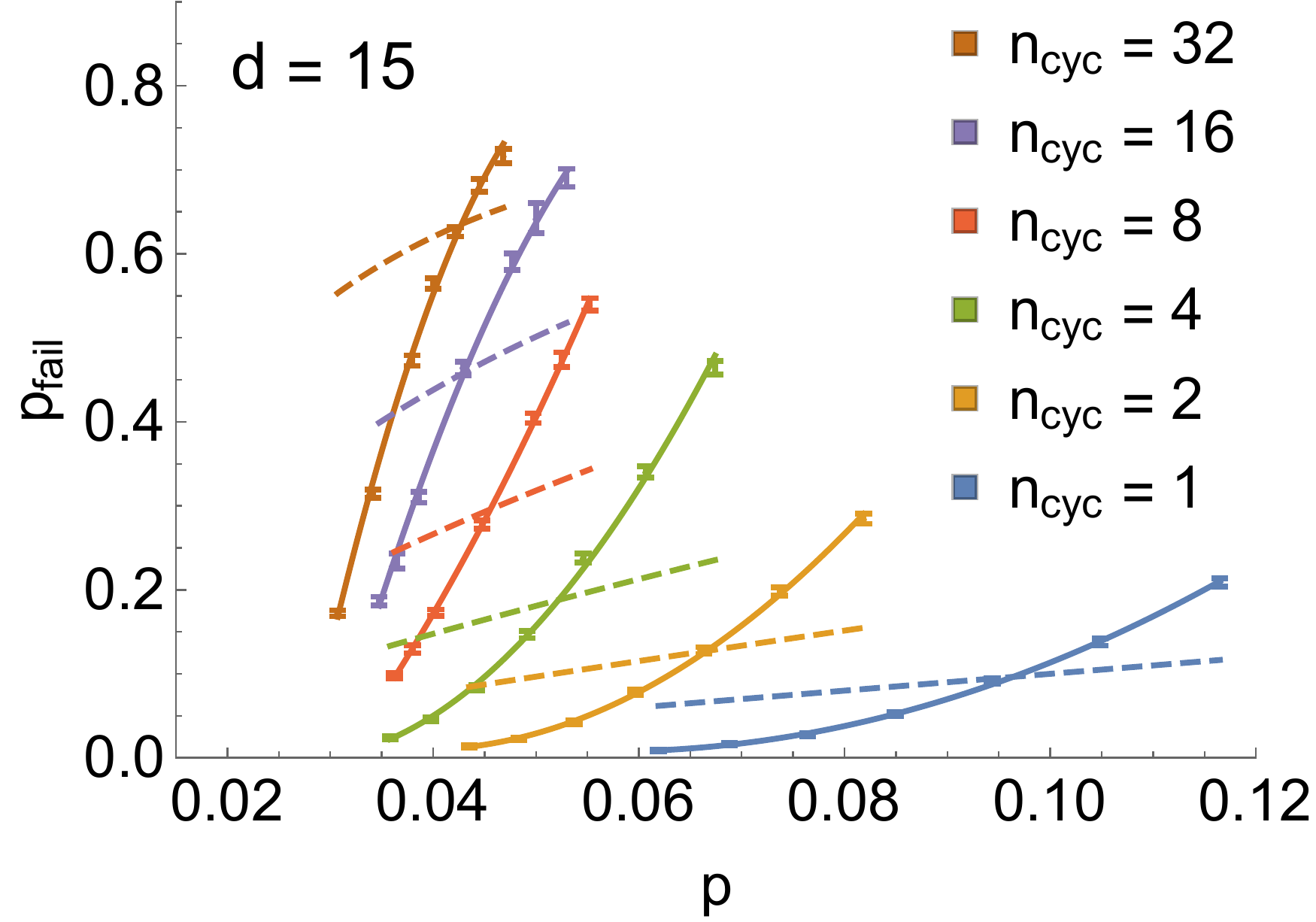}
	\includegraphics[width=.24\textwidth]{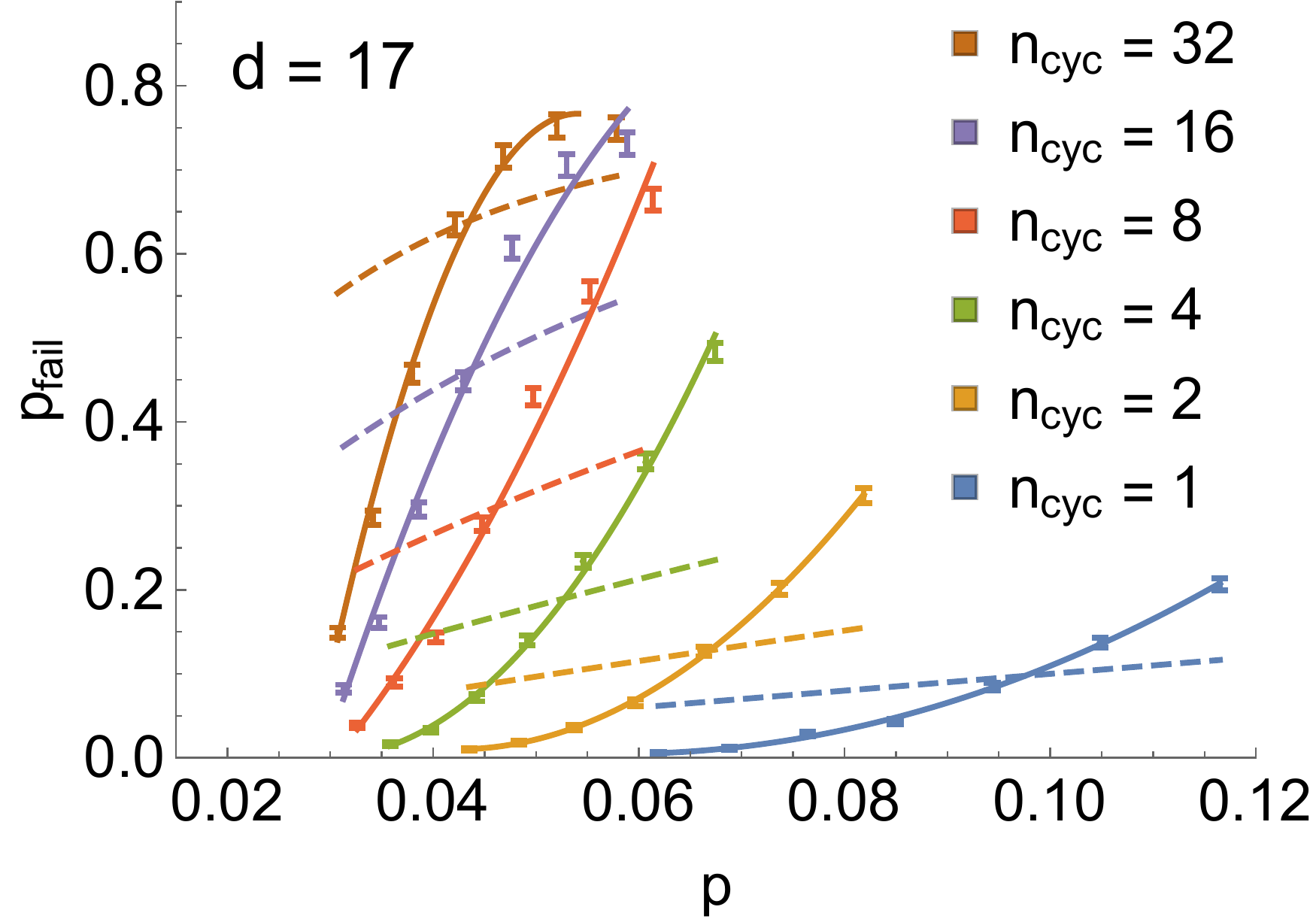}\\\vspace*{5mm}
	(a)\hspace*{-5mm}\includegraphics[width=.45\textwidth]{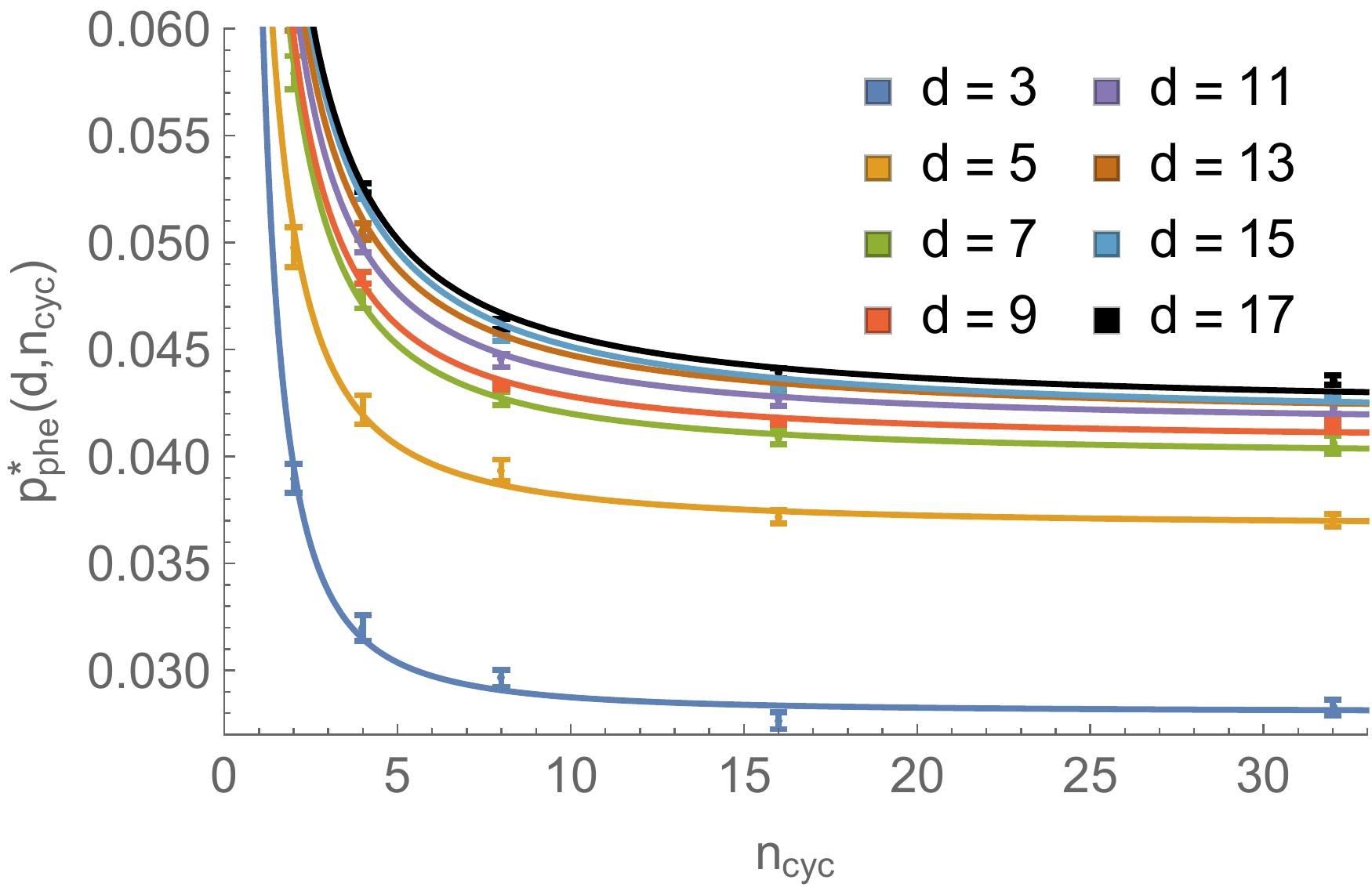}\quad\quad\quad
	(b)\hspace*{-5mm}\includegraphics[width=.45\textwidth]{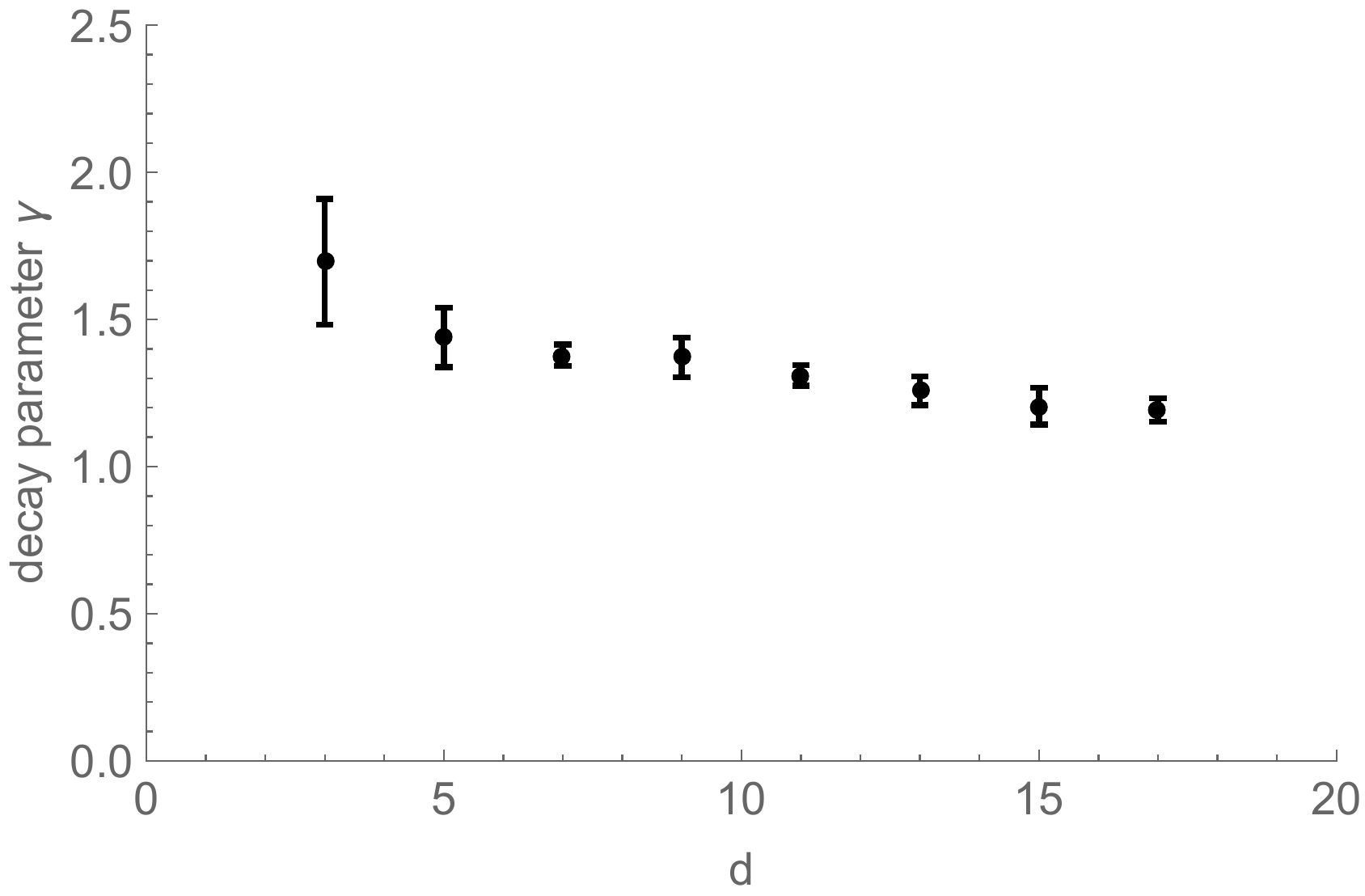}
	\caption{
	(Top two rows) The failure probability of $n_{\text{cyc}}$ QEC cycles for phenomenological noise of strength $p$ given a perfect initial state and final QEC cycle, for various distances $d$.
	We estimate the time-dependent pseudo-threshold $p^*_{\text{phe}}(d,n_{\text{cyc}})$ as an intersection of quadratic fits (solid lines) with the corresponding physical qubit error probability $p_{\text{phy}}(n_{\text{cyc}})$ as defined in \eq{pPhysical} (dashed curves).
	(a) The time-dependent pseudo-threshold $p^*_{\text{phe}}(d,n_{\text{cyc}})$ as a function of $n_\text{cyc}$.
	We estimate the long-time pseudo-thresholds $p^*_{\text{phe}}(d)$ by fitting $p^*_{\text{phe}}(d,n_{\text{cyc}})$ with the ansatz in~\eq{time-dep-pseudothreshold} and in (b) we plot the decay parameter $\gamma$.
	We observe that $\gamma$ stabilizes rapidly with increasing $d$, confirming that the residual noise reaches equilibrium over a time which is independent of system size.	
	} 
	\label{fig:BitflipNoiseAppendix1}
\end{figure}

\begin{figure}[h]
	\includegraphics[width=.24\textwidth]{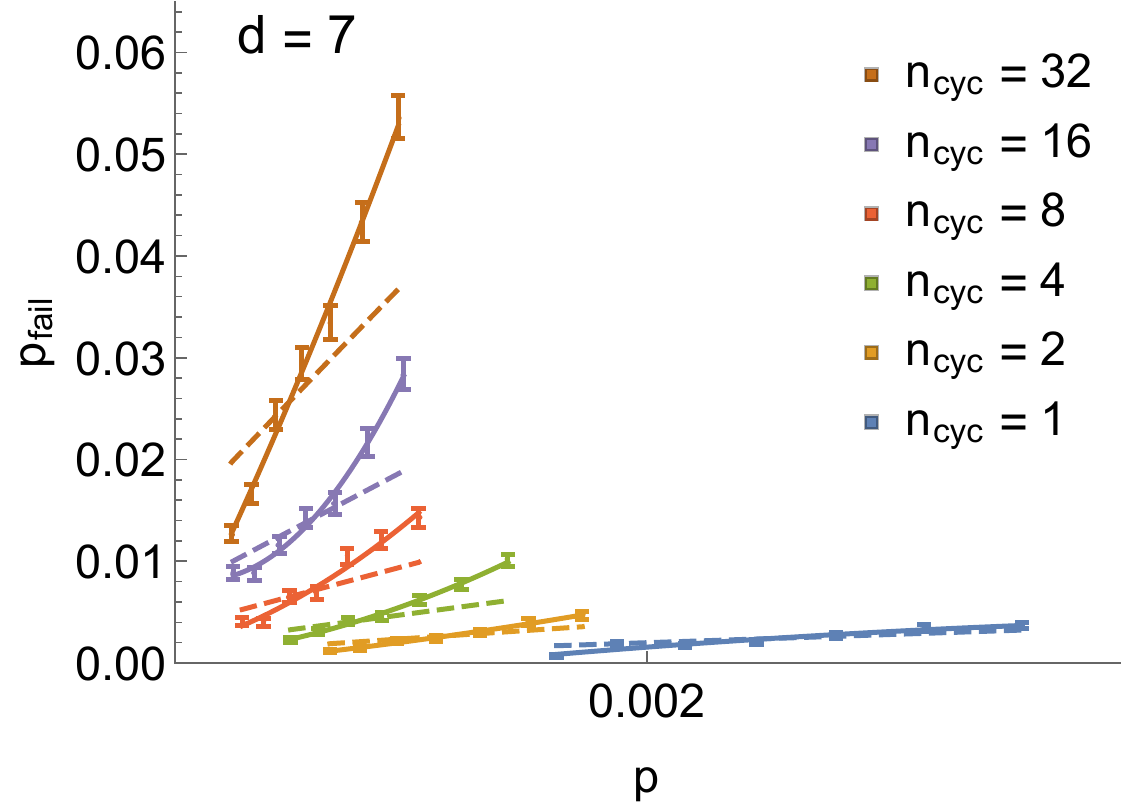}
	\includegraphics[width=.24\textwidth]{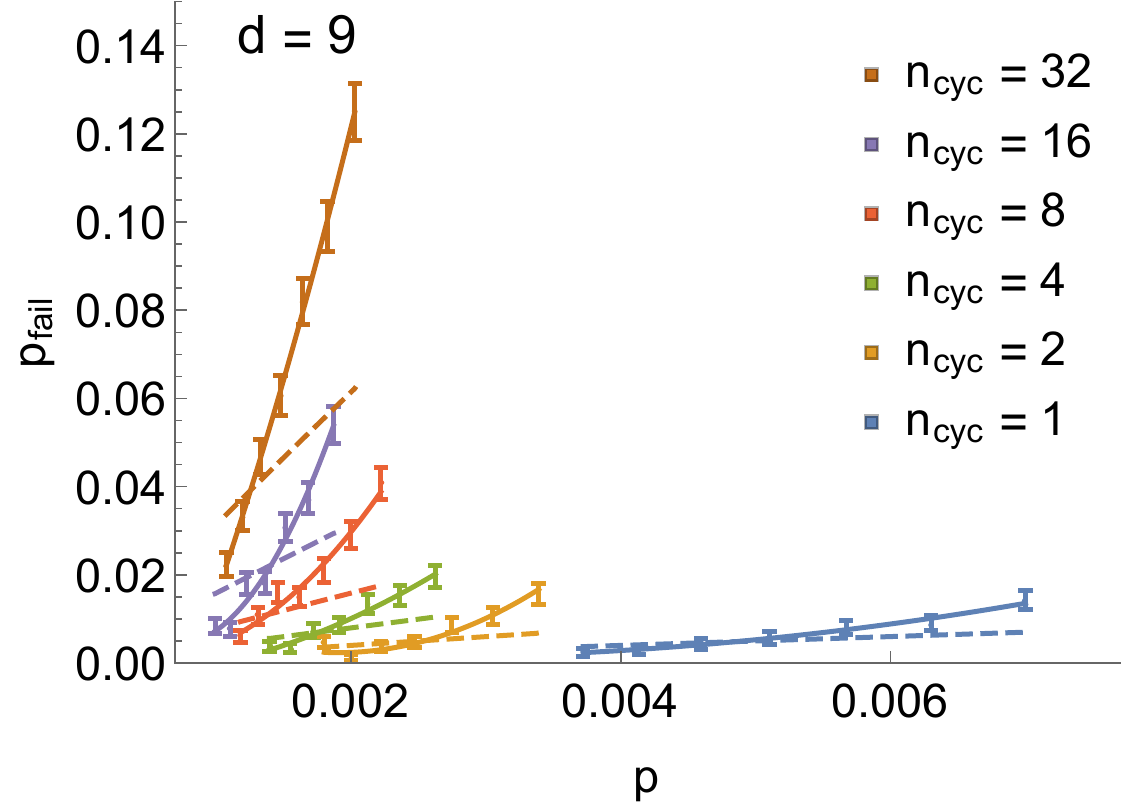}
	\includegraphics[width=.24\textwidth]{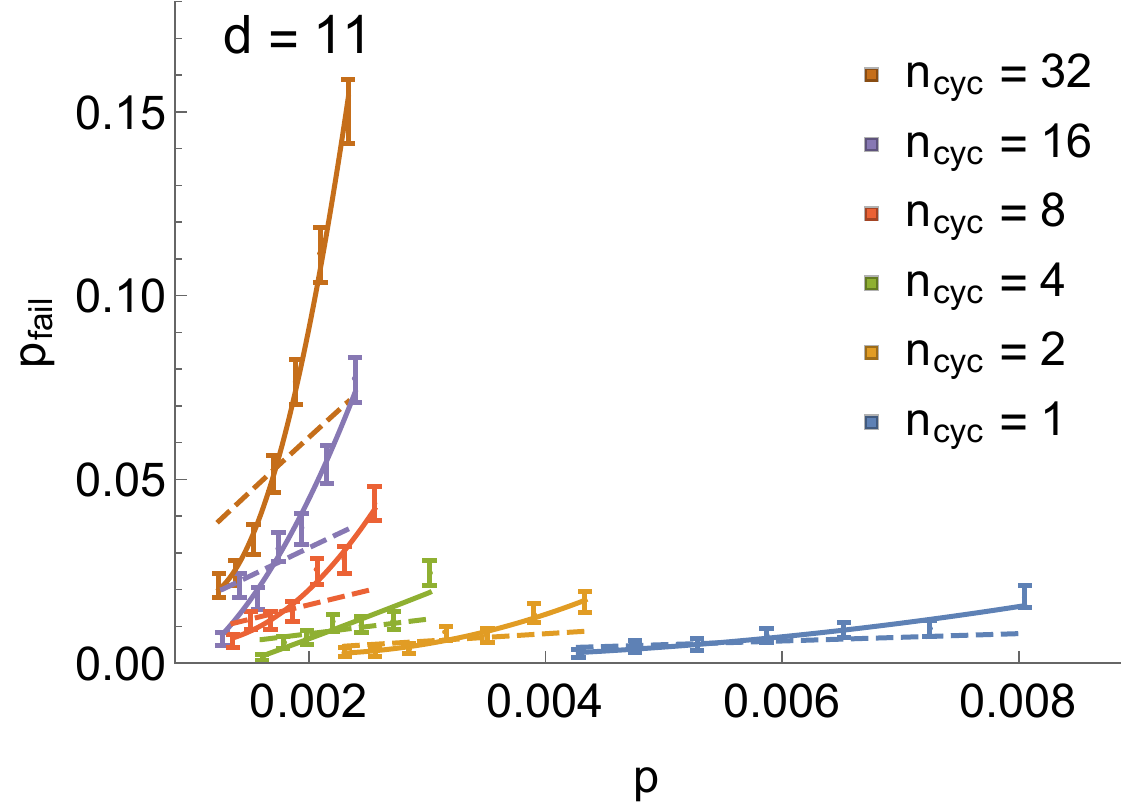}
	\includegraphics[width=.24\textwidth]{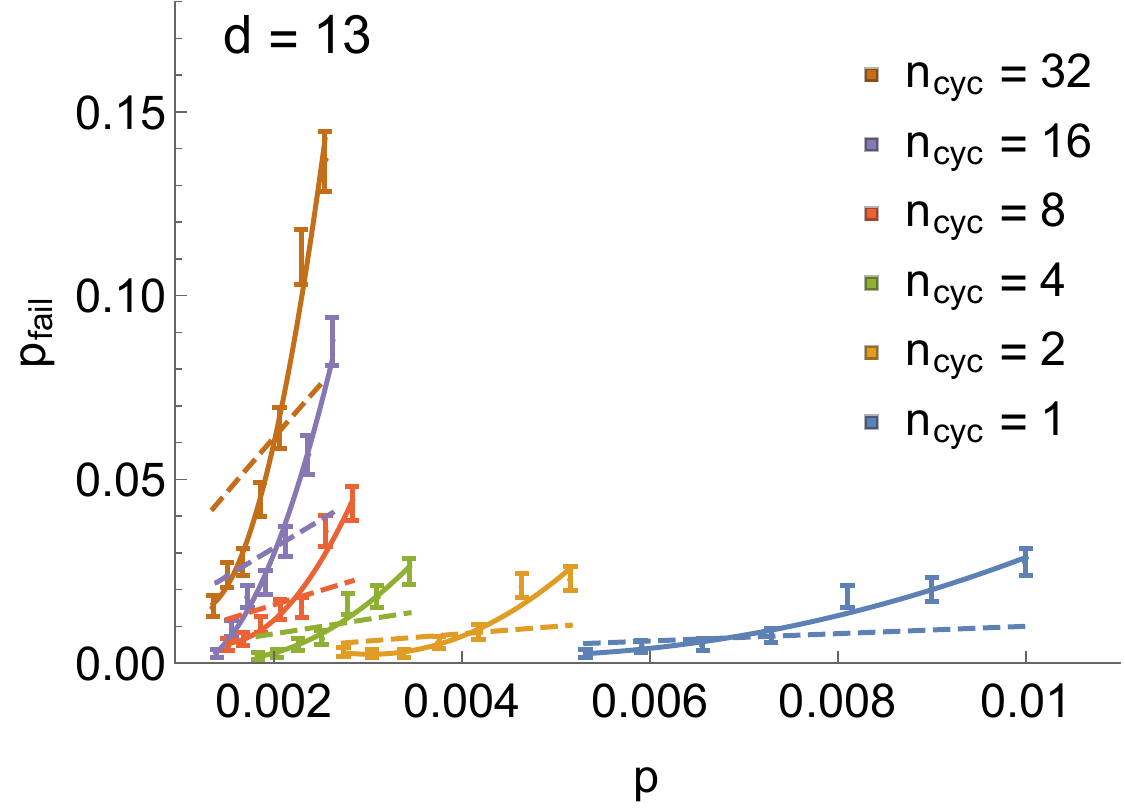}
	\includegraphics[width=.24\textwidth]{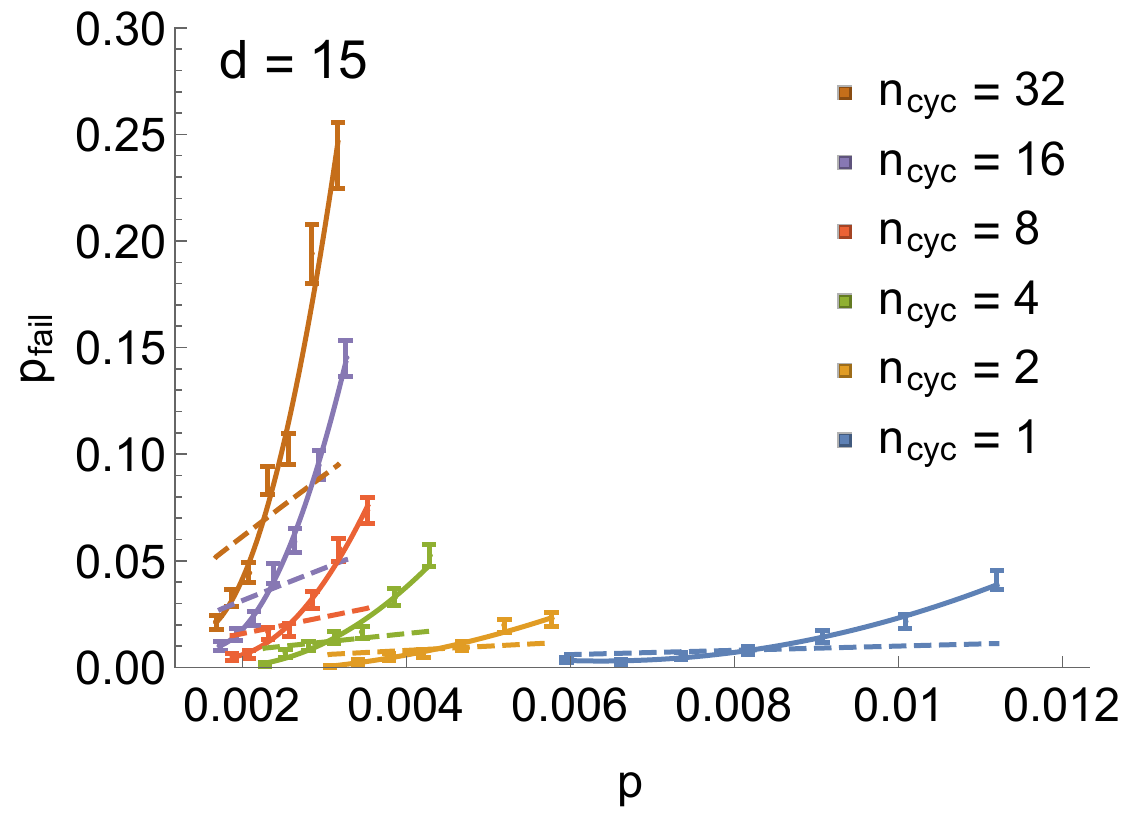}
	\includegraphics[width=.24\textwidth]{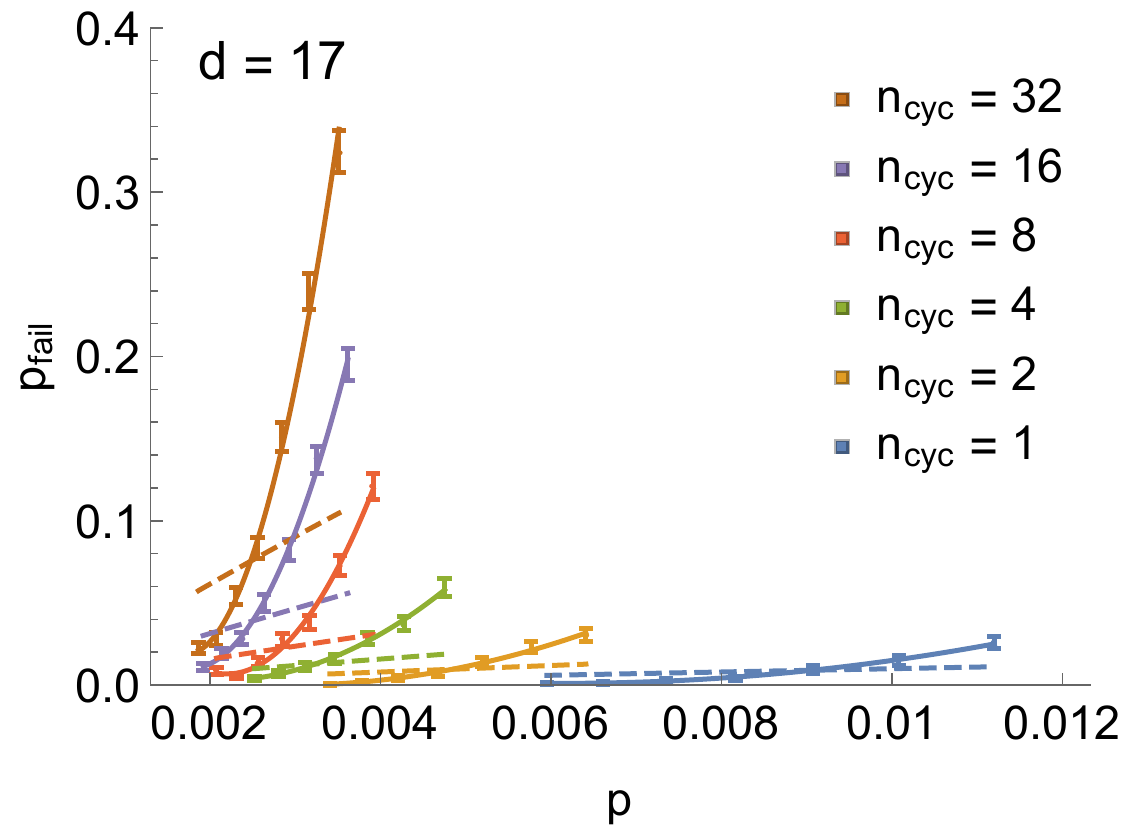}
	\includegraphics[width=.24\textwidth]{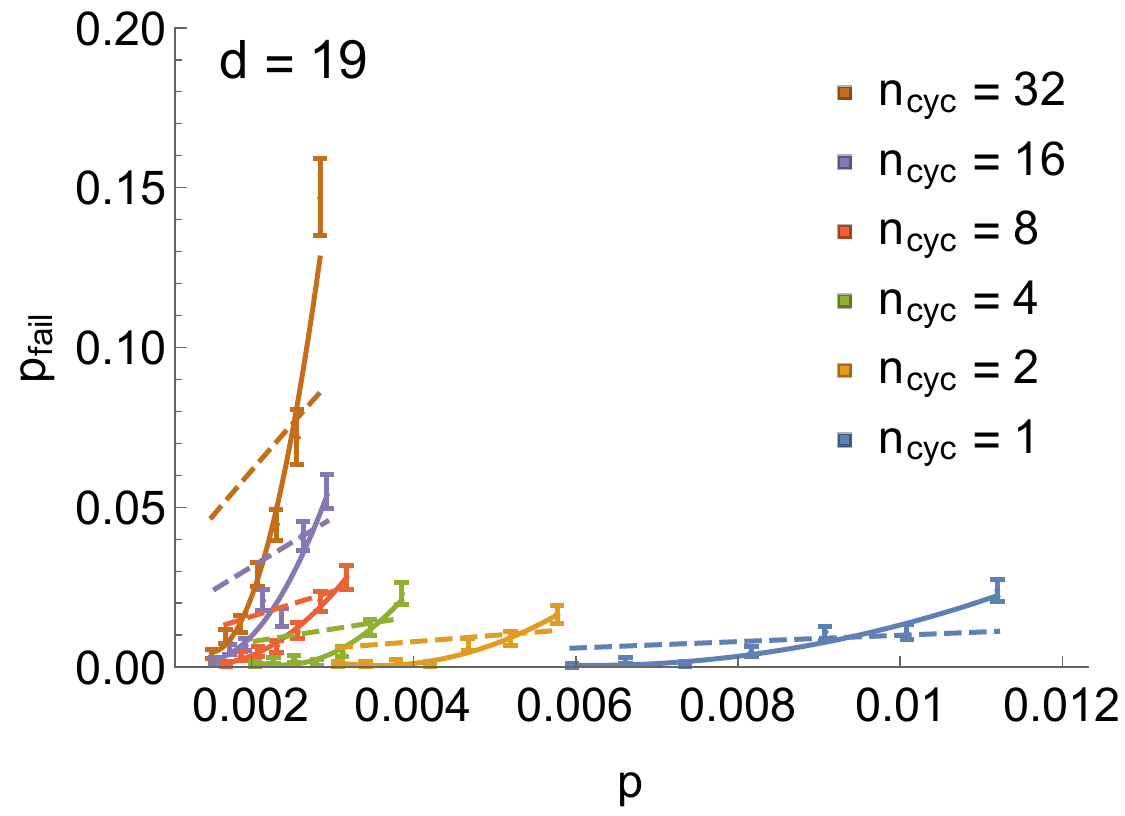}
	\includegraphics[width=.24\textwidth]{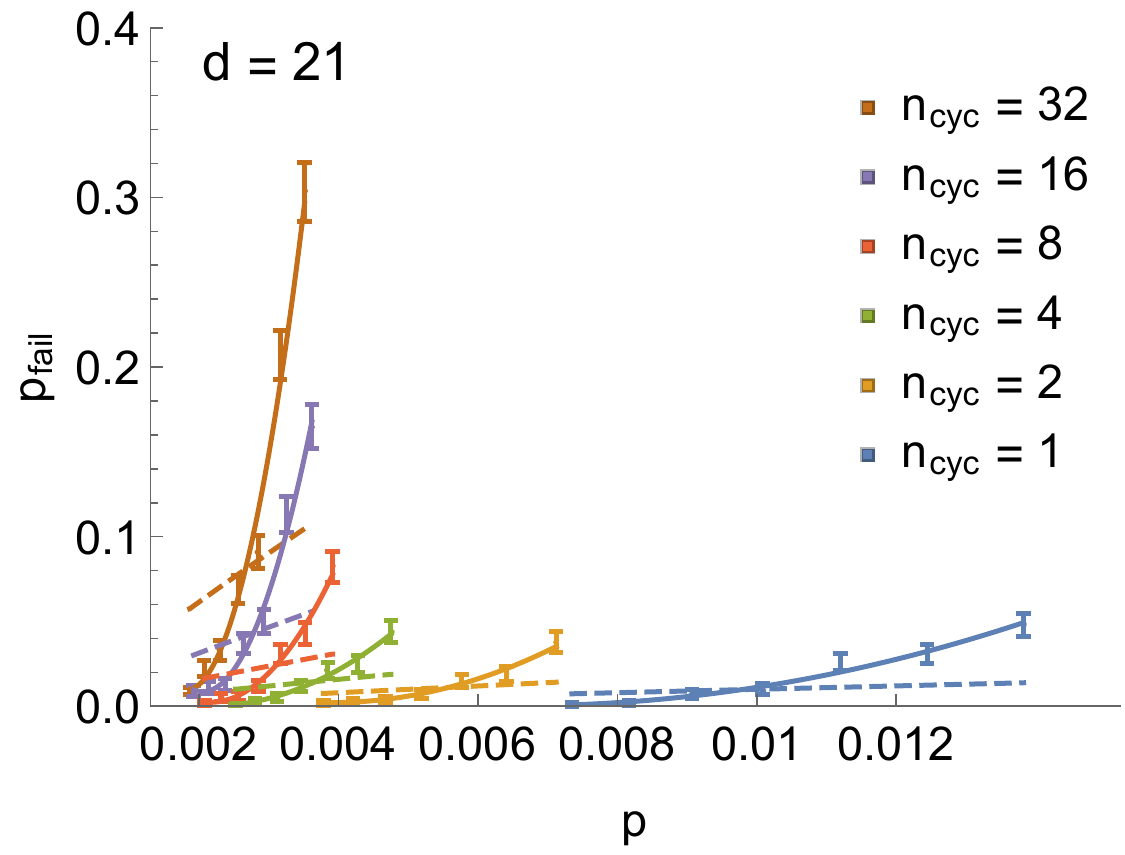}\\
	\vspace*{5mm}
	(a)\hspace*{-5mm}\includegraphics[width=.45\textwidth]{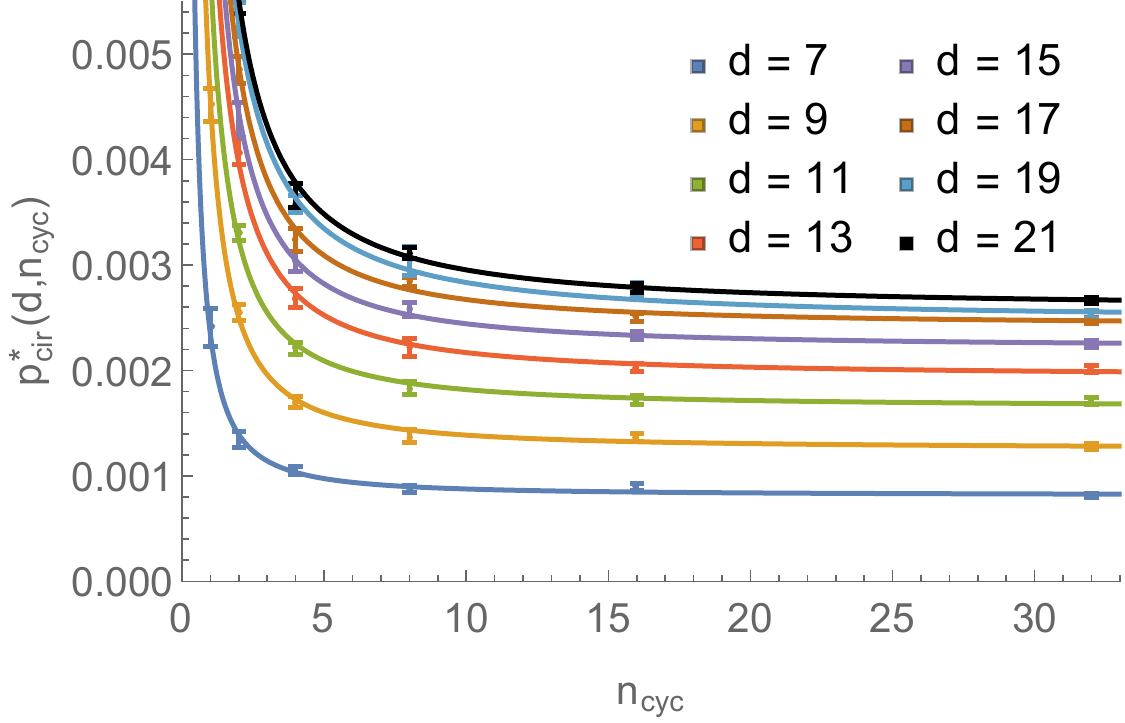}\quad\quad\quad
	(b)\hspace*{-5mm}\includegraphics[width=.45\textwidth]{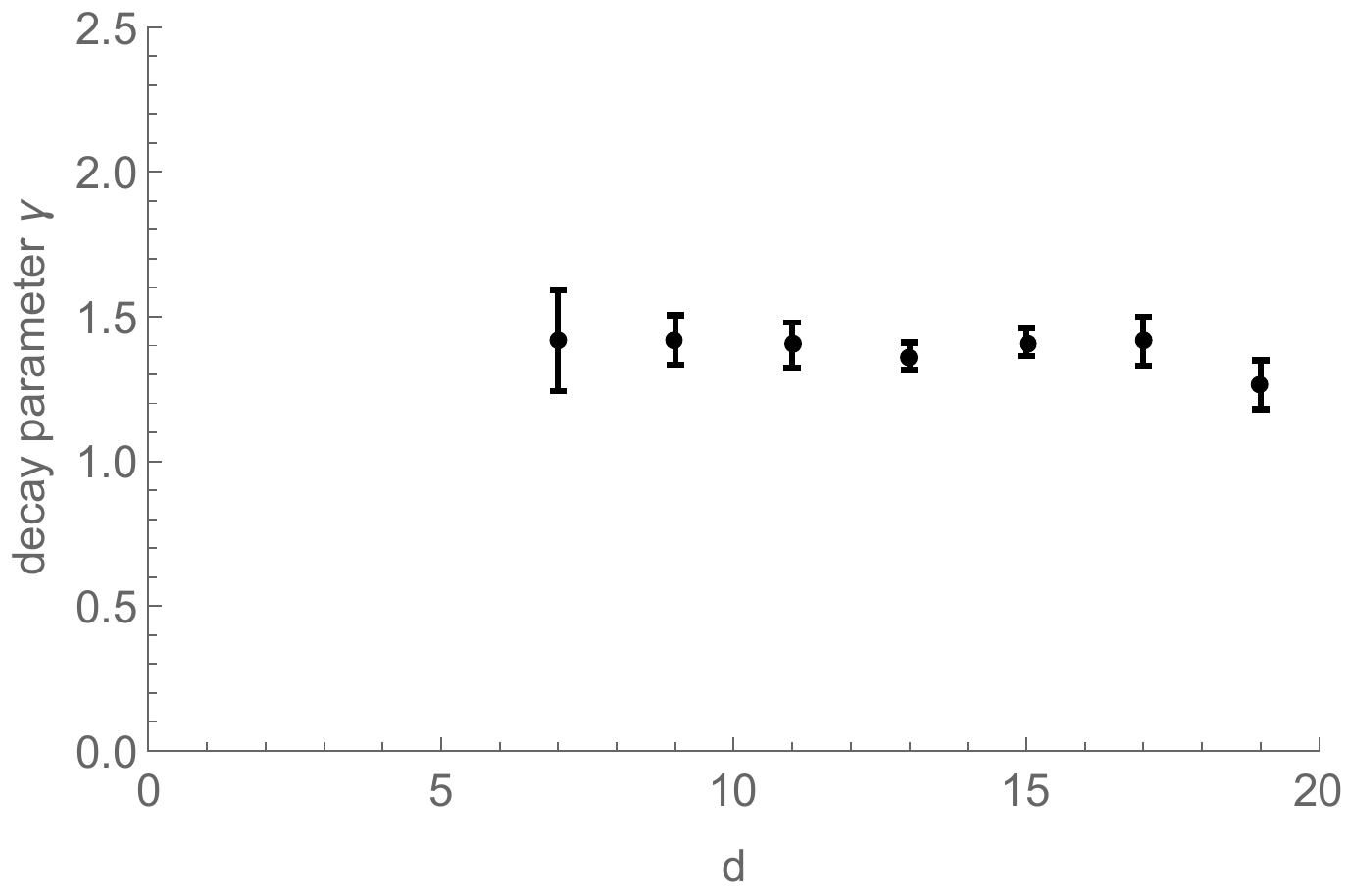}
	\caption{
    (Top two rows) The failure probability of $n_{\text{cyc}}$ QEC cycles for circuit noise of strength $p$ given a perfect initial state and final QEC cycle, for various distances $d$.
	We estimate the time-dependent pseudo-threshold $p^*_{\text{cir}}(d,n_{\text{cyc}})$ as an intersection of quadratic fits (solid lines) with the corresponding physical qubit error probability $p_{\text{phy}}(n_{\text{cyc}})$ as defined in \eq{pPhysical} (dashed curves). 
	(a) The time-dependent pseudo-threshold $p^*_{\text{cir}}(d,n_{\text{cyc}})$ as a function of $n_\text{cyc}$.
	We estimate the long-time pseudo-thresholds $p^*_{\text{cir}}(d)$ by fitting $p^*_{\text{cir}}(d,n_{\text{cyc}})$ with the ansatz in~\eq{time-dep-pseudothreshold} and in (b) we plot the decay parameter $\gamma$.
	We observe that $\gamma$ stabilizes rapidly with increasing $d$, confirming that the residual noise reaches equilibrium over a time which is independent of system size.	
	}
	\label{fig:CircuitNoiseAppendix1}
\end{figure}

\begin{figure}[h]
	\includegraphics[width=.5\textwidth]{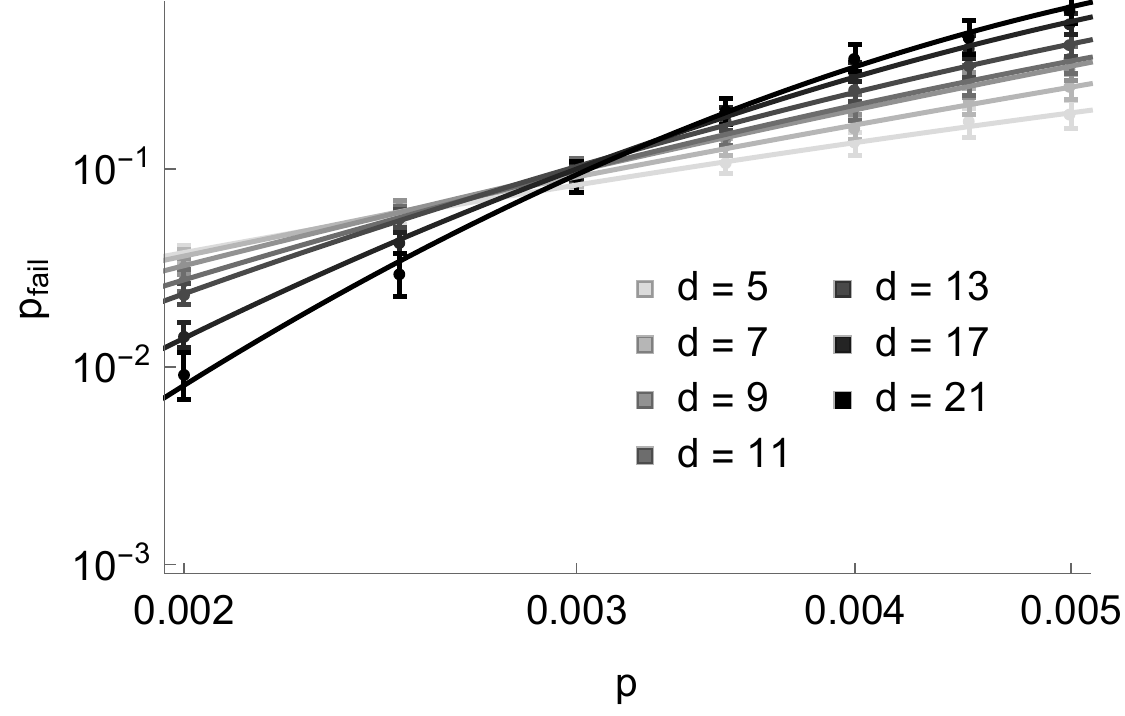}
	\caption{
    The failure probability $p_\text{fail}$ of the noisy-syndrome projection decoder for $n_{\text{cyc}}=d$ rounds of circuit noise and various distances $d$.
    We use this data to find the failure curve crossings $p^\times_{\text{cir}}(d)$, which we plot in \fig{CircuitNoiseThreshold}(b).
	}
	\label{fig:CircuitMeasurementCrossing}
\end{figure}

\begin{figure}[h]
    (a)\hspace*{-5mm}\includegraphics[width=.28\textwidth]{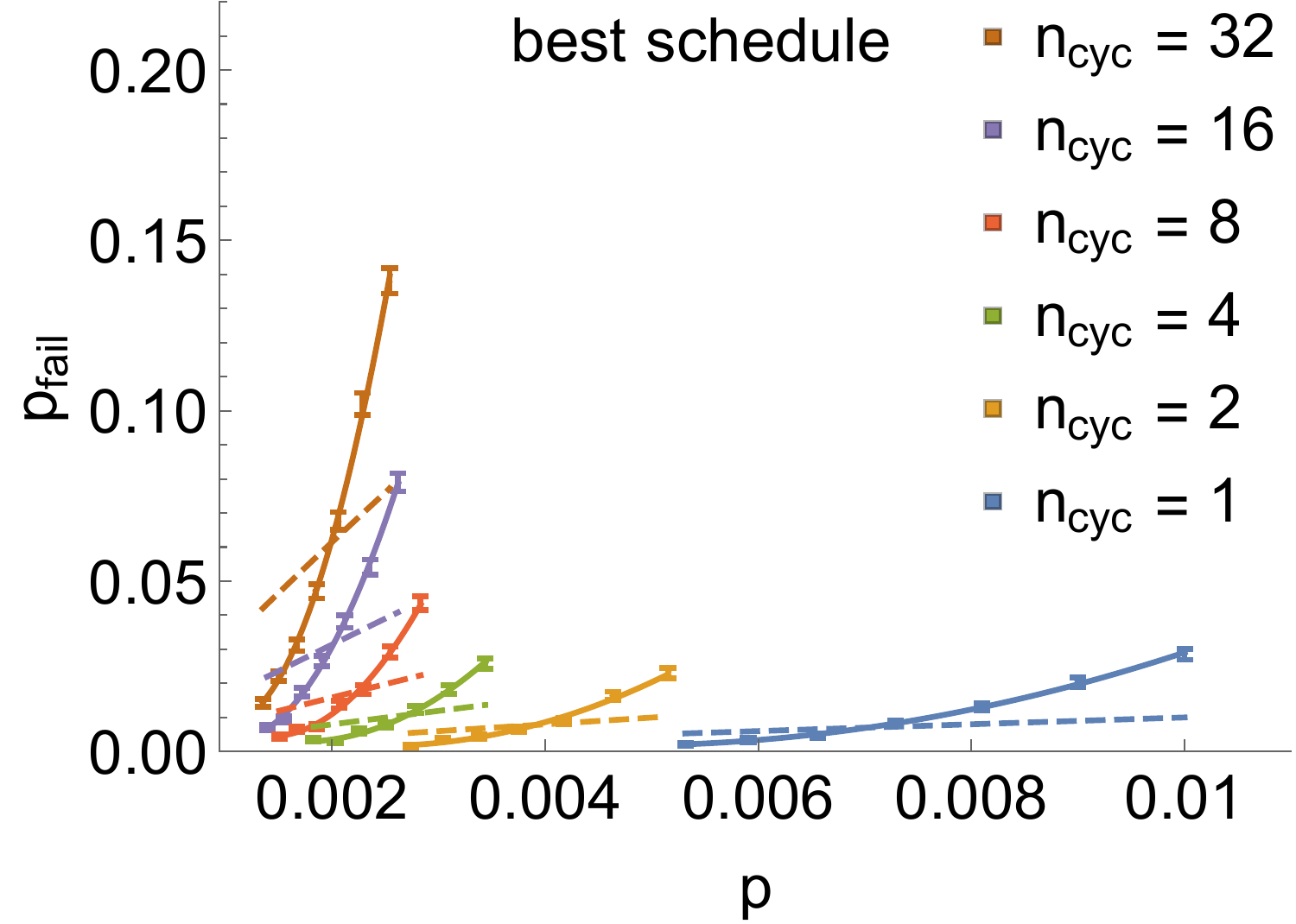}\quad\quad
    (b)\hspace*{-5mm}\includegraphics[width=.28\textwidth]{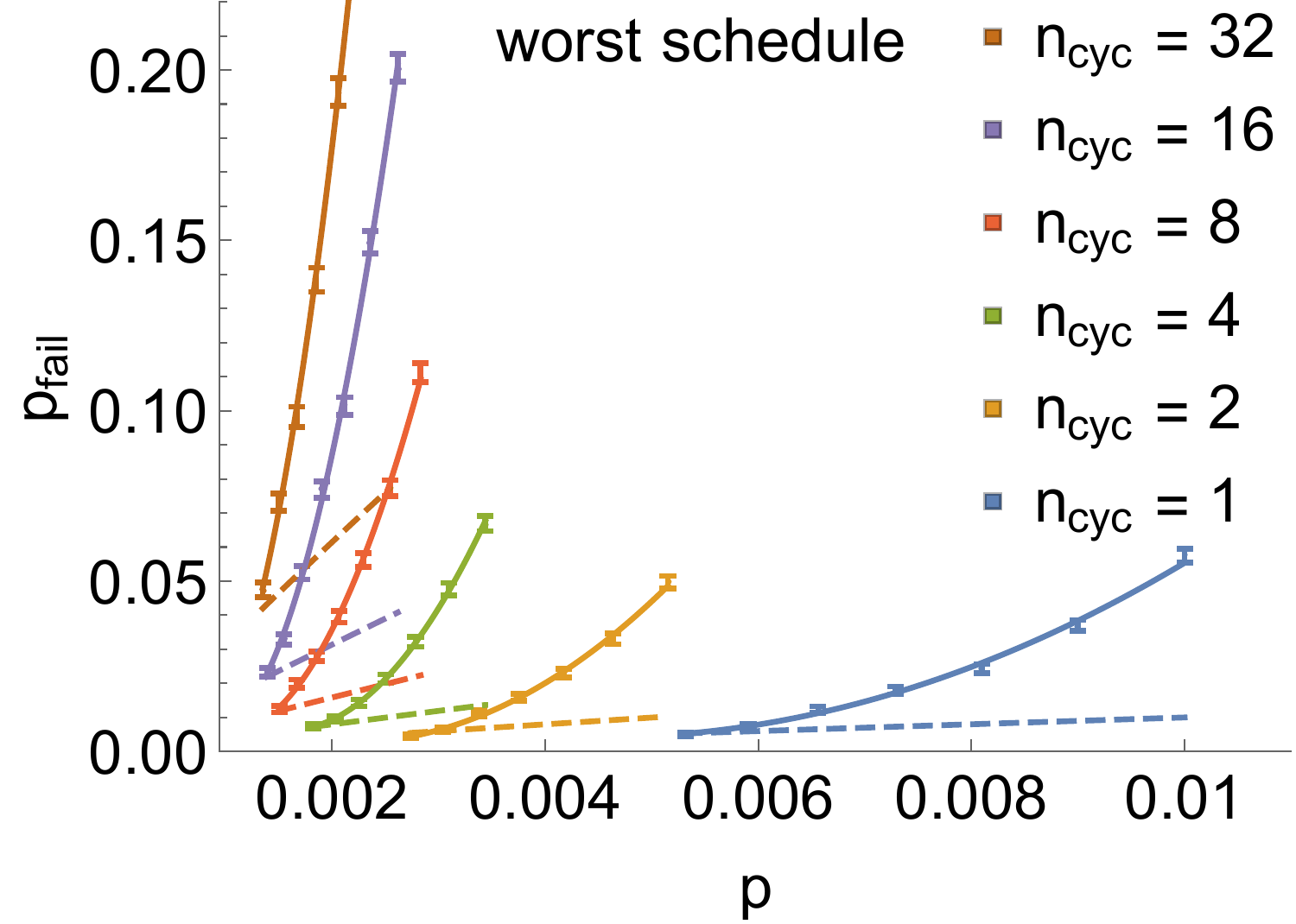}\quad\quad
	(c)\hspace*{-5mm}\includegraphics[width=.34\textwidth]{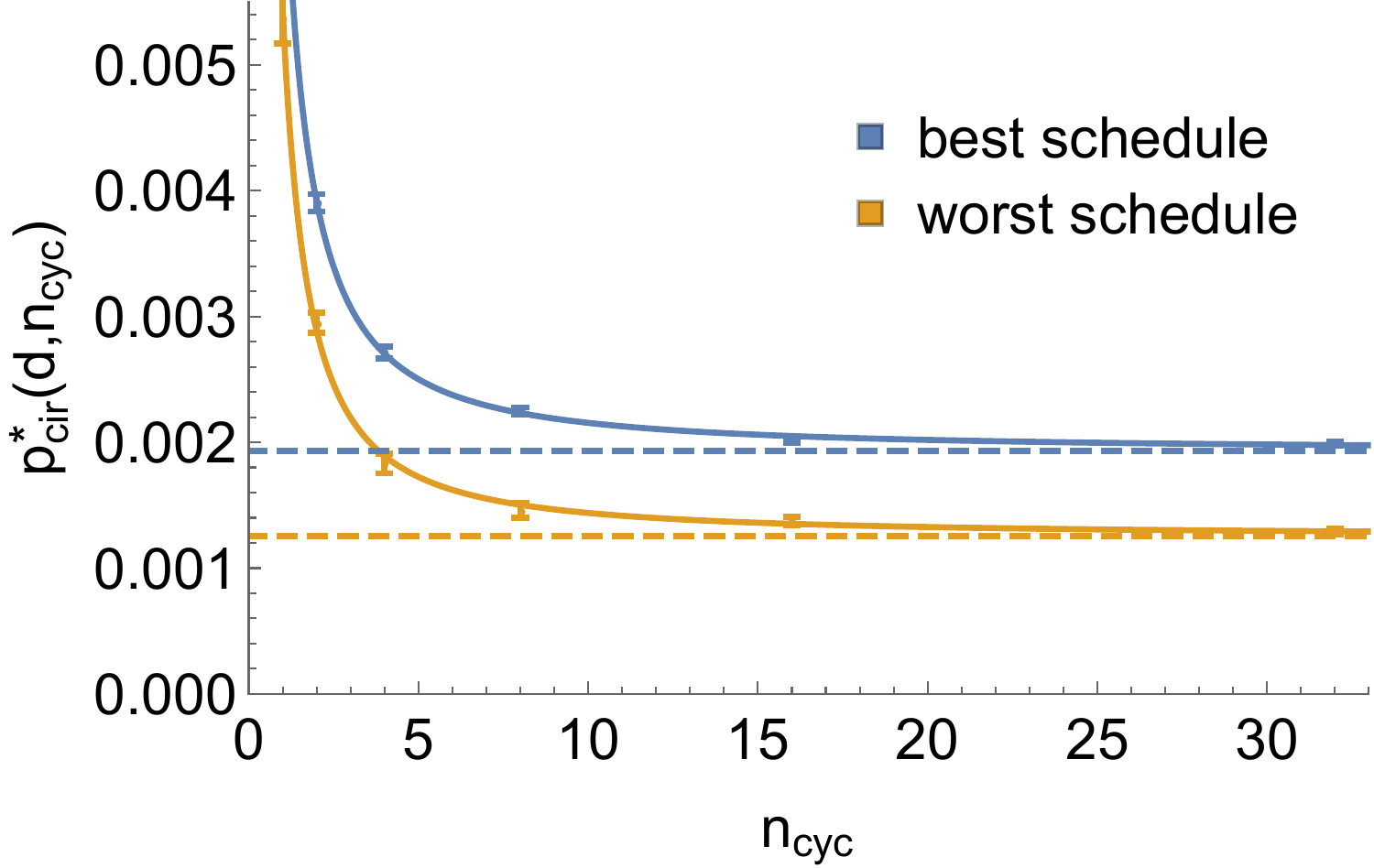}
	\caption{
	Performance comparison of
	(a) the best ranked, and (b) the worst ranked length-7 CNOT schedules from \fig{CircuitNoiseThreshold}(a), namely 
	$\{4, 1, 2, 3, 6, 5 ; 3, 2, 5, 6, 7, 4\}$ and $\{4, 1, 2, 7, 6, 3 ; 1, 6, 7, 4, 5, 2\}$.
	We numerically estimate the failure probability of $n_{\text{cyc}}$ QEC cycles for circuit noise of strength $p$ for $d=13$ given a perfect initial state and final QEC cycle.
	We fit this data with quadratic functions (solid lines), and also show
	the physical qubit error probability $p_{\text{phy}}(n_{\text{cyc}})$ (dashed curves).
	We estimate the pseudo-threshold $p^*_{\text{cir}}(d,n_{\text{cyc}})$ which we plot in sub-figure (c) by identifying the intersection of the solid and dashed curves for a given $d$, $p$ and $n_{\text{cyc}}$.
	We see that the long-time pseudo-threshold for the good circuit is $0.193(2)\%$ (blue, dashed), considerably larger than that of $0.126(3)\%$ for the bad circuit (yellow, dashed).
	}
	\label{fig:cnotscheduleperformance}
\end{figure}

\begin{figure}[h]
	(a)\hspace*{-5mm}\includegraphics[width=.45\textwidth]{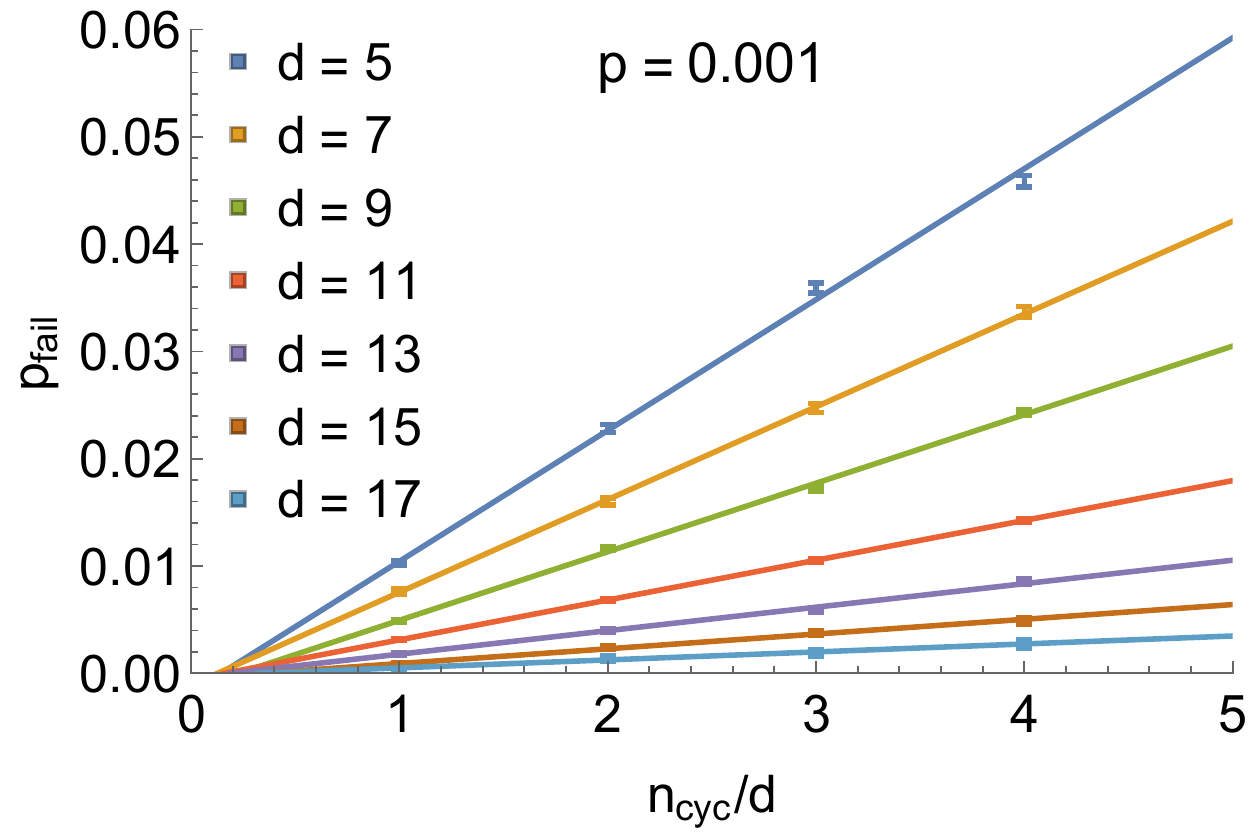}
	\quad\quad\quad
	(b)\hspace*{-5mm}\includegraphics[width=.45\textwidth]{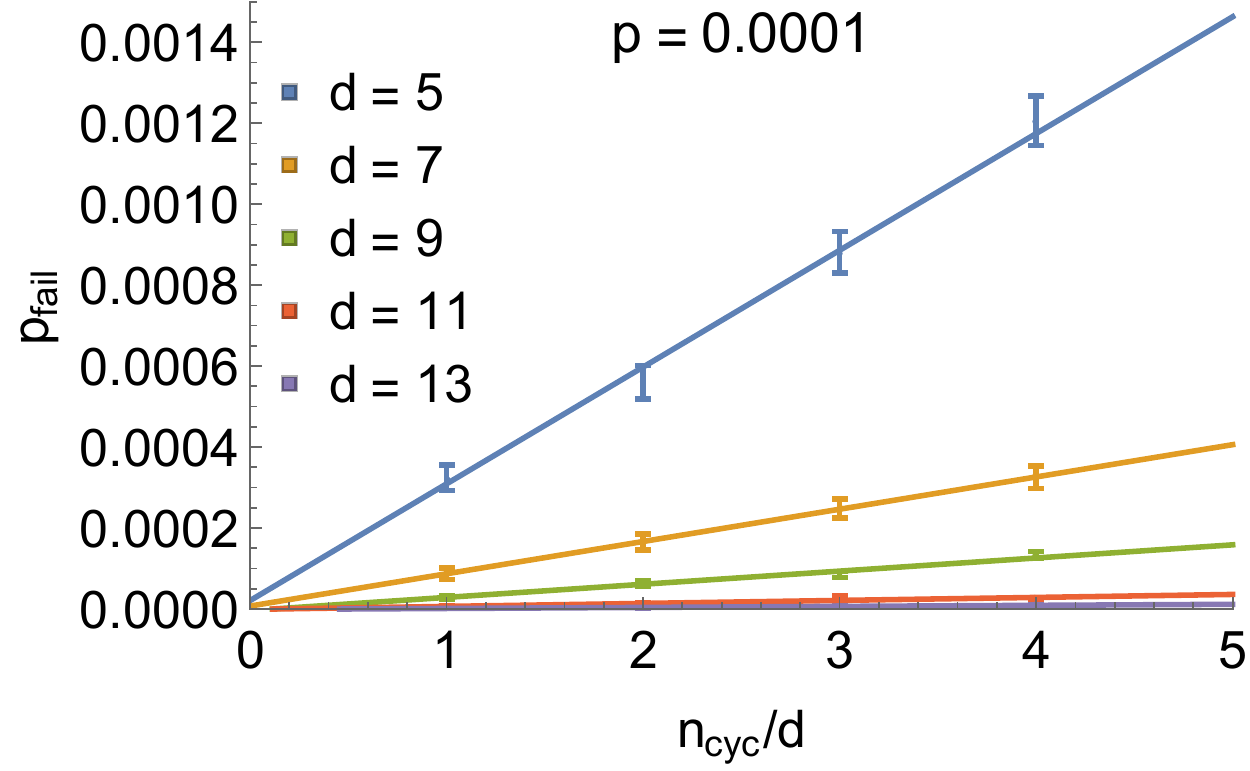}
	\caption{
	The failure probability of the logical idle operation using the best-performing CNOT schedule as a function of $n_{\text{cyc}}/d$ for (a) $p= 10^{-3}$ and (b) $p= 10^{-4}$.
	We use the ansatz in~\eq{ansatz_failure} to find the logical failure rate $\overline{p}(p,d)$.
	We observe that QEC equilibration occurs by the $d$th QEC cycle.
	}
	\label{fig:CircuitNoiseAnalysisAdditionalData}
\end{figure}

Here we provide additional data for the analysis of the performance of the 2D color code using the faulty-measurement projection decoder under phenomenological noise and circuit noise in \sec{2DCCperformance}. 
In \fig{BitflipNoiseAppendix1} we show the performance under phenomenological noise, which we use to produce the long-time pseudo-threshold plot in \fig{BitflipMeasurementThresholdAndStephens}(d) in the main text.
Similarly, in \fig{CircuitNoiseAppendix1} we show the performance under phenomenological noise, which we use to produce the long-time pseudo-threshold in \fig{CircuitNoiseThreshold}(b) in the main text.
We point out that the decay parameter in the fit of the time-dependent pseudo-threshold in \fig{BitflipNoiseAppendix1}(b) and \fig{CircuitNoiseAppendix1}(b) shows that convergence to the long-time pseudo-threshold seems not to depend significantly on the system size.
In \fig{CircuitMeasurementCrossing} we provide the data used to produce the crossings in \fig{CircuitNoiseThreshold}(b).
In \fig{cnotscheduleperformance} we show the impact of optimizing the stabilizer extraction circuit on the performance of error correction under circuit noise.
In \fig{CircuitNoiseAnalysisAdditionalData}, we show additional data from which we extract the logical failure rates for $p=0.001,0.0001$ in \fig{CircuitNoiseAnalysisIdle}(b) in the main text. 

\clearpage
\subsection{Supplementary details for distillation analysis}
\label{app:haah-hastings-distillation}
 Here we provide more details of the distillation analysis in \sec{distillation}. 
 In \tab{distillation-protocols-puncture-optimization} we consider variants of the distillation circuit formed by puncturing different qubits of the quantum Reed-Muller state.
 This helps us identify that puncturing qubit number three is best with the noise model we assume in this paper since it results in the smallest leading order failure probability given that $\overline{p}_{\text{CNOT}} \geq \overline{p}_{\text{prep}} \geq \overline{p}_{\text{idle}}$.

\begin{table}[h]
\begin{center}
 \begin{tabular}{|c || c | c|} 
 \hline
 punctured qubit & failure probability $p_{\text{fail}}^{\text{RM}}$ & rejection probability $p_{\text{rej}}^{\text{RM}}$ \\
 [0.5ex] 
 \hline\hline
 1 & $ 1.98 ~\overline{p}_{\text{CNOT}} + 0.875 ~\overline{p}_{\text{idle}} +  ~\overline{p}_{\text{prep}}$ & $38 ~\overline{p}_{\text{CNOT}} + 75.6 ~\overline{p}_{\text{idle}} + 11.3 ~\overline{p}_{\text{prep}}  $ \\ 
 \hline
 2 & $ 2.2 ~\overline{p}_{\text{CNOT}} + 0.85 ~\overline{p}_{\text{idle}}$ & $38.2 ~\overline{p}_{\text{CNOT}} + 73.3 ~\overline{p}_{\text{idle}} + 12.3 ~\overline{p}_{\text{prep}}  $ \\ 
 \hline
 3 & $ 1.93 ~\overline{p}_{\text{CNOT}} + 0.875 ~\overline{p}_{\text{idle}}$ & $38.2 ~\overline{p}_{\text{CNOT}} + 73 ~\overline{p}_{\text{idle}} + 12.3 ~\overline{p}_{\text{prep}}  $ \\  
 \hline
 4 & $ 2.45 ~\overline{p}_{\text{CNOT}} + 2.31 ~\overline{p}_{\text{idle}} +  \overline{p}_{\text{prep}}$ & $38.1 ~\overline{p}_{\text{CNOT}} + 77.1 ~\overline{p}_{\text{idle}} + 11.8 ~\overline{p}_{\text{prep}}  $\\
 \hline
 5 & $ 1.98 ~\overline{p}_{\text{CNOT}} + 0.875 ~\overline{p}_{\text{idle}}$ & $38 ~\overline{p}_{\text{CNOT}} + 73 ~\overline{p}_{\text{idle}} + 12.3 ~\overline{p}_{\text{prep}}  $ \\    
 \hline
 6 & $ 2.2 ~\overline{p}_{\text{CNOT}} + 0.85 ~\overline{p}_{\text{idle}} + 0.5 ~\overline{p}_{\text{prep}}$ & $38.2 ~\overline{p}_{\text{CNOT}} + 75.5 ~\overline{p}_{\text{idle}} + 12 ~\overline{p}_{\text{prep}}  $ \\     
 \hline
 7 & $ 1.93 ~\overline{p}_{\text{CNOT}} + 0.875 ~\overline{p}_{\text{idle}} + 0.5 ~\overline{p}_{\text{prep}}$ & $38.2 ~\overline{p}_{\text{CNOT}} + 75.3 ~\overline{p}_{\text{idle}} + 12 ~\overline{p}_{\text{prep}}  $ \\    
 \hline
 8 & $ 2.45 ~\overline{p}_{\text{CNOT}} + 3.69 ~\overline{p}_{\text{idle}} + 2 ~\overline{p}_{\text{prep}}$ & $38.1 ~\overline{p}_{\text{CNOT}} + 76.4 ~\overline{p}_{\text{idle}} + 11.3 ~\overline{p}_{\text{prep}}  $\\    
 \hline
 9 & $ 1.98 ~\overline{p}_{\text{CNOT}} + 0.875 ~\overline{p}_{\text{idle}} +  \overline{p}_{\text{prep}}$ & $38 ~\overline{p}_{\text{CNOT}} + 74.8 ~\overline{p}_{\text{idle}} + 11.3 ~\overline{p}_{\text{prep}}  $ \\  
 \hline
 10 & $ 2.2 ~\overline{p}_{\text{CNOT}} + 0.85 ~\overline{p}_{\text{idle}}$ & $38.2 ~\overline{p}_{\text{CNOT}} + 73.3 ~\overline{p}_{\text{idle}} + 12.3 ~\overline{p}_{\text{prep}}  $ \\  
 \hline
 11 & $ 1.93 \overline{p}_{\text{CNOT}} + 0.875 \overline{p}_{\text{idle}}$ & $38.2 \overline{p}_{\text{CNOT}} + 73 \overline{p}_{\text{idle}} + 12.3 \overline{p}_{\text{prep}}  $ \\  
 \hline
 12 & $ 2.45 ~\overline{p}_{\text{CNOT}} + 2.31 ~\overline{p}_{\text{idle}} +  \overline{p}_{\text{prep}}$ & $38.1 ~\overline{p}_{\text{CNOT}} + 77.1 ~\overline{p}_{\text{idle}} + 11.8 ~\overline{p}_{\text{prep}}  $ \\  
 \hline
 13 & $ 1.98 ~\overline{p}_{\text{CNOT}} + 0.875 ~\overline{p}_{\text{idle}}$ & $38 ~\overline{p}_{\text{CNOT}} + 73 ~\overline{p}_{\text{idle}} + 12.3 ~\overline{p}_{\text{prep}}  $ \\  
 \hline
 14 & $ 2.2 ~\overline{p}_{\text{CNOT}} + 0.85 ~\overline{p}_{\text{idle}} + 0.5 ~\overline{p}_{\text{prep}}$ & $38.2 ~\overline{p}_{\text{CNOT}} + 75.9 ~\overline{p}_{\text{idle}} + 12 ~\overline{p}_{\text{prep}}  $ \\   
 \hline
 15 & $ 1.93 ~\overline{p}_{\text{CNOT}} + 0.875 ~\overline{p}_{\text{idle}} + 0.5 ~\overline{p}_{\text{prep}}$ & $38.2 ~\overline{p}_{\text{CNOT}} + 75.5 ~\overline{p}_{\text{idle}} + 12 ~\overline{p}_{\text{prep}}  $ \\   
 \hline 
16 & $ 2.45 ~\overline{p}_{\text{CNOT}} + 2.59 ~\overline{p}_{\text{idle}} + 2 ~\overline{p}_{\text{prep}}$ & $38.1 ~\overline{p}_{\text{CNOT}} + 77 ~\overline{p}_{\text{idle}} + 11.3 ~\overline{p}_{\text{prep}}  $ \\  
 [1ex] 
 \hline
\end{tabular}
 \caption{
 The leading-order contributions to the probability of failure $p_{\text{fail}}^{\text{RM}}$ and rejection $p_{\text{rej}}^{\text{RM}}$ of the 15-to-1 distillation scheme from the production of the Reed-Muller state for each choice of punctured qubit.
 }
 \label{tab:distillation-protocols-puncture-optimization}
\end{center}
\end{table}

Now we elaborate on the procedure to initialize a $T$ state in a distance $d$ color code described in \sec{TStateInitialization}, and explain some structure that we exploit to simplify the noise analysis.
First note that even in the absence of noise, the stabilizers which are measured in the initialization protocol have uncertain outcomes because the state is not a code state of the 2D color code at the start of the protocol. 
Let $A$ and $B$ be the two sets of qubits prepared in $\ket{0}$ and $\ket{+}$, respectively; see~\fig{color-code-magic-initialization}(a).
The protocol involves taking the observed syndrome $\sigma$, and producing a Pauli operator $P(\sigma) = P^X P^Z$, where $P^X$ and $P^Z$ are Pauli $X$ and $Z$ operators supported on qubits in $B$ and $A$, respectively.
The protocol rejects if no such fix exists or if the syndromes from the consecutive rounds of QEC differ .
Note that the fix is unique up to a stabilizer since neither of the disjoint sets $A$ and $B$ support a logical operator.
This implies that for any two syndromes $\sigma$ and $\sigma'$ the Pauli operators $P(\sigma)P(\sigma')$ and $P(\sigma+\sigma')$ are equivalent up to a stabilizer.

To faithfully simulate this protocol in the presence of noise, one can consider the scenario in which a complete set of perfect stabilizer measurements is performed (and their outcomes discarded) before the QEC circuits are applied. 
This allows us to begin the simulation with a state, which is an eigenstate of the stabilizer group.
The simulation can therefore be implemented as follows.
\begin{enumerate}
\item Start with a perfect code state.
\item With some probability $\text{Pr}(\sigma)$, apply a Pauli operator $P(\sigma) = P^X(\sigma) P^Z(\sigma)$, where $P^X(\sigma)$ and $P^Z(\sigma)$ are Pauli $X$ and $Z$ operators supported within $B$ and $A$, respectively.
\item  Run the noisy circuits explicitly as described in \sec{noise}.
A set of faults will propagate to some Pauli error $E$ on the data qubits at the end of the protocol.
The faults could result in the syndromes from the two QEC cycles disagreeing, in which case the protocol is rejected, or both QEC rounds having the same observed syndrome $\widetilde\sigma$, possibly different from the syndrome $\sigma$ of $P$.
Note that $\sigma+\widetilde\sigma$ can only depend on the set of faults, and is independent of $\sigma$.
\item Apply a Pauli operator $P(\widetilde\sigma) = P^X(\widetilde\sigma) P^Z(\widetilde\sigma)$ with the syndrome $\widetilde\sigma$, where $P^X(\widetilde\sigma)$ and $P^Z(\widetilde\sigma)$ are Pauli $X$ and $Z$ operators supported on qubits in $B$ and $A$, or reject if such an operator cannot be found.
\item If the net operator $E P(\sigma) P(\widetilde\sigma)$ is correctable, then we say that the protocol succeeds, and otherwise we say it fails.
\end{enumerate}
However, note that the net operator $E P(\sigma) P(\widetilde\sigma)$ must be independent of $\sigma$, since $P(\widetilde\sigma)$ and $P(\sigma) P(\sigma + \widetilde\sigma)$ are equivalent up to a stabilizer.
In our simulation we therefore set $\sigma$ to be trivial for simplicity without loss of generality.

Using this simulation approach, we analyze all 234 valid CNOT schedules for a variety of distances, and find that 
$\{1, 4, 6, 7, 5, 2 ; 4, 5, 7, 6, 2, 3 \}$
as shown in \fig{color-code-magic-initialization} has the lowest failure pro ability, namely $p_{\text{fail}}^{\text{init}} = 6.07 p$.
For comparison, the worst schedule using $d=3$ was 
$\{1, 4, 6, 7, 5, 2 ; 4, 5, 7, 6, 2, 3 \}$ and would result in a $p_{\text{fail}}^{\text{init}}$ of $12.7p$.

\clearpage
\subsection{Preparing the 3D interior for code switching}
\label{app:residual-noise-sphere-color-code}
\begin{figure}
	\includegraphics[height=.17\textheight]{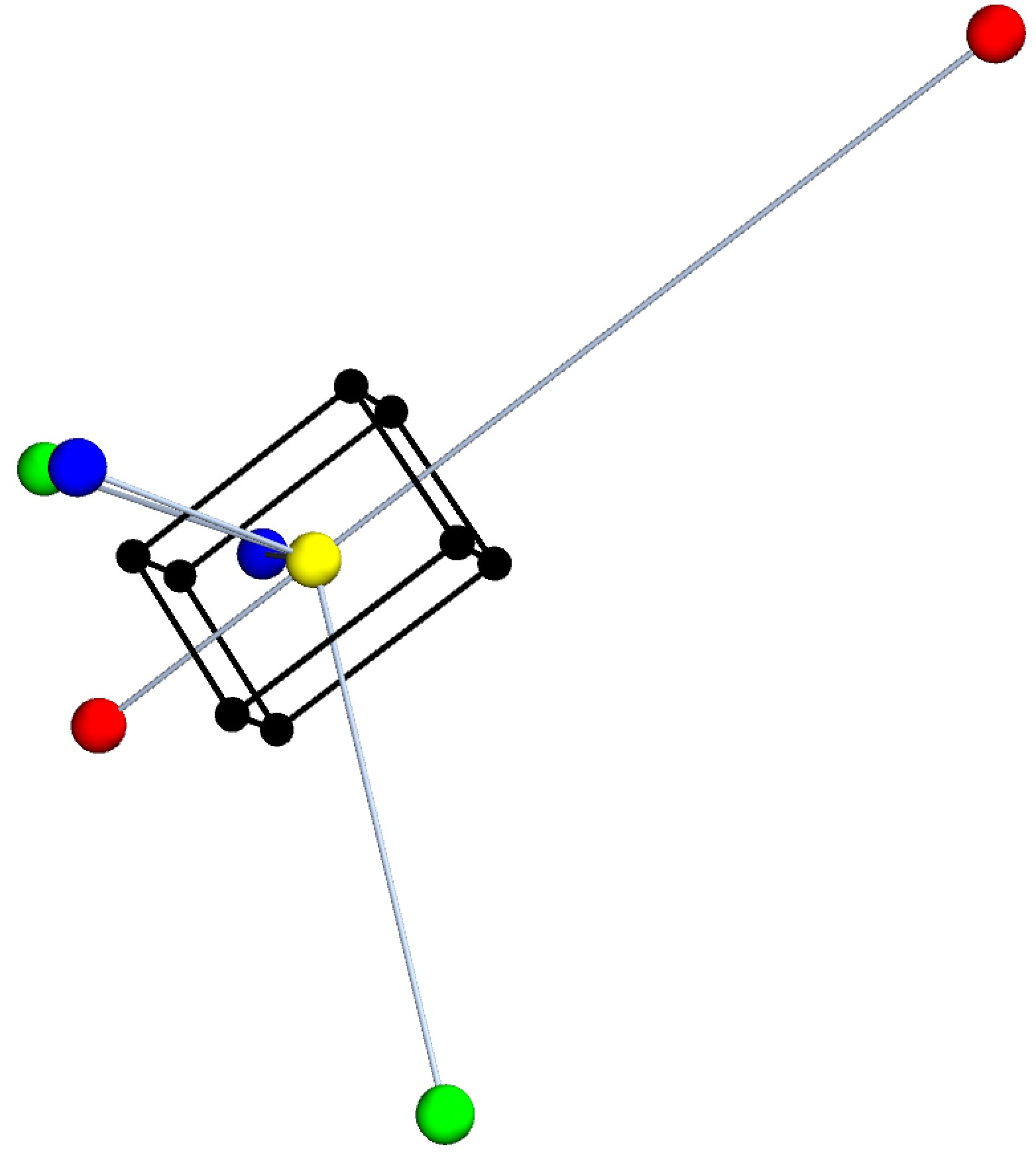}
	\hspace*{-.23\textwidth}(a)\hspace*{.23\textwidth}
	\includegraphics[height=.17\textheight]{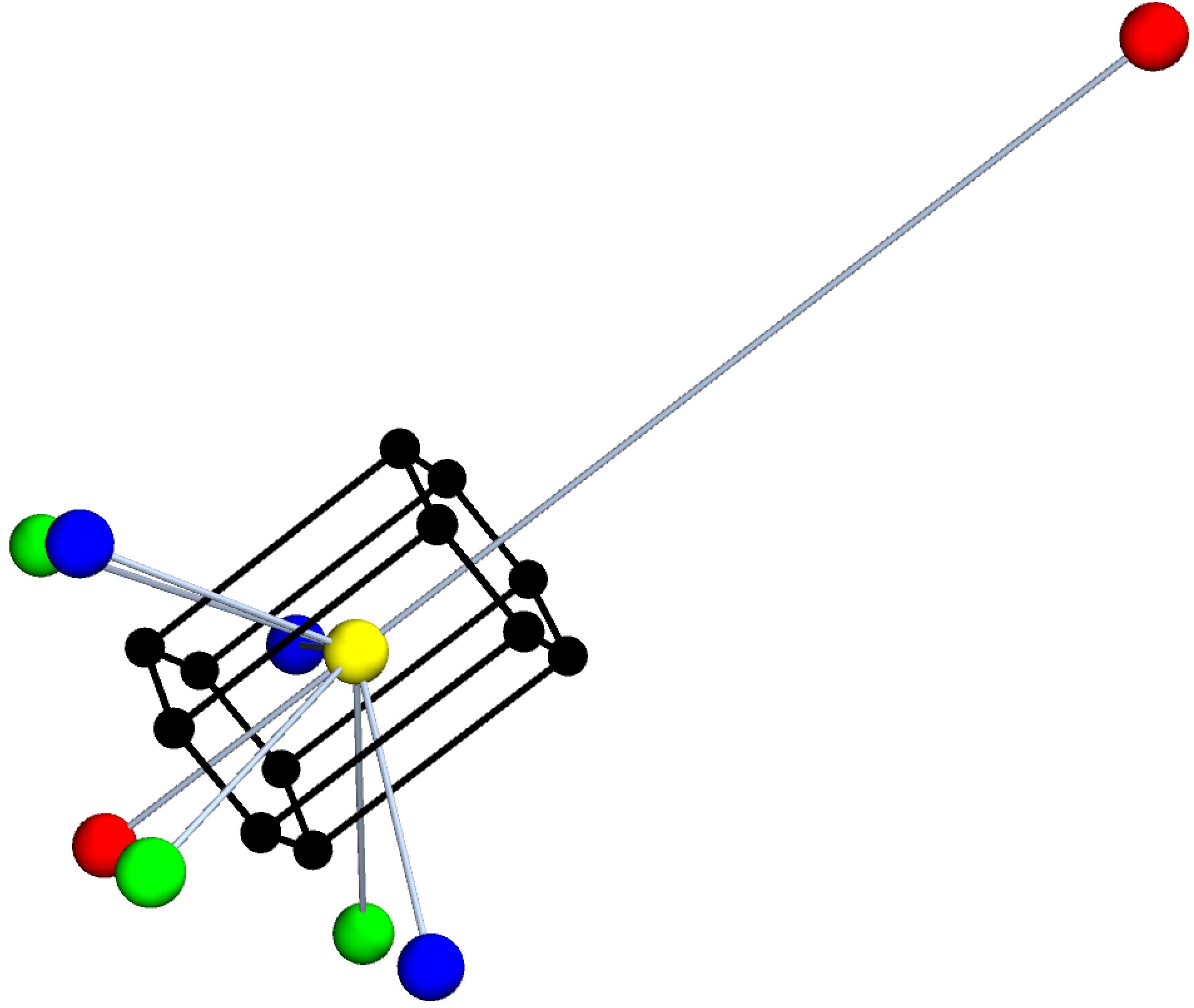}
	\hspace*{-.3\textwidth}(b)\hspace*{.3\textwidth}
    \includegraphics[height=.17\textheight]{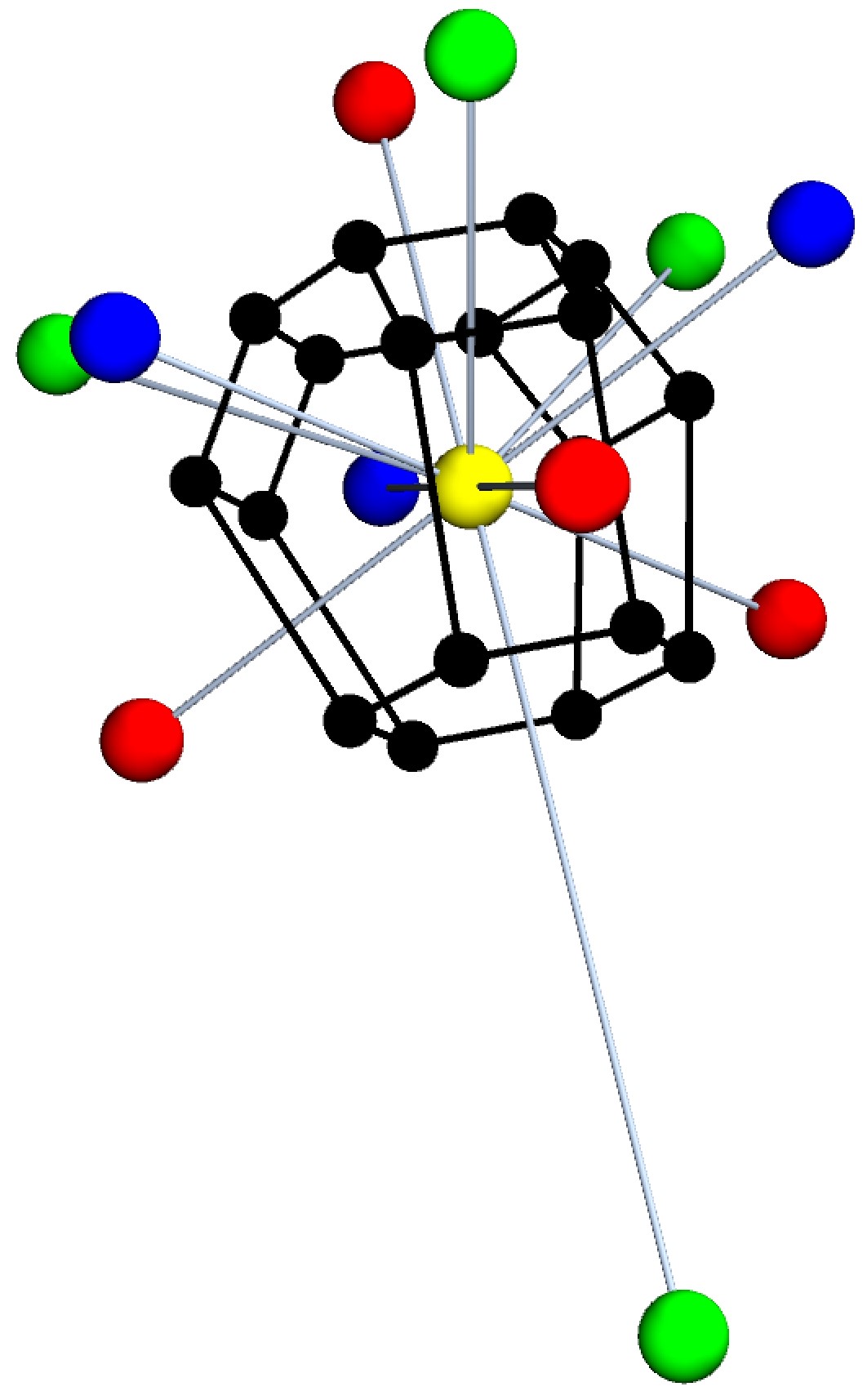}
    \hspace*{-.16\textwidth}(c)\hspace*{.16\textwidth}\quad
    \includegraphics[width=.16\columnwidth]{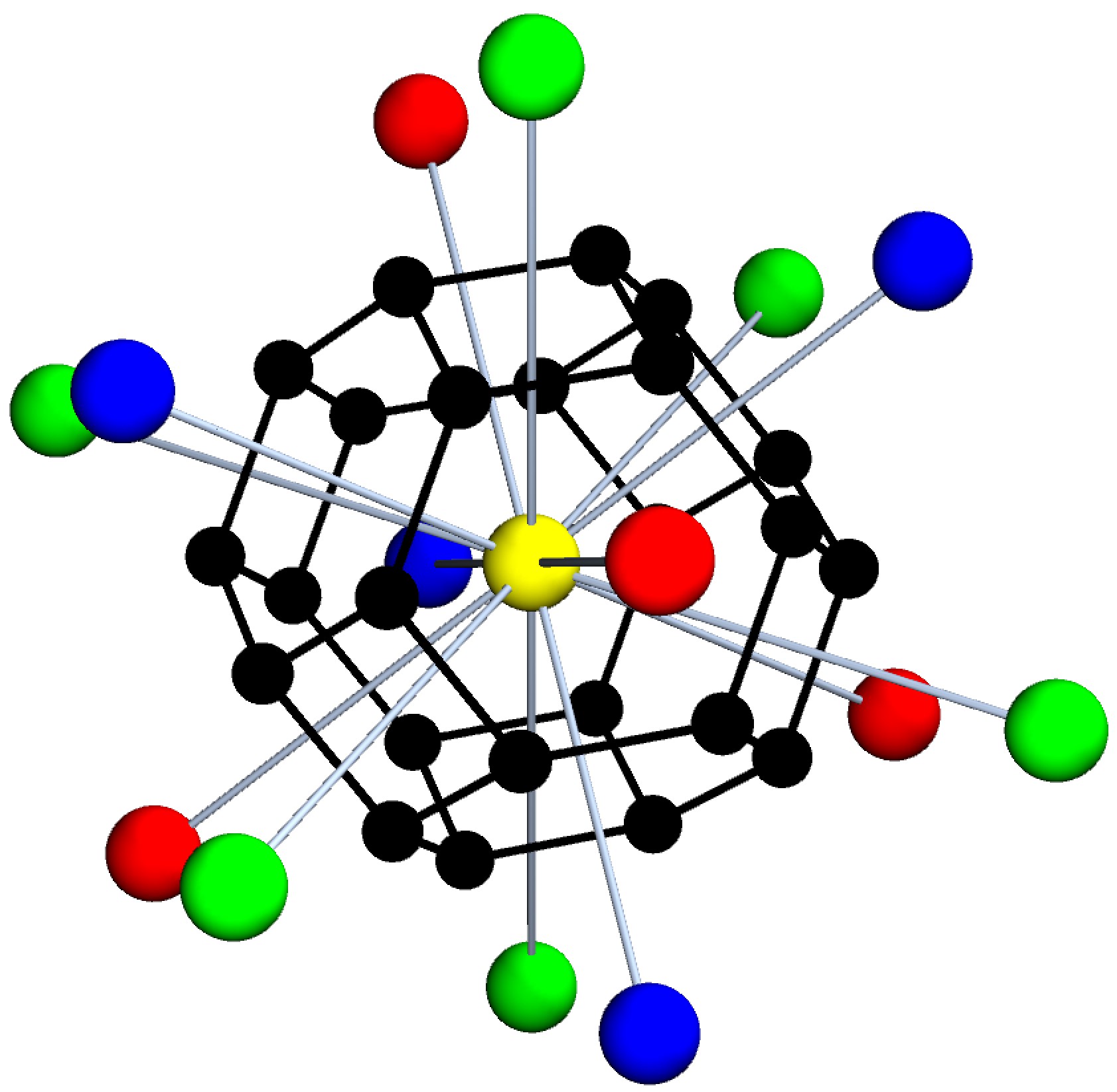}
    \hspace*{-.18\textwidth}(d)\hspace*{.16\textwidth}
	\caption{Dual and primal lattices for the 2D spherical color codes which appear in the preparation of interior of the 3D color code. Qubits are black dots, and stabilizers are grey edges with colored vertices.
	For each, the subset of vertices from $v_{\text{list}}$ in \eq{vList} are: (a) $\{1, 2, 3, 4, 8, 15, 17 \}$, (b) $\{1, 2, 3, 4, 6, 7, 8, 11, 17\}$, (c) $\{1, 2, 3, 4, 5, 8, 9, 10, 12, 13, 14, 15\}$, and (d) $\{1, 2, 3, 4, 5, 6, 7, 8, 9, 10, 11, 12, 13, 14, 16\}$.
	}
	\label{fig:SphericalColorCode}
\end{figure}

Here we consider the 3D interior of the initial state for code switching discussed in \sec{prep-interior}, which consists of a (trivial) 2D color code state on the qubits surrounding each $Y$ interior vertex.
This object can be viewed as a 2D sphere; see \fig{SphericalColorCode}.
This state of each sphere is formed by preparing all data qubits in the $\ket{+}$ state, then measuring the $Z$ stabilizers, each with an ancilla measurement qubit in a single QEC cycle, followed by the application of an appropriate $X$ Pauli to fix incorrect stabilizers.

\textbf{CNOT circuits.---}For each of the four types of spheres, we find an ordering of CNOTs which allows for a minimal-length circuit to measure the stabilizers; see \tab{sphere-measurement-sequence}.
These sequences were found using a greedy algorithm with a random initial sequence and minimizing the cost function of the number of unsatisfied constraints (we require that no qubit is involved in more than one gate per time unit) minus the average time unit when each gate occurs.
The greedy algorithm terminates when all constraints are satisfied and when the cost function cannot be further reduced.
We fixed the total number of time units to be minimal (4 CNOT time units for the 8-qubit sphere, which involves only weight-4 stabilizers, and 6 CNOT time units for the 12-, 18- and 24-qubit spheres, which involve weight-6 stabilizers). 
We maximize the average time unit when each gate occurs since this removes idle time units by not preparing ancillas for lower-weight stabilizer measurements until they are needed.
The sequences we found are optimal with regard to this cost function: all constraints were satisfied using the minimal number of time units, and all weight-$r$ stabilizer generators have CNOTs in the last $r$ time units of each schedule.
There could be many schedules which are optimal in this sense, but we did not attempt to explore among those for schedules which performed better.

To make the representation of the schedules slightly more compact, we specify 17 vertex locations in a list $v_{\text{list}}$ in lexicographical order as follows
 \begin{eqnarray}
 \label{eq:vList}
   v_\text{list} = \bigg\{ &&\left(-1, ~0, ~0\right), \left(-\tfrac{1}{2}, -\tfrac{1}{2}, -\tfrac{1}{2} \right), \left(-\tfrac{1}{2}, -\tfrac{1}{2}, \tfrac{1}{2} \right),
   \left(-\tfrac{1}{2}, ~\tfrac{1}{2}, -\tfrac{1}{2} \right), \left(-\tfrac{1}{2}, ~\tfrac{1}{2}, ~\tfrac{1}{2} \right),\nonumber\\
   &&\left(~0, -1, ~0\right),  
   \left(~0, ~0, -1\right), \left(~0, ~0,  ~0\right), \left(~0, ~0, ~1\right), \left(~0, ~1, ~0\right), \left(~\tfrac{1}{2}, -\tfrac{1}{2}, -\tfrac{1}{2} \right),\nonumber\\
   &&\left(~\tfrac{1}{2}, -\tfrac{1}{2},\tfrac{1}{2} \right),\left(~\tfrac{1}{2}, ~\tfrac{1}{2}, -\tfrac{1}{2} \right), \left(~\tfrac{1}{2}, ~\tfrac{1}{2}, ~\tfrac{1}{2} \right), \left(~1, -1, -1\right), \left(~1, ~0, ~ ~0\right),\left(~\tfrac{3}{2},~\tfrac{3}{2},~\tfrac{3}{2} \right) \bigg\},\quad\quad
 \end{eqnarray}
where the coordinates $\left(~1, -1, -1\right)$ and $\left(~\frac{3}{2},~\frac{3}{2},~\frac{3}{2} \right)$ are for placing boundary vertices $v_G$ and  $v_R$, while the others describe the interior vertices.
The colors of the vertices in
$v_{\text{list}}$ in order are $G$, $R$, $B$, $B$, $R$, $G$, $G$, $Y$, $G$, $G$, $B$, $R$, $R$, $B$, $G$, $G$, $R$. 
The central $Y$ vertex is placed at the origin $(0,0,0)$ and is the eighth entry in $v_{\text{list}}$.
To apply the CNOT schedules we specify in \tab{sphere-measurement-sequence}, one must map the neighborhood of every vertex to one of these four standard configurations.
All interior vertices lie precisely on these standard coordinates relative to their central yellow vertex. 
Boundary vertices connected to their central yellow vertex will not be on the standard coordinates in $v_{\text{list}}$, but can be identified and shifted to the standard coordinate uniquely since there is at most one boundary vertex of each color included in the neighborhood of any yellow vertex.


\begin{table}
\begin{center}
 \begin{tabular}{|c||c|c|c|c|c|c|c|} 
 \hline
 sphere size & edge & time unit 1 & time unit 2 & time unit 3 & time unit 4 & time unit 5 & time unit 6 \\
 \hline
 \hline
 & (8,1) &   (1,2,4,8) & (1,2,3,8) & (1,3,8,17) & (1,4,8,17)&&\\
 & (8,2) &   (1,2,3,8) & (1,2,4,8) & (2,3,8,15) & (2,4,8,15)&&\\
 8& (8,3) &  (3,8,15,17) & (2,3,8,15) & (1,2,3,8) & (1,3,8,17)&&\\
 qubits& (8,4) &  (2,4,8,15) & (4,8,15,17) & (1,4,8,17) & (1,2,4,8)&&\\
 & (8,15) &  (2,3,8,15) & (2,4,8,15) & (4,8,15,17) & (3,8,15,17)&&\\
 & (8,17) &  (1,3,8,17) & (1,4,8,17) & (3,8,15,17) & (4,8,15,17)&&\\
 \hline\hline
 & (8,1) & & & (1,3,8,17) & (1,2,3,8) & (1,2,4,8) & (1,4,8,17)\\
 & (8,2) & (1,2,3,8) & (1,2,4,8) & (2,3,6,8) & (2,6,8,11) & (2,4,7,8) & (2,7,8,11)\\
 & (8,3) & & & (1,2,3,8) & (2,3,6,8) & (1,3,8,17) & (3,6,8,17) \\
 12 & (8,4) & &  & (1,2,4,8) & (2,4,7,8) & (1,4,8,17) & (4,7,8,17)\\
 qubits & (8,6) & & & (6,8,11,17) & (3,6,8,17) & (2,3,6,8) & (2,6,8,11)\\
 & (8,7) & & & (2,7,8,11) & (7,8,11,17) & (4,7,8,17) & (2,4,7,8)\\
 & (8,11) & & & (2,6,8,11) & (2,7,8,11) & (6,8,11,17) & (7,8,11,17)\\
 & (8,17) & (1,3,8,17) & (1,4,8,17) & (3,6,8,17) & (4,7,8,17) & (7,8,11,17) & (6,8,11,17)\\
 [1ex] 
 \hline
 \hline
 & (8,1) & & & (1,2,3,8) & (1,2,4,8) & (1,3,5,8) & (1,4,5,8) \\
 & (8,2) & & & (1,2,4,8) & (1,2,3,8) & (2,3,8,15) & (2,4,8,15) \\
 & (8,3) & (1,2,3,8) & (1,3,5,8) & (3,5,8,9) & (2,3,8,15) & (3,8,9,12) & (3,8,12,15) \\
 & (8,4) & (1,2,4,8) & (1,4,5,8) & (4,5,8,10) & (2,4,8,15) & (4,8,10,13) & (4,8,13,15) \\
18 & (8,5) & (1,3,5,8) & (3,5,8,9) & (1,4,5,8) & (4,5,8,10) & (5,8,9,14) & (5,8,10,14) \\
qubits  & (8,9) & & & (5,8,9,14) & (3,8,9,12) & (3,5,8,9) & (8,9,12,14) \\
 & (8,10) & & & (4,8,10,13) & (5,8,10,14) & (4,5,8,10) & (8,10,13,14) \\
 & (8,12) & & & (3,8,9,12) & (3,8,12,15) & (8,9,12,14) & (8,12,14,15) \\
 & (8,13) & & & (8,13,14,15) & (4,8,13,15) & (8,10,13,14) & (4,8,10,13) \\
 & (8,14) & (5,8,9,14) & (5,8,10,14) & (8,9,12,14) & (8,10,13,14) & (8,12,14,15) & (8,13,14,15) \\
 & (8,15) & (2,4,8,15) & (3,8,12,15) & (4,8,13,15) & (8,12,14,15) & (8,13,14,15) & (2,3,8,15) \\
 [1ex] 
 \hline
 \hline
 & (8,1) & & & (1,2,3,8) & (1,2,4,8) & (1,3,5,8) & (1,4,5,8) \\
 & (8,2) & (1,2,3,8) & (1,2,4,8) & (2,6,8,11) & (2,3,6,8) & (2,4,7,8) & (2,7,8,11) \\
 & (8,3) & (2,3,6,8) & (1,2,3,8) & (1,3,5,8) & (3,5,8,9) & (3,8,9,12) & (3,6,8,12) \\
 & (8,4) & (1,2,4,8) & (2,4,7,8) & (1,4,5,8) & (4,7,8,13) & (4,5,8,10) & (4,8,10,13) \\
 & (8,5) & (1,3,5,8) & (1,4,5,8) & (3,5,8,9) & (4,5,8,10) & (5,8,9,14) & (5,8,10,14) \\
 & (8,6) & & & (2,3,6,8) & (2,6,8,11) & (3,6,8,12) & (6,8,11,12) \\
24 & (8,7) & & & (2,4,7,8) & (2,7,8,11) & (4,7,8,13) & (7,8,11,13) \\
qubits & (8,9) & & & (3,8,9,12) & (5,8,9,14) & (3,5,8,9) & (8,9,12,14) \\
 & (8,10) & & & (4,5,8,10) & (4,8,10,13) & (5,8,10,14) & (8,10,13,14) \\
 & (8,11) & (2,6,8,11) & (2,7,8,11) & (8,11,12,16) & (6,8,11,12) & (7,8,11,13) & (8,11,13,16) \\
 & (8,12) & (3,6,8,12) & (3,8,9,12) & (6,8,11,12) & (8,9,12,14) & (8,11,12,16) & (8,12,14,16) \\
 & (8,13) & (4,7,8,13) & (4,8,10,13) & (7,8,11,13) & (8,10,13,14) & (8,11,13,16) & (8,13,14,16) \\
 & (8,14) & (5,8,10,14) & (8,9,12,14) & (8,10,13,14) & (8,13,14,16) & (8,12,14,16) & (5,8,9,14) \\
 & (8,16) & & & (8,11,13,16) & (8,12,14,16) & (8,13,14,16) & (8,11,12,16) \\
 \hline
\end{tabular}
 \caption{Shortest CNOT sequences to measure each $Z$ stabilizer with a single ancilla for the 8-, 12-, 18- and 24-qubit spherical color codes in \fig{SphericalColorCode}.
 Sets of two and four vertices from $v_{\text{list}}$ are used to label edges (measurement qubits) and tetrahedra (data qubits), which form the target and control of each CNOT which is applied in the specified time unit.
 Note that weight-4 stabilizers have their CNOTs applied in the last 4 time units, avoiding unnecessary idle times in the measurement circuits.
 }
 \label{tab:sphere-measurement-sequence}
\end{center}
\end{table}

\begin{figure}
	\includegraphics[width=.52\columnwidth]{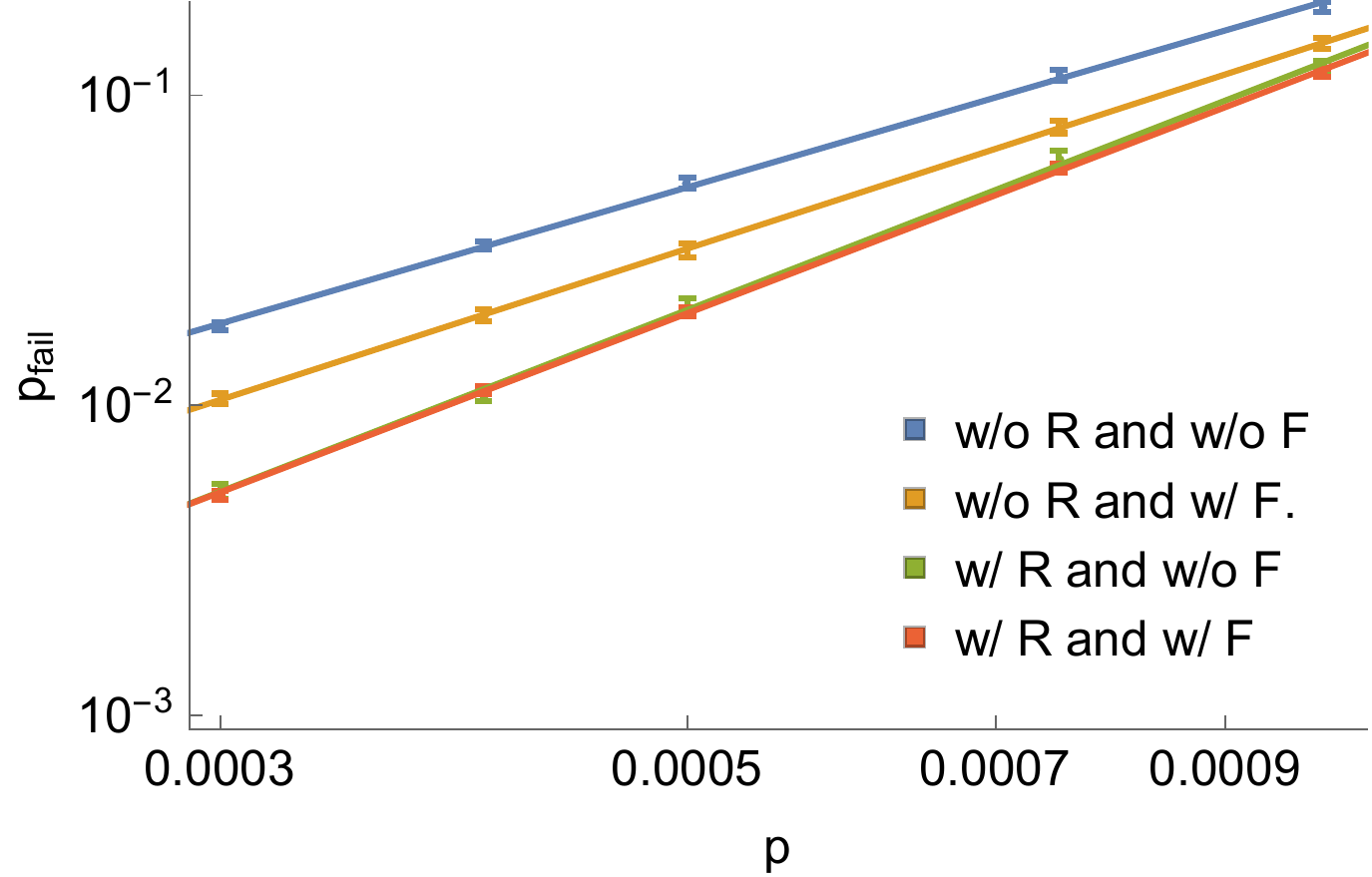}
	\qquad
	\caption{
	    Overall code switching performance of four syndrome fixing approaches for the preparation of 2D spherical color codes. 
	    Each approach either uses or does not use the repeat (R) and flip (F) strategies for distance $d=9$. 
	    In the approach which uses both strategies, we use the repeat strategy, and the flip strategy if the repeat also yields an invalid syndrome.
	    In each case, the stabilizers are measured with the CNOT circuits described in \tab{sphere-measurement-sequence}.
	    Both the repeat and flip strategies improve performance.
	    }
	\label{fig:codeswitchingOptimization}
\end{figure}

\textbf{Syndrome fixing.---}In the absence of error, the syndrome readout of the $Z$-stabilizer measurements for each spherical 2D color code (which will have random outcomes) determines the $X$-type gauge operator to `fix' the outcomes to be $+1$.
Since the spherical 2D color code encodes no logical qubits, any $X$-type operator with the correct syndrome will suffice, and can be found with Gaussian elimination.
However, faults in the measurement circuits can render the syndrome invalid, in the sense that there is no gauge operator with the observed syndrome.
We consider a number of strategies to deal with this. 
Firstly, we consider the \textit{repeat} strategy: if the observed syndrome is invalid, we repeat the preparation once. 
Secondly, we consider the \textit{flip} strategy: if the observed syndrome is invalid, we flip a bit of the syndrome to produce a valid syndrome, and then apply a fixing operator.
We test the impact of these approaches on the failure probability of code switching using a distance 9 code in \fig{codeswitchingOptimization}.
Note that when using the repeat strategy, if the first round is successful, the qubits are left idle in the prepared state to allow for a second round to be applied on spheres that failed the first round.
In practice we find that these idle rounds do not have a lot of impact on the performance in the regime of interest.
We see that both the repeat and flip strategies improve performance, and so we incorporate both into the optimized code switching protocol.

\clearpage
\subsection{Choices of basis for code switching preparation}
\label{app:prep-basis-choice-code-switching}
Here we analyze the effect of the choice of basis in first two steps of the code switching described in \sec{Bell-state-prep} and \sec{prep-interior}. 
The performance of the code may depend on this choice as it could lead to an asymmetry in the $X$ and $Z$ type noise and later steps of the protocol would not treat $X$ and $Z$ error in the same way.

The first choice is for the preparation of the Bell state in the 2D color code patches in \sec{Bell-state-prep}. 
We fault-tolerantly prepare the first patch in $\ket{\overline{+}}$, and the second in $\ket{\overline{0}}$ before applying a transversal CNOT from the first to the second. 
Since the CNOT copies $X$ noise from the first to the second patch, and $Z$ from the second to the first, we can expect that the first patch has more $Z$ noise and the second patch has more $X$ after a faulty preparation.
We choose whether to feed the first or the second patch to be involved in code switching, which we refer to as 2D \textit{patch} $\ket{+}$ and 2D \textit{patch} $\ket{0}$ respectively.

\begin{figure}
	\includegraphics[width=.52\columnwidth]{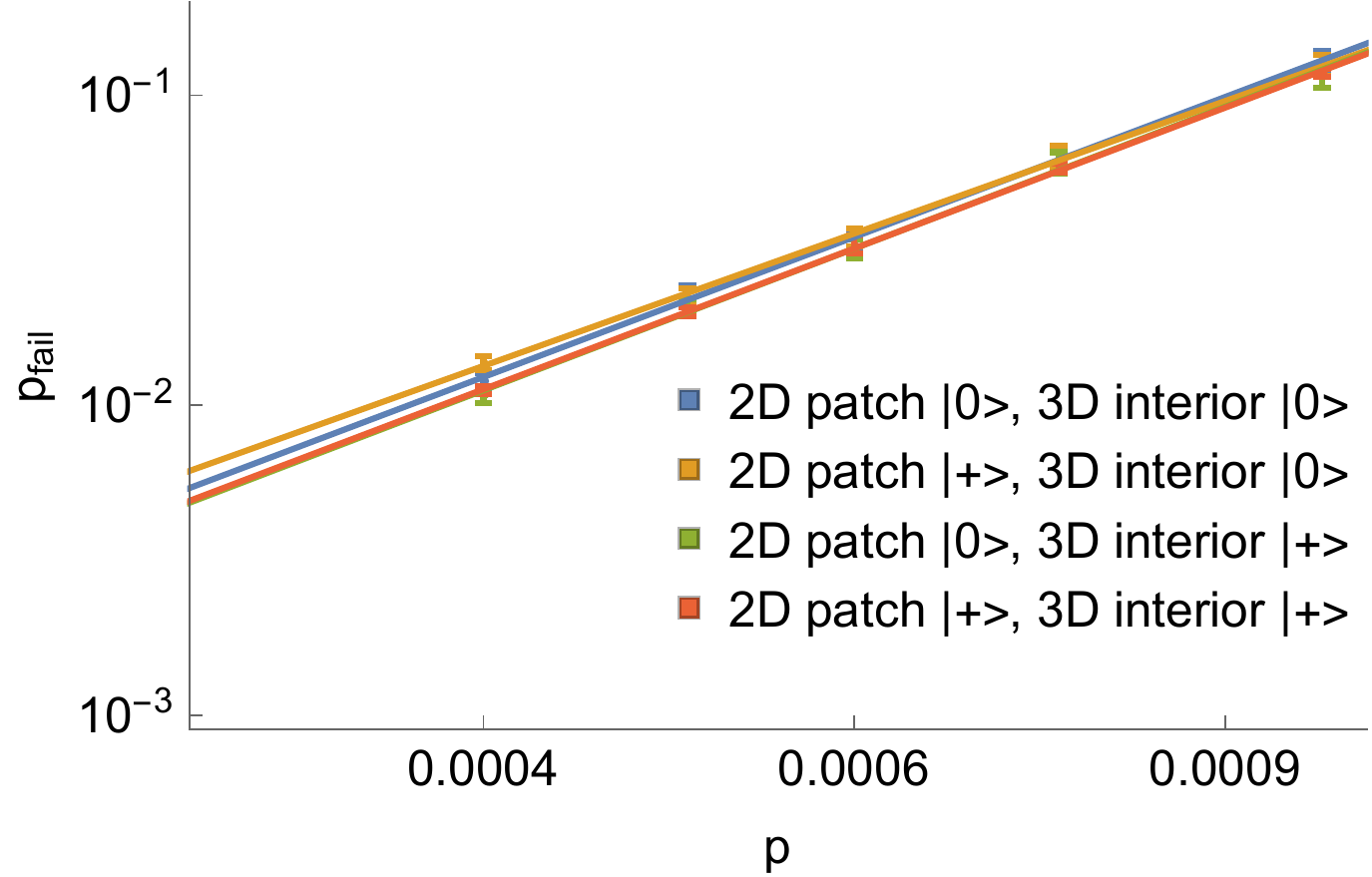}
	\qquad
	\caption{
	    The impact on the failure probability of the code switching protocol from different choices of basis for first two steps of the protocol.
	    The data is for distance $d=9$, and has all other steps as described in \sec{dimjump} for the optimized code switching protocol.
	    }
	\label{fig:codeswitchingBasisOptimization}
\end{figure}

The second choice is for the preparation of the 2D spherical color code in its unique code state around each interior yellow vertex as described in \sec{prep-interior}.
We can do this by either preparing each data qubit in $\ket{0}$ and then measuring the $X$ stabilizers, or by preparing each data qubit in $\ket{+}$ and then measuring the $Z$ stabilizes.
We refer to these cases as 3D \textit{interior} $\ket{0}$ and 3D \textit{interior} $\ket{+}$, respectively, and compare their impact along with the choice for the patch preparation in \fig{codeswitchingBasisOptimization}.
We find that there is very little difference in the performance of each of these choices, and we select 2D \textit{patch} $\ket{+}$ and 3D \textit{interior} $\ket{+}$ for use in the optimized version of the protocol.

\clearpage
\subsection{Gauge measurement circuits for code switching}
\label{app:gauge-measurement-circuits}
Here we discuss the circuits used to measure the $Z$-type gauge generators corresponding to $RG$, $RB$ and $GB$ edges used in \sec{measure-gauge-operators} for code switching.
We find a minimum-length circuit with a single ancilla, which needs 8 time units (including preparation and measurement).
A more naive circuit of 20 time units is constructed by initially preparing ancillas in the first time unit, sequentially applying CNOTs associated with $RG$, $RB$ and $GB$ edges in time units
$2-7$, $8-13$, $14-19$ respectively, and finally measuring ancillas in time unit 20.

To specify the CNOT sequences, we separate edges (measurement qubits) into 20 types by color and orientation and identify the tetrahedra containing them (data qubits) in a systematic way in terms of their spatial position relative to the edge.
The CNOT sequence is then specified for each edge type.
To specify the neighborhood of each edge we label every qubit in its support with some number from 1 to 6 in such a way that all edges of the same type have qubits labelled locally, and are all consistent.

\begin{figure}
	(a)\includegraphics[width=.12\columnwidth]{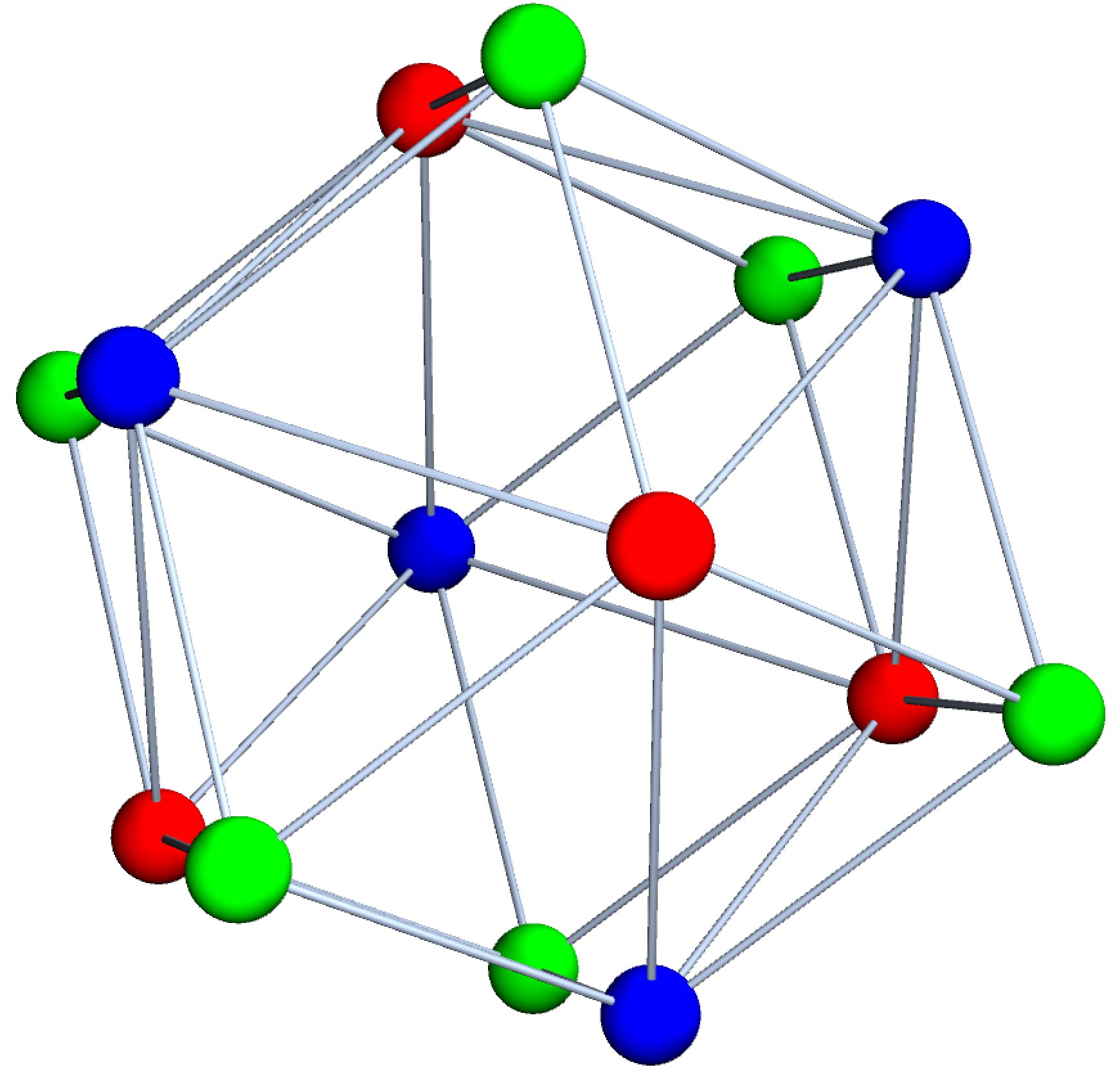}\quad
	(b)\includegraphics[width=.31\columnwidth]{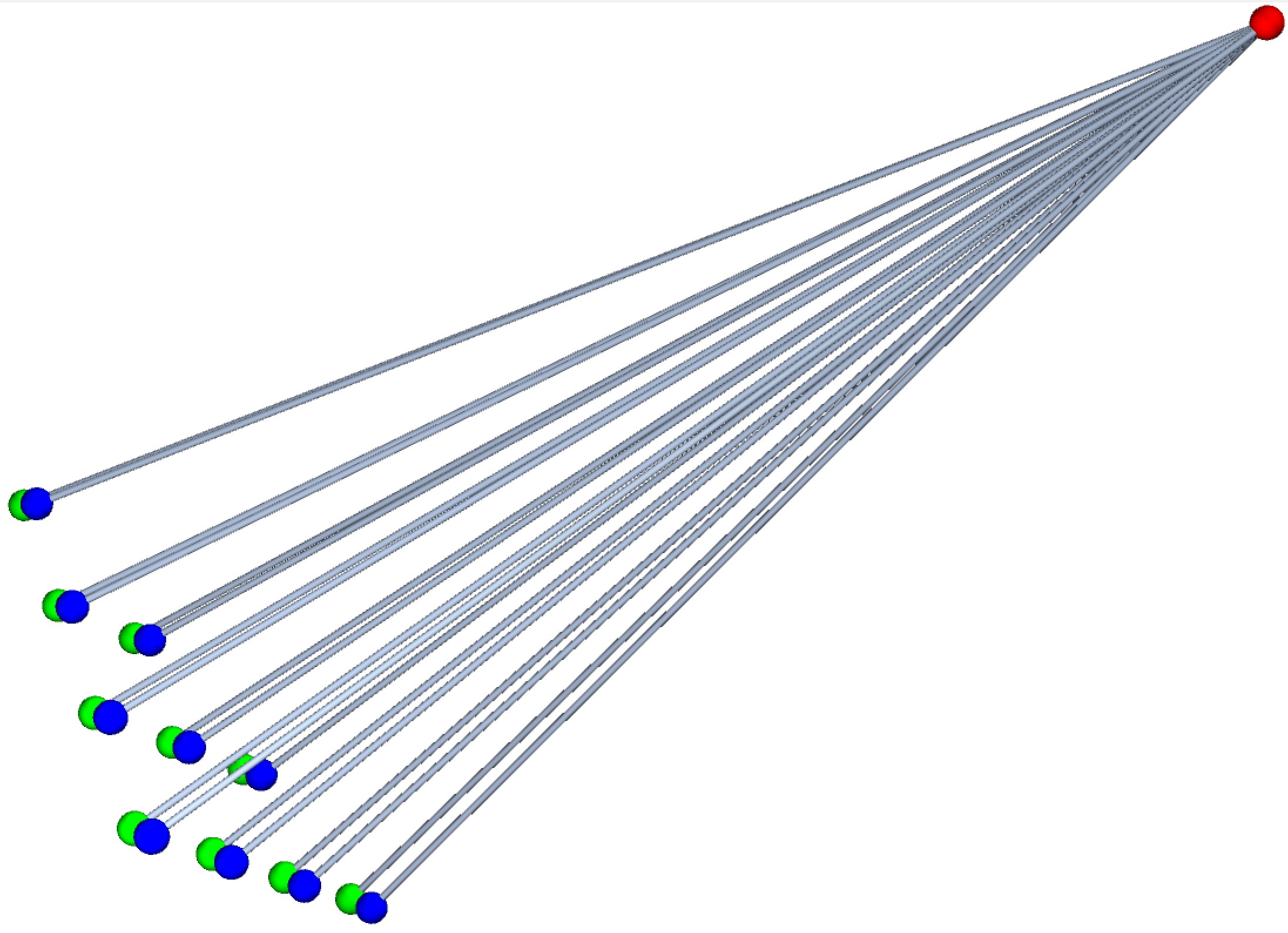}\quad
	(c)\includegraphics[width=.12\columnwidth]{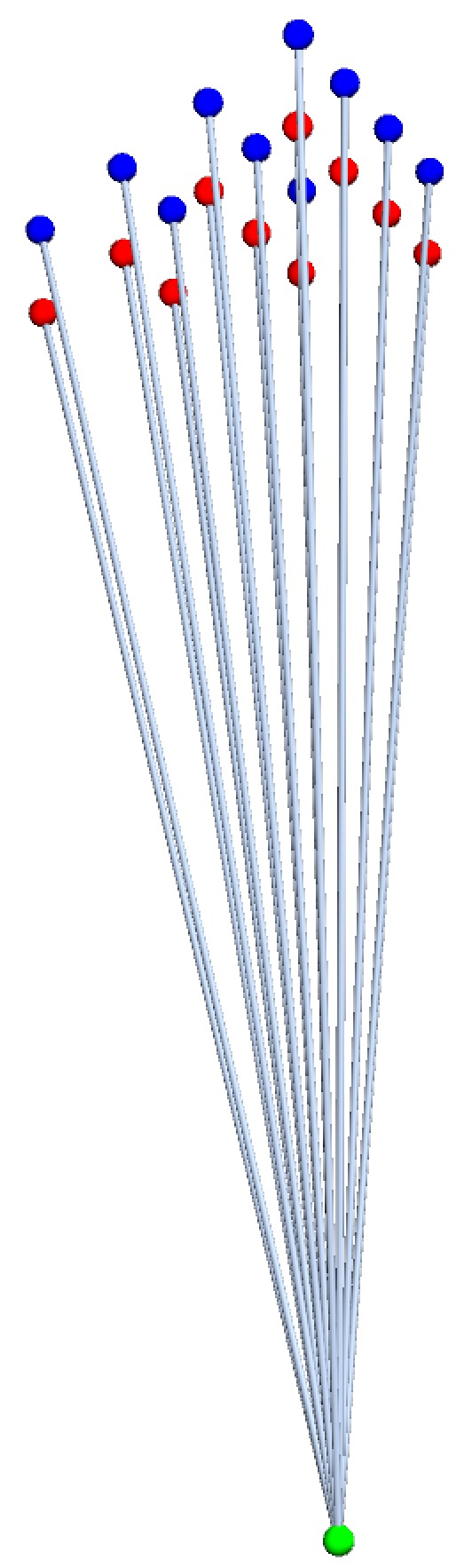}\quad
	(d)\includegraphics[width=.22\columnwidth]{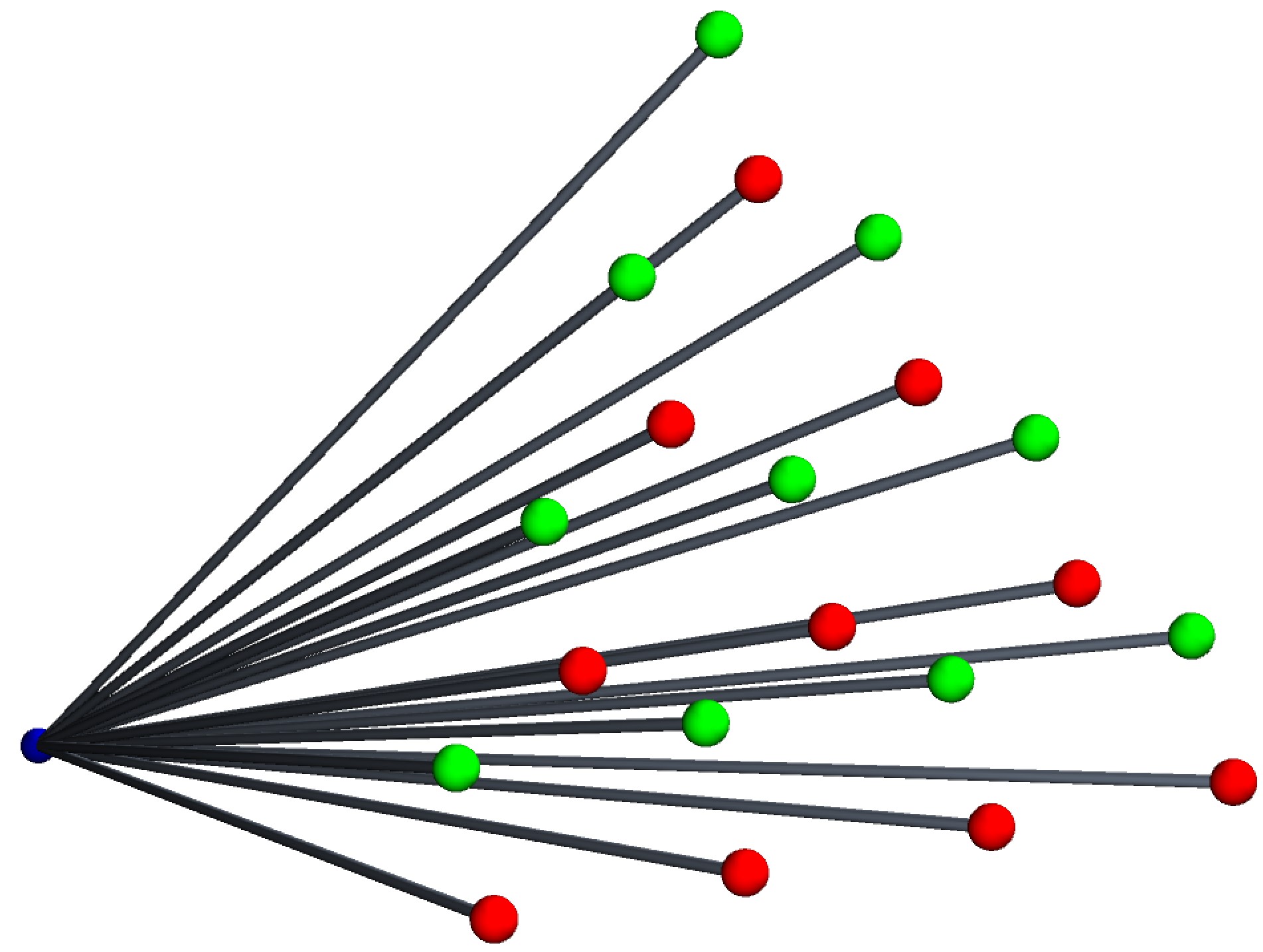}
	\caption{
	The $Z$-type gauge operators that are measured in the protocol to implement code switching are associated with $RG$, $RB$ and $GB$
	edges in the lattice, illustrated for the distance $d=9$ lattice. 
	These are parts of the full lattice in \fig{lattices}(a).
	We separate the edges into 20 cases.
	(a) Edges connecting interior vertices are classified by color and the vector from one vertex to the other.
	(b)-(d) Other edges are classified by the boundary vertex and the color of the interior vertex they are incident to. }
	\label{fig:GaugeFixingEdges}
\end{figure}

For completeness, we now specify our conventions to label the circuits.
To label tetrahedra around an edge with two interior vertices, we specify a local coordinate system.
The $z$ basis vector for an $RG,GB$ or $RB$ edge is the normalized vector $v_{\text{edge}}$ from the first to the second vertex. 
The $x$ basis vector is the normalized vector parallel to $v_{\text{irr}}=(1,\pi,\sqrt{2})$, but perpendicular to $v_{\text{edge}}$, which is guaranteed not to be parallel to any edges.
The $y$ basis vector is then specified uniquely to form a right hand coordinate system $(x,y,z)$.
The tetrahedra around the edge are then labelled in sequence according to the azimuthal angle to their mid-point in this coordinate system.
For edges which contain a boundary vertex, the same applies, but the vector $v_{\text{edge}}$ is specified according to \tab{gauge-measurement-sequence}.
In the bulk, the tetrahedra are positioned regular azimuthal angles for each edge type, but at the boundary we identify the tetrahedra by whichever standard angle it is closest to.
When tetrahedra are `missing' around an edge due to cuts along the boundary, they are simply skipped in the CNOT sequence, with no CNOTs applied during the assigned time unit for that edge.

The sequence in \tab{gauge-measurement-sequence} was found using a greedy algorithm starting with a random initial sequence and minimizing the cost function of the number of unsatisfied constraints (we require that no qubit is involved in more than one gate per time step) minus the average step time for each gate.
The result is the shortest possible measurement circuit using a single measurement qubit per edge, and which involves no idle time units for edges types of weight four.

To simulate the gauge measurements, we first apply an $X$ type gauge operator $g_X$ supported on randomly selected $RB$, $GB$ and $RG$ edges.
Then the specified circuits are applied and produce the outcomes $\tilde{\gamma}$, which in the absence of error would correspond to precisely the $Z$-edges which anti-commute with $g_X$.

\begin{table}
\begin{center}
 \begin{tabular}{|c|c||c|c|c|c|c|c|} 
 \hline
 vertex types & edge vector & time unit 1 & time unit 2 & time unit 3 & time unit 4 & time unit 5 & time unit 6 \\
 [0.5ex] 
 \hline
 \hline
 R-G & $(+\frac{1}{2}, -\frac{1}{2}, -\frac{1}{2})$ & 1 & 3 & 2 & 4 & 5 & 6 \\
 R-G & $(+\frac{1}{2}, +\frac{1}{2}, +\frac{1}{2})$  & 1 & 2 & 3 & 4 & 5 & 6 \\ 
 R-G & $(-\frac{1}{2}, +\frac{1}{2}, -\frac{1}{2})$ & 1 & 2 & 3 & 6 & 5 & 4 \\  
 R-G & $(-\frac{1}{2}, -\frac{1}{2}, +\frac{1}{2})$  & 1 & 2 & 4 & 3 & 5 & 6 \\  
 G-B & $(+\frac{1}{2}, -\frac{1}{2}, -\frac{1}{2})$   & 1 & 2 & 3 & 6 & 4 & 5 \\ 
 G-B & $(+\frac{1}{2}, +\frac{1}{2}, +\frac{1}{2})$  & 1 & 2 & 3 & 6 & 5 & 4 \\ 
 G-B & $(-\frac{1}{2}, +\frac{1}{2}, -\frac{1}{2})$  & 3 & 1 & 2 & 4 & 5 & 6 \\ 
 G-B & $(-\frac{1}{2}, -\frac{1}{2}, +\frac{1}{2})$ & 1 & 2 & 3 & 6 & 5 & 4\\  
 R-B & $(+0, +0, -1)$ & & & 1 & 2 & 3 & 4 \\ 
 R-B & $(+0, +1, +0)$& & & 1 & 2 & 3 & 4 \\ 
 R-B & $(+1, +0, +0)$  & & & 1 & 2 & 3 & 4 \\ 
 R-B & $(-1, +0, +0)$  & & & 1 & 2 & 3 & 4 \\ 
 R-B & $(+0, -1, +0)$ & & & 1 & 2 & 4 & 3 \\
 R-B & $(+0, +0, +1)$  & & & 1 & 2 & 3 & 4 \\ 
 $v_R$-G & $(-1, -1, -1)$ & 1 & 2 & 3 & 4 & 5 & 6 \\ 
 R-$v_G$ & $(+1, -1, -1)$ & 1 & 2 & 3 & 4 & 5 & 6 \\ 
 $v_G$-B & $(-1, +1, +1)$ & 1 & 2 & 3 & 5 & 6 & 4 \\ 
 G-$v_B$ & $(-1, +1, -1)$ & 2 & 1 & 3 & 4 & 5 & 6 \\
 $v_R$-B & $(-1, -1, -1)$ & 1 & 2 & 5 & 3 & 4 & 6 \\ 
 R-$v_B$ & $(-1, +1, -1)$ & 1 & 2 & 3 & 4 & 6 & 5 \\ 
 [0.5ex] 
 \hline
\end{tabular}
 \caption{The measurement circuit specification for $RG$, $RB$ and $GB$ gauge operators.
 Edges connecting interior vertices (first 14 rows) are labelled by the vertex colors (first column) and the vector connecting them (second column).
 Edges connecting a boundary vertex are labelled by that boundary vertex $v_R$, $v_G$ or $v_B$ and the color of the interior vertex. 
 Qubits around an edge are labelled in order of increasing azimuthal angle, and the column in which each data qubit appears specifies which time unit a CNOT occurs from that data qubit to the corresponding mesurement qubit for the edge.
 }
 \label{tab:gauge-measurement-sequence}
\end{center}
\end{table}

\clearpage
\subsection{Optimistic improvements for the 3D color code decoder}
\label{app:3D-color-code-potential-improvements}
Here we justify of our estimate of the impact on code switching performance due to any potential improvements on the 3D color code decoder as used in \sec{DecodingZ3DCC}.
For any error $E$ we define its \textit{stabilizer-reduced weight} to be the minimum of the weight of $s E$ for any stabilizer $s$. 
The estimate rests on the following assumptions.
\begin{enumerate}

    \item We assume that the stabilizer-reduced weight of errors generated by code switching can be approximated by the minimum of the weight of either the error itself or it's correction produced by the modified restriction decoder.

    \item The decoder does not take into account any correlations in the noise produced during the various steps of code switching and its failure probability for errors with stabilizer-reduced weight $w$ produced by code switching is no lower than that of the optimal decoder for iid $Z$ errors of weight $w$. 

    \item The optimal decoder for iid $Z$ noise, when applied to  uniformly drawn weight-$w$ errors has failure probability satisfying
    \begin{eqnarray}
    p_{\text{fail}}^{\text{iid}}(w,d)> p^{(1)}_\text{3DCC}\left(\frac{w/n}{p^{(1)}_\text{3DCC}}\right)^{\frac{d+1}{2}},
    \end{eqnarray}
    where $n$ is the number of data qubits, and $p^{(1)}_\text{3DCC} \simeq 1.9\%$ is the known optimal threshold of the the 3D stabilizer color code decoder for iid $Z$ noise \cite{kubica2018}.
    
\end{enumerate}

We use the stabilizer-reduced weight rather than the actual weight of an error as a proxy for how hard it is to correct.
It is important to do this since errors are equivalent up to the application of a stabilizer when applied to code states. 
Moreover, single faults in a stabilizer measurement circuit can propagate to a high-weight error which is equivalent to a low weight error up to the stabilizer being measured by the circuit.
The first assumption is then made to avoid needing to multiply the error by an exponential number of stabilizers to find its minimum-weight representative. 
This assumption is somewhat justified by the fact that the modified restriction decoder in \sec{3D-color-code-perfect-measurements} seeks to output a low-weight correction.

\begin{figure}
	\includegraphics[width=.44\columnwidth]{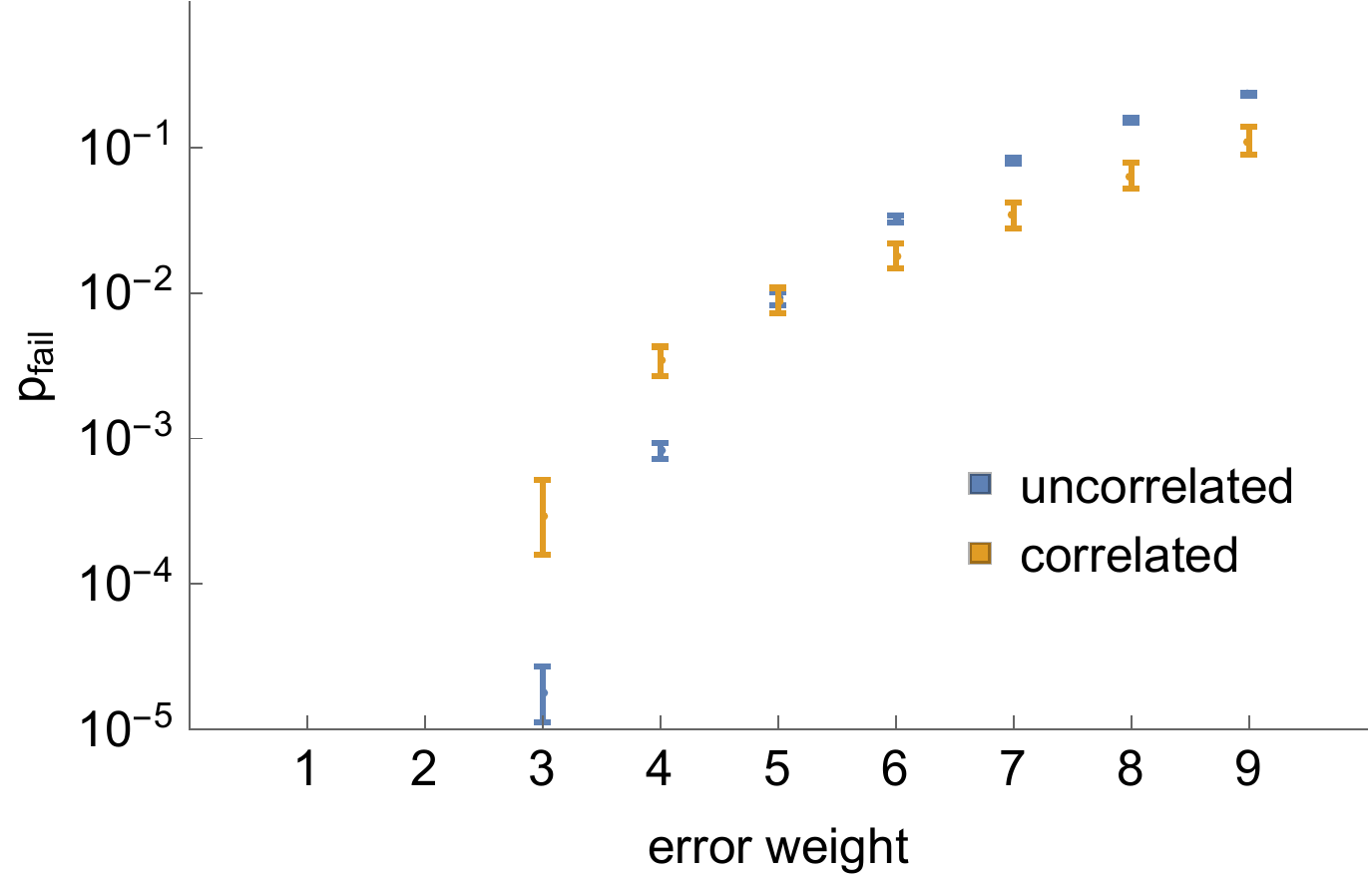}
	\qquad
	\caption{
	Performance of the modified restriction decoder for the 3D color code for distance $d=9$ given various error weights. 
	Uncorrelated $Z$ noise (blue) of weight $w$ is generated by randomly selecting $w$ of the $n$ qubits and applying a $Z$ error.
	Correlated $Z$ noise (yellow) is generated by running the decoder for error rate $p=0.0003$, and for each error generated, estimating its stabilizer-reduced weight $w$ by taking the smaller of the weight of the error itself and its correction.
	The correlated and uncorrelated cases are not identical but are very similar, justifying our assumptions.
	}
	\label{fig:3DColorCodeDecoderPerformanceCorrelatedNoise}
\end{figure}

The second assumption is somewhat crude, since decoders which take correlations into account can outperform those which do not; see e.g. Ref.~\cite{Maskara2018}.
However, we suspect that the assumption holds for typical error patterns, since the decoder will have to correct uncorrelated weight $w$ errors in addition to correlated errors. 
Therefore, we believe it is unlikely that one can design a decoder which significantly outperforms the optimal decoder for iid $Z$ noise in this setting.
We gain further confidence in the first two assumptions by numerically testing the decoder on correlated and uncorrelated noise; see \fig{3DColorCodeDecoderPerformanceCorrelatedNoise}. 
There, we verify that the performance of the modified restriction decoder for noise produced by the code switching protocol with stabilizer-reduced weight $w$ (estimated using the first assumption) is comparable to its performance forr iid $Z$ noise with weight $w$.

The third assumption is based on the heuristic behavior of error correction failure rate $p_{\text{fail}}$ in topological codes for error rate $p$ in the vicinity of their threshold $p^*$ \cite{fowler2012,fowler2013,landahl2011}, i.e., that $p_{\text{fail}}(p,d) \simeq \left(p/p^{*}\right)^{(d+1)/2}$, which is a special case of the ansatz in \eq{ansatz_moregeneral}. 
To form an inequality in terms of the error weight $w$, we first take $w$ as a proxy for the typical error $\langle w \rangle$, which for iid $Z$ noise on $n$ qubits satisfies $\langle w \rangle = p n$. 
We then note that taking $p^{*}$ as a prefactor would correspond to the pseudo-threshold being independent of system size, thereby overestimating error rate for most topological codes.

\clearpage
\subsection{Asymptotic overhead analysis}
\label{app:scaling-calculations}
Here we derive the scaling of the space and space-time overhead of code switching between 2D and 3D codes and state distillation using 2D codes in \eq{scaling-expressions} in \sec{discussion}.  
We consider the regime with $p \ll 1$ and $\log{p_{\text{fin}}}/\log{p} \gg 1$, which allows us to assume the code distances and the number of distillation rounds are large.

First, we consider code switching. 
We assume that the failure probability obeys
\begin{eqnarray}
\overline{p}_{\text{CS}}(p,d)= A_{\text{CS}} \left( \frac{p}{p^*_{\text{CS}}} \right)^{C_{\text{CS}} d}.
\end{eqnarray}
In order to implement code switching we need (to leading order in $d$) $N_{\text{3D}} d^3$ qubits and $\tau_{\text{3D}} d$ time units.
In the implementation presented in \sec{dimjump}, $N_{\text{3D}} = 13/12$ and $\tau_{\text{3D}} = 8$.
To produce a state of infidelity $p_{\text{fin}}$, we find the smallest $d$ for which $\overline{p}_{\text{CS}}(p,d)<p_{\text{fin}}$, arriving at the expressions
\begin{eqnarray}
O_{\text{CS}}^{\text{S}} & \sim & \frac{N_{\text{3D}}}{C_{\text{CS}}^3}\left( 
\frac{\log{p_{\text{fin}}}-\log{A_{\text{CS}}}}{\log{p}-\log{p^*_{\text{CS}}}} \right)^3 , \label{eq:scaling-CS-S-PreLimit}  \\ O_{\text{CS}}^{\text{ST}} & \sim & \frac{N_{\text{3D}} \tau_{\text{3D}}}{C_{\text{CS}}^4}\left( 
\frac{\log{p_{\text{fin}}}-\log{A_{\text{CS}}}}{\log{p}-\log{p^*_{\text{CS}}}} \right)^4.\label{eq:scaling-CS-ST-PreLimit}
\end{eqnarray}

Now we consider distillation. 
The starting point is the simplification that the failure probability of logical operations in the 2D color code is approximately captured by the phenomenological form
\begin{eqnarray}
\overline{p}_{\text{2D}}(p,d)= A_{\text{2D}} \left( \frac{p}{p^*_{\text{2D}}} \right)^{C_{\text{2D}} d}.
\end{eqnarray}
We will assume that the iterative application of an $R$-to-$1$ distillation scheme which (with perfect Clifford operations) reduces the error in a $T$ state from $q$ to $E q^F$ using a Clifford circuit containing $L$ locations and $R_{\text{anc}}$ encoded ancilla qubits.
In the case of the 15-to-1 scheme presented in \sec{distillation}, $R=15$, $E = 35$, and $F=3$, and $L$ is of the order of $10^3$, depending on the precise details of the implementation and the effective noise model.
We assume that the input to the first round is an encoded magic state in a code of fixed size $d_0$ with infidelity $q_0 = G p^H$, where the initialization scheme presented in \sec{magic-state-initialization} has $G=6.07$ and $H=1$. 

We consider $k$ rounds of distillation, with distances $\{d_1, d_2, \dots, d_k \}$, and with outputs of each round having infidelity $\{q_1, q_2, \dots, q_k \}$ satisfying $q_{i+1} = E q_i^F$ in the absence of Clifford noise.
In this analysis, we will assume that for each round the distance $d_i$ is chosen such that $L \overline{p}_{\text{2D}}(p,d_i) \sim q_{i}$, and neglect the effect of Clifford error.
We can use the fact that the final round should output the target infidelity, i.e., $q_k = p_{\text{fin}}$, to relate $q_i$ to $p_{\text{fin}}$ as follows
\begin{eqnarray}
q_{i} = \frac{p_{\text{fin}}^{\alpha(k-i)}}{E^{\beta(k-i)}},\quad\quad\alpha(j) = 1/F^{j},\quad\quad\beta(j) = \sum_{l=1}^j F^{-l} = \frac{1-F^{-j}}{F-1}.
\end{eqnarray}
We solve for the number of rounds $k$ in $q_0 = G p^H= p_{\text{fin}}^{\alpha(k)}/E^{\beta(k)}$, which yields
\begin{eqnarray}
k=\log_F \left( \frac{\log{E} +  (F-1)\log{p_{\text{fin}}}}{\log{E} + (F-1)\log(G p^H)} \right).
\end{eqnarray}
The ratio of the number of qubits needed in round $i+1$ and round $i$ is 
\begin{eqnarray}
\frac{d_{i+1}^2}{R d_{i}^2} = \frac{\left(\log q_{i+1} - \log (A_{\text{2D}} L) \right)^2}{R \left(\log q_{i} - \log (A_{\text{2D}} L) \right)^2} = \frac{\left(\log (E q_i^F) - \log (A_{\text{2D}} L) \right)^2}{R \left(\log q_{i} - \log (A_{\text{2D}} L) \right)^2} \xrightarrow[p\rightarrow 0]{} \frac{F^2}{R},
\end{eqnarray}
as $p\rightarrow 0$ implies $q_i\rightarrow 0$.
Note that this ratio is independent of the round number $i$, implying that the overhead changes monotonically with $i$.
Therefore, if $\log_F R >2$, the first round dominates, whereas if $\log_F R < 2$ then the last round dominates.
We can therefore conclude that the space overhead is dominated by either the first or last round of distillation. 
The distances $d_1$ and $d_k$ for the first and last rounds are
\begin{eqnarray}
d_1 &\sim & \frac{HF}{C_{\text{2D}}}\left( 
\frac{ \log{p}+\frac{1}{HF}\log \left(\frac{E G^F}{A_{\text{2D}} L}\right)}{\log{p}-\log{p^*_{\text{2D}}}} \right),\\
d_k &\sim & \frac{1}{C_{\text{2D}}} \left( \frac{\log{p_{\text{fin}}}-\log(A_{\text{2D}} L)}{\log{p}-\log{p^*_{\text{2D}}}}
 \right).
\end{eqnarray}
There are $\left(\frac{R+R_{\text{anc}}}{R} \right) R^k$ logical qubits needed in the first rounds, and $R+R_{\text{anc}}$ for the last.
To leading order in $d$, there are $N_{\text{2D}} d^2$ qubits needed for a distance $d$ 2D color code, and $\tau_{\text{2D}}d$ time units, where $N_{\text{2D}}=3/2$ and $\tau_{\text{2D}}=8$ in our implementation in \eq{Nqubits2DCC}.
The time cost scales as the sum of the distances, which is between $\tau_{\text{2D}} d_k$ and $k \tau_{\text{2D}}  d_k$ since $d_k$ is the largest distance; see \eq{distillationTime}.
However, we just take the time cost to be $\tau_{\text{2D}} d_k$ (neglecting the prefactor $k$) since $k$ depends doubly logarithmically on both $p$ and $p_{\text{fin}}$.
The space and space-time overhead are then
\begin{eqnarray}
O_{\text{SD}}^{\text{S}}&\sim& N_{\text{2D}}\left(1+R_{\text{anc}/R}\right) \max\left(  R^{k} d_1^2~,~ R d_k^2 \right). \label{eq:scaling-SD-S-PreLimit}\\
O_{\text{SD}}^{\text{ST}}&\sim& \tau_{\text{2D}} O_{\text{SD}}^{\text{S}} d_k. \label{eq:scaling-SD-ST-PreLimit}
\end{eqnarray}

Assuming $p\ll 1$ and $\log{p_{\text{fin}}}/\log{p} \gg 1$ in \eq{scaling-CS-S-PreLimit}, \eq{scaling-CS-ST-PreLimit}, \eq{scaling-SD-S-PreLimit} and \eq{scaling-SD-ST-PreLimit} yields \eq{scaling-expressions} with
\begin{eqnarray}
\Gamma_{\text{CS}} &=& 3,\\ 
\Gamma_{\text{SD}} &=& \max(2,\log_F R), \\
c_{\text{CS}}^{\text{S}} & \sim & \frac{N_{\text{3D}}}{C_{\text{CS}}^3},  \\ c_{\text{CS}}^{\text{ST}} & \sim & \frac{\tau_{\text{3D}} c_{\text{CS}}^{\text{S}}}{C_{\text{CS}}},\\
c_{\text{SD}}^{\text{S}}&\sim&
     \frac{N_{\text{2D}}\left(1+R_{\text{anc}}/R \right)}{C_{\text{2D}}^2}
     \times
     \begin{cases}
      F^2 H^{2-\log_{F}R} &\quad\text{if $\log_{F}R > 2$}\\
      R &\quad\text{otherwise}
    \end{cases},\\
 c_{\text{SD}}^{\text{ST}}&\sim&  \frac{\tau_{\text{2D}} c_{\text{SD}}^{\text{S}}}{C_{\text{2D}}}.
\end{eqnarray}
We expect the asymptotic expressions derived here to apply quite generally for code switching from a 2D to a 3D code, and for state distillation using 2D codes. 


\vfill\eject
\newpage

\bibliography{codeswitchingoverhead}

\begin{thebibliography}{101}%
\makeatletter
\providecommand \@ifxundefined [1]{%
 \@ifx{#1\undefined}
}%
\providecommand \@ifnum [1]{%
 \ifnum #1\expandafter \@firstoftwo
 \else \expandafter \@secondoftwo
 \fi
}%
\providecommand \@ifx [1]{%
 \ifx #1\expandafter \@firstoftwo
 \else \expandafter \@secondoftwo
 \fi
}%
\providecommand \natexlab [1]{#1}%
\providecommand \enquote  [1]{``#1''}%
\providecommand \bibnamefont  [1]{#1}%
\providecommand \bibfnamefont [1]{#1}%
\providecommand \citenamefont [1]{#1}%
\providecommand \href@noop [0]{\@secondoftwo}%
\providecommand \href [0]{\begingroup \@sanitize@url \@href}%
\providecommand \@href[1]{\@@startlink{#1}\@@href}%
\providecommand \@@href[1]{\endgroup#1\@@endlink}%
\providecommand \@sanitize@url [0]{\catcode `\\12\catcode `\$12\catcode
  `\&12\catcode `\#12\catcode `\^12\catcode `\_12\catcode `\%12\relax}%
\providecommand \@@startlink[1]{}%
\providecommand \@@endlink[0]{}%
\providecommand \url  [0]{\begingroup\@sanitize@url \@url }%
\providecommand \@url [1]{\endgroup\@href {#1}{\urlprefix }}%
\providecommand \urlprefix  [0]{URL }%
\providecommand \Eprint [0]{\href }%
\providecommand \doibase [0]{http://dx.doi.org/}%
\providecommand \selectlanguage [0]{\@gobble}%
\providecommand \bibinfo  [0]{\@secondoftwo}%
\providecommand \bibfield  [0]{\@secondoftwo}%
\providecommand \translation [1]{[#1]}%
\providecommand \BibitemOpen [0]{}%
\providecommand \bibitemStop [0]{}%
\providecommand \bibitemNoStop [0]{.\EOS\space}%
\providecommand \EOS [0]{\spacefactor3000\relax}%
\providecommand \BibitemShut  [1]{\csname bibitem#1\endcsname}%
\let\auto@bib@innerbib\@empty
\bibitem [{\citenamefont {Nigg}\ \emph {et~al.}(2014)\citenamefont {Nigg},
  \citenamefont {M{\"{u}}ller}, \citenamefont {Martinez}, \citenamefont
  {Schindler}, \citenamefont {Hennrich}, \citenamefont {Monz}, \citenamefont
  {Martin-Delgado},\ and\ \citenamefont {Blatt}}]{Nigg2014}%
  \BibitemOpen
  \bibfield  {author} {\bibinfo {author} {\bibfnamefont {D.}~\bibnamefont
  {Nigg}}, \bibinfo {author} {\bibfnamefont {M.}~\bibnamefont {M{\"{u}}ller}},
  \bibinfo {author} {\bibfnamefont {E.~A.}\ \bibnamefont {Martinez}}, \bibinfo
  {author} {\bibfnamefont {P.}~\bibnamefont {Schindler}}, \bibinfo {author}
  {\bibfnamefont {M.}~\bibnamefont {Hennrich}}, \bibinfo {author}
  {\bibfnamefont {T.}~\bibnamefont {Monz}}, \bibinfo {author} {\bibfnamefont
  {M.~A.}\ \bibnamefont {Martin-Delgado}}, \ and\ \bibinfo {author}
  {\bibfnamefont {R.}~\bibnamefont {Blatt}},\ }\href {\doibase
  10.1126/science.1253742} {\bibfield  {journal} {\bibinfo  {journal}
  {Science}\ }\textbf {\bibinfo {volume} {345}},\ \bibinfo {pages} {302}
  (\bibinfo {year} {2014})}\BibitemShut {NoStop}%
\bibitem [{\citenamefont {Barends}\ \emph {et~al.}(2014)\citenamefont
  {Barends}, \citenamefont {Kelly}, \citenamefont {Megrant}, \citenamefont
  {Veitia}, \citenamefont {Sank}, \citenamefont {Jeffrey}, \citenamefont
  {White}, \citenamefont {Mutus}, \citenamefont {Fowler}, \citenamefont
  {Campbell}, \citenamefont {Chen}, \citenamefont {Chen}, \citenamefont
  {Chiaro}, \citenamefont {Dunsworth}, \citenamefont {Neill}, \citenamefont
  {O'Malley}, \citenamefont {Roushan}, \citenamefont {Vainsencher},
  \citenamefont {Wenner}, \citenamefont {Korotkov}, \citenamefont {Cleland},\
  and\ \citenamefont {Martinis}}]{Barends2014}%
  \BibitemOpen
  \bibfield  {author} {\bibinfo {author} {\bibfnamefont {R.}~\bibnamefont
  {Barends}}, \bibinfo {author} {\bibfnamefont {J.}~\bibnamefont {Kelly}},
  \bibinfo {author} {\bibfnamefont {A.}~\bibnamefont {Megrant}}, \bibinfo
  {author} {\bibfnamefont {A.}~\bibnamefont {Veitia}}, \bibinfo {author}
  {\bibfnamefont {D.}~\bibnamefont {Sank}}, \bibinfo {author} {\bibfnamefont
  {E.}~\bibnamefont {Jeffrey}}, \bibinfo {author} {\bibfnamefont {T.~C.}\
  \bibnamefont {White}}, \bibinfo {author} {\bibfnamefont {J.}~\bibnamefont
  {Mutus}}, \bibinfo {author} {\bibfnamefont {A.~G.}\ \bibnamefont {Fowler}},
  \bibinfo {author} {\bibfnamefont {B.}~\bibnamefont {Campbell}}, \bibinfo
  {author} {\bibfnamefont {Y.}~\bibnamefont {Chen}}, \bibinfo {author}
  {\bibfnamefont {Z.}~\bibnamefont {Chen}}, \bibinfo {author} {\bibfnamefont
  {B.}~\bibnamefont {Chiaro}}, \bibinfo {author} {\bibfnamefont
  {A.}~\bibnamefont {Dunsworth}}, \bibinfo {author} {\bibfnamefont
  {C.}~\bibnamefont {Neill}}, \bibinfo {author} {\bibfnamefont
  {P.}~\bibnamefont {O'Malley}}, \bibinfo {author} {\bibfnamefont
  {P.}~\bibnamefont {Roushan}}, \bibinfo {author} {\bibfnamefont
  {A.}~\bibnamefont {Vainsencher}}, \bibinfo {author} {\bibfnamefont
  {J.}~\bibnamefont {Wenner}}, \bibinfo {author} {\bibfnamefont {A.~N.}\
  \bibnamefont {Korotkov}}, \bibinfo {author} {\bibfnamefont {A.~N.}\
  \bibnamefont {Cleland}}, \ and\ \bibinfo {author} {\bibfnamefont {J.~M.}\
  \bibnamefont {Martinis}},\ }\href {\doibase 10.1038/nature13171} {\bibfield
  {journal} {\bibinfo  {journal} {Nature}\ }\textbf {\bibinfo {volume} {508}},\
  \bibinfo {pages} {500} (\bibinfo {year} {2014})}\BibitemShut {NoStop}%
\bibitem [{\citenamefont {C{\'{o}}rcoles}\ \emph {et~al.}(2013)\citenamefont
  {C{\'{o}}rcoles}, \citenamefont {Gambetta}, \citenamefont {Chow},
  \citenamefont {Smolin}, \citenamefont {Ware}, \citenamefont {Strand},
  \citenamefont {Plourde},\ and\ \citenamefont {Steffen}}]{Corcoles2013}%
  \BibitemOpen
  \bibfield  {author} {\bibinfo {author} {\bibfnamefont {A.~D.}\ \bibnamefont
  {C{\'{o}}rcoles}}, \bibinfo {author} {\bibfnamefont {J.~M.}\ \bibnamefont
  {Gambetta}}, \bibinfo {author} {\bibfnamefont {J.~M.}\ \bibnamefont {Chow}},
  \bibinfo {author} {\bibfnamefont {J.~A.}\ \bibnamefont {Smolin}}, \bibinfo
  {author} {\bibfnamefont {M.}~\bibnamefont {Ware}}, \bibinfo {author}
  {\bibfnamefont {J.}~\bibnamefont {Strand}}, \bibinfo {author} {\bibfnamefont
  {B.~L.~T.}\ \bibnamefont {Plourde}}, \ and\ \bibinfo {author} {\bibfnamefont
  {M.}~\bibnamefont {Steffen}},\ }\href {\doibase 10.1103/PhysRevA.87.030301}
  {\bibfield  {journal} {\bibinfo  {journal} {Physical Review A}\ }\textbf
  {\bibinfo {volume} {87}},\ \bibinfo {pages} {030301} (\bibinfo {year}
  {2013})}\BibitemShut {NoStop}%
\bibitem [{\citenamefont {Ofek}\ \emph {et~al.}(2016)\citenamefont {Ofek},
  \citenamefont {Petrenko}, \citenamefont {Heeres}, \citenamefont {Reinhold},
  \citenamefont {Leghtas}, \citenamefont {Vlastakis}, \citenamefont {Liu},
  \citenamefont {Frunzio}, \citenamefont {Girvin}, \citenamefont {Jiang},
  \citenamefont {Mirrahimi}, \citenamefont {Devoret},\ and\ \citenamefont
  {Schoelkopf}}]{Ofek2016}%
  \BibitemOpen
  \bibfield  {author} {\bibinfo {author} {\bibfnamefont {N.}~\bibnamefont
  {Ofek}}, \bibinfo {author} {\bibfnamefont {A.}~\bibnamefont {Petrenko}},
  \bibinfo {author} {\bibfnamefont {R.}~\bibnamefont {Heeres}}, \bibinfo
  {author} {\bibfnamefont {P.}~\bibnamefont {Reinhold}}, \bibinfo {author}
  {\bibfnamefont {Z.}~\bibnamefont {Leghtas}}, \bibinfo {author} {\bibfnamefont
  {B.}~\bibnamefont {Vlastakis}}, \bibinfo {author} {\bibfnamefont
  {Y.}~\bibnamefont {Liu}}, \bibinfo {author} {\bibfnamefont {L.}~\bibnamefont
  {Frunzio}}, \bibinfo {author} {\bibfnamefont {S.~M.}\ \bibnamefont {Girvin}},
  \bibinfo {author} {\bibfnamefont {L.}~\bibnamefont {Jiang}}, \bibinfo
  {author} {\bibfnamefont {M.}~\bibnamefont {Mirrahimi}}, \bibinfo {author}
  {\bibfnamefont {M.~H.}\ \bibnamefont {Devoret}}, \ and\ \bibinfo {author}
  {\bibfnamefont {R.~J.}\ \bibnamefont {Schoelkopf}},\ }\href {\doibase
  10.1038/nature18949} {\bibfield  {journal} {\bibinfo  {journal} {Nature}\
  }\textbf {\bibinfo {volume} {536}},\ \bibinfo {pages} {441} (\bibinfo {year}
  {2016})}\BibitemShut {NoStop}%
\bibitem [{\citenamefont {Arute}\ \emph {et~al.}(2019)\citenamefont {Arute}
  \emph {et~al.}}]{Arute2019}%
  \BibitemOpen
  \bibfield  {author} {\bibinfo {author} {\bibfnamefont {F.}~\bibnamefont
  {Arute}} \emph {et~al.},\ }\href {\doibase 10.1038/s41586-019-1666-5}
  {\bibfield  {journal} {\bibinfo  {journal} {Nature}\ }\textbf {\bibinfo
  {volume} {574}},\ \bibinfo {pages} {505} (\bibinfo {year}
  {2019})}\BibitemShut {NoStop}%
\bibitem [{\citenamefont {Preskill}(2018)}]{Preskill2018}%
  \BibitemOpen
  \bibfield  {author} {\bibinfo {author} {\bibfnamefont {J.}~\bibnamefont
  {Preskill}},\ }\href {\doibase 10.22331/q-2018-08-06-79} {\bibfield
  {journal} {\bibinfo  {journal} {Quantum}\ }\textbf {\bibinfo {volume} {2}},\
  \bibinfo {pages} {79} (\bibinfo {year} {2018})}\BibitemShut {NoStop}%
\bibitem [{\citenamefont {Shor}(1996)}]{Shor1996}%
  \BibitemOpen
  \bibfield  {author} {\bibinfo {author} {\bibfnamefont {P.}~\bibnamefont
  {Shor}},\ }in\ \href@noop {} {\emph {\bibinfo {booktitle} {Proceedings of
  37th Conference on Foundations of Computer Science}}}\ (\bibinfo  {publisher}
  {IEEE Comput. Soc. Press},\ \bibinfo {year} {1996})\ pp.\ \bibinfo {pages}
  {56--65}\BibitemShut {NoStop}%
\bibitem [{\citenamefont {Knill}(2005)}]{knill05}%
  \BibitemOpen
  \bibfield  {author} {\bibinfo {author} {\bibfnamefont {E.}~\bibnamefont
  {Knill}},\ }\href {\doibase 10.1103/PhysRevA.71.042322} {\bibfield  {journal}
  {\bibinfo  {journal} {Phys. Rev. A}\ }\textbf {\bibinfo {volume} {71}},\
  \bibinfo {pages} {042322} (\bibinfo {year} {2005})}\BibitemShut {NoStop}%
\bibitem [{\citenamefont {Steane}(1997)}]{steane97}%
  \BibitemOpen
  \bibfield  {author} {\bibinfo {author} {\bibfnamefont {A.~M.}\ \bibnamefont
  {Steane}},\ }\href {\doibase 10.1103/PhysRevLett.78.2252} {\bibfield
  {journal} {\bibinfo  {journal} {Phys. Rev. Lett.}\ }\textbf {\bibinfo
  {volume} {78}},\ \bibinfo {pages} {2252} (\bibinfo {year}
  {1997})}\BibitemShut {NoStop}%
\bibitem [{\citenamefont {Aharonov}\ and\ \citenamefont
  {Ben-Or}(1997)}]{aharonov1997}%
  \BibitemOpen
  \bibfield  {author} {\bibinfo {author} {\bibfnamefont {D.}~\bibnamefont
  {Aharonov}}\ and\ \bibinfo {author} {\bibfnamefont {M.}~\bibnamefont
  {Ben-Or}},\ }in\ \href@noop {} {\emph {\bibinfo {booktitle} {Proceedings of
  the twenty-ninth annual ACM symposium on Theory of computing}}}\ (\bibinfo
  {organization} {ACM},\ \bibinfo {year} {1997})\ pp.\ \bibinfo {pages}
  {176--188}\BibitemShut {NoStop}%
\bibitem [{\citenamefont {Preskill}(1998)}]{Preskill1998}%
  \BibitemOpen
  \bibfield  {author} {\bibinfo {author} {\bibfnamefont {J.}~\bibnamefont
  {Preskill}},\ }\href@noop {} {\bibfield  {journal} {\bibinfo  {journal}
  {Proc. Roy. Soc. Lond.}\ }\textbf {\bibinfo {volume} {454}},\ \bibinfo
  {pages} {385} (\bibinfo {year} {1998})}\BibitemShut {NoStop}%
\bibitem [{\citenamefont {Aliferis}\ \emph {et~al.}(2005)\citenamefont
  {Aliferis}, \citenamefont {Gottesman},\ and\ \citenamefont
  {Preskill}}]{Aliferis2005}%
  \BibitemOpen
  \bibfield  {author} {\bibinfo {author} {\bibfnamefont {P.}~\bibnamefont
  {Aliferis}}, \bibinfo {author} {\bibfnamefont {D.}~\bibnamefont {Gottesman}},
  \ and\ \bibinfo {author} {\bibfnamefont {J.}~\bibnamefont {Preskill}},\
  }\href {http://arxiv.org/abs/quant-ph/0504218} {\bibfield  {journal}
  {\bibinfo  {journal} {Quantum Information and Computation}\ }\textbf
  {\bibinfo {volume} {6}},\ \bibinfo {pages} {097} (\bibinfo {year}
  {2005})}\BibitemShut {NoStop}%
\bibitem [{\citenamefont {Raussendorf}\ and\ \citenamefont
  {Harrington}(2007)}]{Raussendorf2007}%
  \BibitemOpen
  \bibfield  {author} {\bibinfo {author} {\bibfnamefont {R.}~\bibnamefont
  {Raussendorf}}\ and\ \bibinfo {author} {\bibfnamefont {J.}~\bibnamefont
  {Harrington}},\ }\href {\doibase 10.1103/PhysRevLett.98.190504} {\bibfield
  {journal} {\bibinfo  {journal} {Physical Review Letters}\ }\textbf {\bibinfo
  {volume} {98}} (\bibinfo {year} {2007}),\
  10.1103/PhysRevLett.98.190504}\BibitemShut {NoStop}%
\bibitem [{\citenamefont {Raussendorf}\ \emph {et~al.}(2007)\citenamefont
  {Raussendorf}, \citenamefont {Harrington},\ and\ \citenamefont
  {Goyal}}]{Raussendorf2007a}%
  \BibitemOpen
  \bibfield  {author} {\bibinfo {author} {\bibfnamefont {R.}~\bibnamefont
  {Raussendorf}}, \bibinfo {author} {\bibfnamefont {J.}~\bibnamefont
  {Harrington}}, \ and\ \bibinfo {author} {\bibfnamefont {K.}~\bibnamefont
  {Goyal}},\ }\href {\doibase 10.1088/1367-2630/9/6/199} {\bibfield  {journal}
  {\bibinfo  {journal} {New Journal of Physics}\ }\textbf {\bibinfo {volume}
  {9}} (\bibinfo {year} {2007}),\ 10.1088/1367-2630/9/6/199}\BibitemShut
  {NoStop}%
\bibitem [{\citenamefont {Dennis}\ \emph {et~al.}(2002)\citenamefont {Dennis},
  \citenamefont {Kitaev}, \citenamefont {Landahl},\ and\ \citenamefont
  {Preskill}}]{Dennis01}%
  \BibitemOpen
  \bibfield  {author} {\bibinfo {author} {\bibfnamefont {E.}~\bibnamefont
  {Dennis}}, \bibinfo {author} {\bibfnamefont {A.}~\bibnamefont {Kitaev}},
  \bibinfo {author} {\bibfnamefont {A.}~\bibnamefont {Landahl}}, \ and\
  \bibinfo {author} {\bibfnamefont {J.}~\bibnamefont {Preskill}},\ }\href@noop
  {} {\bibfield  {journal} {\bibinfo  {journal} {J. Math. Phys.}\ }\textbf
  {\bibinfo {volume} {43}},\ \bibinfo {pages} {4452} (\bibinfo {year}
  {2002})}\BibitemShut {NoStop}%
\bibitem [{\citenamefont {Cross}\ \emph {et~al.}(2009)\citenamefont {Cross},
  \citenamefont {Divincenzo},\ and\ \citenamefont {Terhal}}]{Cross2009}%
  \BibitemOpen
  \bibfield  {author} {\bibinfo {author} {\bibfnamefont {A.~W.}\ \bibnamefont
  {Cross}}, \bibinfo {author} {\bibfnamefont {D.~P.}\ \bibnamefont
  {Divincenzo}}, \ and\ \bibinfo {author} {\bibfnamefont {B.~M.}\ \bibnamefont
  {Terhal}},\ }\href@noop {} {\bibfield  {journal} {\bibinfo  {journal}
  {Quantum Info. Comput.}\ }\textbf {\bibinfo {volume} {9}},\ \bibinfo {pages}
  {541–572} (\bibinfo {year} {2009})}\BibitemShut {NoStop}%
\bibitem [{\citenamefont {Brown}\ \emph {et~al.}(2016)\citenamefont {Brown},
  \citenamefont {Loss}, \citenamefont {Pachos}, \citenamefont {Self},\ and\
  \citenamefont {Wootton}}]{Brown2014}%
  \BibitemOpen
  \bibfield  {author} {\bibinfo {author} {\bibfnamefont {B.~J.}\ \bibnamefont
  {Brown}}, \bibinfo {author} {\bibfnamefont {D.}~\bibnamefont {Loss}},
  \bibinfo {author} {\bibfnamefont {J.~K.}\ \bibnamefont {Pachos}}, \bibinfo
  {author} {\bibfnamefont {C.~N.}\ \bibnamefont {Self}}, \ and\ \bibinfo
  {author} {\bibfnamefont {J.~R.}\ \bibnamefont {Wootton}},\ }\href {\doibase
  10.1103/RevModPhys.88.045005} {\bibfield  {journal} {\bibinfo  {journal}
  {Reviews of Modern Physics}\ }\textbf {\bibinfo {volume} {88}},\ \bibinfo
  {pages} {045005} (\bibinfo {year} {2016})}\BibitemShut {NoStop}%
\bibitem [{\citenamefont {Duclos-Cianci}\ and\ \citenamefont
  {Poulin}(2010)}]{Duclos-Cianci2010}%
  \BibitemOpen
  \bibfield  {author} {\bibinfo {author} {\bibfnamefont {G.}~\bibnamefont
  {Duclos-Cianci}}\ and\ \bibinfo {author} {\bibfnamefont {D.}~\bibnamefont
  {Poulin}},\ }\href {\doibase 10.1103/PhysRevLett.104.050504} {\bibfield
  {journal} {\bibinfo  {journal} {Physical Review Letters}\ }\textbf {\bibinfo
  {volume} {104}},\ \bibinfo {pages} {050504} (\bibinfo {year}
  {2010})}\BibitemShut {NoStop}%
\bibitem [{\citenamefont {Anwar}\ \emph {et~al.}(2014)\citenamefont {Anwar},
  \citenamefont {Brown}, \citenamefont {Campbell},\ and\ \citenamefont
  {Browne}}]{Anwar2013}%
  \BibitemOpen
  \bibfield  {author} {\bibinfo {author} {\bibfnamefont {H.}~\bibnamefont
  {Anwar}}, \bibinfo {author} {\bibfnamefont {B.~J.}\ \bibnamefont {Brown}},
  \bibinfo {author} {\bibfnamefont {E.~T.}\ \bibnamefont {Campbell}}, \ and\
  \bibinfo {author} {\bibfnamefont {D.~E.}\ \bibnamefont {Browne}},\ }\href
  {\doibase 10.1088/1367-2630/16/6/063038} {\bibfield  {journal} {\bibinfo
  {journal} {New Journal of Physics}\ }\textbf {\bibinfo {volume} {16}},\
  \bibinfo {pages} {063038} (\bibinfo {year} {2014})}\BibitemShut {NoStop}%
\bibitem [{\citenamefont {Duclos-Cianci}\ and\ \citenamefont
  {Poulin}(2013)}]{Duclos-Cianci2013}%
  \BibitemOpen
  \bibfield  {author} {\bibinfo {author} {\bibfnamefont {G.}~\bibnamefont
  {Duclos-Cianci}}\ and\ \bibinfo {author} {\bibfnamefont {D.}~\bibnamefont
  {Poulin}},\ }\href {\doibase 10.1103/PhysRevA.87.062338} {\bibfield
  {journal} {\bibinfo  {journal} {Physical Review A}\ }\textbf {\bibinfo
  {volume} {87}},\ \bibinfo {pages} {062338} (\bibinfo {year}
  {2013})}\BibitemShut {NoStop}%
\bibitem [{\citenamefont {Bravyi}\ and\ \citenamefont
  {Haah}(2013)}]{Bravyi2013a}%
  \BibitemOpen
  \bibfield  {author} {\bibinfo {author} {\bibfnamefont {S.}~\bibnamefont
  {Bravyi}}\ and\ \bibinfo {author} {\bibfnamefont {J.}~\bibnamefont {Haah}},\
  }\href {\doibase 10.1103/PhysRevLett.111.200501} {\bibfield  {journal}
  {\bibinfo  {journal} {Physical Review Letters}\ }\textbf {\bibinfo {volume}
  {111}},\ \bibinfo {pages} {200501} (\bibinfo {year} {2013})}\BibitemShut
  {NoStop}%
\bibitem [{\citenamefont {Duivenvoorden}\ \emph {et~al.}(2019)\citenamefont
  {Duivenvoorden}, \citenamefont {Breuckmann},\ and\ \citenamefont
  {Terhal}}]{Duivenvoorden2017}%
  \BibitemOpen
  \bibfield  {author} {\bibinfo {author} {\bibfnamefont {K.}~\bibnamefont
  {Duivenvoorden}}, \bibinfo {author} {\bibfnamefont {N.~P.}\ \bibnamefont
  {Breuckmann}}, \ and\ \bibinfo {author} {\bibfnamefont {B.~M.}\ \bibnamefont
  {Terhal}},\ }\href {\doibase 10.1109/TIT.2018.2879937} {\bibfield  {journal}
  {\bibinfo  {journal} {IEEE Transactions on Information Theory}\ }\textbf
  {\bibinfo {volume} {65}},\ \bibinfo {pages} {2545} (\bibinfo {year}
  {2019})}\BibitemShut {NoStop}%
\bibitem [{\citenamefont {Kubica}\ and\ \citenamefont
  {Preskill}(2019)}]{Kubica2018toom}%
  \BibitemOpen
  \bibfield  {author} {\bibinfo {author} {\bibfnamefont {A.}~\bibnamefont
  {Kubica}}\ and\ \bibinfo {author} {\bibfnamefont {J.}~\bibnamefont
  {Preskill}},\ }\href {\doibase 10.1103/PhysRevLett.123.020501} {\bibfield
  {journal} {\bibinfo  {journal} {Physical Review Letters}\ }\textbf {\bibinfo
  {volume} {123}} (\bibinfo {year} {2019}),\
  10.1103/PhysRevLett.123.020501}\BibitemShut {NoStop}%
\bibitem [{\citenamefont {Vasmer}\ \emph {et~al.}(2020)\citenamefont {Vasmer},
  \citenamefont {Browne},\ and\ \citenamefont {Kubica}}]{Vasmer2020}%
  \BibitemOpen
  \bibfield  {author} {\bibinfo {author} {\bibfnamefont {M.}~\bibnamefont
  {Vasmer}}, \bibinfo {author} {\bibfnamefont {D.~E.}\ \bibnamefont {Browne}},
  \ and\ \bibinfo {author} {\bibfnamefont {A.}~\bibnamefont {Kubica}},\ }\href
  {http://arxiv.org/abs/2004.07247} {\bibfield  {journal} {\bibinfo  {journal}
  {arXiv:2004.07247}\ } (\bibinfo {year} {2020})}\BibitemShut {NoStop}%
\bibitem [{\citenamefont {Bravyi}\ \emph {et~al.}(2014)\citenamefont {Bravyi},
  \citenamefont {Suchara},\ and\ \citenamefont {Vargo}}]{Bravyi2014}%
  \BibitemOpen
  \bibfield  {author} {\bibinfo {author} {\bibfnamefont {S.}~\bibnamefont
  {Bravyi}}, \bibinfo {author} {\bibfnamefont {M.}~\bibnamefont {Suchara}}, \
  and\ \bibinfo {author} {\bibfnamefont {A.}~\bibnamefont {Vargo}},\ }\href
  {\doibase 10.1103/PhysRevA.90.032326} {\bibfield  {journal} {\bibinfo
  {journal} {Physical Review A - Atomic, Molecular, and Optical Physics}\
  }\textbf {\bibinfo {volume} {90}} (\bibinfo {year} {2014}),\
  10.1103/PhysRevA.90.032326}\BibitemShut {NoStop}%
\bibitem [{\citenamefont {Darmawan}\ and\ \citenamefont
  {Poulin}(2018)}]{Darmawan2018}%
  \BibitemOpen
  \bibfield  {author} {\bibinfo {author} {\bibfnamefont {A.~S.}\ \bibnamefont
  {Darmawan}}\ and\ \bibinfo {author} {\bibfnamefont {D.}~\bibnamefont
  {Poulin}},\ }\href {\doibase 10.1103/PhysRevE.97.051302} {\bibfield
  {journal} {\bibinfo  {journal} {Physical Review E}\ }\textbf {\bibinfo
  {volume} {97}} (\bibinfo {year} {2018}),\
  10.1103/PhysRevE.97.051302}\BibitemShut {NoStop}%
\bibitem [{\citenamefont {Nickerson}\ and\ \citenamefont
  {Brown}(2019)}]{Nickerson2017}%
  \BibitemOpen
  \bibfield  {author} {\bibinfo {author} {\bibfnamefont {N.~H.}\ \bibnamefont
  {Nickerson}}\ and\ \bibinfo {author} {\bibfnamefont {B.~J.}\ \bibnamefont
  {Brown}},\ }\href {\doibase 10.22331/q-2019-04-08-131} {\bibfield  {journal}
  {\bibinfo  {journal} {Quantum}\ }\textbf {\bibinfo {volume} {3}},\ \bibinfo
  {pages} {131} (\bibinfo {year} {2019})}\BibitemShut {NoStop}%
\bibitem [{\citenamefont {Maskara}\ \emph {et~al.}(2019)\citenamefont
  {Maskara}, \citenamefont {Kubica},\ and\ \citenamefont
  {Jochym-O'Connor}}]{Maskara2018}%
  \BibitemOpen
  \bibfield  {author} {\bibinfo {author} {\bibfnamefont {N.}~\bibnamefont
  {Maskara}}, \bibinfo {author} {\bibfnamefont {A.}~\bibnamefont {Kubica}}, \
  and\ \bibinfo {author} {\bibfnamefont {T.}~\bibnamefont {Jochym-O'Connor}},\
  }\href {\doibase 10.1103/PhysRevA.99.052351} {\bibfield  {journal} {\bibinfo
  {journal} {Physical Review A}\ }\textbf {\bibinfo {volume} {99}} (\bibinfo
  {year} {2019}),\ 10.1103/PhysRevA.99.052351}\BibitemShut {NoStop}%
\bibitem [{\citenamefont {Chamberland}\ and\ \citenamefont
  {Ronagh}(2018)}]{Chamberland2018}%
  \BibitemOpen
  \bibfield  {author} {\bibinfo {author} {\bibfnamefont {C.}~\bibnamefont
  {Chamberland}}\ and\ \bibinfo {author} {\bibfnamefont {P.}~\bibnamefont
  {Ronagh}},\ }\href {\doibase 10.1088/2058-9565/aad1f7} {\bibfield  {journal}
  {\bibinfo  {journal} {Quantum Science and Technology}\ }\textbf {\bibinfo
  {volume} {3}} (\bibinfo {year} {2018}),\
  10.1088/2058-9565/aad1f7}\BibitemShut {NoStop}%
\bibitem [{\citenamefont {Delfosse}(2014)}]{delfosse2014}%
  \BibitemOpen
  \bibfield  {author} {\bibinfo {author} {\bibfnamefont {N.}~\bibnamefont
  {Delfosse}},\ }\href@noop {} {\bibfield  {journal} {\bibinfo  {journal}
  {Phys. Rev. A}\ }\textbf {\bibinfo {volume} {89}},\ \bibinfo {pages} {012317}
  (\bibinfo {year} {2014})}\BibitemShut {NoStop}%
\bibitem [{\citenamefont {Kubica}\ and\ \citenamefont
  {Delfosse}(2019)}]{kubica2019}%
  \BibitemOpen
  \bibfield  {author} {\bibinfo {author} {\bibfnamefont {A.}~\bibnamefont
  {Kubica}}\ and\ \bibinfo {author} {\bibfnamefont {N.}~\bibnamefont
  {Delfosse}},\ }\href@noop {} {\bibfield  {journal} {\bibinfo  {journal}
  {arXiv}\ } (\bibinfo {year} {2019})},\ \Eprint
  {http://arxiv.org/abs/arXiv:1905.07393} {arXiv:1905.07393} \BibitemShut
  {NoStop}%
\bibitem [{\citenamefont {Kitaev}(1997)}]{Kitaev1997}%
  \BibitemOpen
  \bibfield  {author} {\bibinfo {author} {\bibfnamefont {A.~Y.}\ \bibnamefont
  {Kitaev}},\ }\href@noop {} {\bibfield  {journal} {\bibinfo  {journal} {Russ.
  Math. Surv.}\ }\textbf {\bibinfo {volume} {52}},\ \bibinfo {pages} {1191}
  (\bibinfo {year} {1997})}\BibitemShut {NoStop}%
\bibitem [{\citenamefont {Bravyi}\ and\ \citenamefont
  {Kitaev}(1998)}]{Bravyi98}%
  \BibitemOpen
  \bibfield  {author} {\bibinfo {author} {\bibfnamefont {S.}~\bibnamefont
  {Bravyi}}\ and\ \bibinfo {author} {\bibfnamefont {A.~Y.}\ \bibnamefont
  {Kitaev}},\ }\href@noop {} {\bibfield  {journal} {\bibinfo  {journal}
  {arXiv}\ } (\bibinfo {year} {1998})},\ \bibinfo {note}
  {arXiv:9811052}\BibitemShut {NoStop}%
\bibitem [{\citenamefont {Bombin}\ and\ \citenamefont
  {Martin-Delgado}(2006)}]{Bombin2006}%
  \BibitemOpen
  \bibfield  {author} {\bibinfo {author} {\bibfnamefont {H.}~\bibnamefont
  {Bombin}}\ and\ \bibinfo {author} {\bibfnamefont {M.}~\bibnamefont
  {Martin-Delgado}},\ }\href@noop {} {\bibfield  {journal} {\bibinfo  {journal}
  {Physical Review Letters}\ }\textbf {\bibinfo {volume} {97}},\ \bibinfo
  {pages} {180501} (\bibinfo {year} {2006})}\BibitemShut {NoStop}%
\bibitem [{\citenamefont {Fowler}\ \emph {et~al.}(2012)\citenamefont {Fowler},
  \citenamefont {Mariantoni}, \citenamefont {Martinis},\ and\ \citenamefont
  {Cleland}}]{fowler2012}%
  \BibitemOpen
  \bibfield  {author} {\bibinfo {author} {\bibfnamefont {A.~G.}\ \bibnamefont
  {Fowler}}, \bibinfo {author} {\bibfnamefont {M.}~\bibnamefont {Mariantoni}},
  \bibinfo {author} {\bibfnamefont {J.~M.}\ \bibnamefont {Martinis}}, \ and\
  \bibinfo {author} {\bibfnamefont {A.~N.}\ \bibnamefont {Cleland}},\
  }\href@noop {} {\bibfield  {journal} {\bibinfo  {journal} {Physical Review
  A}\ }\textbf {\bibinfo {volume} {86}},\ \bibinfo {pages} {032324} (\bibinfo
  {year} {2012})}\BibitemShut {NoStop}%
\bibitem [{\citenamefont {Chamberland}\ \emph
  {et~al.}(2020{\natexlab{a}})\citenamefont {Chamberland}, \citenamefont {Zhu},
  \citenamefont {Yoder}, \citenamefont {Hertzberg},\ and\ \citenamefont
  {Cross}}]{Chamberland2020b}%
  \BibitemOpen
  \bibfield  {author} {\bibinfo {author} {\bibfnamefont {C.}~\bibnamefont
  {Chamberland}}, \bibinfo {author} {\bibfnamefont {G.}~\bibnamefont {Zhu}},
  \bibinfo {author} {\bibfnamefont {T.~J.}\ \bibnamefont {Yoder}}, \bibinfo
  {author} {\bibfnamefont {J.~B.}\ \bibnamefont {Hertzberg}}, \ and\ \bibinfo
  {author} {\bibfnamefont {A.~W.}\ \bibnamefont {Cross}},\ }\href {\doibase
  10.1103/PhysRevX.10.011022} {\bibfield  {journal} {\bibinfo  {journal}
  {Physical Review X}\ }\textbf {\bibinfo {volume} {10}} (\bibinfo {year}
  {2020}{\natexlab{a}}),\ 10.1103/PhysRevX.10.011022}\BibitemShut {NoStop}%
\bibitem [{\citenamefont {Chamberland}\ \emph
  {et~al.}(2020{\natexlab{b}})\citenamefont {Chamberland}, \citenamefont
  {Kubica}, \citenamefont {Yoder},\ and\ \citenamefont
  {Zhu}}]{Chamberland2020}%
  \BibitemOpen
  \bibfield  {author} {\bibinfo {author} {\bibfnamefont {C.}~\bibnamefont
  {Chamberland}}, \bibinfo {author} {\bibfnamefont {A.}~\bibnamefont {Kubica}},
  \bibinfo {author} {\bibfnamefont {T.~J.}\ \bibnamefont {Yoder}}, \ and\
  \bibinfo {author} {\bibfnamefont {G.}~\bibnamefont {Zhu}},\ }\href {\doibase
  10.1088/1367-2630/ab68fd} {\bibfield  {journal} {\bibinfo  {journal} {New
  Journal of Physics}\ }\textbf {\bibinfo {volume} {22}} (\bibinfo {year}
  {2020}{\natexlab{b}}),\ 10.1088/1367-2630/ab68fd}\BibitemShut {NoStop}%
\bibitem [{\citenamefont {Karzig}\ \emph {et~al.}(2017)\citenamefont {Karzig},
  \citenamefont {Knapp}, \citenamefont {Lutchyn}, \citenamefont {Bonderson},
  \citenamefont {Hastings}, \citenamefont {Nayak}, \citenamefont {Alicea},
  \citenamefont {Flensberg}, \citenamefont {Plugge}, \citenamefont {Oreg},
  \citenamefont {Marcus},\ and\ \citenamefont {Freedman}}]{Karzig2017}%
  \BibitemOpen
  \bibfield  {author} {\bibinfo {author} {\bibfnamefont {T.}~\bibnamefont
  {Karzig}}, \bibinfo {author} {\bibfnamefont {C.}~\bibnamefont {Knapp}},
  \bibinfo {author} {\bibfnamefont {R.~M.}\ \bibnamefont {Lutchyn}}, \bibinfo
  {author} {\bibfnamefont {P.}~\bibnamefont {Bonderson}}, \bibinfo {author}
  {\bibfnamefont {M.~B.}\ \bibnamefont {Hastings}}, \bibinfo {author}
  {\bibfnamefont {C.}~\bibnamefont {Nayak}}, \bibinfo {author} {\bibfnamefont
  {J.}~\bibnamefont {Alicea}}, \bibinfo {author} {\bibfnamefont
  {K.}~\bibnamefont {Flensberg}}, \bibinfo {author} {\bibfnamefont
  {S.}~\bibnamefont {Plugge}}, \bibinfo {author} {\bibfnamefont
  {Y.}~\bibnamefont {Oreg}}, \bibinfo {author} {\bibfnamefont {C.~M.}\
  \bibnamefont {Marcus}}, \ and\ \bibinfo {author} {\bibfnamefont {M.~H.}\
  \bibnamefont {Freedman}},\ }\href {\doibase 10.1103/PhysRevB.95.235305}
  {\bibfield  {journal} {\bibinfo  {journal} {Phys. Rev. B}\ }\textbf {\bibinfo
  {volume} {95}},\ \bibinfo {pages} {235305} (\bibinfo {year}
  {2017})}\BibitemShut {NoStop}%
\bibitem [{\citenamefont {Chao}\ \emph {et~al.}(2020)\citenamefont {Chao},
  \citenamefont {Beverland}, \citenamefont {Delfosse},\ and\ \citenamefont
  {Haah}}]{Chao2020}%
  \BibitemOpen
  \bibfield  {author} {\bibinfo {author} {\bibfnamefont {R.}~\bibnamefont
  {Chao}}, \bibinfo {author} {\bibfnamefont {M.~E.}\ \bibnamefont {Beverland}},
  \bibinfo {author} {\bibfnamefont {N.}~\bibnamefont {Delfosse}}, \ and\
  \bibinfo {author} {\bibfnamefont {J.}~\bibnamefont {Haah}},\ }\href {\doibase
  10.22331/q-2020-10-28-352} {\bibfield  {journal} {\bibinfo  {journal}
  {{Quantum}}\ }\textbf {\bibinfo {volume} {4}},\ \bibinfo {pages} {352}
  (\bibinfo {year} {2020})}\BibitemShut {NoStop}%
\bibitem [{\citenamefont {Eastin}\ and\ \citenamefont
  {Knill}(2009)}]{Eastin2009}%
  \BibitemOpen
  \bibfield  {author} {\bibinfo {author} {\bibfnamefont {B.}~\bibnamefont
  {Eastin}}\ and\ \bibinfo {author} {\bibfnamefont {E.}~\bibnamefont {Knill}},\
  }\href@noop {} {\bibfield  {journal} {\bibinfo  {journal} {Phys. Rev. Lett.}\
  }\textbf {\bibinfo {volume} {102}},\ \bibinfo {pages} {110502} (\bibinfo
  {year} {2009})}\BibitemShut {NoStop}%
\bibitem [{\citenamefont {Zeng}\ \emph {et~al.}(2011)\citenamefont {Zeng},
  \citenamefont {Cross},\ and\ \citenamefont {Chuang}}]{zeng2011}%
  \BibitemOpen
  \bibfield  {author} {\bibinfo {author} {\bibfnamefont {B.}~\bibnamefont
  {Zeng}}, \bibinfo {author} {\bibfnamefont {A.}~\bibnamefont {Cross}}, \ and\
  \bibinfo {author} {\bibfnamefont {I.~L.}\ \bibnamefont {Chuang}},\
  }\href@noop {} {\bibfield  {journal} {\bibinfo  {journal} {IEEE Transactions
  on Information Theory}\ }\textbf {\bibinfo {volume} {57}},\ \bibinfo {pages}
  {6272} (\bibinfo {year} {2011})}\BibitemShut {NoStop}%
\bibitem [{\citenamefont {Jochym-O'Connor}\ \emph {et~al.}(2018)\citenamefont
  {Jochym-O'Connor}, \citenamefont {Kubica},\ and\ \citenamefont
  {Yoder}}]{Jochym-OConnor2018}%
  \BibitemOpen
  \bibfield  {author} {\bibinfo {author} {\bibfnamefont {T.}~\bibnamefont
  {Jochym-O'Connor}}, \bibinfo {author} {\bibfnamefont {A.}~\bibnamefont
  {Kubica}}, \ and\ \bibinfo {author} {\bibfnamefont {T.~J.}\ \bibnamefont
  {Yoder}},\ }\href {\doibase 10.1103/PhysRevX.8.021047} {\bibfield  {journal}
  {\bibinfo  {journal} {Physical Review X}\ }\textbf {\bibinfo {volume} {8}},\
  \bibinfo {pages} {21047} (\bibinfo {year} {2018})}\BibitemShut {NoStop}%
\bibitem [{\citenamefont {Bravyi}\ and\ \citenamefont
  {K\"{o}nig}(2013)}]{Bravyi2013}%
  \BibitemOpen
  \bibfield  {author} {\bibinfo {author} {\bibfnamefont {S.}~\bibnamefont
  {Bravyi}}\ and\ \bibinfo {author} {\bibfnamefont {R.}~\bibnamefont
  {K\"{o}nig}},\ }\href@noop {} {\bibfield  {journal} {\bibinfo  {journal}
  {Phys. Rev. Lett.}\ }\textbf {\bibinfo {volume} {110}},\ \bibinfo {pages}
  {170503} (\bibinfo {year} {2013})}\BibitemShut {NoStop}%
\bibitem [{\citenamefont {Pastawski}\ and\ \citenamefont
  {Yoshida}(2014)}]{PastawskiYoshida14}%
  \BibitemOpen
  \bibfield  {author} {\bibinfo {author} {\bibfnamefont {F.}~\bibnamefont
  {Pastawski}}\ and\ \bibinfo {author} {\bibfnamefont {B.}~\bibnamefont
  {Yoshida}},\ }\href@noop {} {\enquote {\bibinfo {title} {Fault-tolerant
  logical gates in quantum error-correcting codes},}\ } (\bibinfo {year}
  {2014}),\ \bibinfo {note} {{ arXiv:1408.1720}}\BibitemShut {NoStop}%
\bibitem [{\citenamefont {Beverland}\ \emph {et~al.}(2016)\citenamefont
  {Beverland}, \citenamefont {Buerschaper}, \citenamefont {Koenig},
  \citenamefont {Pastawski}, \citenamefont {Preskill},\ and\ \citenamefont
  {Sijher}}]{Beverland2014}%
  \BibitemOpen
  \bibfield  {author} {\bibinfo {author} {\bibfnamefont {M.~E.}\ \bibnamefont
  {Beverland}}, \bibinfo {author} {\bibfnamefont {O.}~\bibnamefont
  {Buerschaper}}, \bibinfo {author} {\bibfnamefont {R.}~\bibnamefont {Koenig}},
  \bibinfo {author} {\bibfnamefont {F.}~\bibnamefont {Pastawski}}, \bibinfo
  {author} {\bibfnamefont {J.}~\bibnamefont {Preskill}}, \ and\ \bibinfo
  {author} {\bibfnamefont {S.}~\bibnamefont {Sijher}},\ }\href@noop {}
  {\bibfield  {journal} {\bibinfo  {journal} {Journal of Mathematical Physics}\
  }\textbf {\bibinfo {volume} {57}},\ \bibinfo {pages} {022201} (\bibinfo
  {year} {2016})}\BibitemShut {NoStop}%
\bibitem [{\citenamefont {Webster}\ \emph {et~al.}(2020)\citenamefont
  {Webster}, \citenamefont {Vasmer}, \citenamefont {Scruby},\ and\
  \citenamefont {Bartlett}}]{Webster2020}%
  \BibitemOpen
  \bibfield  {author} {\bibinfo {author} {\bibfnamefont {P.}~\bibnamefont
  {Webster}}, \bibinfo {author} {\bibfnamefont {M.}~\bibnamefont {Vasmer}},
  \bibinfo {author} {\bibfnamefont {T.~R.}\ \bibnamefont {Scruby}}, \ and\
  \bibinfo {author} {\bibfnamefont {S.~D.}\ \bibnamefont {Bartlett}},\ }\href
  {http://arxiv.org/abs/2012.05260} {\bibfield  {journal} {\bibinfo  {journal}
  {arXiv:2012.05260}\ } (\bibinfo {year} {2020})}\BibitemShut {NoStop}%
\bibitem [{\citenamefont {Bombin}\ and\ \citenamefont
  {Martin-Delgado}(2007)}]{Bombin2007}%
  \BibitemOpen
  \bibfield  {author} {\bibinfo {author} {\bibfnamefont {H.}~\bibnamefont
  {Bombin}}\ and\ \bibinfo {author} {\bibfnamefont {M.}~\bibnamefont
  {Martin-Delgado}},\ }\href@noop {} {\bibfield  {journal} {\bibinfo  {journal}
  {Physical Review B}\ }\textbf {\bibinfo {volume} {75}},\ \bibinfo {pages}
  {075103} (\bibinfo {year} {2007})}\BibitemShut {NoStop}%
\bibitem [{\citenamefont {Kubica}\ \emph {et~al.}(2015)\citenamefont {Kubica},
  \citenamefont {Yoshida},\ and\ \citenamefont {Pastawski}}]{kubica2015}%
  \BibitemOpen
  \bibfield  {author} {\bibinfo {author} {\bibfnamefont {A.}~\bibnamefont
  {Kubica}}, \bibinfo {author} {\bibfnamefont {B.}~\bibnamefont {Yoshida}}, \
  and\ \bibinfo {author} {\bibfnamefont {F.}~\bibnamefont {Pastawski}},\ }\href
  {\doibase 10.1088/1367-2630/17/8/083026} {\bibfield  {journal} {\bibinfo
  {journal} {New Journal of Physics}\ }\textbf {\bibinfo {volume} {17}},\
  \bibinfo {pages} {083026} (\bibinfo {year} {2015})}\BibitemShut {NoStop}%
\bibitem [{\citenamefont {Vasmer}\ and\ \citenamefont
  {Browne}(2019)}]{Vasmer2019}%
  \BibitemOpen
  \bibfield  {author} {\bibinfo {author} {\bibfnamefont {M.}~\bibnamefont
  {Vasmer}}\ and\ \bibinfo {author} {\bibfnamefont {D.~E.}\ \bibnamefont
  {Browne}},\ }\href {\doibase 10.1103/PhysRevA.100.012312} {\bibfield
  {journal} {\bibinfo  {journal} {Physical Review A}\ }\textbf {\bibinfo
  {volume} {100}},\ \bibinfo {pages} {012312} (\bibinfo {year}
  {2019})}\BibitemShut {NoStop}%
\bibitem [{\citenamefont {Bravyi}\ and\ \citenamefont
  {Kitaev}(2005)}]{Bravyi2005}%
  \BibitemOpen
  \bibfield  {author} {\bibinfo {author} {\bibfnamefont {S.}~\bibnamefont
  {Bravyi}}\ and\ \bibinfo {author} {\bibfnamefont {A.}~\bibnamefont
  {Kitaev}},\ }\href@noop {} {\bibfield  {journal} {\bibinfo  {journal} {Phys.
  Rev. A}\ }\textbf {\bibinfo {volume} {71}},\ \bibinfo {pages} {022316}
  (\bibinfo {year} {2005})}\BibitemShut {NoStop}%
\bibitem [{\citenamefont {Knill}(2004{\natexlab{a}})}]{knill2004a}%
  \BibitemOpen
  \bibfield  {author} {\bibinfo {author} {\bibfnamefont {E.}~\bibnamefont
  {Knill}},\ }\href@noop {} {\bibfield  {journal} {\bibinfo  {journal} {arXiv
  preprint arXiv:0404104}\ } (\bibinfo {year}
  {2004}{\natexlab{a}})}\BibitemShut {NoStop}%
\bibitem [{\citenamefont {Knill}(2004{\natexlab{b}})}]{knill2004b}%
  \BibitemOpen
  \bibfield  {author} {\bibinfo {author} {\bibfnamefont {E.}~\bibnamefont
  {Knill}},\ }\href@noop {} {\bibfield  {journal} {\bibinfo  {journal} {arXiv
  preprint arXiv:0402171}\ } (\bibinfo {year}
  {2004}{\natexlab{b}})}\BibitemShut {NoStop}%
\bibitem [{\citenamefont {Litinski}(2019)}]{Litinski2019}%
  \BibitemOpen
  \bibfield  {author} {\bibinfo {author} {\bibfnamefont {D.}~\bibnamefont
  {Litinski}},\ }\href {\doibase 10.22331/q-2019-12-02-205} {\bibfield
  {journal} {\bibinfo  {journal} {{Quantum}}\ }\textbf {\bibinfo {volume}
  {3}},\ \bibinfo {pages} {205} (\bibinfo {year} {2019})}\BibitemShut {NoStop}%
\bibitem [{\citenamefont {Paetznick}\ and\ \citenamefont
  {Reichardt}(2013)}]{paetznick2013}%
  \BibitemOpen
  \bibfield  {author} {\bibinfo {author} {\bibfnamefont {A.}~\bibnamefont
  {Paetznick}}\ and\ \bibinfo {author} {\bibfnamefont {B.~W.}\ \bibnamefont
  {Reichardt}},\ }\href@noop {} {\bibfield  {journal} {\bibinfo  {journal}
  {Phys. Rev. Lett.}\ }\textbf {\bibinfo {volume} {111}},\ \bibinfo {pages}
  {090505} (\bibinfo {year} {2013})}\BibitemShut {NoStop}%
\bibitem [{\citenamefont {Anderson}\ \emph {et~al.}(2014)\citenamefont
  {Anderson}, \citenamefont {Duclos-Cianci},\ and\ \citenamefont
  {Poulin}}]{anderson2014}%
  \BibitemOpen
  \bibfield  {author} {\bibinfo {author} {\bibfnamefont {J.~T.}\ \bibnamefont
  {Anderson}}, \bibinfo {author} {\bibfnamefont {G.}~\bibnamefont
  {Duclos-Cianci}}, \ and\ \bibinfo {author} {\bibfnamefont {D.}~\bibnamefont
  {Poulin}},\ }\href@noop {} {\bibfield  {journal} {\bibinfo  {journal} {Phys.
  Rev. Lett.}\ }\textbf {\bibinfo {volume} {113}},\ \bibinfo {pages} {080501}
  (\bibinfo {year} {2014})}\BibitemShut {NoStop}%
\bibitem [{\citenamefont {Bombin}(2015{\natexlab{a}})}]{bombin2015}%
  \BibitemOpen
  \bibfield  {author} {\bibinfo {author} {\bibfnamefont {H.}~\bibnamefont
  {Bombin}},\ }\href@noop {} {\bibfield  {journal} {\bibinfo  {journal} {New
  Journal of Physics}\ }\textbf {\bibinfo {volume} {17}},\ \bibinfo {pages}
  {083002} (\bibinfo {year} {2015}{\natexlab{a}})}\BibitemShut {NoStop}%
\bibitem [{\citenamefont {Bomb{\'\i}n}(2016)}]{bombin2016}%
  \BibitemOpen
  \bibfield  {author} {\bibinfo {author} {\bibfnamefont {H.}~\bibnamefont
  {Bomb{\'\i}n}},\ }\href@noop {} {\bibfield  {journal} {\bibinfo  {journal}
  {New Journal of Physics}\ }\textbf {\bibinfo {volume} {18}},\ \bibinfo
  {pages} {043038} (\bibinfo {year} {2016})}\BibitemShut {NoStop}%
\bibitem [{\citenamefont {Bombin}(2015{\natexlab{b}})}]{Bombin2015b}%
  \BibitemOpen
  \bibfield  {author} {\bibinfo {author} {\bibfnamefont {H.}~\bibnamefont
  {Bombin}},\ }\href@noop {} {\bibfield  {journal} {\bibinfo  {journal}
  {Physical Review X}\ }\textbf {\bibinfo {volume} {5}},\ \bibinfo {pages}
  {031043} (\bibinfo {year} {2015}{\natexlab{b}})}\BibitemShut {NoStop}%
\bibitem [{\citenamefont {Bravyi}\ and\ \citenamefont
  {Haah}(2012)}]{bravyi2012}%
  \BibitemOpen
  \bibfield  {author} {\bibinfo {author} {\bibfnamefont {S.}~\bibnamefont
  {Bravyi}}\ and\ \bibinfo {author} {\bibfnamefont {J.}~\bibnamefont {Haah}},\
  }\href@noop {} {\bibfield  {journal} {\bibinfo  {journal} {Physical Review
  A}\ }\textbf {\bibinfo {volume} {86}},\ \bibinfo {pages} {052329} (\bibinfo
  {year} {2012})}\BibitemShut {NoStop}%
\bibitem [{\citenamefont {Haah}\ \emph
  {et~al.}(2017{\natexlab{a}})\citenamefont {Haah}, \citenamefont {Hastings},
  \citenamefont {Poulin},\ and\ \citenamefont {Wecker}}]{haah2017}%
  \BibitemOpen
  \bibfield  {author} {\bibinfo {author} {\bibfnamefont {J.}~\bibnamefont
  {Haah}}, \bibinfo {author} {\bibfnamefont {M.~B.}\ \bibnamefont {Hastings}},
  \bibinfo {author} {\bibfnamefont {D.}~\bibnamefont {Poulin}}, \ and\ \bibinfo
  {author} {\bibfnamefont {D.}~\bibnamefont {Wecker}},\ }\href {\doibase
  10.22331/q-2017-10-03-31} {\bibfield  {journal} {\bibinfo  {journal}
  {{Quantum}}\ }\textbf {\bibinfo {volume} {1}},\ \bibinfo {pages} {31}
  (\bibinfo {year} {2017}{\natexlab{a}})}\BibitemShut {NoStop}%
\bibitem [{\citenamefont {Haah}\ and\ \citenamefont
  {Hastings}(2018)}]{Haah2018a}%
  \BibitemOpen
  \bibfield  {author} {\bibinfo {author} {\bibfnamefont {J.}~\bibnamefont
  {Haah}}\ and\ \bibinfo {author} {\bibfnamefont {M.~B.}\ \bibnamefont
  {Hastings}},\ }\href {\doibase 10.22331/q-2018-06-07-71} {\bibfield
  {journal} {\bibinfo  {journal} {{Quantum}}\ }\textbf {\bibinfo {volume}
  {2}},\ \bibinfo {pages} {71} (\bibinfo {year} {2018})}\BibitemShut {NoStop}%
\bibitem [{\citenamefont {Brooks}(2013)}]{brooks2013}%
  \BibitemOpen
  \bibfield  {author} {\bibinfo {author} {\bibfnamefont {P.}~\bibnamefont
  {Brooks}},\ }\href@noop {} {\bibfield  {journal} {\bibinfo  {journal}
  {Caltech Ph.D Thesis}\ } (\bibinfo {year} {2013})}\BibitemShut {NoStop}%
\bibitem [{\citenamefont {Jochym-O'Connor}\ \emph {et~al.}(2013)\citenamefont
  {Jochym-O'Connor}, \citenamefont {Yu}, \citenamefont {Helou},\ and\
  \citenamefont {Laflamme}}]{jochym2013}%
  \BibitemOpen
  \bibfield  {author} {\bibinfo {author} {\bibfnamefont {T.}~\bibnamefont
  {Jochym-O'Connor}}, \bibinfo {author} {\bibfnamefont {Y.}~\bibnamefont {Yu}},
  \bibinfo {author} {\bibfnamefont {B.}~\bibnamefont {Helou}}, \ and\ \bibinfo
  {author} {\bibfnamefont {R.}~\bibnamefont {Laflamme}},\ }\href@noop {}
  {\bibfield  {journal} {\bibinfo  {journal} {Quantum Information and
  Computation}\ }\textbf {\bibinfo {volume} {13}},\ \bibinfo {pages} {0361}
  (\bibinfo {year} {2013})}\BibitemShut {NoStop}%
\bibitem [{\citenamefont {Jones}(2013{\natexlab{a}})}]{jones2013b}%
  \BibitemOpen
  \bibfield  {author} {\bibinfo {author} {\bibfnamefont {N.~C.}\ \bibnamefont
  {Jones}},\ }\href@noop {} {\enquote {\bibinfo {title} {Logic synthesis for
  fault-tolerant quantum computers},}\ } (\bibinfo {year}
  {2013}{\natexlab{a}}),\ \Eprint {http://arxiv.org/abs/1310.7290} {PhD Thesis
  (Stanford):1310.7290 [quant-ph]} \BibitemShut {NoStop}%
\bibitem [{\citenamefont {Bravyi}\ and\ \citenamefont
  {Cross}(2015)}]{Bravyi2015}%
  \BibitemOpen
  \bibfield  {author} {\bibinfo {author} {\bibfnamefont {S.}~\bibnamefont
  {Bravyi}}\ and\ \bibinfo {author} {\bibfnamefont {A.}~\bibnamefont {Cross}},\
  }\href {http://arxiv.org/abs/1509.03239} {\bibfield  {journal} {\bibinfo
  {journal} {arXiv:1509.03239}\ } (\bibinfo {year} {2015})}\BibitemShut
  {NoStop}%
\bibitem [{\citenamefont {Jochym-O'Connor}\ and\ \citenamefont
  {Bartlett}(2016)}]{Jochym-OConnor2016}%
  \BibitemOpen
  \bibfield  {author} {\bibinfo {author} {\bibfnamefont {T.}~\bibnamefont
  {Jochym-O'Connor}}\ and\ \bibinfo {author} {\bibfnamefont {S.~D.}\
  \bibnamefont {Bartlett}},\ }\href {\doibase 10.1103/PhysRevA.93.022323}
  {\bibfield  {journal} {\bibinfo  {journal} {Physical Review A}\ }\textbf
  {\bibinfo {volume} {93}} (\bibinfo {year} {2016}),\
  10.1103/PhysRevA.93.022323}\BibitemShut {NoStop}%
\bibitem [{\citenamefont {Jones}\ \emph {et~al.}(2016)\citenamefont {Jones},
  \citenamefont {Brooks},\ and\ \citenamefont {Harrington}}]{Jones2016}%
  \BibitemOpen
  \bibfield  {author} {\bibinfo {author} {\bibfnamefont {C.}~\bibnamefont
  {Jones}}, \bibinfo {author} {\bibfnamefont {P.}~\bibnamefont {Brooks}}, \
  and\ \bibinfo {author} {\bibfnamefont {J.}~\bibnamefont {Harrington}},\
  }\href {\doibase 10.1103/PhysRevA.93.052332} {\bibfield  {journal} {\bibinfo
  {journal} {Physical Review A}\ }\textbf {\bibinfo {volume} {93}} (\bibinfo
  {year} {2016}),\ 10.1103/PhysRevA.93.052332}\BibitemShut {NoStop}%
\bibitem [{\citenamefont {Bombin}(2018{\natexlab{a}})}]{Bombin2018a}%
  \BibitemOpen
  \bibfield  {author} {\bibinfo {author} {\bibfnamefont {H.}~\bibnamefont
  {Bombin}},\ }\href {http://arxiv.org/abs/1810.09571} {\bibfield  {journal}
  {\bibinfo  {journal} {arXiv:1810.09571}\ } (\bibinfo {year}
  {2018}{\natexlab{a}})}\BibitemShut {NoStop}%
\bibitem [{\citenamefont {Brown}(2020)}]{Brown2020}%
  \BibitemOpen
  \bibfield  {author} {\bibinfo {author} {\bibfnamefont {B.~J.}\ \bibnamefont
  {Brown}},\ }\href {\doibase 10.1126/sciadv.aay4929} {\bibfield  {journal}
  {\bibinfo  {journal} {Science Advances}\ }\textbf {\bibinfo {volume} {6}}
  (\bibinfo {year} {2020}),\ 10.1126/sciadv.aay4929}\BibitemShut {NoStop}%
\bibitem [{\citenamefont {Scruby}\ \emph {et~al.}(2020)\citenamefont {Scruby},
  \citenamefont {Browne}, \citenamefont {Webster},\ and\ \citenamefont
  {Vasmer}}]{Scruby2020}%
  \BibitemOpen
  \bibfield  {author} {\bibinfo {author} {\bibfnamefont {T.~R.}\ \bibnamefont
  {Scruby}}, \bibinfo {author} {\bibfnamefont {D.~E.}\ \bibnamefont {Browne}},
  \bibinfo {author} {\bibfnamefont {P.}~\bibnamefont {Webster}}, \ and\
  \bibinfo {author} {\bibfnamefont {M.}~\bibnamefont {Vasmer}},\ }\href
  {http://arxiv.org/abs/2012.08536} {\bibfield  {journal} {\bibinfo  {journal}
  {arXiv:2012.08536}\ } (\bibinfo {year} {2020})}\BibitemShut {NoStop}%
\bibitem [{\citenamefont {Iverson}\ and\ \citenamefont
  {Kubica}(2021)}]{Iverson2020}%
  \BibitemOpen
  \bibfield  {author} {\bibinfo {author} {\bibfnamefont {J.}~\bibnamefont
  {Iverson}}\ and\ \bibinfo {author} {\bibfnamefont {A.}~\bibnamefont
  {Kubica}},\ }\href@noop {} {}\bibinfo {howpublished} {in preparation}
  (\bibinfo {year} {2021})\BibitemShut {NoStop}%
\bibitem [{\citenamefont {Hill}\ \emph {et~al.}(2013)\citenamefont {Hill},
  \citenamefont {Fowler}, \citenamefont {Wang},\ and\ \citenamefont
  {Hollenberg}}]{Hill2013}%
  \BibitemOpen
  \bibfield  {author} {\bibinfo {author} {\bibfnamefont {C.~D.}\ \bibnamefont
  {Hill}}, \bibinfo {author} {\bibfnamefont {A.~G.}\ \bibnamefont {Fowler}},
  \bibinfo {author} {\bibfnamefont {D.~S.}\ \bibnamefont {Wang}}, \ and\
  \bibinfo {author} {\bibfnamefont {L.~C.~L.}\ \bibnamefont {Hollenberg}},\
  }\href@noop {} {\bibfield  {journal} {\bibinfo  {journal} {Quantum Info.
  Comput.}\ }\textbf {\bibinfo {volume} {13}},\ \bibinfo {pages} {439–451}
  (\bibinfo {year} {2013})}\BibitemShut {NoStop}%
\bibitem [{\citenamefont {Yoder}\ \emph {et~al.}(2016)\citenamefont {Yoder},
  \citenamefont {Takagi},\ and\ \citenamefont {Chuang}}]{yoder2016}%
  \BibitemOpen
  \bibfield  {author} {\bibinfo {author} {\bibfnamefont {T.~J.}\ \bibnamefont
  {Yoder}}, \bibinfo {author} {\bibfnamefont {R.}~\bibnamefont {Takagi}}, \
  and\ \bibinfo {author} {\bibfnamefont {I.~L.}\ \bibnamefont {Chuang}},\
  }\href@noop {} {\bibfield  {journal} {\bibinfo  {journal} {Physical Review
  X}\ }\textbf {\bibinfo {volume} {6}},\ \bibinfo {pages} {031039} (\bibinfo
  {year} {2016})}\BibitemShut {NoStop}%
\bibitem [{\citenamefont {Jochym-O'Connor}\ and\ \citenamefont
  {Laflamme}(2014)}]{jochym2014}%
  \BibitemOpen
  \bibfield  {author} {\bibinfo {author} {\bibfnamefont {T.}~\bibnamefont
  {Jochym-O'Connor}}\ and\ \bibinfo {author} {\bibfnamefont {R.}~\bibnamefont
  {Laflamme}},\ }\href@noop {} {\bibfield  {journal} {\bibinfo  {journal}
  {Physical Review Letters}\ }\textbf {\bibinfo {volume} {112}},\ \bibinfo
  {pages} {010505} (\bibinfo {year} {2014})}\BibitemShut {NoStop}%
\bibitem [{\citenamefont {Chamberland}\ and\ \citenamefont
  {Cross}(2019)}]{Chamberland2019}%
  \BibitemOpen
  \bibfield  {author} {\bibinfo {author} {\bibfnamefont {C.}~\bibnamefont
  {Chamberland}}\ and\ \bibinfo {author} {\bibfnamefont {A.~W.}\ \bibnamefont
  {Cross}},\ }\href {\doibase 10.22331/q-2019-05-20-143} {\bibfield  {journal}
  {\bibinfo  {journal} {{Quantum}}\ }\textbf {\bibinfo {volume} {3}},\ \bibinfo
  {pages} {143} (\bibinfo {year} {2019})}\BibitemShut {NoStop}%
\bibitem [{\citenamefont {Chamberland}\ and\ \citenamefont
  {Noh}(2020{\natexlab{a}})}]{Chamberland2020c}%
  \BibitemOpen
  \bibfield  {author} {\bibinfo {author} {\bibfnamefont {C.}~\bibnamefont
  {Chamberland}}\ and\ \bibinfo {author} {\bibfnamefont {K.}~\bibnamefont
  {Noh}},\ }\href {\doibase 10.1038/s41534-020-00319-5} {\bibfield  {journal}
  {\bibinfo  {journal} {npj Quantum Information}\ }\textbf {\bibinfo {volume}
  {6}},\ \bibinfo {pages} {91} (\bibinfo {year}
  {2020}{\natexlab{a}})}\BibitemShut {NoStop}%
\bibitem [{\citenamefont {Fowler}(2013)}]{fowler2013}%
  \BibitemOpen
  \bibfield  {author} {\bibinfo {author} {\bibfnamefont {A.~G.}\ \bibnamefont
  {Fowler}},\ }\href {\doibase 10.1103/PhysRevA.87.040301} {\bibfield
  {journal} {\bibinfo  {journal} {Phys. Rev. A}\ }\textbf {\bibinfo {volume}
  {87}},\ \bibinfo {pages} {040301} (\bibinfo {year} {2013})}\BibitemShut
  {NoStop}%
\bibitem [{\citenamefont {Landahl}\ \emph {et~al.}(2011)\citenamefont
  {Landahl}, \citenamefont {Anderson},\ and\ \citenamefont
  {Rice}}]{landahl2011}%
  \BibitemOpen
  \bibfield  {author} {\bibinfo {author} {\bibfnamefont {A.~J.}\ \bibnamefont
  {Landahl}}, \bibinfo {author} {\bibfnamefont {J.~T.}\ \bibnamefont
  {Anderson}}, \ and\ \bibinfo {author} {\bibfnamefont {P.~R.}\ \bibnamefont
  {Rice}},\ }\href@noop {} {\bibfield  {journal} {\bibinfo  {journal} {arXiv
  preprint arXiv:1108.5738}\ } (\bibinfo {year} {2011})}\BibitemShut {NoStop}%
\bibitem [{\citenamefont {Kubica}\ and\ \citenamefont
  {Beverland}(2015)}]{kubica2015a}%
  \BibitemOpen
  \bibfield  {author} {\bibinfo {author} {\bibfnamefont {A.}~\bibnamefont
  {Kubica}}\ and\ \bibinfo {author} {\bibfnamefont {M.}~\bibnamefont
  {Beverland}},\ }\href {\doibase 10.1103/PhysRevA.91.032330} {\bibfield
  {journal} {\bibinfo  {journal} {Physical Review A}\ }\textbf {\bibinfo
  {volume} {91}},\ \bibinfo {pages} {032330} (\bibinfo {year}
  {2015})}\BibitemShut {NoStop}%
\bibitem [{\citenamefont {Kubica}(2018)}]{Kubicathesis}%
  \BibitemOpen
  \bibfield  {author} {\bibinfo {author} {\bibfnamefont {A.}~\bibnamefont
  {Kubica}},\ }\emph {\bibinfo {title} {{The ABCs of the color code: A study of
  topological quantum codes as toy models for fault-tolerant quantum
  computation and quantum phases of matter}}},\ \href
  {https://thesis.library.caltech.edu/10955/} {Ph.D. thesis},\ \bibinfo
  {school} {Caltech} (\bibinfo {year} {2018})\BibitemShut {NoStop}%
\bibitem [{\citenamefont {Landahl}\ and\ \citenamefont
  {Ryan-Anderson}(2014)}]{landahl2014}%
  \BibitemOpen
  \bibfield  {author} {\bibinfo {author} {\bibfnamefont {A.~J.}\ \bibnamefont
  {Landahl}}\ and\ \bibinfo {author} {\bibfnamefont {C.}~\bibnamefont
  {Ryan-Anderson}},\ }\href@noop {} {\enquote {\bibinfo {title} {Quantum
  computing by color-code lattice surgery},}\ } (\bibinfo {year} {2014}),\
  \Eprint {http://arxiv.org/abs/1407.5103} {arXiv:1407.5103 [quant-ph]}
  \BibitemShut {NoStop}%
\bibitem [{\citenamefont {Landahl}\ and\ \citenamefont
  {Cesare}(2013)}]{Landahl2013}%
  \BibitemOpen
  \bibfield  {author} {\bibinfo {author} {\bibfnamefont {A.~J.}\ \bibnamefont
  {Landahl}}\ and\ \bibinfo {author} {\bibfnamefont {C.}~\bibnamefont
  {Cesare}},\ }\href@noop {} {\bibfield  {journal} {\bibinfo  {journal}
  {arXiv}\ } (\bibinfo {year} {2013})},\ \bibinfo {note} {{
  arXiv:1302.3240}}\BibitemShut {NoStop}%
\bibitem [{\citenamefont {Hastings}\ and\ \citenamefont
  {Haah}(2018)}]{Haah2018b}%
  \BibitemOpen
  \bibfield  {author} {\bibinfo {author} {\bibfnamefont {M.~B.}\ \bibnamefont
  {Hastings}}\ and\ \bibinfo {author} {\bibfnamefont {J.}~\bibnamefont
  {Haah}},\ }\href {\doibase 10.1103/PhysRevLett.120.050504} {\bibfield
  {journal} {\bibinfo  {journal} {Phys. Rev. Lett.}\ }\textbf {\bibinfo
  {volume} {120}},\ \bibinfo {pages} {050504} (\bibinfo {year}
  {2018})}\BibitemShut {NoStop}%
\bibitem [{\citenamefont {Reichardt}(2005)}]{reichardt2005}%
  \BibitemOpen
  \bibfield  {author} {\bibinfo {author} {\bibfnamefont {B.~W.}\ \bibnamefont
  {Reichardt}},\ }\href@noop {} {\bibfield  {journal} {\bibinfo  {journal}
  {Quantum Information Processing}\ }\textbf {\bibinfo {volume} {4}},\ \bibinfo
  {pages} {251} (\bibinfo {year} {2005})}\BibitemShut {NoStop}%
\bibitem [{\citenamefont {Meier}\ \emph {et~al.}(2013)\citenamefont {Meier},
  \citenamefont {Eastin},\ and\ \citenamefont {Knill}}]{meier2013}%
  \BibitemOpen
  \bibfield  {author} {\bibinfo {author} {\bibfnamefont {A.~M.}\ \bibnamefont
  {Meier}}, \bibinfo {author} {\bibfnamefont {B.}~\bibnamefont {Eastin}}, \
  and\ \bibinfo {author} {\bibfnamefont {E.}~\bibnamefont {Knill}},\
  }\href@noop {} {\bibfield  {journal} {\bibinfo  {journal} {Quantum
  Information and Computation}\ }\textbf {\bibinfo {volume} {13}},\ \bibinfo
  {pages} {0195} (\bibinfo {year} {2013})}\BibitemShut {NoStop}%
\bibitem [{\citenamefont {Jones}(2013{\natexlab{b}})}]{jones2013}%
  \BibitemOpen
  \bibfield  {author} {\bibinfo {author} {\bibfnamefont {C.}~\bibnamefont
  {Jones}},\ }\href@noop {} {\bibfield  {journal} {\bibinfo  {journal}
  {Physical Review A}\ }\textbf {\bibinfo {volume} {87}},\ \bibinfo {pages}
  {042305} (\bibinfo {year} {2013}{\natexlab{b}})}\BibitemShut {NoStop}%
\bibitem [{\citenamefont {Duclos-Cianci}\ and\ \citenamefont
  {Poulin}(2015)}]{duclos2015}%
  \BibitemOpen
  \bibfield  {author} {\bibinfo {author} {\bibfnamefont {G.}~\bibnamefont
  {Duclos-Cianci}}\ and\ \bibinfo {author} {\bibfnamefont {D.}~\bibnamefont
  {Poulin}},\ }\href@noop {} {\bibfield  {journal} {\bibinfo  {journal}
  {Physical Review A}\ }\textbf {\bibinfo {volume} {91}},\ \bibinfo {pages}
  {042315} (\bibinfo {year} {2015})}\BibitemShut {NoStop}%
\bibitem [{\citenamefont {Campbell}\ and\ \citenamefont
  {O’Gorman}(2016)}]{campbell2016}%
  \BibitemOpen
  \bibfield  {author} {\bibinfo {author} {\bibfnamefont {E.~T.}\ \bibnamefont
  {Campbell}}\ and\ \bibinfo {author} {\bibfnamefont {J.}~\bibnamefont
  {O’Gorman}},\ }\href@noop {} {\bibfield  {journal} {\bibinfo  {journal}
  {Quantum Science and Technology}\ }\textbf {\bibinfo {volume} {1}},\ \bibinfo
  {pages} {015007} (\bibinfo {year} {2016})}\BibitemShut {NoStop}%
\bibitem [{\citenamefont {Haah}\ \emph
  {et~al.}(2017{\natexlab{b}})\citenamefont {Haah}, \citenamefont {Hastings},
  \citenamefont {Poulin},\ and\ \citenamefont {Wecker}}]{haah2017b}%
  \BibitemOpen
  \bibfield  {author} {\bibinfo {author} {\bibfnamefont {J.}~\bibnamefont
  {Haah}}, \bibinfo {author} {\bibfnamefont {M.~B.}\ \bibnamefont {Hastings}},
  \bibinfo {author} {\bibfnamefont {D.}~\bibnamefont {Poulin}}, \ and\ \bibinfo
  {author} {\bibfnamefont {D.}~\bibnamefont {Wecker}},\ }\href@noop {}
  {\bibfield  {journal} {\bibinfo  {journal} {arXiv preprint arXiv:1709.02789}\
  } (\bibinfo {year} {2017}{\natexlab{b}})}\BibitemShut {NoStop}%
\bibitem [{\citenamefont {Haah}(2018)}]{haah2018c}%
  \BibitemOpen
  \bibfield  {author} {\bibinfo {author} {\bibfnamefont {J.}~\bibnamefont
  {Haah}},\ }\href {\doibase 10.1103/PhysRevA.97.042327} {\bibfield  {journal}
  {\bibinfo  {journal} {Phys. Rev. A}\ }\textbf {\bibinfo {volume} {97}},\
  \bibinfo {pages} {042327} (\bibinfo {year} {2018})}\BibitemShut {NoStop}%
\bibitem [{\citenamefont {Wang}\ \emph {et~al.}(2011)\citenamefont {Wang},
  \citenamefont {Fowler},\ and\ \citenamefont {Hollenberg}}]{Wang2011}%
  \BibitemOpen
  \bibfield  {author} {\bibinfo {author} {\bibfnamefont {D.~S.}\ \bibnamefont
  {Wang}}, \bibinfo {author} {\bibfnamefont {A.~G.}\ \bibnamefont {Fowler}}, \
  and\ \bibinfo {author} {\bibfnamefont {L.~C.~L.}\ \bibnamefont
  {Hollenberg}},\ }\href {\doibase 10.1103/PhysRevA.83.020302} {\bibfield
  {journal} {\bibinfo  {journal} {Phys. Rev. A}\ }\textbf {\bibinfo {volume}
  {83}},\ \bibinfo {pages} {020302} (\bibinfo {year} {2011})}\BibitemShut
  {NoStop}%
\bibitem [{\citenamefont {Stephens}(2014)}]{stephens2014}%
  \BibitemOpen
  \bibfield  {author} {\bibinfo {author} {\bibfnamefont {A.~M.}\ \bibnamefont
  {Stephens}},\ }\href@noop {} {\bibfield  {journal} {\bibinfo  {journal}
  {arXiv preprint arXiv:1402.3037}\ } (\bibinfo {year} {2014})}\BibitemShut
  {NoStop}%
\bibitem [{\citenamefont {Brown}\ \emph {et~al.}(2015)\citenamefont {Brown},
  \citenamefont {Nickerson},\ and\ \citenamefont {Browne}}]{Brown2015}%
  \BibitemOpen
  \bibfield  {author} {\bibinfo {author} {\bibfnamefont {B.~J.}\ \bibnamefont
  {Brown}}, \bibinfo {author} {\bibfnamefont {N.~H.}\ \bibnamefont
  {Nickerson}}, \ and\ \bibinfo {author} {\bibfnamefont {D.~E.}\ \bibnamefont
  {Browne}},\ }\href {\doibase 10.1038/ncomms12302} {\bibfield  {journal}
  {\bibinfo  {journal} {Nature Communications}\ }\textbf {\bibinfo {volume}
  {7}},\ \bibinfo {pages} {4} (\bibinfo {year} {2015})},\ \Eprint
  {http://arxiv.org/abs/1503.08217} {arXiv:1503.08217} \BibitemShut {NoStop}%
\bibitem [{\citenamefont {Terhal}(2015)}]{Terhal2015}%
  \BibitemOpen
  \bibfield  {author} {\bibinfo {author} {\bibfnamefont {B.~M.}\ \bibnamefont
  {Terhal}},\ }\href {\doibase 10.1103/RevModPhys.87.307} {\bibfield  {journal}
  {\bibinfo  {journal} {Rev. Mod. Phys.}\ }\textbf {\bibinfo {volume} {87}},\
  \bibinfo {pages} {307} (\bibinfo {year} {2015})}\BibitemShut {NoStop}%
\bibitem [{\citenamefont {Kubica}\ \emph {et~al.}(2018)\citenamefont {Kubica},
  \citenamefont {Beverland}, \citenamefont {Brand\~ao}, \citenamefont
  {Preskill},\ and\ \citenamefont {Svore}}]{kubica2018}%
  \BibitemOpen
  \bibfield  {author} {\bibinfo {author} {\bibfnamefont {A.}~\bibnamefont
  {Kubica}}, \bibinfo {author} {\bibfnamefont {M.~E.}\ \bibnamefont
  {Beverland}}, \bibinfo {author} {\bibfnamefont {F.}~\bibnamefont
  {Brand\~ao}}, \bibinfo {author} {\bibfnamefont {J.}~\bibnamefont {Preskill}},
  \ and\ \bibinfo {author} {\bibfnamefont {K.~M.}\ \bibnamefont {Svore}},\
  }\href {\doibase 10.1103/PhysRevLett.120.180501} {\bibfield  {journal}
  {\bibinfo  {journal} {Phys. Rev. Lett.}\ }\textbf {\bibinfo {volume} {120}},\
  \bibinfo {pages} {180501} (\bibinfo {year} {2018})}\BibitemShut {NoStop}%
\bibitem [{\citenamefont {Fowler}(2020)}]{fowler2020}%
  \BibitemOpen
  \bibfield  {author} {\bibinfo {author} {\bibfnamefont {A.}~\bibnamefont
  {Fowler}},\ }\href@noop {} {}\bibinfo {howpublished} {private communication}
  (\bibinfo {year} {2020})\BibitemShut {NoStop}%
\bibitem [{\citenamefont {Li}(2015)}]{Li2015}%
  \BibitemOpen
  \bibfield  {author} {\bibinfo {author} {\bibfnamefont {Y.}~\bibnamefont
  {Li}},\ }\href {\doibase 10.1088/1367-2630/17/2/023037} {\bibfield  {journal}
  {\bibinfo  {journal} {New Journal of Physics}\ }\textbf {\bibinfo {volume}
  {17}},\ \bibinfo {pages} {023037} (\bibinfo {year} {2015})}\BibitemShut
  {NoStop}%
\bibitem [{\citenamefont {Chamberland}\ and\ \citenamefont
  {Noh}(2020{\natexlab{b}})}]{Chamberland2020a}%
  \BibitemOpen
  \bibfield  {author} {\bibinfo {author} {\bibfnamefont {C.}~\bibnamefont
  {Chamberland}}\ and\ \bibinfo {author} {\bibfnamefont {K.}~\bibnamefont
  {Noh}},\ }\href@noop {} {\bibfield  {journal} {\bibinfo  {journal} {arXiv}\ }
  (\bibinfo {year} {2020}{\natexlab{b}})},\ \bibinfo {note}
  {arXiv:2003.03049}\BibitemShut {NoStop}%
\bibitem [{\citenamefont {Bombin}(2018{\natexlab{b}})}]{Bombin2018}%
  \BibitemOpen
  \bibfield  {author} {\bibinfo {author} {\bibfnamefont {H.}~\bibnamefont
  {Bombin}},\ }\href@noop {} {\bibfield  {journal} {\bibinfo  {journal}
  {arXiv}\ } (\bibinfo {year} {2018}{\natexlab{b}})},\ \Eprint
  {http://arxiv.org/abs/arXiv:1810.09575} {arXiv:1810.09575} \BibitemShut
  {NoStop}%
\bibitem [{\citenamefont {Turner}\ \emph {et~al.}(2020)\citenamefont {Turner},
  \citenamefont {Hanish}, \citenamefont {Blanchard}, \citenamefont {Davis},\
  and\ \citenamefont {Cour}}]{Turner2020}%
  \BibitemOpen
  \bibfield  {author} {\bibinfo {author} {\bibfnamefont {S.}~\bibnamefont
  {Turner}}, \bibinfo {author} {\bibfnamefont {J.}~\bibnamefont {Hanish}},
  \bibinfo {author} {\bibfnamefont {E.}~\bibnamefont {Blanchard}}, \bibinfo
  {author} {\bibfnamefont {N.}~\bibnamefont {Davis}}, \ and\ \bibinfo {author}
  {\bibfnamefont {B.~L.}\ \bibnamefont {Cour}},\ }\href@noop {} {\bibfield
  {journal} {\bibinfo  {journal} {arXiv}\ } (\bibinfo {year} {2020})},\ \Eprint
  {http://arxiv.org/abs/arXiv:2003.11602} {arXiv:2003.11602} \BibitemShut
  {NoStop}%
\bibitem [{\citenamefont {Bruzewicz}\ \emph {et~al.}(2019)\citenamefont
  {Bruzewicz}, \citenamefont {Chiaverini}, \citenamefont {McConnell},\ and\
  \citenamefont {Sage}}]{Bruzewicz2019}%
  \BibitemOpen
  \bibfield  {author} {\bibinfo {author} {\bibfnamefont {C.~D.}\ \bibnamefont
  {Bruzewicz}}, \bibinfo {author} {\bibfnamefont {J.}~\bibnamefont
  {Chiaverini}}, \bibinfo {author} {\bibfnamefont {R.}~\bibnamefont
  {McConnell}}, \ and\ \bibinfo {author} {\bibfnamefont {J.~M.}\ \bibnamefont
  {Sage}},\ }\href {\doibase 10.1063/1.5088164} {\bibfield  {journal} {\bibinfo
   {journal} {Applied Physics Reviews}\ }\textbf {\bibinfo {volume} {6}},\
  \bibinfo {pages} {021314} (\bibinfo {year} {2019})},\ \Eprint
  {http://arxiv.org/abs/https://doi.org/10.1063/1.5088164}
  {https://doi.org/10.1063/1.5088164} \BibitemShut {NoStop}%
\end{thebibliography}%

\end{document}